\documentclass[12pt, a4paper, twoside]{book}
\usepackage{a4wide}

\usepackage[Sonny]{fncychap}
\usepackage{fancyhdr}
\usepackage[latin1]{inputenc}
\usepackage[dutch,english]{babel}
\usepackage[small,bf,flushleft]{caption}
\usepackage{amsmath}
\usepackage{amsfonts}
\usepackage{amssymb}
\usepackage{graphicx}
\usepackage{tocbibind}
\usepackage{emptypage}
\usepackage{braket}
\usepackage{caption}
\usepackage{multirow}
\usepackage{rotating}
\usepackage[usenames,dvipsnames]{color}
\usepackage[colorlinks=true, urlcolor=blue, citecolor=OliveGreen]{hyperref}
\usepackage[numbers, sort&compress]{natbib}
\newcommand*{\doi}[1]{\href{http://dx.doi.org/#1}{doi: #1}}

\usepackage{pdfpages}
\usepackage{multirow}

\author{ir. Sebastian Wouters}
\title{Accurate variational electronic structure calculations with the density matrix renormalization group}

\usepackage{makeidx}
\makeindex

\usepackage{color}

\graphicspath{{figures/}}

\makeatletter
\newenvironment{chapquote}[2][2em]
  {\setlength{\@tempdima}{#1}%
   \def\chapquote@author{#2}%
   \parshape 1 \@tempdima \dimexpr\textwidth-2\@tempdima\relax%
   \itshape}
  {\par\normalfont\hfill--\ \chapquote@author\hspace*{\@tempdima}\par\bigskip}
\makeatother

\begin{document}
\includepdf[pages=1,landscape,fitpaper=true]{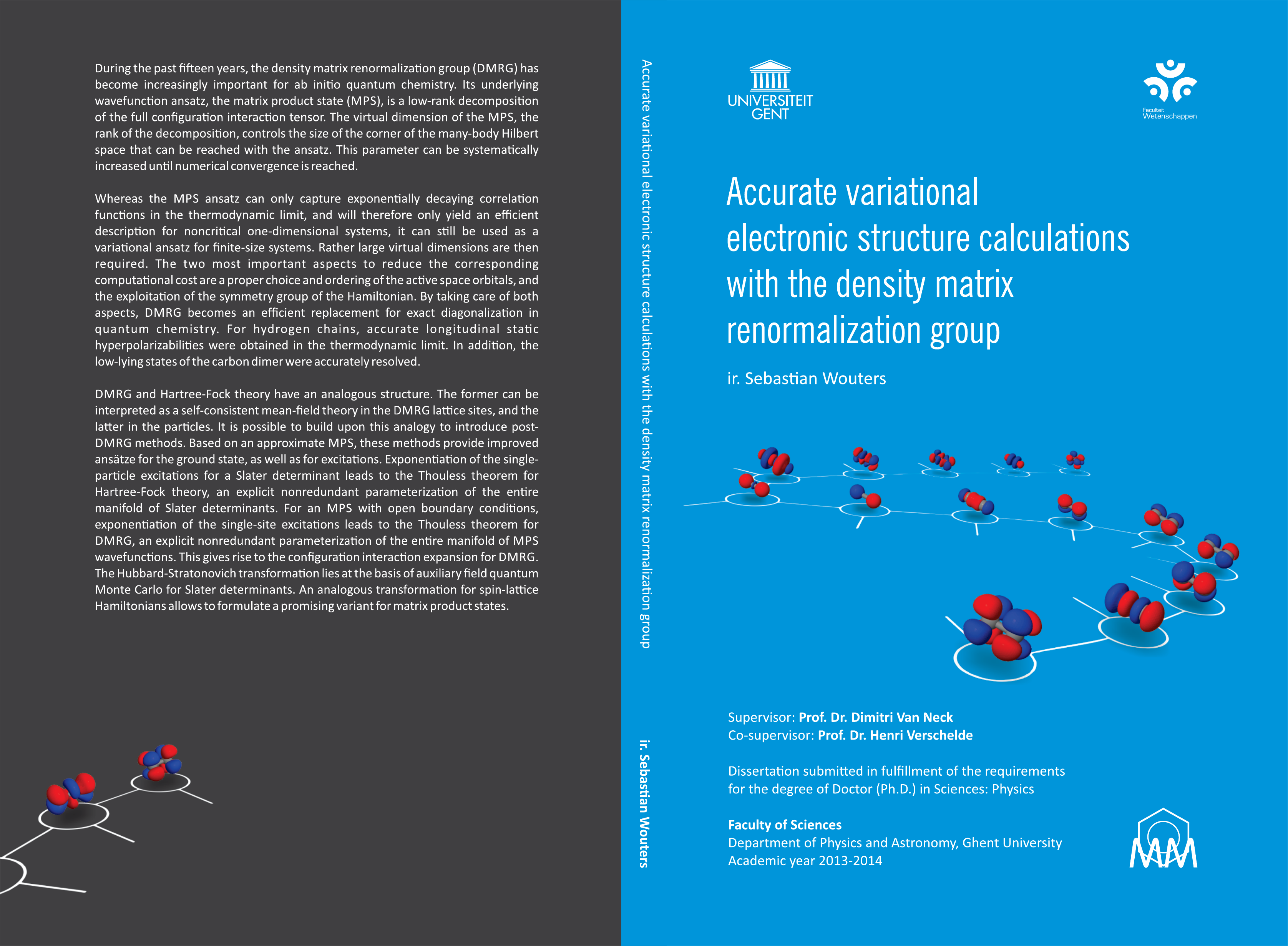}

\newpage
\thispagestyle{empty}
\hbox{}

\includepdf[pages={1}]{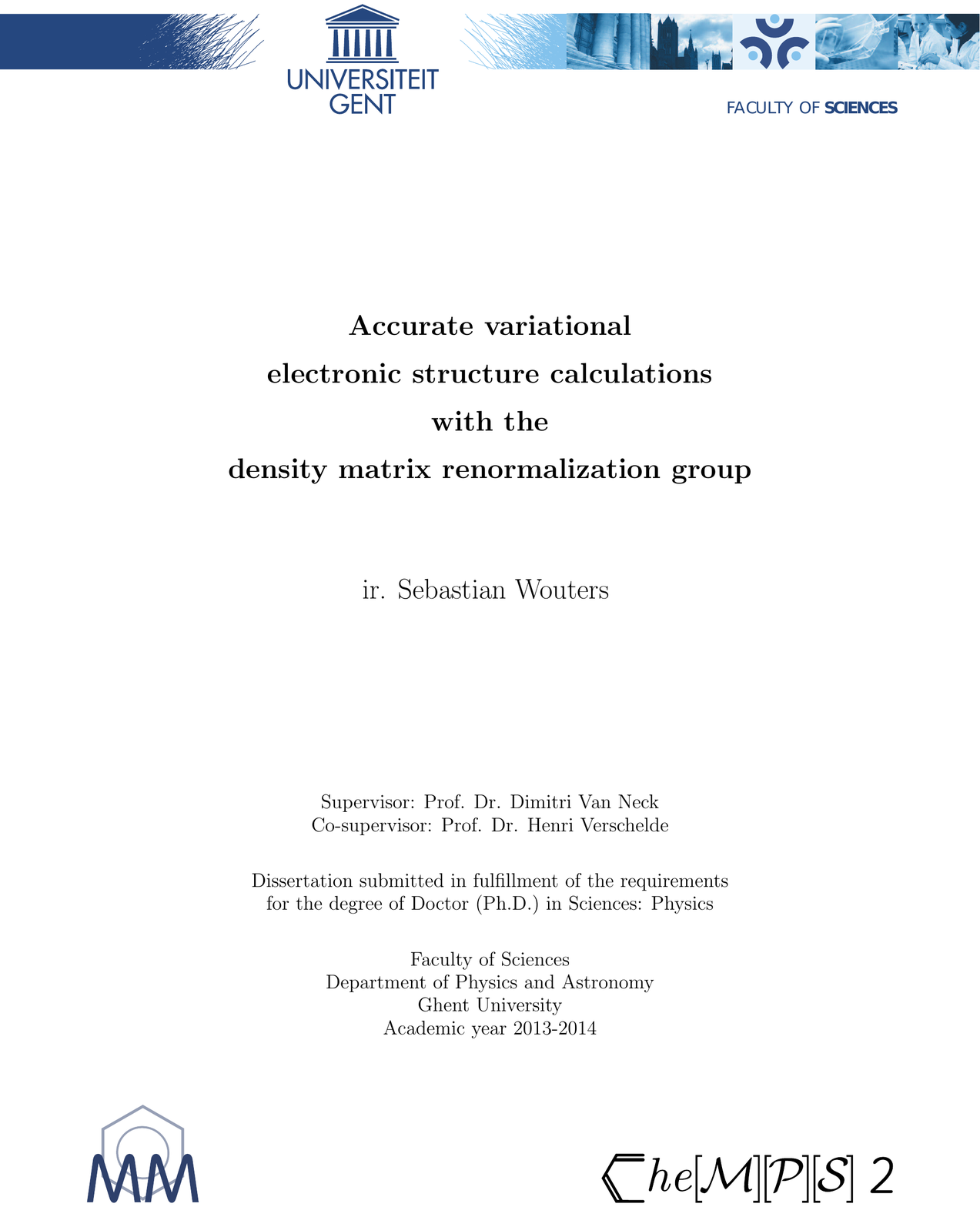}

\frontmatter

\chapter*{Abstract}
\addcontentsline{toc}{chapter}{Abstract}
During the past fifteen years, the density matrix renormalization group (DMRG) has become increasingly important for ab initio quantum chemistry. Its underlying wavefunction ansatz, the matrix product state (MPS), is a low-rank decomposition of the full configuration interaction tensor. The virtual dimension of the MPS, the rank of the decomposition, controls the size of the corner of the many-body Hilbert space that can be reached with the ansatz. This parameter can be systematically increased until numerical convergence is reached.

Chapter \ref{DMRG-QC-chapter} of this Ph.D. thesis contains a literature study about DMRG for ab initio quantum chemistry (QC-DMRG). The chapter starts by assessing DMRG and the MPS ansatz from the viewpoint of quantum information theory. DMRG works well for noncritical one-dimensional systems, as the MPS ansatz only captures exponentially decaying correlation functions in the thermodynamic limit. The active orbital spaces studied in quantum chemistry are often far from one-dimensional, and therefore relatively large virtual dimensions are required. The QC-DMRG algorithm, its computational cost, and its properties are discussed. Special attention is given to the orbital choice and ordering, as they influence the convergence behaviour significantly.

The symmetry group of a Hamiltonian allows to make it block-diagonal. In chapter \ref{SYMM-chapter}, an MPS wavefunction is constructed which is a symmetry eigenstate of this group. The Wigner-Eckart theorem allows to factorize this MPS ansatz in Clebsch-Gordan coefficients and reduced tensors. This introduces block-sparsity in the ansatz. For non-abelian groups, this encompasses information compression as well. Both lead to a decrease in required memory and computational time. The QC-DMRG implementation of the author, \textsc{CheMPS2}, exploits $\mathsf{SU(2)}$ spin symmetry, $\mathsf{U(1)}$ particle-number symmetry, and the abelian point groups $\mathsf{P}$ with real-valued character tables. The exploitation of non-abelian spatial symmetries is also briefly touched upon.

Chapters \ref{HCHAIN-chapter} and \ref{C2-chapter} contain a review of two applications. The Coulomb interaction in hydrogen chains is effectively local due to the mutual screening of electrons and nuclei. QC-DMRG therefore only requires a small virtual dimension to reach numerical convergence, and accurate longitudinal response properties were obtained for this system. The ground state of the carbon dimer has significant multireference character, and many crossings and avoided crossings
occur between its low-lying states. Due to the exploitation of symmetry in \textsc{CheMPS2}, it was possible to accurately resolve the low-lying states per symmetry sector.

DMRG and Hartree-Fock theory have an analogous structure. The former can be interpreted as a self-consistent mean-field theory in the DMRG lattice sites, and the latter in the particles. Chapters \ref{Thouless-chapter} and \ref{DMC-MPS-chapter} build upon this analogy to introduce post-DMRG methods. Based on an approximate MPS, these methods provide improved ans\"atze for the ground state, as well as for excitations. Exponentiation of the single-particle excitations for a Slater determinant leads to the Thouless theorem for Hartree-Fock theory, an explicit nonredundant parameterization of the entire manifold of Slater determinants. For an MPS with open boundary conditions, exponentiation of the single-site excitations leads to the Thouless theorem for DMRG, an explicit nonredundant parameterization of the entire manifold of MPS wavefunctions. This gives rise to the configuration interaction expansion for DMRG. The Hubbard-Stratonovich transformation lies at the basis of auxiliary field quantum Monte Carlo for Slater determinants. An analogous transformation for spin-lattice Hamiltonians allows to formulate a promising variant for matrix product states.

\chapter*{Acknowledgements}
\addcontentsline{toc}{chapter}{Acknowledgements}
\begin{chapquote}{Richard P. Feynman, 1974}
The first principle is that you must not fool yourself - and you are the easiest person to fool.
\end{chapquote}

Can there be a better solution to adhere to the first principle than to surround oneself with knowledgeable people?\\

For scientific knowledge, these include my mentors, to whom I would like to express my sincere gratitude and appreciation. To Michel Waroquier, for teaching me the principles of quantum mechanics. To Veronique Van Speybroeck, for guiding me through the jungle of electronic structure methods. To my supervisor Dimitri Van Neck, for exposing many-body theory. To my co-supervisor Henri Verschelde, for teaching me quantum field theory. To Paul Ayers, for sharing his immense wisdom on theoretical chemistry. To Garnet Chan, for the numerous long, deep, and stimulating conversations; for his inexhaustible patience; and for his hospitality both in the past and in the future.\\

I want to thank my colleagues at the Center for Molecular Modelling for all the science, fun, and unforgettable moments. Especially Brecht Verstichel and Matthias Degroote for pointing me in the direction of DMRG, Ward Poelmans for his magic computer skills, Stijn De Baerdemacker, Andy Van Yperen-De Deyne, Thierry De Meyer, Paul Johnson, and Mario Van Raemdonck for the numerous chats, as well as the Friday afternoon coffee-and-\textit{pateekes} team.\\

The Chan group was my second scientific home. I want to thank my overseas colleagues Mark Watson, Sandeep Sharma, 
Roberto Olivares-Amaya, George Booth, Tom Watson, Gerald Knizia, James McClain, Qiming Sun, Qiaoni Chen, Barbara Sandh\"ofer, Michael Roemelt, Naoki Nakatani, Jun Yang, Weifeng Hu, Elvira Sayfutyarova, and Boxiao Zheng for their hospitality, barbeques, and dinners; for showing me the ins and outs of Princeton; and for the joy of running.\\

Wetenschappelijke kennis alleen volstaat niet. In de voorbije jaren heb ik ook altijd kunnen rekenen op mijn vrienden en familie. Hun vriendschap, steun en menselijke kennis waren onontbeerlijk. Om te beginnen horen ook alle collega's in dit lijstje thuis. Daarnaast wil ik ook speciaal Thomas bedanken om mij te stimuleren om een BAEF aanvraag in te dienen; Sven, Brenda en Jense om de vreugde en het verdriet in het leven te delen; Benoit en Sofie voor een vriendschap die al langer meegaat dan het bestaan van DMRG; Ellen, Timothy en Bernd voor de pastateams; Christine om mij zo goed te verzorgen in de VS; Poenki, Rune en Fluweel om ten gepaste tijde over mijn toetsenbord te wandelen; Mathilde, Pol, Bernadette en Florent voor hun doorgedreven introductie in het Leedse dialect; Bart, Elise, Martine, Albert en Maxim om leven in de brouwerij te brengen; Ben en Barbara voor de kansen die ze mij hebben gegeven; en Jan en Marleen om minstens even goede ouders te zijn.\\

Tot slot wil ik de allerbelangrijkste persoon bedanken, mijn vrouw Kathleen. Zonder haar aanmoediging, geduld, liefde, steun, en zorg zou ik niet zijn wie ik ben, en was dit werk nooit tot stand kunnen komen. Danku schat.\\

\vspace{2cm}

\begin{flushright}
Sebastian,\\
Ghent, March 2014
\end{flushright}

\vspace{5.2cm}

This work was supported by a Ph.D. fellowship of the Research Foundation Flanders (Aspirant Fonds Wetenschappelijk Onderzoek Vlaanderen); and was carried out using the Stevin Supercomputer Infrastructure at Ghent University, funded by Ghent University, the Hercules Foundation and the Flemish Government - department EWI.

The Belgian American Educational Foundation is acknowledged for a Gustave Bo\"el-Sofina postdoctoral fellowship to work next academic year (2014-15) in Garnet Chan's group at Princeton University.

\chapter*{List of papers}
\addcontentsline{toc}{chapter}{List of papers}
\begin{enumerate}
\item Sebastian Wouters, Peter A. Limacher, Dimitri Van Neck, and Paul W. Ayers, \textit{Longitudinal static optical properties of hydrogen chains: Finite field extrapolations of matrix product state calculations}, The Journal of Chemical Physics \textbf{136}, 134110 (2012), \href{http://dx.doi.org/10.1063/1.3700087}{\nolinkurl{doi:10.1063/1.3700087}}
      
\item Brecht Verstichel, Helen van Aggelen, Ward Poelmans, Sebastian Wouters, and Dimitri Van Neck, \textit{Extensive v2DM study of the one-dimensional Hubbard model for large lattice sizes: Exploiting translational invariance and parity}, Computational and Theoretical Chemistry \textbf{1003}, 12 (2013), \href{http://dx.doi.org/10.1016/j.comptc.2012.09.014}{\nolinkurl{doi:10.1016/j.comptc.2012.09.014}}
      
\item Sebastian Wouters, Naoki Nakatani, Dimitri Van Neck, and Garnet K.-L. Chan, \textit{Thouless theorem for matrix product states and subsequent post density matrix renormalization group methods}, Physical Review B \textbf{88}, 075122 (2013), \href{http://dx.doi.org/10.1103/PhysRevB.88.075122}{\nolinkurl{doi:10.1103/PhysRevB.88.075122}}

\item Naoki Nakatani, Sebastian Wouters, Dimitri Van Neck, and Garnet K.-L. Chan, \textit{Linear response theory for the density matrix renormalization group: Efficient algorithms for strongly correlated excited states}, The Journal of Chemical Physics \textbf{140}, 024108 (2014), \href{http://dx.doi.org/10.1063/1.4860375}{\nolinkurl{doi:10.1063/1.4860375}}
      
\item Brecht Verstichel, Ward Poelmans, Stijn De Baerdemacker, Sebastian Wouters, and Dimitri Van Neck, \textit{Variational optimization of the 2DM: approaching three-index accuracy using extended cluster constraints}, The European Physical Journal B \textbf{87}, 59 (2014), \href{http://dx.doi.org/10.1140/epjb/e2014-40788-x}{\nolinkurl{doi:10.1140/epjb/e2014-40788-x}}

\item Sebastian Wouters, Ward Poelmans, Paul W. Ayers, and Dimitri Van Neck, \textit{CheMPS2: a free open-source spin-adapted implementation of the density matrix renormalization group for ab initio quantum chemistry}, Computer Physics Communications \textbf{185}, 1501 (2014), \href{http://dx.doi.org/10.1016/j.cpc.2014.01.019}{\nolinkurl{doi:10.1016/j.cpc.2014.01.019}}

\item Sebastian Wouters, Brecht Verstichel, Dimitri Van Neck, and Garnet K.-L. Chan, \textit{Projector quantum Monte Carlo with matrix product states}, ArXiv e-prints (2014), \href{http://arxiv.org/abs/1403.3125}{\nolinkurl{arXiv:1403.3125}}

\end{enumerate}








\tableofcontents

\mainmatter

\chapter{Introduction} \label{Intro-chapter}
\begin{chapquote}{Paul A. M. Dirac, 1929}
The general theory of quantum mechanics is now almost complete, the imperfections that still remain being in connection with the exact fitting in of the theory with relativity ideas. These give rise to difficulties only when high-speed particles are involved, and are therefore of no importance in the consideration of atomic and molecular structure and ordinary chemical reactions, in which it is, indeed, usually sufficiently accurate if one neglects relativity variation of mass with velocity and assumes only Coulomb forces between the various electrons and atomic nuclei. The underlying physical laws necessary for the mathematical theory of a large part of physics and the whole of chemistry are thus completely known, and the difficulty is only that the exact application of these laws leads to equations much too complicated to be soluble. It therefore becomes desirable that approximate practical methods of applying quantum mechanics should be developed, which can lead to an explanation of the main features of complex atomic systems without too much computation.
\end{chapquote}

The twentieth century was a thriving period for physics. The theories of special relativity and quantum mechanics were invented. They were unified in quantum field theory, a framework to study particles and their electroweak and strong interactions. The invariance principle of mechanics in special relativity was later extended to all physical laws in general relativity, the framework for gravitation and acceleration.

This thesis deals with quantum chemistry, the nonrelativistic quantum mechanical description of electrostatically interacting particles, more specifically electrons and atomic nuclei. Section \ref{many-body-problem-sec} introduces quantum mechanics for identical fermions from a historical perspective. Section \ref{ab-initio-qc-section} discusses quantum chemistry: its approximations, the terminology, and how the density matrix renormalization group fits in.

\section{The quantum mechanics of identical fermions} \label{many-body-problem-sec}
Schr\"odinger was able to rederive Bohr's semiclassical energy spectrum for hydrogenlike atoms within the framework of quantum mechanics \cite{Schrodinger}. He obtained that each eigenstate can be uniquely labeled by three quantum numbers. To explain the spectra of more complicated atoms, Pauli introduced a yet unknown fourth degree of freedom, and stated that two electrons can never have the same four quantum numbers \cite{Pauli}. This principle is currently known as Pauli's exclusion principle. Uhlenbeck and Goudsmit identified Pauli's fourth quantum number as the spin projection of the electron \cite{Uhlenbeck}.

In order to explain the occurrence of para- and ortho-Helium, Heisenberg explored many-body quantum mechanics \cite{Heisenberg}. For indistinguishable particles, the Hamiltonian is invariant to particle interchange. Its eigenfunctions can hence be separated into corresponding symmetry classes. Symmetric eigenfunctions with respect to particle interchange are not connected to antisymmetric ones by the Hamiltonian. Moreover, only fully antisymmetric eigenfunctions comply with Pauli's exclusion principle. This does not provide a rigorous proof that for identical particles which obey Pauli's exclusion principle, the wavefunction has to be fully antisymmetric. Heisenberg could however explain the correction factor $n!$ in Bose-Einstein statistics, which had to be introduced to make the entropy extensive \cite{Bose1}, with his wavefunction proposal for systems of identical particles \cite{Heisenberg}:
\begin{equation}
\phi = \frac{1}{\sqrt{n!}} \sum\limits_{k=1}^{n!} (\pm 1)^{\delta_k} \phi_1(m_{\alpha}^{k}) \phi_2(m_{\beta}^{k}) ... \phi_n(m_{\nu}^{k}) \label{HFwfn}
\end{equation}
where $\delta_k$ denotes the permutation order of the phase space variables $m$. A fully symmetric wavefunction arises for $(+1)$ and a fully antisymmetric one for $(-1)$. All possible distributions of the phase space variables $m$ over the single particle states $\phi_j$ contribute equally to $\phi$, with prefactors determined by the (anti)symmetry. According to Heisenberg, the phase space size in Bose-Einstein statistics has to be reduced with a factor $n!$ because a single term in Eq. \eqref{HFwfn} provides by itself no physical wavefunction, only the total sum does. The antisymmetric wavefunction of Eq. \eqref{HFwfn} will later be given the name Slater determinant \cite{PhysRev.34.1293}, the variational ansatz for Hartree-Fock (HF) theory \cite{PSP.1733252, PSP.1733252bis, PhysRev.32.339, Fock1}.

Heisenberg's feeling that identical particles which obey Pauli's exclusion principle should form antisymmetric wavefunctions, was confirmed with the advent of quantum field theory, in the so-called spin-statistics theorem \cite{Fierz, PhysRev.58.716}. Just like electrons have spin-$\frac{1}{2}$, all particles have spin, either integer or half-integer. The particles with integer spin are called bosons. They obey Bose-Einstein statistics. The wavefunction for a system of identical bosons is symmetric with respect to the interchange of any two particles. The particles with half-integer spin are called fermions. They obey Fermi-Dirac statistics. The wavefunction for a system of identical fermions is antisymmetric with respect to the interchange of any two particles. Fermions therefore obey Pauli's exclusion principle.

Any linear combination of Slater determinants is still an antisymmetric wavefunction, and hence provides a better variational ansatz for fermions. This is the basis of the configuration interaction method \cite{PhysRev.34.1293}. In this method, one needs to keep track of the occupation of single particle states $\phi_j$ with certain fermions $m_{\delta}$ and the corresponding phase prefactors $(-1)^{\delta_k}$. Dirac and Fock established a nice bookkeeping device which has exactly this functionality, called second quantization \cite{Dirac01031927, Fock2ndquant}. In what follows, a short introduction of second quantization for fermions is given.

Consider a set of orthonormal single particle states $\phi_j$:
\begin{equation}
  \int dm ~ \phi_i^{*}(m) \phi_j(m) = \delta_{ij}
\end{equation}
 Creation $\hat{a}_k^{\dagger}$ and annihilation $\hat{a}_k$ operators are introduced, which obey anticommutation relations:
\begin{eqnarray}
 \left\{ \hat{a}_l, \hat{a}_k^{\dagger} \right\} & = & \hat{a}_l \hat{a}_k^{\dagger} +  \hat{a}_k^{\dagger} \hat{a}_l = \delta_{kl},\\
 \left\{ \hat{a}_l^{\dagger}, \hat{a}_k^{\dagger} \right\} & = & 0 \quad \& \quad \text{hermitian conjugate}. \label{anti-comm-relation-two-crea}
\end{eqnarray}
When $\hat{a}_k^{\dagger}$ acts on the vacuum $\ket{-}$, which contains no particles, the single particle state $\phi_k$ is filled:
\begin{equation}
 \hat{a}_k^{\dagger} \ket{-} = \ket{\phi_k}.
\end{equation}
The annihilation operator $\hat{a}_k$ destroys the vacuum:
\begin{equation}
 \hat{a}_k \ket{-} = 0.
\end{equation}
The anticommutation relation \eqref{anti-comm-relation-two-crea} ensures that a single particle state cannot be filled with more than one fermion, in accordance with Pauli's exclusion principle:
\begin{equation}
 \hat{a}_k^{\dagger} \hat{a}_k^{\dagger} \ket{-} = \frac{1}{2} \left\{ \hat{a}_k^{\dagger} , \hat{a}_k^{\dagger} \right\} \ket{-} = 0.
\end{equation}
The same anticommutation relation also ensures antisymmetry for multiple fermions:
\begin{equation}
 \hat{a}_k^{\dagger} \hat{a}_l^{\dagger} = - \hat{a}_l^{\dagger} \hat{a}_k^{\dagger}.
\end{equation}
The antisymmetric $n$-particle state of Eq. \eqref{HFwfn} is for example represented by
\begin{equation}
   \ket{\phi} = \hat{a}^{\dagger}_1 \hat{a}^{\dagger}_2 ... \hat{a}^{\dagger}_n \ket{-}. \label{HFin2ndQuant}
\end{equation}
In this thesis, the occupation number representation is often used. A fixed order is given to the $L$ single particle states under consideration: $\phi_1; \phi_2; \phi_3; ...; \phi_L$. With the notation
\begin{equation}
\ket{n_1 n_2 ... n_L} = \left( \hat{a}^{\dagger}_1 \right)^{n_1} \left( \hat{a}^{\dagger}_2 \right)^{n_2} ...  \left( \hat{a}^{\dagger}_L \right)^{n_L} \ket{-} \label{occ-num-represent}
\end{equation}
the global sign of the wavefunction is well-defined. Due to Pauli's exclusion principle $n_j$ can be either 0 or 1, but not larger than one. Second quantization is useful, because it allows to express both the Hilbert space basis vectors and the Hamiltonian in a convenient way. For pairwise and number-conserving interactions, the Hamiltonian can be expressed as \cite{Fock2ndquant}:
\begin{equation}
\hat{H} = E_0 + \sum\limits_{ij} (i | \hat{T} | j ) \hat{a}_i^{\dagger} \hat{a}_j + \frac{1}{2} \sum\limits_{ijkl} (ij | \hat{V} | kl ) \hat{a}_i^{\dagger} \hat{a}_j ^{\dagger} \hat{a}_l \hat{a}_k. \label{Ham}
\end{equation}
For $N$ identical fermions:
\begin{equation}
 \sum\limits_j n_j = N,
\end{equation}
there are $\binom{L}{N}$ orthonormal states $\ket{n_1 n_2 ... n_L}$. The exact diagonalization of the Hamiltonian \eqref{Ham} in the basis \eqref{occ-num-represent} is hence NP-complete (in the number of single particle states $L$).

Monte Carlo methods allow to efficiently sample large spaces, if a positive-semidefinite probability distribution can be associated to it \cite{Ulam, Metropolis2}. This is the case for bosonic systems, for which the wavefunction is symmetric. For fermionic systems, the wavefunction is antisymmetric, and except for a few marginal cases, one always ends up with indefinite distributions. Unfortunately, the latter cannot be interpreted as a probability. Workarounds do exist for fermionic systems, but they suffer from the fermion sign problem, which is NP-hard \cite{PhysRevLett.94.170201}. Chapter \ref{DMC-MPS-chapter} deals with one particular flavour of quantum Monte Carlo, diffusion Monte Carlo, which introduces a controllable systematic bias to deal with the sign problem.

No exact solution methods are known which scale polynomially with $L$. We therefore have to resort to approximate solution methods.

\section{Ab initio quantum chemistry} \label{ab-initio-qc-section}
In the first paragraph of ``Quantum Mechanics of Many-Electron Systems", the opening quote of this chapter, Dirac gives his perspective on the status of the field \cite{Dirac06041929}. His comments are still valid. To study chemistry on a computer, several approximations need to be made. Not all physical interactions and effects are required to provide an accurate description of chemistry. The infite set of orthonormal single particle states which span the whole of space needs to be reduced to a finite set. Approximate solution methods are required.

\subsection{The relevant physics}
Currently, there is no single theory available to describe all observed phenomena in nature. Quantum field theory and general relativity are two distinct theories, and much effort is put into a possible unification. On the energy and distance scales relevant to chemistry, the gravitational, weak, and strong interactions are negligible compared to electromagnetism. We therefore have to resort to quantum electrodynamics \cite{Tomonaga01081946, PhysRev.73.416, PhysRev.74.1439, PhysRev.76.769, PhysRev.76.749, PhysRev.80.440, PhysRev.75.486, PhysRev.75.1736} to study chemistry.

To obtain a workable theory, relativistic effects are initially neglected. Instead of a field theory, where electrons can be created and annihilated, and where they interact by exchanging photons, the particle number is fixed and all charged particles interact instantaneously. The mass of the particles is assumed to be velocity-independent. Spin-orbit coupling and the Darwin term are neglected. The last three corrections can be understood in terms of Dirac's equation for hydrogenlike atoms \cite{Dirac01021928}. Relativistic effects become important in heavy atoms, where they can be treated in perturbation \cite{Pyykko}. Direct treatment is also possible, with four-component electronic structure theories \cite{Swirles15111935, four-comp-overview}.

This leaves us with the nonrelativistic Hamiltonian:
\begin{equation}
\hat{H} = - \sum\limits_{i} \frac{\nabla^2_{i}}{2} - \sum\limits_{\alpha} \frac{\nabla^2_{\alpha}}{2 M_{\alpha}} + \frac{1}{2} \sum\limits_{\alpha \neq \beta} \frac{Z_{\alpha} Z_{\beta}}{\mid \vec{R}_{\alpha} - \vec{R}_{\beta}\mid } - \sum\limits_{\alpha  i} \frac{Z_{\alpha}}{\mid \vec{R}_{\alpha} - \vec{r}_{i} \mid} + \frac{1}{2} \sum\limits_{i \neq j} \frac{1}{\mid \vec{r}_{i} - \vec{r}_{j} \mid}. \label{chem-ham-eq}
\end{equation}
$Z_{\alpha}$, $M_{\alpha}$ and $\vec{R}_{\alpha}$ refer respectively to the charge, mass, and position of atomic nucleus $\alpha$. $\vec{r}_i$ refers to the position of electron $i$. Atomic units are used: mass, charge, action, and dielectric constant are expressed as multiples of respectively the electron mass $m_e$, the electron charge $e$, the reduced Planck constant $\hbar$, and $4 \pi \epsilon_0$ with $\epsilon_0$ the electric permittivity of free space. All other atomic units can be derived from these four, e.g. \cite{RevModPhys.84.1527}
\begin{eqnarray}
\text{Bohr radius (length)} & \qquad & a_0 = \frac{4 \pi \epsilon_0 \hbar^2}{m_e e^2} = 5.2917721092(17) \times 10^{-11} \text{m} \\
\text{Hartree (energy)} & \qquad & E_h = \frac{m_e e^4}{\left( 4 \pi \epsilon_0 \hbar \right)^2} = 4.35974434(19) \times 10^{-18} \text{J}.
\end{eqnarray}
The Hamiltonian \eqref{chem-ham-eq} can be rewritten as
\begin{equation}
 \hat{H} = \hat{H}_e(\vec{\mathbf{R}}) - \sum\limits_{\alpha} \frac{\nabla^2_{\alpha}}{2 M_{\alpha}}.
\end{equation}
Because the nuclei are much heavier than the electrons, the motion of the latter can be regarded instantaneous. This is the basis of the Born-Oppenheimer approximation \cite{BornOppenheimerCit}. The electronic structure is solved for fixed nuclear positions:
\begin{equation}
\hat{H}_e(\vec{\mathbf{R}}) \Phi_e \left( \vec{\mathbf{r}} \mid \vec{\mathbf{R}} \right) = E_e( \vec{\mathbf{R}} ) \Phi_e \left( \vec{\mathbf{r}} \mid \vec{\mathbf{R}} \right) \label{BO-Ham}
\end{equation}
and the nuclear motion is subsequently treated in the potential energy surface (PES) $E_e( \vec{\mathbf{R}} )$:
\begin{equation}
 \left( - \sum\limits_{\alpha} \frac{\nabla^2_{\alpha}}{2 M_{\alpha}} + E_e( \vec{\mathbf{R}} ) \right) \Xi_n \left( \vec{\mathbf{R}}   \right) = E_{\text{total}} \Xi_n \left( \vec{\mathbf{R}}   \right). \label{nucl-motion_BO-eq}
\end{equation}
Equation \eqref{nucl-motion_BO-eq} yields the vibrational, rotational, and translational motion of the nuclei. The total wavefunction is hence factorized in the Born-Oppenheimer approximation:
\begin{equation}
 \Psi \left( \vec{\mathbf{r}} ; \vec{\mathbf{R}} \right) = \Phi_e \left( \vec{\mathbf{r}} \mid \vec{\mathbf{R}} \right) \Xi_n \left( \vec{\mathbf{R}}   \right).
\end{equation}
This is a good approximation if the electronic PESs are well separated:
\begin{equation}
 \forall \vec{\mathbf{R}} : E_e^0( \vec{\mathbf{R}} ) \ll E_e^1( \vec{\mathbf{R}} ) \ll E_e^2( \vec{\mathbf{R}} ) \ll ...
\end{equation}

\subsection{A finite basis set} \label{subsec-finite-basis-set-and-symmetry}
The Hamiltonian $\hat{H}_e(\vec{\mathbf{R}})$ leads to a partial differential equation (PDE) for $\Phi_e(\vec{\mathbf{r}} \mid \vec{\mathbf{R}})$. In the HF method, when a Slater determinant ansatz is used, the Schr\"odinger equation can be rewritten as a set of coupled PDEs for the HF single particle states \cite{PSP.1733252, PSP.1733252bis, PhysRev.32.339, Fock1}. Electrons can have spin projection up ($\alpha$, $\uparrow$) or down ($\beta$, $\downarrow$). It is therefore useful to introduce spin-orbitals as the single particle states:
\begin{equation}
\phi_j \left( \vec{\mathbf{r}} ; \sigma \right) = \phi^{\sigma}_j \left( \vec{ \mathbf{r} } \right) \ket{\sigma}
\end{equation}
with $\sigma$ either up or down. Roothaan was the first one to point out that the set of HF PDEs can be rewritten as an algebraic equation \cite{RevModPhys.23.69}:
\begin{equation}
 \mathbf{F} C = E \mathbf{S} C
\end{equation}
by introducing a fixed and (for practical reasons) finite basis set $\left\{ \gamma_{\kappa} \left( \vec{\mathbf{r}} \right) \right\}$ for the spin-orbitals:
\begin{equation}
 \phi_j \left( \vec{\mathbf{r}} ; \sigma \right) = \sum\limits_{\kappa} C_{\kappa j}^{\sigma} \gamma_{\kappa} \left( \vec{\mathbf{r}} \right) \ket{\sigma}.
\end{equation}
Boys noted that the required integrals $(i | \hat{T} | j )$ and $(ij | \hat{V} | kl )$ can be evaluated analytically if gaussian basis functions are used \cite{Boys22021950}:
\begin{equation}
\gamma_{\kappa} \left( \vec{\mathbf{r}} \right) = P(x,y,z)e^{-\alpha r^2}
\end{equation}
with $P(x,y,z)$ a polynomial in $x$, $y$, and $z$. This led to the advent of computational quantum chemistry, with Pople's Gaussian-70 program, and the development of a plethora of gaussian basis sets \cite{PopleIniPaper, Pople1978161}.

With spin-orbitals, Eq. \eqref{occ-num-represent} becomes
\begin{equation}
\ket{n_{1\uparrow} n_{1\downarrow} n_{2\uparrow} ... n_{L \uparrow} n_{L \downarrow}} = \left( \hat{a}^{\dagger}_{1 \uparrow} \right)^{n_{1\uparrow}} \left( \hat{a}^{\dagger}_{1 \downarrow} \right)^{n_{1\downarrow}} \left( \hat{a}^{\dagger}_{2\uparrow} \right)^{n_{2\uparrow}} ...  \left( \hat{a}^{\dagger}_{L\uparrow} \right)^{n_{L \uparrow}} \left( \hat{a}^{\dagger}_{L\downarrow} \right)^{n_{L \downarrow}}  \ket{-}. \label{occupation-number-representation-spin-orbs}
\end{equation}
The number of antisymmetric $N$-particle states scales as $\binom{2L}{N}$. For spin-independent spatial orbitals $\phi_i(\vec{\mathbf{r}}) = \phi_i^{\uparrow}(\vec{\mathbf{r}}) = \phi_i^{\downarrow}(\vec{\mathbf{r}})$, Eq. \eqref{Ham} can be written as
\begin{equation}
\hat{H}_e = E_0 + \sum\limits_{ij} (i | \hat{T} | j ) \sum\limits_{\sigma} \hat{a}_{i\sigma}^{\dagger} \hat{a}_{j\sigma} + \frac{1}{2} \sum\limits_{ijkl} (ij | \hat{V} | kl ) \sum\limits_{\sigma\tau} \hat{a}_{i\sigma}^{\dagger} \hat{a}_{j\tau} ^{\dagger} \hat{a}_{l\tau} \hat{a}_{k\sigma}, \label{QC-ham}
\end{equation}
because $\hat{H}_e(\vec{\mathbf{R}})$ is spin-independent. The Latin letters denote spatial orbitals and the Greek letters electron spin projections. The possible orbital fillings are then $\ket{n_i} = \ket{-}$, $\ket{\uparrow}$, $\ket{\downarrow}$, or $\ket{\uparrow\downarrow}$.

The symmetry group of this Hamiltonian is $\mathsf{SU(2)} \otimes \mathsf{U(1)} \otimes \mathsf{P}$, or total electronic spin, particle number, and molecular point group symmetry. By defining the operators
\begin{eqnarray}
\hat{S}^{+} & = & \sum\limits_i \hat{a}_{i \uparrow}^{\dagger} \hat{a}_{i \downarrow} \\
\hat{S}^{-} & = & \left( \hat{S}^{+} \right)^{\dagger} = \sum\limits_i \hat{a}_{i \downarrow}^{\dagger} \hat{a}_{i \uparrow} \\
\hat{S}^{z} & = & \frac{1}{2} \sum\limits_i \left( \hat{a}_{i \uparrow}^{\dagger} \hat{a}_{i \uparrow} - \hat{a}_{i \downarrow}^{\dagger} \hat{a}_{i \downarrow} \right) \\
\hat{N} & = & \sum\limits_i \left( \hat{a}_{i \uparrow}^{\dagger} \hat{a}_{i \uparrow} + \hat{a}_{i \downarrow}^{\dagger} \hat{a}_{i \downarrow} \right) \\
\hat{S}^2 & = & \frac{\hat{S}^+ \hat{S}^- + \hat{S}^- \hat{S}^+}{2} + \hat{S}^z \hat{S}^z,
\end{eqnarray}
it can be easily checked that $\hat{H}_e$, $\hat{S}^2$, $\hat{S}^z$, and $\hat{N}$ form a set of commuting observables. This constitutes the $\mathsf{SU(2)}$ total electronic spin and $\mathsf{U(1)}$ particle-number symmetries. For fixed particle number $N$, Eq. \eqref{QC-ham} can also be written as
\begin{eqnarray}
\hat{H}_e & = & E_0 + \frac{1}{2} \sum\limits_{ijkl} h_{ij;kl} \sum\limits_{\sigma\tau} \hat{a}_{i\sigma}^{\dagger} \hat{a}_{j\tau} ^{\dagger} \hat{a}_{l\tau} \hat{a}_{k\sigma} \label{QC-ham-2} \\
h_{ij;kl} & = & (ij | \hat{V} | kl ) + \frac{1}{N-1} \left[ (i | \hat{T} | k) \delta_{j,l} + (j | \hat{T} | l ) \delta_{i,k} \right].
\end{eqnarray}

The molecular point group symmetry $\mathsf{P}$ consists of the rotations, reflections, and inversions which leave the external potential due to the nuclei invariant. These symmetry operations map nuclei with equal charges onto each other. The point group symmetry has implications for the spatial orbitals. Linear combinations of the gaussian basis functions $\gamma_{\kappa} \left( \vec{\mathbf{r}} \right)$ can be constructed which transform according to a particular row of a particular irreducible representation (irrep) of $\mathsf{P}$ \cite{BookCornwell}. As the Hamiltonian transforms according to the trivial irrep $I_0$ of $\mathsf{P}$, $h_{ij;kl}$ can only be nonzero if the reductions of  $I_i \otimes I_j$ and $I_k \otimes I_l$ have at least one irrep in common. Most molecular electronic structure programs make use of the abelian point groups with real-valued character tables.

\subsection{Approximate solution methods}
Exact diagonalization of the quantum chemistry Hamiltonian \eqref{QC-ham-2} scales nonpolynomial with the number of single particle states. Exact eigenstates can hence only be obtained for small system sizes $L$. For larger system sizes, approximate solution methods need to be used. There are many methods available, providing a delicate trade-off between desired accuracy and available computational time. They can be divided into several categories: classical vs. quantum mechanical, semi-empirical vs. ab initio, single reference vs. multireference... This section attempts to provide a minimal overview.

For small systems, highly accurate PESs $E_e(\vec{\mathbf{R}})$ can be obtained. They can be used to fit the parameters of so-called force fields \cite{FF1, FF2, FF3}. Force fields provide a simplified classical model of intra- and intermolecular interactions. They try to divide the entire PES $E_e(\vec{\mathbf{R}})$ into specific contributions such as bond stretching, bond rotations, electrostatic repulsion, and van der Waals interaction, each with its own classical functional form. Obtaining the optimal force-field parameters is a separate area of specialization, as it is nontrivial to accurately mimic the surface $E_e(\vec{\mathbf{R}})$ with a limited number of parameters over a wide range of nuclear positions $\vec{\mathbf{R}}$. Once the force-field parameters are chosen, the model can be used in molecular mechanics simulations to study the thermodynamics of large systems.

In semi-empirical methods, some parameters are fitted to experiments or more accurate calculations, e.g. in force fields. In ab initio methods, one starts with the Hamiltonian \eqref{QC-ham-2} and an approximate wavefunction ansatz, e.g. the Slater determinant. The orbitals in the latter are optimized to yield the minimal energy. The exact ground state has contributions from many orthogonal Slater determinants. The difference in energy between the HF solution with a single Slater determinant reference and the exact (nonrelativistic) ground state is the correlation energy \cite{PhysRev.97.1509}. This energy is often (ambiguously) divided into two contributions: static (or nondynamic) correlation and dynamic correlation \cite{helgaker2}. When near-degeneracies between determinants occur, and more than one determinant is needed to describe the qualitative behaviour of a molecule, it is said to have static correlation. This type of correlation often arises in transition metal complexes or $\pi$-conjugated systems, as well as for geometries far from equilibrium. It is typically resolved with only a few determinants. The Coulomb repulsion results in a nonzero occupancy of virtual HF orbitals in the true ground state. This effect is called dynamic correlation and constitutes the remainder of the energy gap.

All static and dynamic correlation can in principle be retrieved at HF cost with density functional theory (DFT). Hohenberg and Kohn have shown that the electron density provides sufficient information to determine all ground state properties, and that there exists a unique universal functional of the electron density which can be used to obtain the exact ground state density \cite{PhysRev.136.B864}. Kohn and Sham rewrote the universal functional as the sum of the kinetic energy of a noninteracting system and an exchange-correlation functional \cite{PhysRev.140.A1133}. This allows to represent the electron density as a Slater determinant, which immediately ensures correct N-representability. Unfortunately, the universal functional is unknown. Many approximate semi-empirical exchange-correlation functionals of various complexity have been proposed. They each have their limited area of applicability, which renders DFT a separate  area of specialization. Because the exact exchange-correlation functional is unknown, not all static and dynamic correlation is retrieved with current DFT methods. It can even be stated that DFT is rather bad in capturing static correlation \cite{BeckeQuote}.

Dynamic correlation can also be captured with ab initio post-HF methods. These start from the optimized HF orbitals and the corresponding Slater determinant $\ket{\text{HF}}$, and build in dynamic correlation on top of the single determinant reference. Commonly known are M\o{}ller-Plesset (Rayleigh-Schr\"odinger) perturbation theory \cite{PhysRev.46.618}, the configuration interaction (CI) expansion \cite{PhysRev.34.1293, PhysRev.36.1121}, and coupled cluster (CC) theory \cite{Coester1958421, Coester1960477, CizekCC}. These methods are truncated in their perturbation or expansion order. For example, the CI and CC ansatzes with single and double excitations (CISD and CCSD) for a spin singlet system can resp.\ be written with second quantization as
\begin{eqnarray}
\ket{\text{CISD}} & = & \left( x + \sum\limits_{vo \sigma} y^{vo} \hat{a}^{\dagger}_{v\sigma} \hat{a}_{o\sigma} + \frac{1}{2} \sum\limits_{vwop \sigma\tau} z^{vw;op} \hat{a}^{\dagger}_{v\sigma} \hat{a}^{\dagger}_{w\tau} \hat{a}_{p\tau} \hat{a}_{o\sigma} \right) \ket{\text{HF}}\\
\ket{\text{CCSD}} & = & \exp\left( \sum\limits_{vo \sigma} y^{vo} \hat{a}^{\dagger}_{v\sigma} \hat{a}_{o\sigma} + \frac{1}{2} \sum\limits_{vwop \sigma\tau} z^{vw;op} \hat{a}^{\dagger}_{v\sigma} \hat{a}^{\dagger}_{w\tau} \hat{a}_{p\tau} \hat{a}_{o\sigma} \right) \ket{\text{HF}},
\end{eqnarray}
where $v,w$ denote virtual or empty HF orbitals and $o,p$ denote occupied HF orbitals. An important property of ansatz wave functions is their size consistency: the fact that for two noninteracting subsystems, the compound wave function should be multiplicatively separable and the total energy additively separable. CISD is not size consistent if there are more than two electrons in the compound system, whereas CCSD is always size consistent because of the exponential ansatz \cite{helgaker2}. Because these post-HF methods start from a single determinant reference, they have difficulty building in static correlation. Mostly, very large expansion orders are required to retrieve static correlation.

It is therefore better to resort to multireference (MR) methods for systems with pronounced static correlation. For such systems, the subset of important orbitals (the active space), in which the occupation changes over the relevant determinants, is often rather small. This allows for a particular MR solution method: the complete active space (CAS) self-consistent field (SCF) method \cite{Roos1, Roos2, Roos3}. From the HF solution, a subset of occupied and virtual orbitals is selected as active space. While the remaining occupied and virtual orbitals are kept frozen at HF level, the electronic structure in the active space is solved exactly (the CAS-part). Subsequently, the occupied, active, and virtual spaces are rotated to further minimize the energy. This two-step cycle, which is sometimes implemented together, is repeated until convergence is reached (the SCF-part). CASSCF resolves the static correlation in the system. Dynamic correlation can be built in on top of the CASSCF reference wavefunction by perturbation theory (CASPT2) \cite{CASPT2-1,CASPT2-2}, a CI expansion (MRCI or CASCI) \cite{MRCIfirst, MRCIsecond, CASCI-1, CASCI-2, CASCI-3}, or CC theory (MRCC or CASCC) \cite{MRCC,Stolarczyk19941}. For the latter, approximate schemes such as canonical transformation (CT) theory \cite{CT-first} are often used.

\subsection{The density matrix renormalization group} \label{sec1p4-in-chap1}
An eigenstate of the Hamiltonian \eqref{QC-ham-2} can be written as
\begin{equation}
\ket{\Psi} = \sum_{\{n_{j} \}} C^{n_{1} n_{2} ... n_{L}} \ket{n_{1} n_{2} ... n_{L}}, \label{asdfhasdgjhsajcucahsgasgsjkkskkkk}
\end{equation}
with $\ket{n_i} = \ket{n_{i\uparrow} n_{i\downarrow}}$. The full CI (FCI) tensor can be exactly decomposed into the following contracted matrix product:
\begin{equation}
C^{n_{1} n_{2} ... n_{L}} = \sum_{\{ \alpha_k \}} A[1]^{n_{1}}_{\alpha_1} A[2]^{n_{2}}_{\alpha_1 ; \alpha_2} A[3]^{n_{3}}_{\alpha_2 ; \alpha_3} ... A[L-1]^{n_{L-1}}_{\alpha_{L-2} ; \alpha_{L-1}} A[L]^{n_{{L}}}_{\alpha_{L-1}}, \label{MPS-A-tensor}
\end{equation}
for example by successive singular value decompositions (SVD). Since no assumptions are made about the FCI tensor, the dimension of the indices $\{\alpha_k\}$ has to grow exponentially towards the middle of this contracted product:
\begin{equation}
 \text{dim}\left( \alpha_j \right) = \min \left( 4^j , 4^{L-j} \right).
\end{equation}
This is solely due to the increasing matrix dimensions in the successive SVDs. Instead of variationally optimizing over the FCI tensor, one may as well optimize over the tensors of its decomposition \eqref{MPS-A-tensor}. To make Eq. \eqref{MPS-A-tensor} of practical use, its dimensions can be truncated:
\begin{equation}
 \text{dim}\left( \alpha_j \right) = \min \left( 4^j , 4^{L-j}, D \right). \label{MPSapproxSizes}
\end{equation}
The corresponding ansatz is called a matrix product state (MPS) with open boundary conditions and bond (or virtual) dimension $D$. It can be optimized by the density matrix renormalization group (DMRG) algorithm \cite{PhysRevLett.69.2863, PhysRevB.48.10345, WhiteQCDMRG}, yielding a variational upper bound for the ground state energy. Historically, DMRG was invented first, and its underlying MPS ansatz was discovered only later \cite{PhysRevLett.75.3537, PhysRevB.55.2164}. A thorough discussion of the DMRG algorithm is given in chapter \ref{DMRG-QC-chapter}. Chapter \ref{SYMM-chapter} deals with the exploitation of the symmetry group $\mathsf{SU(2)} \otimes \mathsf{U(1)} \otimes \mathsf{P}$ of the Hamiltonian \eqref{QC-ham-2} in the DMRG algorithm, and in particular in \textsc{CheMPS2} \cite{CheMPS2github, 2013arXiv1312.2415W}, the implementation of the author.

In the large-$D$ regime, the DMRG ground state energy and its corresponding MPS become numerically exact. For most systems, this already happens for moderate values of $D$, and DMRG is therefore an efficient route to exact diagonalization accuracy. In methods which rely on a FCI solver, such as CASSCF and CASPT2 for example, the FCI solver can be replaced with DMRG. Example calculations in the large-$D$ regime are presented in chapters \ref{HCHAIN-chapter} and \ref{C2-chapter}.

Just like HF theory can be interpreted as a mean-field theory for particles, DMRG can be interpreted as a mean-field theory for sites. In analogy to a Slater determinant in HF, the MPS can then be treated as a zeroth order reference, on top of which excitations and/or dynamic correlation can be built. This is the subject of chapters \ref{Thouless-chapter} and \ref{DMC-MPS-chapter}.

\chapter{DMRG for ab initio quantum chemistry} \label{DMRG-QC-chapter}
\begin{chapquote}{Kenneth G. Wilson, 1975}
If one finds this prospect {\upshape [RG results depend on the specific setup]} discouraging, one should remember that the successful tricks of one generation become the more formal and more easily learned mathematical methods of the next generation.
\end{chapquote}

\section{Introduction}
The density matrix renormalization group (DMRG) was invented in 1992 by White in the field of condensed matter theory \cite{PhysRevLett.69.2863}. \"Ostlund and Rommer discovered in 1995 its underlying variational ansatz, the matrix product state (MPS) \cite{PhysRevLett.75.3537}. The area law for one-dimensional quantum systems was proven by Hastings in 2007 \cite{1742-5468-2007-08-P08024}, and constituted a hard proof that an MPS is very efficient in representing the ground state of gapped one-dimensional quantum systems. The discovery of the MPS ansatz and the understanding provided by quantum information theory induced the development of a plethora of new variational renormalization group ansatzes in subsequent years.

MPSs were in fact discovered earlier, under various names. Nishino found that they were used in statistical physics as a variational optimization technique \cite{NishinoHistory}: in 1941 by Kramers and Wannier \cite{PhysRev.60.263} and in 1968 by Baxter \cite{Baxter}. Nightingale and Bl\"ote recycled Baxter's ansatz in 1986 to approximate quantum eigenstates \cite{PhysRevB.33.659}. In 1987, Affleck, Kennedy, Lieb and Tasaki constructed the exact valence-bond ground state of a particular next-nearest-neighbour spin chain \cite{PhysRevLett.59.799}. They obtained an MPS with bond dimension 2. In mathematics, the translationally invariant valence-bond state is known as a finitely correlated state \cite{0295-5075-10-7-005, FCSmath}, and in the context of information compression, an MPS is known as a tensor train \cite{TensorTrains, TensorTrainsNMRspinSystem}.

The concept of a renormalization group was first used in quantum electrodynamics. The coarse-grained view of a point-like electron breaks down at small distance scales (i.e. large energy scales). The electron itself consists of electrons, positrons, and photons. The mass and charge contributions from this fine structure lead to infinities. These were successfully resolved by Tomonaga, Schwinger, and Feynman \cite{Tomonaga01081946, PhysRev.73.416, PhysRev.74.1439, PhysRev.76.769, PhysRev.76.749, PhysRev.80.440}. Later, Wilson used a numerical renormalization group (NRG) to solve the long-standing Kondo problem \cite{RevModPhys.47.773}. He turned the coupling of the impurity to the conduction band into a half-infinite lattice problem by discretizing the conduction band in momentum space. For increasing lattice sizes, only the lowest energy states are kept at each renormalization step. These are necessary and (numerically) sufficient to study the low-temperature thermodynamics of the impurity system. Although very successful for impurity systems, NRG fails for real-space lattice systems such as the discretized particle-in-a-box, spin-lattice, and Hubbard models. For these systems, the low energy states of a small subsystem are often irrelevant for the ground state of the total system \cite{PhysRevLett.68.3487}. Consider for example the ground state of the particle-in-a-box problem. By concatenating the solution of two smaller sized boxes, an unphysical node is introduced in the approximation of the ground state of the larger problem. It was White who pointed out this problem and resolved it with his DMRG method \cite{PhysRevLett.69.2863}. Instead of selecting the degrees of freedom with lowest energy, the most relevant degrees of freedom should be selected.

\section{Entanglement and the von Neumann entropy} \label{entanglement-section}
This section attempts to clarify the broader context of DMRG. A brief introduction to quantum entanglement, the von Neumann entropy, and the so-called area laws is given. More information on the first two subjects can be found in Nielsen and Chuang \cite{Nielsen}.

\begin{figure}[h!]
\centering
\includegraphics[width=0.50\textwidth]{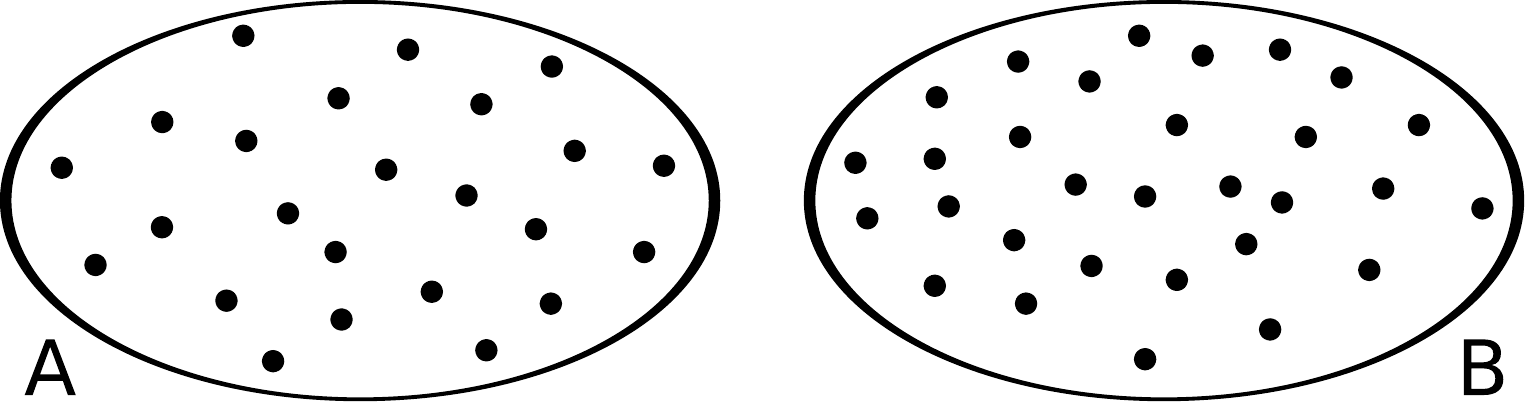}
\caption{\label{Bipartition-plot} Bipartition of the $L$ single-particle states.}
\end{figure}
Consider the bipartition of $L$ orthonormal single-particle states in two subsystems $A$ and $B$ in Fig. \ref{Bipartition-plot}. Suppose $\{ \ket{A_i} \}$ and $\{ \ket{B_j} \}$ are the orthonormal basis states of the many-body Hilbert spaces of resp.\ subsystem $A$ and $B$. The Hilbert space of the composite system is spanned by the product space $\{ \ket{A_i} \} \otimes \{ \ket{B_j} \}$, and a general quantum many-body state $\ket{\Psi}$ of the composite system can be written as
\begin{equation}
\ket{\Psi} = \sum_{ij} C_{ij} \ket{A_i} \ket{B_j}.\label{chap2-eq21-genrealstate}
\end{equation}
The Schmidt decomposition of $\ket{\Psi}$ is obtained by performing an SVD on $C_{ij}$ and by rotating the orthonormal bases $\{ \ket{A_i} \}$ and $\{ \ket{B_j} \}$ with the unitary matrices $U$ and $V$:
\begin{equation}
\ket{\Psi} = \sum_{ij} C_{ij} \ket{A_i} \ket{B_j} = \sum_{ijk} U_{ik} \sigma_k V^{\dagger}_{kj} \ket{A_i} \ket{B_j} = \sum\limits_k \sigma_k \ket{\widetilde{A}_k} \ket{\widetilde{B}_k} .\label{Schmidt-decomp}
\end{equation}
For normalized $\ket{\Psi}$:
\begin{equation}
\braket{\Psi \mid \Psi} = \sum\limits_k \sigma_k^2 = 1.
\end{equation}
For the given bipartition, the optimal approximation $\ket{\widetilde{\Psi}}$ of $\ket{\Psi}$ in least squares sense $\| \ket{\widetilde{\Psi}} - \ket{\Psi} \|_2$, with a smaller number of terms in the summation \eqref{chap2-eq21-genrealstate}, is obtained by keeping the states with the largest Schmidt numbers $\sigma_k$ in Eq. \eqref{Schmidt-decomp}. This fact will be of key importance for the DMRG algorithm (see section \ref{subsec-micro-iteraions}).

In classical theories, the sum over $k$ can contain only one nonzero value $\sigma_k$. A measurement in subsystem $A$ then does not influence the outcome in subsystem $B$, and the two subsystems are not entangled. In quantum theories, the sum over $k$ can contain many nonzero values $\sigma_k$. State $\ket{\widetilde{A}_k}$ in subsytem $A$ occurs with probability $\sigma_{k}^2$, as can be observed from the reduced density matrix (RDM) of subsystem $A$:
\begin{equation}
\hat{\rho}^A = \text{Tr}_{B} \ket{\Psi} \bra{\Psi} = \sum\limits_{j} \braket{B_j \mid \Psi} \braket{\Psi \mid B_j} = \sum\limits_{ijl} \ket{A_i} C_{ij} C^{\dagger}_{jl} \bra{A_l} = \sum\limits_{k} \ket{\widetilde{A}_k} \sigma_k^2 \bra{\widetilde{A}_k}.
\end{equation}
Analogously the RDM of subsystem $B$ can be constructed:
\begin{equation}
\hat{\rho}^B = \sum\limits_{k} \ket{\widetilde{B}_k} \sigma_k^2 \bra{\widetilde{B}_k}.
\end{equation}

From \eqref{Schmidt-decomp}, it follows that the measurement of $\ket{\widetilde{A}_k}$ in subsystem $A$ implies the measurement of $\ket{\widetilde{B}_k}$ in subsystem $B$ with probability 1. Measurements in $A$ and $B$ are hence not independent, and the two subsystems are said to be entangled.

Consider for example two singly occupied orbitals $A$ and $B$ in the spin-0 singlet state:
\begin{equation}
\ket{\Psi} = \frac{\ket{\uparrow_A \downarrow_B} - \ket{\downarrow_A \uparrow_B}}{\sqrt{2}}.
\end{equation}
The measurements of the spin projections of the electrons are not independent. Each possible spin projection of the electron in $A$ can be measured with probability $\frac{1}{2}$, but the simultaneous measurement of both spin projections will always yield
\begin{equation}
\braket{\Psi \mid \hat{S}^z_A \hat{S}^z_B \mid \Psi} = -\frac{1}{4}
\end{equation}
with probability 1. The two electron spins are maximally entangled.

The RDMs $\hat{\rho}^A$ and $\hat{\rho}^B$ allow to define the von Neumann entanglement entropy \cite{Neumann1927}:
\begin{equation}
S_{A \mid B} = - \text{Tr}_A ~ \hat{\rho}^A \ln \hat{\rho}^A = - \text{Tr}_B ~ \hat{\rho}^B \ln \hat{\rho}^B = - \sum\limits_k  \sigma_k^2 \ln \sigma_k^2. \label{von-neumann-entropy-eq}
\end{equation}
This quantum analogue of the Shannon entropy is a measure of how entangled subsystems $A$ and $B$ are. If they are not entangled, $\sigma_1=1$ and $\forall k \geq 2 : \sigma_k = 0$, which implies $S_{A \mid B} = 0$. If they are maximally entangled, $\forall k,l: \sigma_k = \sigma_l$, which implies $S_{A \mid B} = \ln(Z)$, with $Z$ the minimum of the sizes of the many-body Hilbert spaces of $A$ and $B$.

A $K$-dimensional quantum lattice system in the thermodynamic limit is called local if there exists a distance cutoff beyond which the interaction terms decay at least exponentially.
Consider the \textit{ground state} $\ket{\Psi_0}$ of a gapped $K$-dimensional quantum system in the thermodynamic limit, and select as subsystem a hypercube with side $L$ and volume $L^K$. The von Neumann entropy is believed to obey an area law \cite{PhysRevLett.94.060503, RevModPhys.82.277, PhysRevLett.111.170501}:
\begin{equation}
S_{\text{hypercube}} \propto L^{K-1}.
\end{equation}
This is the result of a finite correlation length, as only lattice sites in the immediate vicinity of the hypercube's boundary are then correlated with lattice sites on the other side of the boundary. This is a theorem for one-dimensional systems \cite{1742-5468-2007-08-P08024} and a conjecture in higher dimensions \cite{RevModPhys.82.277}, supported by numerical examples and theoretical arguments \cite{PhysRevLett.111.170501}. For critical quantum systems, with a closed excitation gap, there can be logarithmic corrections to the area law \cite{PhysRevLett.90.227902, RevModPhys.82.277}.

For gapped one-dimensional systems, consider as subsystem a line segment of length $L$. Its boundary consists of two points. Due to the finite correlation length in the ground state, the entanglement of the subsystem does not increase with $L$, if $L$ is significantly larger than the correlation length. The von Neumann entropy is then a constant independent of $L$, and the ground state $\ket{\Psi_0}$ can be well represented by retaining only a finite number of states $D$ in the Schmidt decomposition of any bipartition of the lattice in two semi-infinite line segments. This is the reason why the MPS ansatz and the corresponding DMRG algorithm work very well to study the ground states of gapped one-dimensional systems.

\begin{figure}[h!]
\centering
\includegraphics[width=0.55\textwidth]{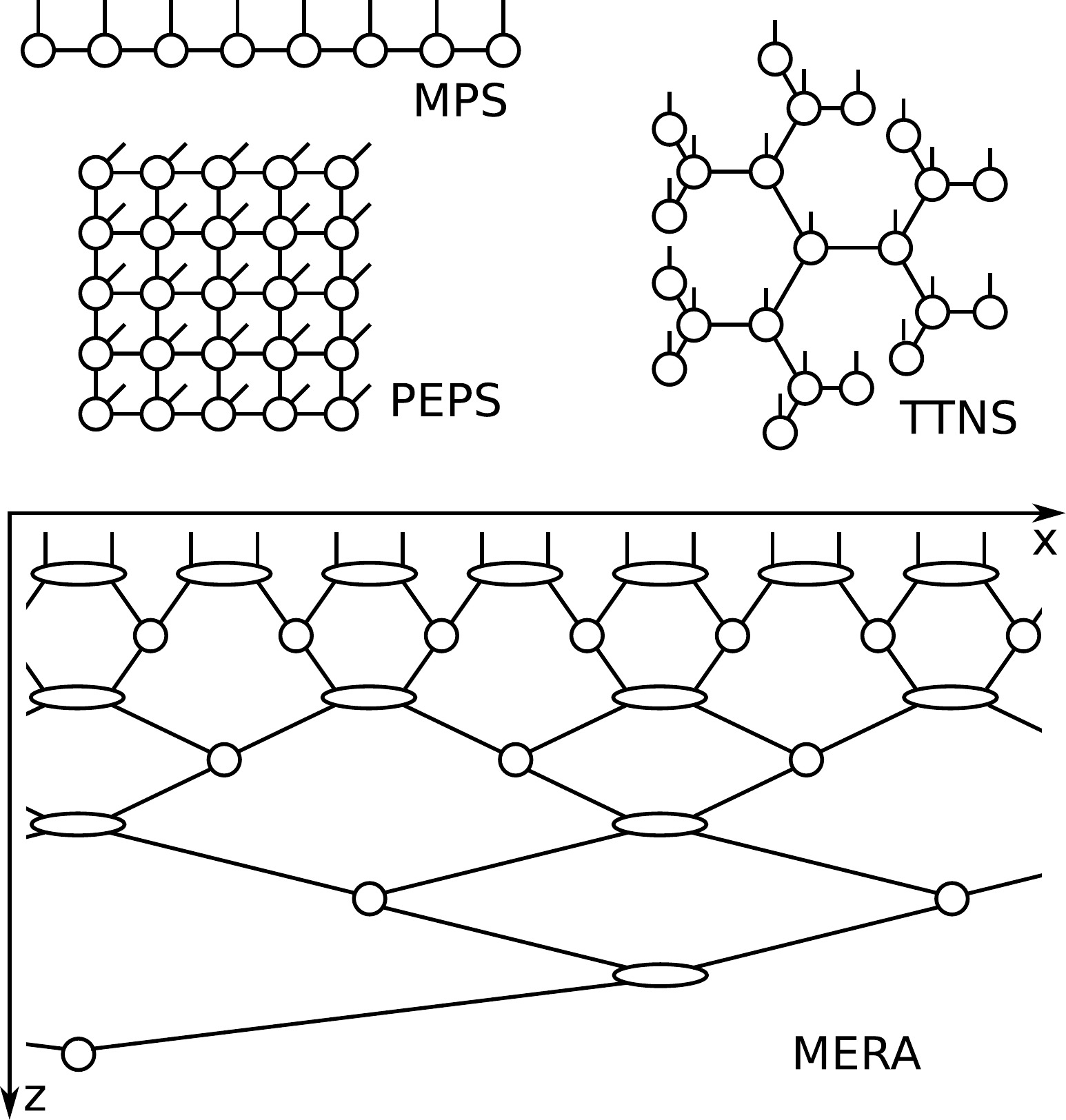}
\caption{\label{TNS-plot} Several tensor network states. Tensors are represented by circles, physical indices by open lines, and virtual indices by connected lines. The graph hence represents how the ansatz decomposes the FCI tensor.}
\end{figure}
The MPS ansatz
\begin{equation}
\ket{\Psi} = \sum_{\{n_{j} \} \{ \alpha_k \}} A[1]^{n_{1}}_{\alpha_1} A[2]^{n_{2}}_{\alpha_1 ; \alpha_2} ... A[L-1]^{n_{L-1}}_{\alpha_{L-2} ; \alpha_{L-1}} A[L]^{n_{{L}}}_{\alpha_{L-1}} \ket{n_1 n_2 ... n_{L-1} n_L}, \label{MPSansatz}
\end{equation}
is shown pictorially in Fig. \ref{TNS-plot}. Except for the first and last orbital (or site), the MPS ansatz introduces a rank-3 tensor per site. One of its indices corresponds to the physical index $n_i$, the other two to the virtual indices $\alpha_{i-1}$ and $\alpha_i$. In Fig. \ref{TNS-plot}, tensors are represented by circles, physical indices by open lines, and virtual indices by connected lines. The graph hence represents how the ansatz decomposes the FCI tensor. The finite size $D$ of the virtual indices can capture finite-length correlations along the one-dimensional chain. Stated more rigorously: for a system in the thermodynamic limit, all correlation functions $C_{\text{MPS}}(\Delta x)$ measured in an MPS ansatz with finite $D$ decay exponentially with increasing site distance $\Delta x$ \cite{FCSmath, TNSoverview}:
\begin{equation}
C_{\text{MPS}}(\Delta x) \propto e^{-\alpha \Delta x}. \label{MPSexpodecaycorrfuncchap2}
\end{equation}

Unless the lattice size is reasonably small \cite{2D-DMRG-citation}, an MPS is not efficient to represent the ground state of higher dimensional or critical systems. Fortunately, efficient tensor network states (TNS) for higher dimensional and critical lattice systems, which do obey the correct entanglement scaling laws, have been developed \cite{TNSoverview}. There even exists a continuous MPS ansatz for quantum fields \cite{PhysRevLett.104.190405}.

The ansatz for gapped two-dimensional systems is called the projected entangled pair state (PEPS) \cite{PEPS-arxiv}, see Fig. \ref{TNS-plot}. Instead of two virtual indices, each tensor now has four virtual indices,  which allows to arrange the sites in a square lattice. A finite virtual dimension $D$ still introduces a finite correlation length, but due to the topology of the PEPS, this is sufficient for gapped two-dimensional systems, even in the thermodynamic limit. Analogous extensions exist for other lattice topologies (other than the square lattice).

The ansatz for critical one-dimensional systems is called the multi-scale entanglement renormalization ansatz (MERA) \cite{PhysRevLett.99.220405}, see Fig. \ref{TNS-plot}. This ansatz has two axes: $x$ along the physical one-dimensional lattice and $z$ along the renormalization direction. Consider two sites separated by $\Delta x$ along $x$. The number of virtual bonds between those sites is only of order $\Delta z \propto \ln \Delta x$. With finite $D$, all correlation functions $C_{\text{MERA}}(\Delta x)$ measured in a MERA decay exponentially with increasing renormalization distance $\Delta z$:
\begin{equation}
 C_{\text{MERA}}(\Delta x) \propto e^{-\alpha \Delta z} \propto e^{- \beta \ln \Delta x} = (\Delta x)^{-\beta},
\end{equation}
and therefore only polynomially with increasing lattice distance $\Delta x$ \cite{PhysRevLett.99.220405, TNSoverview}.

An inconvenient property of the PEPS, MERA, and MPS with periodic boundary conditions \cite{PhysRevLett.93.227205}, is the introduction of loops in the network. This results in the inability to exploit the TNS gauge invariance to work with orthonormal renormalized environment states, see sections \ref{subsec-can-form-2.3.2} and \ref{subsec-micro-iteraions}. One particular network which avoids such loops, but which is still able to capture polynomially decaying correlation functions is the tree TNS (TTNS) \cite{TTNS, PhysRevB.87.125139}, see Fig. \ref{TNS-plot}. From a central tensor with $z$ virtual bonds, $Y$ consecutive onion-like layers are built of tensors with also $z$ virtual bonds. The last layer consists of tensors with only 1 virtual bond. An MPS is hence a TTNS with $z=2$. For $z \geq 3$, the number of sites $L$ increases as \cite{PhysRevB.82.205105, LegezaTTNS}:
\begin{equation}
 L = 1 + z \sum\limits_{k=1}^{Y} (z-1)^{k-1} = \frac{z (z-1)^{Y} - 2}{z-2}
\end{equation}
and thus $Y \propto \ln(L)$ for $z \geq 3$. The maximum number of virtual bonds between any two sites is $2Y$. The correlation functions in a TTNS with finite $D$ and $z \geq 3$ decrease exponentially with increasing separation $Y$:
\begin{equation}
 C_{\text{TTNS}}(L) \propto e^{-\alpha Y} \propto e^{- \beta \ln L} = L^{-\beta}.
\end{equation}
and therefore only polynomially with increasing number of sites $L$ \cite{TTNS, PhysRevB.87.125139}.

For higher-dimensional or critical systems, DMRG can still be useful \cite{2D-DMRG-citation}. The virtual dimension $D$ then has to be increased to a rather large size to obtain numerical convergence. In the case of multiple dimensions, the question arises if one should work in real or momentum space, and how the corresponding single-particle degrees of freedom should be mapped to the one-dimensional lattice \cite{PhysRevB.53.R10445}. Ab initio quantum chemistry can be considered as a higher-dimensional system, due to the full-rank two-body interaction in the Hamiltonian \eqref{QC-ham-2}, and the often compact spatial extent of molecules. Nevertheless, DMRG turned out to be very useful for ab initio quantum chemistry (QC-DMRG) \cite{WhiteQCDMRG, QUA:QUA1, mitrushenkov:6815, Chan2002, PhysRevB.67.125114, chan:8551, DMRG_LiF, mitrushenkov:4148, PhysRevB.68.195116, chan:3172, chan:6110, PhysRevB.70.205118, moritz:024107, chan:204101, moritz:184105, moritz:034103, hachmann:144101, Rissler2006519, moritz:244109, dorando:084109, hachmann:134309, marti:014104, zgid:014107, zgid:144115, zgid:144116, ghosh:144117, ChanB805292C, ChanQUA:QUA22099, dorando:184111,kurashige:234114,yanai:024105, neuscamman:024106, RichardsonControlReiher, PhysRevB.81.235129, mizukami:091101, 1367-2630-12-10-103008, PhysRevB.82.205105, Marti:C0CP01883J, PhysRevA.83.012508, boguslawski:224101, kurashige:094104, QUA:QUA23173, sharma:124121, woutersJCP1, JCTCspindens, C2CP23767A, JPCLentanglement, JCTCgrapheneNano, nakatani:134113, JCTCbondForm, naturechem, ma:224105, saitow:044118, Spiropyran, C3CP53975J, NaokiLRTpaper, Knecht-4c-DMRG, 2013arXiv1312.2415W, Harris2014, 2014arXiv1401.5437M, newKura, LegezaTTNS, 1.4867383}.

An excellent description of QC-DMRG in terms of renormalization transformations is given in Chan and Head-Gordon \cite{Chan2002}. Section \ref{DMRG-sec-chapt-2} contains a description in terms of the underlying MPS ansatz, because this approach will be used in chapter \ref{SYMM-chapter} to introduce $\mathsf{SU(2)} \otimes \mathsf{U(1)} \otimes \mathsf{P}$ symmetry (see section \ref{subsec-finite-basis-set-and-symmetry}) in the DMRG algorithm. The properties of the DMRG algorithm are discussed in section \ref{DMRG-prop}. Several convergence strategies are listed in section \ref{conv_strat_sec}. An overview of the strategies to choose and order orbitals is given in section \ref{DMRG-orb}. As mentioned earlier, a converged DMRG calculation can be the starting point of other methods. These methods are summarized in section \ref{DMRG-algos}. Section \ref{DMRG-systems} gives an overview of the currently existing QC-DMRG codes, and the systems which have been studied with them. The reader can also find several QC-DMRG reviews in the literature \cite{ChanRevFrontiers, Chan2009149, MartiReiherOldenburg2, chan:annurevphys, WCMS:WCMS1095, KurashigeMolPhys, Accuracy-Keller}.

\section{The DMRG algorithm} \label{DMRG-sec-chapt-2}
\subsection{The MPS ansatz} \label{TheMPSansatzSectionInChapterTwo}
DMRG can be formulated as the variational optimization of an MPS ansatz \cite{PhysRevLett.75.3537, PhysRevB.55.2164}. The MPS ansatz in Eq. \eqref{MPSansatz} has open boundary conditions, because sites 1 and L only have one virtual index. To be of practical use, the virtual dimensions $\alpha_j$ are truncated to $D$: $\text{dim}(\alpha_j) = \min(4^j, 4^{L-j}, D)$. The sites are assumed to be orbitals, which have 4 possible occupancies (see section \ref{sec1p4-in-chap1}). With increasing $D$, the MPS ansatz spans a larger region of the full Hilbert space, but it is of course not useful to make $D$ larger than $4^{\lfloor \frac{L}{2} \rfloor}$ as the MPS ansatz then spans the whole Hilbert space.

In a Slater determinant, there is gauge freedom: a rotation in the occupied orbital space alone, or a rotation in the virtual orbital space alone, does not change the physical wavefunction. Only occupied-virtual rotations change the wavefunction. In an MPS, there is gauge freedom as well. If for two neighbouring sites $i$ and $i+1$, the left MPS tensors are right-multiplied with the non-singular matrix $G$
\begin{equation}
\tilde{A}[i]^{n_i}_{\alpha_{i-1};\alpha_i} = \sum\limits_{\beta_i} A[i]^{n_i}_{\alpha_{i-1};\beta_i} G_{\beta_i;\alpha_i}
\end{equation}
and the right MPS tensors are left-multiplied with the inverse of $G$
\begin{equation}
\tilde{A}[i+1]^{n_{i+1}}_{\alpha_{i};\alpha_{i+1}} = \sum\limits_{\beta_i} G^{-1}_{\alpha_{i};\beta_i} A[i+1]^{n_{i+1}}_{\beta_i;\alpha_{i+1}}
\end{equation}
the wavefunction does not change, i.e. $\forall n_i,n_{i+1},\alpha_{i-1},\alpha_{i+1}$:
\begin{equation}
\sum\limits_{\alpha_i} \tilde{A}[i]^{n_i}_{\alpha_{i-1};\alpha_i} \tilde{A}[i+1]^{n_{i+1}}_{\alpha_{i};\alpha_{i+1}} = \sum\limits_{\alpha_i} A[i]^{n_i}_{\alpha_{i-1};\alpha_i} A[i+1]^{n_{i+1}}_{\alpha_{i};\alpha_{i+1}}.
\end{equation}

\subsection{Canonical forms} \label{subsec-can-form-2.3.2}
The two-site DMRG algorithm consists of consecutive sweeps or macro-iterations, where at each sweep step the rank-3 MPS tensors of two neighbouring sites are optimized in the micro-iteration. Suppose these sites are $i$ and $i+1$. The gauge freedom of the MPS is used to bring it in a particular canonical form. For all sites to the left of $i$, the MPS tensors are left-normalized:
\begin{equation}
\sum\limits_{\alpha_{k-1}, n_k} \left(A[k]^{n_k}\right)^{\dagger}_{\alpha_k; \alpha_{k-1}} A[k]^{n_k}_{\alpha_{k-1};\beta_k} = \delta_{\alpha_k, \beta_k} \label{left-normalized}
\end{equation}
and for all sites to the right of $i+1$, the MPS tensors are right-normalized:
\begin{equation}
\sum\limits_{\alpha_{k}, n_k} A[k]^{n_k}_{\alpha_{k-1};\alpha_k} \left(A[k]^{n_k}\right)^{\dagger}_{\alpha_k; \beta_{k-1}} = \delta_{\alpha_{k-1}, \beta_{k-1}}. \label{right-normalized}
\end{equation}
Left-normalization can be performed with consecutive QR-decompositions:
\begin{equation}
A[k]^{n_k}_{\alpha_{k-1};\alpha_k} = A[k]_{(\alpha_{k-1} n_k) ; \alpha_k} = \sum\limits_{\beta_k} Q[k]_{( \alpha_{k-1} n_k ) ; \beta_k} R_{\beta_k ; \alpha_k} = \sum\limits_{\beta_k} Q[k]^{n_k}_{\alpha_{k-1} ; \beta_k} R_{\beta_k ; \alpha_k}.
\end{equation}
The MPS tensor $Q[k]$ is now left-normalized. The $R$-matrix is multiplied into $A[k+1]$. From site 1 to $i-1$, the MPS tensors are left-normalized this way, without changing the wavefunction. Right-normalization occurs with LQ-decompositions. In section \ref{subsec-macro-it}, it will become clear that this normalization procedure only needs to occur at the start of the DMRG algorithm.

At this point, it is instructive to make the analogy to the renormalization group formulation of the DMRG algorithm. Define the following vectors:
\begin{eqnarray}
\ket{\alpha_{i-1}^L} & = & \sum_{\{n_{j} \} \{ \alpha_1 ... \alpha_{i-2} \}} A[1]^{n_{1}}_{\alpha_1} A[2]^{n_{2}}_{\alpha_1 ; \alpha_2} ... A[i-1]^{n_{i-1}}_{\alpha_{i-2} ; \alpha_{i-1}} \ket{n_1 n_2 ... n_{i-1}},\\
\ket{\alpha_{i+1}^R} & = & \sum_{\{n_{j} \} \{ \alpha_{i+2} ... \alpha_{L-1} \}} A[i+2]^{n_{i+2}}_{\alpha_{i+1} ; \alpha_{i+2}} ... A[L]^{n_{L}}_{\alpha_{L-1}} \ket{n_{i+2} ... n_{L}}.
\end{eqnarray}
Due to the left- and right-normalization described above, these vectors are orthonormal:
\begin{eqnarray}
\braket{\alpha_{i-1}^L \mid \beta_{i-1}^L } & = & \delta_{\alpha_{i-1}, \beta_{i-1}}, \\
\braket{\alpha_{i+1}^R \mid \beta_{i+1}^R } & = & \delta_{\alpha_{i+1}, \beta_{i+1}}.
\end{eqnarray}
$\{\ket{\alpha_{i-1}^L}\}$ and $\{\ket{\alpha_{i+1}^R}\}$ are renormalized bases of the many-body Hilbert spaces spanned by resp.\ orbitals 1 to $i-1$ and orbitals $i+2$ to $L$. Consider for example the left side. For site $k$ from 1 to $i-2$, the orbital basis is augmented by one orbital and subsequently truncated again to at most $D$ renormalized basis states:
\begin{equation}
 \{ \ket{\alpha_{k-1}^L} \} \otimes \{ \ket{n_k} \} \rightarrow \ket{\alpha_{k}^L} = \sum\limits_{\alpha_{k-1},n_k} A[k]^{n_k}_{\alpha_{k-1};\alpha_k}\ket{\alpha_{k-1}^L} \ket{n_k}.
\end{equation}
DMRG is hence a renormalization group for increasing many-body Hilbert spaces. The next section addresses how this renormalization transformation is chosen.

\subsection{Micro-iterations} \label{subsec-micro-iteraions}
\begin{figure}
\centering
\includegraphics[width=0.65\textwidth]{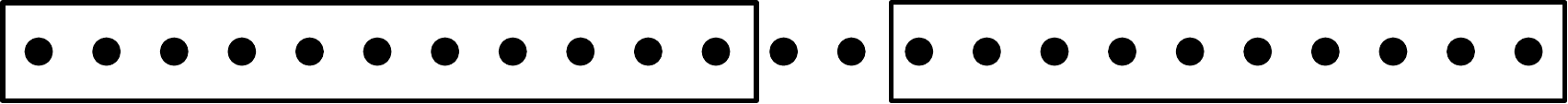}
\put(-235,24){$\ket{\alpha^L_{i-1}}$}
\put(-160,24){$\ket{n_i n_{i+1}}$}
\put(-70,24){$\ket{\alpha^R_{i+1}}$}
\caption{\label{Sweeps-plot} Optimization of the MPS tensors at sites $i$ and $i+1$ in the two-site DMRG algorithm. The effective Hamiltonian equation \eqref{effHameq} obtained by variation of the Lagrangian \eqref{Lagrangian_eq} can be interpreted as the approximate diagonalization of the exact Hamiltonian $\hat{H}$ in the orthonormal basis $\{ \ket{ \alpha_{i-1}^L } \} \otimes \{ \ket{ n_i } \} \otimes \{ \ket{ n_{i+1} } \} \otimes \{ \ket{ \alpha_{i+1}^R } \}$.}
\end{figure}

Combine the MPS tensors of the two sites under consideration into a single two-site tensor:
\begin{equation}
\sum\limits_{\alpha_i} A[i]^{n_i}_{\alpha_{i-1};\alpha_i} A[i+1]^{n_{i+1}}_{\alpha_{i};\alpha_{i+1}} = B[i]_{\alpha_{i-1};\alpha_{i+1}}^{n_i;n_{i+1}}. \label{two-site-object-not-reduced}
\end{equation}
At the current micro-iteration of the DMRG algorithm, $\mathbf{B}[i]$ (the flattened column form of the tensor $B[i]$) is used as an initial guess for the effective Hamiltonian equation. This equation is obtained by variation of the Lagrangian \cite{ChanB805292C}
\begin{equation}
\mathcal{L} = \braket{\Psi(\mathbf{B}[i]) \mid \hat{H} \mid \Psi(\mathbf{B}[i])} - E_i \braket{\Psi(\mathbf{B}[i]) \mid \Psi(\mathbf{B}[i])} \label{Lagrangian_eq}
\end{equation}
with respect to the complex conjugate of $\mathbf{B}[i]$:
\begin{equation}
\mathbf{H}[i]^{\text{eff}} \mathbf{B}[i] = E_i \mathbf{B}[i]. \label{effHameq}
\end{equation}
The canonical form in Eqs. (\ref{left-normalized})-(\ref{right-normalized}) ensured that no overlap matrix is present in this effective Hamiltonian equation. In the DMRG language, this equation can be interpreted as the approximate diagonalization of the exact Hamiltonian $\hat{H}$ in the orthonormal basis $\{ \ket{ \alpha_{i-1}^L } \} \otimes \{ \ket{ n_i } \} \otimes \{ \ket{ n_{i+1} } \} \otimes \{ \ket{ \alpha_{i+1}^R } \}$, see Fig. \ref{Sweeps-plot}. Because of the underlying MPS ansatz, DMRG is variational: $E_i$ is always an upper bound to the energy of the true ground state.

The lowest eigenvalue and corresponding eigenvector of the effective Hamiltonian are then searched with iterative sparse eigensolvers. Typical choices are the Lanczos or Davidson algorithms \cite{Lanczos, Davidson197587}. Once $\mathbf{B}[i]$ is found, it is decomposed with an SVD:
\begin{equation}
B[i]_{ \left( \alpha_{i-1} n_i \right) ; \left( n_{i+1} \alpha_{i+1} \right)} = \sum\limits_{\beta_i} U[i]_{ \left( \alpha_{i-1} n_i \right) ; \beta_i} \kappa[i]_{\beta_i} V[i]_{ \beta_i ; \left( n_{i+1} \alpha_{i+1} \right)} \label{SVDofBsolution}
\end{equation}
Note that $U[i]$ is hence left-normalized and $V[i]$ right-normalized. The sum over $\beta_i$ is truncated if there are more than $D$ nonzero Schmidt values $\kappa[i]_{\beta_i}$, thereby keeping the $D$ largest ones. This is the optimal approximation for the bipartition of $\{ \ket{ \alpha_{i-1}^L } \} \otimes \{ \ket{ n_i } \} \otimes \{ \ket{ n_{i+1} } \} \otimes \{ \ket{ \alpha_{i+1}^R } \}$ into $A = \{ \ket{ \alpha_{i-1}^L } \} \otimes \{ \ket{n_i} \}$ and $B = \{ \ket{n_{i+1}} \} \otimes \ket{ \alpha_{i+1}^R } \}$. In the original DMRG algorithm, $U[i]$ and $V[i]$ were obtained as the eigenvectors of resp.\ $\hat{\rho}^A$ and $\hat{\rho}^B$.

A discarded weight can be associated to the truncation of the sum over $\beta_i$:
\begin{equation}
 w[i]^{\text{disc}}_D = \sum\limits_{\beta_i > D} \kappa[i]^2_{\beta_i}.
\end{equation}
This is the probability to measure one of the discarded states in the subsystems $A$ or $B$.
The approximation introduced by the truncation becomes beter with increasingly small discarded weight.
Instead of working with a fixed $D$, one could also choose $D$ dynamically in order to keep $w[i]^{\text{disc}}_D$ below a preset threshold, as is done in Legeza's dynamic block state selection approach \cite{PhysRevB.67.125114}.

\subsection{Macro-iterations} \label{subsec-macro-it}

So far, we have looked at a micro-iteration of the DMRG algorithm. This micro-iteration happens during left or right sweeps. During a left sweep, $B[i]$ is constructed, the corresponding effective Hamiltonian equation solved, the solution $B[i]$ decomposed, the Schmidt spectrum truncated, $\kappa[i]$ is contracted into $U[i]$, $A[i]$ is set to this contraction $U[i] \times \kappa[i]$, $A[i+1]$ is set to $V[i]$, and $i$ is decreased by 1. Note that $A[i+1]$ is right-normalized for the next micro-iteration as required. This stepping to the left occurs until $i=1$, and then the sweep direction is reversed from left to right. Based on energy differences, or wavefunction overlaps, between consecutive sweeps, a convergence criterium is triggered, and the sweeping stops.

DMRG can be regarded as a self-consistent field method: at convergence the neighbours of an MPS tensor generate the field which yields the local solution, and this local solution generates the field for its neighbours \cite{Chan2002, hachmann:144101, ChanB805292C}.

\subsection{Renormalized operators and their complements}
The effective Hamiltonian in Eq. (\ref{effHameq}) is too large to be fully constructed. Only its action on a particular guess $\mathbf{B}[i]$ is available as a function. In order to construct $\mathbf{H}[i]^{\text{eff}} \mathbf{B}[i]$ efficiently for general quantum chemistry Hamiltonians, several tricks are used. Suppose that a right sweep is performed and that the MPS tensors of sites $i$ and $i+1$ are about to be optimized.

Renormalized operators such as $\braket{\alpha_{i-1}^L \mid \hat{a}_{k\sigma}^{\dagger} \hat{a}_{l\tau} \mid \beta_{i-1}^L}$ with $k,l \leq i-1$ are constructed and stored on disk \cite{WhiteQCDMRG, Chan2002, kurashige:234114}. The renormalized operators needed for the previous micro-iteration can be recycled to this end. Suppose $k,l \leq i-2$:
\begin{equation}
\braket{\alpha_{i-1}^L \mid \hat{a}_{k\sigma}^{\dagger} \hat{a}_{l\tau} \mid \beta_{i-1}^L} = \sum\limits_{\alpha_{i-2} \beta_{i-2} n_{i-1}} \left(A[i-1]^{n_{i-1}}\right)^{\dagger}_{\alpha_{i-1} ; \alpha_{i-2}} \braket{\alpha_{i-2}^L \mid \hat{a}_{k\sigma}^{\dagger} \hat{a}_{l\tau} \mid \beta_{i-2}^L} A[i-1]^{n_{i-1}}_{\beta_{i-2} ; \beta_{i-1}}. \label{operator_tfo}
\end{equation}
Note that no phases appear because an even number of second-quantized operators was transformed. For an odd number, there should be an additional phase $(-1)^{n_{(i-1)\uparrow} + n_{(i-1)\downarrow}}$ at the right-hand side (RHS) due to the Jordan-Wigner transformation \cite{JordanWignerTfo}. Renormalized operators to the right of $B[i]$ can be loaded from disk, as they have been saved during the previous left sweep.

Once three second-quantized operators are on one side of $B[i]$, they are multiplied with the matrix elements $h_{kl;mn}$, and a summation is performed over the common indices to construct complementary renormalized operators \cite{PhysRevB.53.R10445, WhiteQCDMRG, Chan2002, kurashige:234114}:
\begin{equation}
\braket{\alpha_{i-1}^L \mid \hat{Q}_{n\tau} \mid \beta_{i-1}^L} = \sum\limits_{\sigma} \sum\limits_{k,l,m<i} h_{kl;mn} \braket{\alpha_{i-1}^L \mid \hat{a}_{k\sigma}^{\dagger} \hat{a}^{\dagger}_{l\tau} \hat{a}_{m\sigma} \mid \beta_{i-1}^L}. \label{Q_complement_op}
\end{equation}
For two, three, and four second-quantized operators on one side of $B[i]$, these complementary renormalized operators are constructed. A bare renormalized operator (without matrix elements) is only constructed for one or two second-quantized operators.

Hermitian conjugation and commutation relations, i.e.
\begin{equation}
\braket{\alpha_{i-1}^L \mid \hat{a}_{k\sigma}^{\dagger} \hat{a}^{\dagger}_{l\tau} \mid \beta_{i-1}^L} = \braket{\beta_{i-1}^L \mid \hat{a}_{l\tau} \hat{a}_{k\sigma} \mid \alpha_{i-1}^L}^{\dagger} = - \braket{\alpha_{i-1}^L \mid \hat{a}^{\dagger}_{l\tau} \hat{a}_{k\sigma}^{\dagger} \mid \beta_{i-1}^L},
\end{equation}
are also used to further limit the storage requirements for the (complementary) renormalized operators. A few examples of renormalized operators and the fermion sign handling will be given in section \ref{CheMPS2-renorm-op-subsection} in conjunction with symmetry handling.

\subsection{Computational cost}
This section describes the cost of the DMRG algorithm per sweep in terms of memory, disk, and computational time \cite{WhiteQCDMRG, Chan2002, kurashige:234114}. To analyze this cost, let us first look at the cost per micro-iteration. A micro-iteration consists of three steps: solving the effective Hamiltonian equation \eqref{effHameq}, performing an SVD of the solution \eqref{SVDofBsolution}, and constructing the (complementary) renormalized operators for the next micro-iteration.

To solve the effective Hamiltonian equation with Davidson's algorithm, a set of $N_{vec}$ trial vectors $\{ \mathbf{B}[i] \}$ are kept in memory, as well as $\mathbf{H}[i]^{\text{eff}} \{ \mathbf{B}[i] \}$. To construct $\mathbf{H}[i]^{\text{eff}} \{ \mathbf{B}[i] \}$, (complementary) renormalized operators should also be stored in memory. The latter have at most two site indices. The total memory cost is hence $\mathcal{O}((N_{vec} + L^2) D^2)$.

The action of $\mathbf{H}[i]^{\text{eff}}$ on $\mathbf{B}[i]$ is divided into several contributions. Each contribution consists of the joint action of a renormalized operator and the corresponding complementary renormalized operator. For each contribution, two matrix-matrix multiplications need to be performed, of computational cost $\mathcal{O}(D^3)$. In total there are $\mathcal{O}(L^2)$ contributions, because complementary renormalized operators have at most two site indices. The total computational cost is hence $\mathcal{O}(N_{\text{vec}} L^2 D^3)$ for the multiplications, and $\mathcal{O}(N_{\text{vec}} L^2 D^2)$ for the summation of the different contributions.

The SVD of the solution $\mathbf{B}[i]$ and its subsequent truncation take $\mathcal{O}(D^3)$ computational time and $\mathcal{O}(D^2)$ memory.

The construction of one particular renormalized operator takes $\mathcal{O}(D^3)$ computational time and $\mathcal{O}(D^2)$ memory, and there are $\mathcal{O}(L^2)$ such operators. The most tedious part to analyze is the construction of the two-site complementary renormalized operators:
\begin{equation}
\braket{\alpha_{i-1}^L \mid \hat{F}_{m \sigma ; n\tau} \mid \beta_{i-1}^L} = \sum\limits_{k,l<i} h_{kl;mn} \braket{\alpha_{i-1}^L \mid \hat{a}_{k\sigma}^{\dagger} \hat{a}^{\dagger}_{l\tau} \mid \beta_{i-1}^L},
\end{equation}
which takes at first sight $\mathcal{O}(L^2 D^2)$ computational time and $\mathcal{O}(D^2)$ memory per operator. There are $\mathcal{O}(L^2)$ such operators, and a naive implementation would hence result in a computational cost of $\mathcal{O}(L^4 D^2)$ per micro-iteration. However, this summation needs to be performed only once for each operator, at the moment when the second second-quantized operator is added:
\begin{equation}
\braket{\alpha_{i-1}^L \mid \hat{F}_{m \sigma ; n\tau} \mid \beta_{i-1}^L} = \sum\limits_{k<i} h_{k(i-1);mn} \braket{\alpha_{i-1}^L \mid \hat{a}_{k\sigma}^{\dagger} \hat{a}^{\dagger}_{(i-1)\tau} \mid \beta_{i-1}^L}. \label{smart-two-index-complements}
\end{equation}
From then on, this operator can be transformed as in Eq. \eqref{operator_tfo}. The total computational cost per micro-iteration is hence reduced to $\mathcal{O}(L^3 D^2)$ for the summation (there are three variable site indices in Eq. \eqref{smart-two-index-complements}), and $\mathcal{O}(L^2 D^3)$ for the transformation (there are $\mathcal{O}(L^2)$ operators to be transformed). The one-site complementary renormalized operator (the complement of three second-quantized operators) can be constructed from the two-site complementary renormalized operators at the moment when the third second-quantized operator is added. From then on, this operator can also be transformed as in Eq. \eqref{operator_tfo}.

As mentioned earlier, the (complementary) renormalized operators are stored to disk, as well as the MPS site tensors, in order to be recycled when the sweep direction is reversed. An overview of the resulting \textit{total cost per macro-iteration} is given in Tab. \ref{compuReqDMRG}.

\begin{table}
\centering
\caption{\label{compuReqDMRG} Computational requirements per \textit{macro-iteration} for the DMRG algorithm.}
\begin{tabular}{|l|rrrr|}
\hline
$\mathcal{O}(\text{task})$ & time & memory & & disk \\
\hline
$\mathbf{H}[i]^{\text{eff}} \{ \mathbf{B}[i] \}$ & $N_{vec} L^3 D^3$    & $N_{vec} D^2~^{(a)}$ & & - \\
SVD and basis truncation                          & $L D^3$                & $D^2$              & \multirow{3}{*}{$\hspace{-0.25cm} \left\} \begin{array}{c} \\ \\ \\ \end{array} \hspace{-0.45cm} \xrightarrow{L\text{ sites}} \right\{ \hspace{-0.3cm}$} & $L D^2$ \\
Renormalized operators                            & $L^3 D^3$            & $L^2 D^2$            & & $L^3 D^2$ \\
Complementary renorm. op.              & $L^4 D^2 + L^3 D^3$  & $L^2 D^2$            & & $L^3 D^2$ \\
\hline
Total              & $L^4 D^2 + N_{vec} L^3 D^3$ & $(N_{\text{vec}} + L^2) D^2$ & & $L^3 D^2$ \\
\hline
\end{tabular}

{$^{(a)}$ The (complementary) renormalized operators are mentioned separately.}
\end{table}

As long as the required $D$ to yield numerical convergence is not exponentially large, the DMRG algorithm is of polynomial cost in $L$. The computational requirements in Tab. \ref{compuReqDMRG} are upper bounds if the symmetry group of the Hamiltonian is exploited. Then the MPS tensors and corresponding (complementary) renormalized operators become block-sparse, and $h_{kl;mn}$ is not full rank. This will be discussed in chapter \ref{SYMM-chapter}.

\section{Properties} \label{DMRG-prop}
\subsection{DMRG is variational}
The DMRG algorithm is variational, because it can be formulated as the optimization of an MPS ansatz. All energies obtained during all micro-iterations are therefore upper bounds to the true ground state energy. These energies do not go down monotonically however, because the basis $\{ \ket{ \alpha_{i-1}^L } \} \otimes \{ \ket{ n_i } \} \otimes \{ \ket{ n_{i+1} } \} \otimes \{ \ket{ \alpha_{i+1}^R } \}$ in which $\hat{H}$ is diagonalized changes between different micro-iterations due to the truncation of the Schmidt spectrum \cite{Chan2002}.

\subsection{Energy extrapolation}
With increasing virtual dimension $D$, the MPS ansatz spans an increasing part of the many-body Hilbert space. Call $E_D$ the minimum energy encountered in Eq. \eqref{effHameq} during the micro-iterations for a given virtual dimension $D$. Several calculations with increasing $D$ can be performed, in order to assess the convergence. This even allows to make an extrapolation of the energy to the FCI limit. Several extrapolation schemes have been suggested. Note that $E_{\text{FCI}}$ and $\{ C_i, p_j, q_k \}$ below are parameters to be fitted. The maximum discarded weight encountered during the last sweep before convergence is abbreviated as:
\begin{equation}
w^{\text{disc}}_D = \max\limits_i \left\{ w[i]^{\text{disc}}_D \right\}.
\end{equation}
The initial assumption of exponential convergence \cite{WhiteQCDMRG}
\begin{equation}
\ln \left( E_D - E_{\text{FCI}} \right) \propto C_1 + C_2 D \label{ExtrapolExpon}
\end{equation}
was rapidly abandoned for the relation \cite{PhysRevB.53.14349,Chan2002,PhysRevB.67.125114}
\begin{equation}
E_D - E_{\text{FCI}} = C_3 w^{\text{disc}}_D, \label{Eextrapol1}
\end{equation}
because the energy is a linear function of the RDM \cite{Chan2002}. The tail of the distribution of RDM eigenvalues scales as \cite{AyersRDMcalues, Chan2002}
\begin{equation}
\kappa[i]^2_{\beta_i} \propto \exp \left\{ - C_4 \left( \ln \beta_i \right)^2 \right\}. \label{schmidtspectrumdecayrelationcroot}
\end{equation}
Substituting this relation in Eq. \eqref{Eextrapol1} yields an improved version of Eq. \eqref{ExtrapolExpon} \cite{Chan2002}:
\begin{equation}
\ln \left( E_D - E_{\text{FCI}} \right) \propto C_5 - C_4 \left( \ln D \right)^2. \label{Eextrapol2}
\end{equation}
Eqs. \eqref{Eextrapol1} and \eqref{Eextrapol2} are still the most widely used extrapolation schemes in QC-DMRG. Three other relations have been proposed, but they have not been used except in their introduction papers. A relation for incremental energies $\Delta E_{D_1} = E_{D_1} - E_{D_0}$ has been suggested \cite{mitrushenkov:4148}:
\begin{equation}
\Delta E_{D} = \frac{C_6 + C_7 E_D}{\sqrt{L^3D^2 + 2L^2D^3}},
\end{equation}
but the extrapolated $E_{\text{FCI}}$ often violates the variational principle. An alternative relation based on the discarded weight has also been proposed \cite{mitrushenkov:4148}:
\begin{equation}
\ln \left( E_D - E_{\text{FCI}} \right) = C_8 - C_9 \left( w^{\text{disc}}_D \right)^{-\frac{1}{2}},
\end{equation}
as well as a Richardson-type extrapolation scheme, based on the assumption that the energy is an analytic function of $w^{\text{disc}}_D$ \cite{RichardsonControlReiher}:
\begin{equation}
E^{(\mu\nu)}(w^{\text{disc}}_D) = \frac{p_0 + p_1 w^{\text{disc}}_D + ... + p_{\mu} \left( w^{\text{disc}}_D \right)^{\mu}}{q_0 + q_1 w^{\text{disc}}_D + ... + q_{\nu} \left( w^{\text{disc}}_D \right)^{\nu}}.
\end{equation}

\subsection{The CI content of the wavefunction}
To analyze the MPS wavefunction \eqref{MPSansatz}, suppose that the $L$ orthonormal orbitals are the HF single-particle states. An important difference with traditional post-HF methods such as CI expansions, is that no FCI coefficients are a priori zero. An MPS hence captures CI coefficients of any particle-excitation rank relative to HF \cite{chan:6110, hachmann:144101}. A small virtual dimension implies little information content in the FCI coefficient tensor, or equivalently that the many nonzero FCI coefficients are in fact highly correlated. This has to be contrasted with CI expansions, which are truncated in their particle-excitation rank and therefore set many FCI coefficients a priori to zero. The nonzero FCI coefficients are however not a priori correlated in a CI expansion: they are entirely free to be variationally optimized.

\subsection{Size-consistency}
Is DMRG size-consistent? For noninteracting subsystems $A$ and $B$, the compound wavefunction should be multiplicatively separable $\ket{\Psi} = \ket{A} \ket{B}$ and the energy additively separable $E = E_A + E_B$. From the discussion of the Schmidt decomposition above, it follows immediately that an MPS is size-consistent if the orbitals of subsystems $A$ and $B$ do not overlap, and if they are separated into two groups on the one-dimensional DMRG lattice \cite{Chan2002, chan:annurevphys}. The latter is for example realized if orbitals $1$ to $k$ correspond to subsystem $A$ and orbitals $k+1$ to $L$ correspond to subsystem $B$. DMRG will then automatically generate the solution with virtual dimension 1 on the $A$-$B$ boundary: $\text{dim}(\alpha_k) = 1$.

\subsection{DMRG is not FCI}
An accurate variational energy does not necessarily imply that the wavefunction is good. Suppose we have an orthonormal MPS $\ket{\Psi_{\text{MPS}}}$ with virtual dimension $D$ which has been variationally optimized to approximate the true ground state $\ket{\Psi_{\text{true}}}$. Suppose that
\begin{equation}
\ket{\Psi_{\text{MPS}}} = \sqrt{1 - \epsilon^2} \ket{\Psi_{\text{true}}} + \epsilon \ket{\Psi_{\text{error}}}
\end{equation}
with $\braket{ \Psi_{\text{true}} \mid \Psi_{\text{error}} } = 0$. Then
\begin{equation}
\| \ket{\Psi_{\text{MPS}}} - \ket{\Psi_{\text{true}}} \|_2 = \sqrt{\left(\sqrt{1-\epsilon^2} - 1\right)^2 + \epsilon^2} = \epsilon + \mathcal{O}(\epsilon^3)
\end{equation}
and
\begin{equation}
\braket{\Psi_{\text{MPS}} \mid \hat{H} \mid \Psi_{\text{MPS}}} - E_{\text{true}} = \epsilon^2 \left( \braket{\Psi_{\text{error}} \mid \hat{H} \mid \Psi_{\text{error}}} - E_{\text{true}} \right).
\end{equation}
The energy converges quadratically in the wavefunction error! Most DMRG convergence criteria rely on energy convergence ($\epsilon^2 \approx 0$). An important implication is that, except for tremendously large virtual dimensions $D$ where $\epsilon \approx 0$, the MPS wavefunction is not invariant to orbital rotations. The orbital choice and their ordering on a one-dimensional lattice also influence the convergence rate with $D$. Strategies to choose and order orbitals are discussed in section \ref{DMRG-orb}. Sparse iterative FCI eigensolvers converge the FCI tensor to a predefined threshold instead of the energy. A FCI solution can therefore be considered invariant to orbital rotations.

\section{Convergence strategies} \label{conv_strat_sec}
The DMRG algorithm can get stuck in a local minimum or a limit cycle, if $D$ is insufficiently large \cite{Chan2002}. The chance of occurrence is larger for inconvenient orbital choices and orderings. Because the virtual dimension $D$ cannot be increased indefinitely in practice, it is important to choose the set of orbitals and their ordering well, see section \ref{DMRG-orb}. Additional considerations to enhance convergence are described here.

\subsection{The number of sites to be optimized in a micro-iteration}
It is better to use the two-site DMRG algorithm than the one-site version \cite{PhysRevB.72.180403}. In the one-site version, the Hamiltonian $\hat{H}$ is diagonalized during the micro-iterations in the basis $\{ \ket{ \alpha_{i-1}^L } \} \otimes \{ \ket{ n_i } \} \otimes \{ \ket{ \alpha_{i}^R } \}$ instead of $\{ \ket{ \alpha_{i-1}^L } \} \otimes \{ \ket{ n_i } \} \otimes \{ \ket{ n_{i+1} } \} \otimes \{ \ket{ \alpha_{i+1}^R } \}$. Because of the larger variational freedom in the two-site DMRG algorithm, lower energy solutions are obtained, and the algorithm is less likely to get stuck \cite{zgid:144115}. It might therefore be worthwhile to optimize three or more MPS tensors simultaneously in a micro-iteration, or to group several orbitals into a single DMRG lattice site \cite{WhiteQCDMRG}.

The two-site algorithm has another important advantage, when the symmetry group of the Hamiltonian is exploited. The virtual dimension $D$ is then distributed over several symmetry sectors, see chapter \ref{SYMM-chapter}. In the one-site algorithm, the virtual dimension of a symmetry sector has to be changed ``manually'' during the sweeps \cite{zgid:144115}, while the SVD \eqref{SVDofBsolution} in the two-site algorithm automatically picks the best distribution.

\subsection{Perturbative corrections and noise}
White suggested to add perturbative corrections to the RDM in order to enhance convergence \cite{PhysRevB.72.180403}. Instead of using perturbative corrections, one can also add noise to the RDM prior to diagonalization or to $B[i]$ prior to SVD \cite{Chan2002}. The corrections or noise help to reintroduce lost symmetry sectors (lost quantum numbers) in the renormalized basis, which are important for the true ground state. Instead of adding noise or perturbative corrections, one can also reserve a certain percentage of the virtual dimension $D$ to distribute equally over all symmetry sectors \cite{chan:3172}.

\subsection{Getting started}
The wavefunction from which QC-DMRG starts has influence on the converged energy (getting stuck in a local minimum) and on the rate of convergence \cite{PhysRevB.67.125114, PhysRevB.68.195116, moritz:034103}. The effect of the starting guess is estimated to be an order of magnitude smaller than the effect of the choice and ordering of the orbitals \cite{moritz:034103}. Nevertheless, it deserves attention.

One possibility is to choose a small active space to start from, and subsequently augment this active space stepwise with previously frozen orbitals \cite{mitrushenkov:6815}, in analogy to the infinite-system DMRG algorithm \cite{PhysRevLett.69.2863}. Natural orbitals from a small CASSCF calculation or HF orbitals can be used to this end \cite{moritz:034103}. An alternative is to make an a priori guess of how correlated the orbitals are. This can be done with a DMRG calculation with small virtual dimension $D$, from which the approximate single-orbital entropies can be obtained, see section \ref{subsec-entang-meas}. The subsystem $A$ is then chosen to be a single orbital in Eq. \eqref{von-neumann-entropy-eq}. The larger the single-orbital entropy, the more it is correlated. The active space can then be chosen and dynamically extended based on the single-orbital entropies \cite{PhysRevA.83.012508}.

One can also decompose a cheap CISD calculation into an MPS to start from \cite{Chan2002, moritz:034103}. The author has found that distributing $D$ equally over the symmetry sectors, and filling the so-obtained MPS with noise, retrieves energies below the HF energy well within the first macro-iteration \cite{2013arXiv1312.2415W}.

To achieve a very accurate MPS quickly, it is also best to start from calculations with relatively small virtual dimension $D$, and to enlarge it stepwise \cite{Chan2002, moritz:034103, PhysRevLett.77.3633}.

\section{Orbital choice and ordering} \label{DMRG-orb}
The opening quote of this chapter refers to section \ref{conv_strat_sec} and this section. There are plenty of ways to set up an RG flow, and the specific setup influences the outcome. One consideration of key importance in QC-DMRG is the choice and ordering of orbitals. Most molecules or active spaces are far from one-dimensional. By placing the orbitals on a one-dimensional lattice, and by assuming an MPS ansatz with modest $D$, an artifical correlation length is introduced in the system, which can be a bad approximation. Over time, several rules of thumb have been established to choose and order the orbitals.

\subsection{Elongated molecules}
Quantum information theory learns that locality is an important concept (see section \ref{entanglement-section}). The Coulomb interaction, however, is long-ranged. On the other hand, the mutual screening of electrons and nuclei can result in an effectively local interaction. For elongated molecules such as hydrogen chains \cite{Chan2002, hachmann:144101, QUA:QUA23173, woutersJCP1, nakatani:134113, ma:224105}, polyenes \cite{Chan2002, chan:204101, hachmann:144101, ghosh:144117, yanai:024105}, or acenes \cite{hachmann:134309, dorando:084109, JCTCgrapheneNano}, which are more or less one-dimensional, choosing a spatially local basis has turned out to be very beneficial. There are roughly three ways to choose a local basis: symmetric orthogonalization as it lies closest to the original gaussian basis functions \cite{hachmann:134309, dorando:084109, QUA:QUA23173, woutersJCP1, ma:224105, PhysRev.105.102}, explicit localization procedures such as Pipek-Mezey or Edmiston-Ruedenberg \cite{ghosh:144117, JCTCgrapheneNano, Pipek-Mezey, RevModPhys.35.457}, and working in a biorthogonal basis \cite{chan:204101, QUA:QUA23173}. For the latter, the effective Hamiltonian is not hermitian anymore. The DMRG algorithm should then be correspondingly adapted \cite{Mitru_arxiv, chan:204101, QUA:QUA23173}. The adapted algorithm is slower and prone to convergence issues, and it is therefore better to use one of the other two localized bases \cite{chan:204101, QUA:QUA23173}.

\subsection{Hamiltonian measures} \label{sec-integral-measures}
If the topology of the molecule does not provide hints for choosing and ordering orbitals, it was investigated whether the Hamiltonian \eqref{QC-ham} can be of use. Several integral measures have been proposed, for which a minimal bandwidth is believed to yield a good orbital order. Chan and Head-Gordon proposed to minimize the bandwidth of the one-electron integral matrix $(i | \hat{T} | j)$ of the HF orbitals \cite{Chan2002}. In quantum chemistry, it is often stated that the one-electron integrals are an order of magnitude larger than the two-electron integrals, and that quantum chemistry therefore corresponds to the small-$U$ limit of the Hubbard model \cite{PhysRevB.67.125114, PhysRevA.83.012508, Hubbard26111963}. On the other hand, there are many two-electron integrals, and they may become important due to their number. When other orbitals than the HF orbitals are used, it may therefore be interesting to minimize the bandwidth of the Fock matrix \cite{DMRG_LiF}:
\begin{equation}
F_{ij} = (i | \hat{T} | j) + \sum\limits_{k \in \text{occ}} \left( 4 (ik | \hat{V} | jk) - 2 (ik | \hat{V} | kj) \right).
\end{equation}
Other proposed integral measures are the MP2-inspired matrix \cite{mitrushenkov:4148}:
\begin{equation}
G_{ij} = \frac{ (ii | \hat{V} | jj)^2}{| \epsilon_i - \epsilon_j |}
\end{equation}
where $\{ \epsilon_i \}$ are the HF single-particle energies, as well as several measures in Ref. \cite{moritz:024107}. These are the Coulomb matrix $J_{ij} = (ij | \hat{V} | ij)$, the exchange matrix $K_{ij} = (ij | \hat{V} | ji)$, the mean-field matrix $M_{ij} = \left( 2 J_{ij} - K_{ij} \right)$, and two derived quantities:
\begin{eqnarray}
J_{ij}^{'} & = & e^{-J_{ij}} \\
M_{ij}^{'} & = & e^{-M_{ij}}.
\end{eqnarray}
While the one-electron integrals $(i | \hat{T} | j)$ vanish when orbitals $i$ and $j$ belong to different molecular point group irreps, $J_{ij}$ and $K_{ij}$ do not. Ref. \cite{moritz:024107} used a genetic algorithm to find the optimal HF orbital ordering, in order to assess the proposed integral measures. This genetic algorithm was expensive, which limited its usage to small test systems. It favoured $K_{ij}$ bandwidth minimization, although no definite conclusions were drawn \cite{moritz:024107}. The exchange matrix $K_{ij}$ was recently used in two DMRG studies \cite{JCTCgrapheneNano,nakatani:134113} in conjunction with localized orbitals, because it then directly reflects their overlaps and distances.

\subsection{Entanglement measures} \label{subsec-entang-meas}
DMRG can be analyzed by means of the underlying MPS ansatz and quantum information theory. Can the latter tell us something more than locality? Legeza and S\'olyom proposed to use the single-orbital entropies to find an optimal ordering \cite{PhysRevB.68.195116}. Subsystem A is then chosen to be a single orbital $k$ in Eq. \eqref{von-neumann-entropy-eq}, and its entropy is denoted by $S_1(k)$. It can be efficiently calculated in the DMRG algorithm, because the corresponding RDM $\hat{\rho}^k$ can be built from the following expectation values \cite{Rissler2006519}:
\begin{equation}
\hat{\rho}^k = \left( \begin{array}{cccc}
\braket{(1-\hat{n}_{k\uparrow})(1- \hat{n}_{k\downarrow})} & 0 & 0 & 0 \\
0 & \braket{\hat{n}_{k\uparrow} (1-\hat{n}_{k\downarrow})} & 0 & 0 \\
0 & 0 & \braket{(1-\hat{n}_{k\uparrow}) \hat{n}_{k\downarrow}} & 0 \\
0 & 0 & 0 & \braket{\hat{n}_{k\uparrow} \hat{n}_{k\downarrow}}
\end{array} \right) \label{one-orb-RDM}
\end{equation}
with $\hat{n}_{k\sigma} = \hat{a}_{k\sigma}^{\dagger}\hat{a}_{k\sigma}$, hence without reordering any orbitals. The larger the single-orbital entropy $S_1(k)$, the more orbital $k$ is correlated. Legeza and S\'olyom proposed to perform a small-$D$ DMRG calculation to estimate $S_1(k)$, and to place the orbitals with large $S_1(k)$ in the center of the chain, and the ones with small $S_1(k)$ near the edges. They reasoned that orbitals close to the Fermi surface are more entangled and therefore have a larger single-orbital entropy. Because DMRG only captures local correlations, these orbitals should lie close to each other.

Rissler, Noack and White proposed to use the two-orbital mutual information $I_{k,l}$ to order the orbitals \cite{Rissler2006519}. In addition to the single-orbital entropies $S_1(k)$ and $S_1(l)$, the two-orbital entropy $S_2(k,l)$ is also needed to calculate $I_{k,l}$. It can be obtained by choosing for subsystem $A$ the two orbitals $k$ and $l$. $S_2(k,l)$ can again be efficiently calculated in the DMRG algorithm, as its RDM can be built from expectation values of operators acting on at most two sites \cite{Rissler2006519}. Although a $16 \times 16$ RDM needs to be constructed, many of its entries are zero due to symmetry considerations, as was the case in Eq. \eqref{one-orb-RDM}. The so-called subadditivity property of the entanglement entropy dictates that:
\begin{equation}
S_2(k,l) \leq S_1(k) + S_1(l).
\end{equation}
Any entanglement between orbitals $k$ and $l$ reduces $S_2(k,l)$ with respect to $S_1(k) + S_1(l)$. The two-orbital mutual information is defined by:
\begin{equation}
I_{k,l} = \frac{1}{2} \left( S_1(k) + S_1(l) - S_2(k,l) \right)(1 - \delta_{k,l}) \geq 0,
\end{equation}
and is thus a symmetric measure of the correlation between orbitals $k$ and $l$. Its bandwidth can be minimized, for example based on cost functions such as
\begin{equation}
I = \sum\limits_{k,l} I_{k,l} |k-l|^{\eta}.
\end{equation}
Rissler, Noack and White found no clear correspondence between $I_{k,l}$ and the integral measures of section \ref{sec-integral-measures}. They observed that $I_{k,l}$ is large between orbitals which belong to the same molecular point group irrep, as well as between corresponding bonding and anti-bonding orbitals with large partial occupations (far from empty or doubly occupied) \cite{Rissler2006519}. Later studies of various groups supported this finding and corresponding ordering \cite{kurashige:234114, yanai:024105, PhysRevA.83.012508, JPCLentanglement, ma:224105, 2013arXiv1312.2415W}. For small molecules such as dimers, it is best to group orbitals of the same molecular point group irrep into blocks, and place irrep blocks of bonding and anti-bonding type next to each other. If in addition natural orbitals (NO) are used, the orbitals within an irrep block should be reordered so that the ones with NO occupation number (NOON) closest to one, are nearest to the block of their bonding or anti-bonding colleagues \cite{ma:224105}.

The gradient and Hessian of $I_{k,l}$ with respect to orbital rotations can be calculated by resp. three- and four-point correlation functions on the one-dimensional DMRG lattice \cite{2013arXiv1312.2415W}. These can still be obtained efficiently \cite{zgid:144115}. With a corresponding Newton-Raphson algorithm, $I_{k,l}$ might not only yield the optimal ordering of a given set of orbitals, but also the optimal choice of orbitals.

\section{Variations on QC-DMRG} \label{DMRG-algos}
\subsection{Quadratic scaling DMRG} \label{QS-DMRG-section}
For elongated molecules, when the active space is studied in a localized basis,
\begin{equation}
(ij | \hat{V} | kl) = \int d\vec{r}_1 d\vec{r}_2 \frac{\phi^*_i(\vec{r}_1) \phi_k(\vec{r}_1) \phi^*_j(\vec{r}_2) \phi_l(\vec{r}_2)}{| \vec{r}_1 - \vec{r}_2 |} \label{physics-notation-two-body-matrix-elements}
\end{equation}
vanishes exponentially with the separation of orbitals $i$ and $k$, and the separation of orbitals $j$ and $l$. By defining a threshold, below which these two-body matrix elements can be neglected, one can reduce the cost of the DMRG algorithm in Tab. \ref{compuReqDMRG} to $\mathcal{O}(L^2D^3)$ computational time, $\mathcal{O}(LD^2)$ memory, and $\mathcal{O}(L^2D^2)$ disk \cite{WhiteQCDMRG, hachmann:144101, dorando:084109}. Quadratic scaling DMRG (QS-DMRG) is not variational anymore because the Hamiltonian is altered, but the error can be controlled with the threshold. At present, QC-DMRG can achieve FCI energy accuracy for about 40 electrons in 40 orbitals \cite{sharma:124121, 2013arXiv1312.2415W}. With QS-DMRG, one can achieve FCI energy accuracy for 100 electrons in 100 orbitals \cite{hachmann:144101}, and maybe more. It should however be repeated, that this method relies on the topology of the molecule, and exploits the fact that DMRG works very well for one-dimensional systems.

\subsection{Building-in dynamic correlation}
QC-DMRG can at present achieve FCI energy accuracy for about 40 electrons in 40 orbitals. The static correlation in active spaces up to this size can hence be resolved, while dynamic correlation has to be treated a posteriori. Luckily, QC-DMRG allows for an efficient extraction of the two-body RDM (2-RDM) \cite{zgid:144115, ghosh:144117}. The 2-RDM is not only required to calculate analytic nuclear gradients \cite{Chan2002, Spiropyran}, but also to compute the gradient and the Hessian in CASSCF \cite{Roos3}. It is therefore natural to introduce a CASSCF variant with DMRG as active space solver, DMRG-CASSCF or DMRG-SCF \cite{zgid:144116, ghosh:144117, ChanQUA:QUA22099}. Static correlation can be treated with DMRG-SCF. To add dynamic correlation as well, three methods have been introduced.

With a little more effort, the 3-RDM and some specific contracted 4-RDMs can be extracted from DMRG as well. These are required to apply second-order perturbation theory to a CASSCF wavefunction, called CASPT2, in internally contracted form. The DMRG variant is called DMRG-CASPT2 \cite{kurashige:094104,ma:224105,Spiropyran}.

Based on a CASSCF wavefunction, a configuration interaction expansion can be introduced, called MRCI. Recently, an internally contracted MRCI variant was proposed, which only requires the 4-RDM \cite{saitow:044118}. By approximating the 4-RDM with a cumulant reconstruction from lower-rank RDMs, DMRG-MRCI was made possible \cite{saitow:044118}.

Yet another way is to perform a canonical transformation (CT) on top of an MR wavefunction, in internally contracted form. When an MPS is used as MR wavefunction, the method is called DMRG-CT \cite{yanai:024105, neuscamman:024106, C2CP23767A}.

\subsection{Excited states} \label{DMRG-ExcitedStatesSectionInChatperTwo}
In addition to ground states, DMRG can also find excited states. By projecting out lower-lying eigenstates \cite{2013arXiv1312.2415W}, or by targeting a specific energy with the harmonic Davidson algorithm \cite{dorando:084109}, DMRG solves for a particular excited state. In these state-specific algorithms, the whole renormalized basis is used to represent one single eigenstate. In state-averaged DMRG, several eigenstates are targeted at once to prevent root-flipping. Their RDMs are weighted and summed to perform the DMRG renormalization step \cite{HallbergBook}. The renormalized basis then represents several eigenstates simultaneously.

DMRG linear response theory (DMRG-LRT) \cite{dorando:184111} allows to calculate response properties, as well as excited states. Once the ground state has been found, the MPS tangent vectors to this optimized point can be used as an (incomplete) variational basis to approximate excited states \cite{dorando:184111, PhysRevB.85.035130, PhysRevB.85.100408, PhysRevB.88.075122, PhysRevB.88.075133, NaokiLRTpaper}, see chapter \ref{Thouless-chapter}. As the tangent vectors to an optimized Slater determinant yield the configuration interaction with singles (CIS), also called the Tamm-Dancoff approximation (TDA), for HF theory \cite{helgaker2}, the same names are used for DMRG: DMRG-CIS or DMRG-TDA. The variational optimization in an (incomplete) basis of MPS tangent vectors can be extended to higher-order tangent spaces as well. DMRG-CISD, or DMRG configuration interaction with singles and doubles, is a variational approximation to target both ground and excited states in the space spanned by the MPS reference and its single and double tangent spaces \cite{PhysRevB.88.075122}.

By linearizing the time-dependent variational principle for matrix product states \cite{PhysRevLett.107.070601}, the DMRG random phase approximation (DMRG-RPA) is found \cite{2011arXiv1103.2155K, PhysRevB.88.075122, PhysRevB.88.075133, NaokiLRTpaper}, again in complete analogy with RPA for HF theory.

\subsection{Other ansatzes}
Two other related ansatzes have been employed in quantum chemistry: the TTNS \cite{PhysRevB.82.205105, nakatani:134113, LegezaTTNS} and the complete-graph TNS (CGTNS) \cite{1367-2630-12-10-103008, Marti:C0CP01883J}:
\begin{equation}
\ket{\Psi} = \sum\limits_{\{ n_k \}} \left( \prod\limits_{ i < j} C[i,j]^{n_i n_j} \right) \ket{n_1 ... n_L}. \label{CGTNS}
\end{equation}
The latter is an example of a correlator product state (CPS) \cite{1367-2630-11-8-083026}, in which multiple tensors can have the same physical index. The TTNS requires a smaller virtual dimension than DMRG to achieve the same accuracy. The accuracy of the CGTNS is limited by the number of correlated orbitals in each cluster (two in Eq. \eqref{CGTNS}). The optimization algorithms for TTNSs and CGTNSs are less efficient than QC-DMRG for an MPS, and as a result an MPS is currently still the preferred choice for ab initio quantum chemistry.

There is also a QC-DMRG algorithm for the relativistic many-body four-component Dirac equation \cite{Knecht-4c-DMRG}.

\section{QC-DMRG codes and studied systems} \label{DMRG-systems}
Tab. \ref{QC_DMRG-codes} gives an overview of the currently existing QC-DMRG codes. Two of them are freely available, \textsc{Block} and \textsc{CheMPS2}. Four codes have $\mathsf{SU(2)}$ spin symmetry: Zgid's code, \textsc{Rego}, \textsc{Block}, and \textsc{CheMPS2}. The former two explicitly retain entire multiplets at each virtual bond, while the latter two exploit the Wigner-Eckart theorem to work with a reduced renormalized basis and reduced renormalized operators, see chapter \ref{SYMM-chapter}.

Two message-passing interface (MPI) strategies are currently used: processes can become responsible of certain site indices of the (complementary) renormalized operators \cite{chan:3172}, or of certain symmetry blocks in the virtual bonds \cite{kurashige:234114}.

\begin{table}
\centering
\caption{\label{QC_DMRG-codes} Overview of QC-DMRG codes. This list may be incomplete. All codes known to the author are listed.}
\begin{tabular}{|l|l|l|}
\hline
Name                      & Authors             & Selected papers \\
\hline
                          & White               & \cite{WhiteQCDMRG,Rissler2006519} \\
                          & Mitrushenkov        & \cite{mitrushenkov:6815,QUA:QUA23173} \\
\textsc{Block}$^{(a)}$    & Chan \& Sharma      & \cite{Chan2002,sharma:124121}\\
\textsc{Qc-Dmrg-Budapest} & Legeza              & \cite{PhysRevB.67.125114,JCTCbondForm}\\
\textsc{Qc-Dmrg-Eth}      & Reiher              & \cite{RichardsonControlReiher, JCTCspindens}\\
                          & Zgid                & \cite{zgid:014107,zgid:144116}\\
                          & Xiang               & \cite{PhysRevB.81.235129}\\
\textsc{Rego}             & Kurashige \& Yanai  & \cite{kurashige:234114, naturechem}\\
\textsc{CheMPS2}$^{(b)}$  & Wouters             & \cite{woutersJCP1,2013arXiv1312.2415W}\\
\hline
\end{tabular}

$^{(a)}$Freely available from \cite{BlockCodeChan}.\\
$^{(b)}$Freely available from \cite{CheMPS2github} or \cite{2013arXiv1312.2415W}.
\end{table}

Many properties of many systems have been studied. QC-DMRG is of course able to calculate the ground state energy, but also excited state energies \cite{PhysRevB.67.125114, DMRG_LiF, moritz:184105, dorando:084109, ghosh:144117, Spiropyran, NaokiLRTpaper,2013arXiv1312.2415W, 1.4867383}, avoided crossings \cite{DMRG_LiF, moritz:184105, 2013arXiv1312.2415W, LegezaTTNS}, spin splittings \cite{hachmann:134309, marti:014104, zgid:014107, neuscamman:024106, mizukami:091101, 1367-2630-12-10-103008, sharma:124121, woutersJCP1, C3CP53975J, 2013arXiv1312.2415W, Harris2014}, polyradical character by means of the NOON spectrum \cite{hachmann:134309,ChanQUA:QUA22099,JCTCgrapheneNano}, static and dynamic polarizabilities \cite{dorando:184111, woutersJCP1}, static second hyperpolarizabilities \cite{woutersJCP1}, particle-particle, spin-spin, and singlet diradical correlation functions \cite{hachmann:134309,sharma:124121,JCTCgrapheneNano, saitow:044118}, as well as expectation values based on the 1- or 2-RDM such as spin densities \cite{JCTCspindens, newKura} and dipole moments \cite{DMRG_LiF}.

The systems which have been studied range from atoms and first-row dimers to large transition metal clusters and $\pi$-conjugated hydrocarbons. Several of them have repeatedly received attention in the QC-DMRG community:
\begin{itemize}
\item H$_2$O \cite{WhiteQCDMRG, Chan2002, PhysRevB.67.125114, chan:8551, PhysRevB.68.195116, chan:3172, PhysRevB.70.205118, RichardsonControlReiher, PhysRevB.81.235129, C2CP23767A, nakatani:134113, NaokiLRTpaper} was already the subject of several FCI studies, due to its natural abundance and small number of electrons.
\item Hydrogen chains \cite{Chan2002, hachmann:144101, zgid:144115, zgid:144116, QUA:QUA23173, woutersJCP1, nakatani:134113, ma:224105}: these one-dimensional systems exhibit large static correlation at stretched geometries. They are optimal testcases for QC-DMRG.
\item All-trans polyenes \cite{Chan2002, chan:204101, hachmann:144101, ghosh:144117, yanai:024105, saitow:044118, NaokiLRTpaper}: they are also one-dimensional, with a large MR character.
\item N$_2$ \cite{mitrushenkov:6815, Chan2002, mitrushenkov:4148, PhysRevB.68.195116, chan:6110, Rissler2006519, moritz:244109, C2CP23767A, nakatani:134113, JCTCbondForm, ma:224105, saitow:044118} was already the subject of several FCI studies, due to its MR character at stretched bond lengths and its small number of electrons.
\item Cr$_2$ \cite{mitrushenkov:6815, moritz:024107, moritz:034103, kurashige:234114, kurashige:094104, sharma:124121, nakatani:134113, ma:224105} is only found to be bonding at the CASPT2 level. A complete basis set extrapolation of DMRG-CASPT2 calculations in the cc-pwCV(T,Q,5)Z basis, correlating 12 electrons in 28 orbitals, was needed to retrieve an acceptable dissociation energy \cite{kurashige:094104}.
\item $\left[\text{Cu}_2\text{O}_2\right]^{2+}$ \cite{marti:014104, kurashige:234114, yanai:024105, PhysRevA.83.012508} requires accurate descriptions of both static and dynamic correlation along its isomerization coordinate. DMRG-CT, correlating 28 electrons in 32 orbitals, showed that the bis($\mu$-oxo) isomer is more stable than the $\mu-\eta^2:\eta^2$ peroxo isomer \cite{yanai:024105}.
\end{itemize}
Other QC-DMRG studies treat
\begin{itemize}
\item the avoided crossings in LiF \cite{DMRG_LiF, LegezaTTNS}, CsH \cite{moritz:184105, JCTCbondForm}, and C$_2$ \cite{2013arXiv1312.2415W}
\item the static correlation due to $\pi$-conjugation in acenes \cite{dorando:084109,hachmann:134309,JCTCgrapheneNano}, poly(phenyl) carbenes \cite{ChanQUA:QUA22099, mizukami:091101}, perylene \cite{C2CP23767A}, graphene nanoribbons \cite{JCTCgrapheneNano}, free base porphyrin \cite{neuscamman:024106, saitow:044118}, and spiropyran \cite{Spiropyran}
\item transition metal clusters such as $\left[\text{Fe}_2\text{S}_2(\text{SCH}_3)_4\right]^{2-}$ \cite{sharma:124121, NaokiLRTpaper}, $\left[\text{Fe(NO)}\right]^{2+}$ \cite{JCTCspindens, JPCLentanglement}, Mn$_4$CaO$_5$ in photosystem II \cite{naturechem}, and the two dinuclear oxo-bridged complexes $\left[\text{Fe}_2\text{O}\text{Cl}_6\right]^{2-}$ and $\left[\text{Cr}_2\text{O}(\text{NH}_3)_{10}\right]^{4+}$ \cite{Harris2014}
\item molecules with heavy elements, for which relativistic effects become important, such as CsH \cite{moritz:184105, JCTCbondForm}, the complexation of CUO with four Ne or Ar atoms \cite{C3CP53975J}, and the binding energy of TlH \cite{Knecht-4c-DMRG}
\end{itemize}
Many more molecules were, are, and will be studied, which renders this list incomplete.

\chapter{Symmetry-adapted DMRG and \textsc{CheMPS2}} \label{SYMM-chapter}
\begin{chapquote}{Wolfgang E. Pauli}
Die Gruppenpest!
\end{chapquote}

\section{Introduction}

The symmetry group of a Hamiltonian can be used to reduce the dimensionality of the exact diagonalization problem \cite{Weyl, Wigner1939}. The Hamiltonian does not connect states which belong to different irreps or to different rows of the same irrep. By choosing a basis of symmetry eigenvectors, the Hamiltonian becomes block diagonal, and each block can be diagonalized separately. The blocks which belong to different rows of the same irrep are closely related, and yield the same energies. In chapter \ref{DMRG-QC-chapter}, it was discussed how locality leads to low-entanglement wavefunctions. These allow to reduce the dimensionality of the exact diagonalization problem as well, at least for ground and low-lying eigenstates. Symmetry and locality can be combined, which is shown in this chapter for DMRG.

From the very beginning, the abelian particle-number and spin-projection symmetries were incorporated in QC-DMRG \cite{WhiteQCDMRG,mitrushenkov:6815,Chan2002}. Abelian point group symmetry followed quickly \cite{chan:6110, PhysRevB.68.195116}. These symmetries are easy to implement, because they commute with the DMRG RDM. For $\mathsf{SU(2)}$ spin symmetry this is not the case, which is why its implementation took longer.

Sierra and Nishino first introduced exact $\mathsf{SU(2)}$ spin symmetry into DMRG with the interaction-round-a-face DMRG method \cite{Sierra1997505}. McCulloch and Gul\'acsi later found an easier way, based on a quasi-RDM \cite{McCulloch1, McCUlloch2, 0295-5075-57-6-852}, see section \ref{theo-symm-sec}. For the underlying MPS, this boils down to assuming that the rank-three MPS tensors are irreducible tensor operators of the symmetry group \cite{1742-5468-2007-10-P10014}. This opened the path to implement multiplicity-free non-Abelian symmetries also in TNSs \cite{1367-2630-12-3-033029, PhysRevA.82.050301, PhysRevB.86.195114}. The spin-adapted DMRG method of McCulloch and Gul\'acsi was later introduced in nuclear structure calculations \cite{PhysRevC.73.014301, PhysRevLett.97.110603, PhysRevC.78.041303}, where it is known as angular momentum DMRG or JDMRG, as well as in QC-DMRG \cite{zgid:014107, sharma:124121, woutersJCP1, 2013arXiv1312.2415W}. Non-multiplicity-free symmetries can also be exploited in DMRG, but require special considerations \cite{Weichselbaum20122972}.

Before the introduction of exact $\mathsf{SU(2)}$ symmetry in QC-DMRG, several tricks were employed. Legeza used a spin-reflection operator to distinguish even- and odd-spin states based on their spin parity \cite{PhysRevB.56.14449,PhysRevB.67.125114,DMRG_LiF}. A level shift operator \cite{moritz:184105, marti:014104, ghosh:144117,1367-2630-12-10-103008}
\begin{eqnarray}
\hat{H} & = & \hat{H}_0 + \alpha \hat{S}^- \hat{S}^+ \\
\hat{H} & = & \hat{H}_0 + \alpha \hat{S}^2 
\end{eqnarray}
can also be used to raise higher spin states in energy. Zgid and Nooijen \cite{zgid:014107} used the quasi-RDM to impose exact $\mathsf{SU(2)}$ spin symmetry in QC-DMRG, but they retained all states of a multiplet explicitly in the renormalized basis. In the works of Sharma and Chan \cite{sharma:124121} and the author \cite{woutersJCP1, 2013arXiv1312.2415W}, the Wigner-Eckart theorem was exploited to work with reduced renormalized basis states instead of entire multiplets.

\section{Spin-adapted DMRG} \label{theo-symm-sec}
McCulloch's quasi-RDM method \cite{McCulloch1, McCUlloch2, 0295-5075-57-6-852, 1742-5468-2007-10-P10014} is reviewed in this section.

\subsection{The quasi-RDM} \label{subsec-quasi-rdm}
Consider the bases $\{ \ket{j_A j^z_A \alpha_A} \}$ and $\{ \ket{j_B j^z_B \alpha_B} \}$ for subsystems $A$ and $B$ respectively, which have good spin $j$ and spin projection $j^z$ quantum numbers. $\alpha$ keeps track of the number of basis states with symmetry $(j,j^z)$. The wavefunction for the compound system with spin $S$ and spin projection $S^z$ can be written as
\begin{equation}
\ket{\Psi} = \sum\limits_{j_A j^z_A \alpha_A j_B j^z_B \alpha_B} \Psi^{S S^z}_{(j_A j^z_A \alpha_A) ; (j_B j^z_B \alpha_B)} \ket{j_A j^z_A \alpha_A} \ket{j_B j^z_B \alpha_B}. \label{McCullochWfn}
\end{equation}
The coefficients $\Psi^{S S^z}_{(j_A j^z_A \alpha_A) ; (j_B j^z_B \alpha_B)}$ are not completely independent, but are related to each other by Clebsch-Gordan coefficients. The triangle condition for angular momentum and the sum rule for spin projections have to be fulfilled for example:
\begin{eqnarray}
|j_A - j_B| & \leq & S \leq j_A + j_B, \label{triangleRelation}\\
j_A^z + j_B^z & = & S^z.
\end{eqnarray}
Only if the compound wavefunction is a spin singlet, $j_A$ and $j_B$ are constrained to be equal in the summation. This implies that the RDM $\hat{\rho}^A$ for subsystem $A$ is in general not block-diagonal with respect to $j_A$, except if $\ket{\Psi}$ is a singlet:
\begin{equation}
\hat{\rho}^A = \sum\limits_{j_A j^z_A \alpha_A \widetilde{j}_A \widetilde{\alpha}_A}  \ket{j_A j^z_A \alpha_A} \left( \sum\limits_{j_B j^z_B \alpha_B} \Psi^{S S^z}_{(j_A j^z_A \alpha_A) ; (j_B j^z_B \alpha_B)} \Psi^{S S^z *}_{(\widetilde{j}_A j^z_A \widetilde{\alpha}_A) ; (j_B j^z_B \alpha_B)} \right) \bra{\widetilde{j}_A j^z_A \widetilde{\alpha}_A}.
\end{equation}
The eigenvectors of $\hat{\rho}^A$ will then not be spin eigenvectors, i.e. of $\hat{S}^2$. One way to obtain a renormalized basis of spin eigenvectors, is by using the quasi-RDM. It can be obtained from $\hat{\rho}^A$ by setting the off-diagonal blocks, which connect different spin symmetry sectors, to zero:
\begin{equation}
\hat{\rho}^A_{\text{quasi}} = \sum\limits_{j_A j^z_A \alpha_A \widetilde{\alpha}_A}  \ket{j_A j^z_A \alpha_A} \left( \sum\limits_{j_B j^z_B \alpha_B} \Psi^{S S^z}_{(j_A j^z_A \alpha_A) ; (j_B j^z_B \alpha_B)} \Psi^{S S^z *}_{(j_A j^z_A \widetilde{\alpha}_A) ; (j_B j^z_B \alpha_B)} \right) \bra{j_A j^z_A \widetilde{\alpha}_A}.
\end{equation}
The eigenvectors of $\hat{\rho}^A_{\text{quasi}}$ are spin eigenvectors, and their probability of occurrence in subsystem $A$ is given by the corresponding eigenvalues of $\hat{\rho}^A_{\text{quasi}}$. Quasi-RDMs can be constructed analogously for other non-Abelian symmetries.

\subsection{Reduced basis states}
A performance gain in memory and computer time can be obtained by working with reduced basis states. If for all multiplets $(j, \alpha)$, all spin projections $j^z$ are present, a Clebsch-Gordan coefficient can be factorized from the coefficient tensor in Eq. \eqref{McCullochWfn} due to the Wigner-Eckart theorem:
\begin{equation}
\ket{\Psi} = \sum\limits_{j_A j^z_A \alpha_A j_B j^z_B \alpha_B} \braket{j_A j_A^z j_B j_B^z \mid S S^z} \Psi^{S}_{(j_A \alpha_A) ; (j_B \alpha_B)} \ket{j_A j^z_A \alpha_A} \ket{j_B j^z_B \alpha_B}, \label{FullWfnButRedCoeffTensor}
\end{equation}
or in reduced form:
\begin{equation}
\left| \Ket{\Psi} \right. = \sum\limits_{j_A \alpha_A j_B \alpha_B} \Psi^{S}_{(j_A \alpha_A) ; (j_B \alpha_B)} \left| \ket{j_A \alpha_A} \right. \left|\ket{j_B \alpha_B}\right. . \label{reducedWfn}
\end{equation}

The DMRG renormalization tranformation to augment the left renormalized basis with one site (containing one spin) can analogously be written as
\begin{equation}
\ket{j_{i} j_{i}^z \alpha_{i}} = \sum\limits_{ j_{i-1} j_{i-1}^z \alpha_{i-1} s_i s^z_i} A[i]^{(s_i s_i^z)}_{ (j_{i-1} j_{i-1}^z \alpha_{i-1}) ; (j_{i} j_{i}^z \alpha_{i})} \ket{j_{i-1} j_{i-1}^z \alpha_{i-1}} \ket{s_i s^z_i},
\end{equation}
or in reduced form as
\begin{equation}
\left| \ket{j_{i} \alpha_{i}} \right. = \sum\limits_{ j_{i-1} \alpha_{i-1} s_i} T[i]^{(s_i)}_{ (j_{i-1} \alpha_{i-1}) ; (j_{i} \alpha_{i})} \left|\ket{j_{i-1} \alpha_{i-1}}\right. \left|\ket{s_i}\right. ,
\end{equation}
with
\begin{equation}
A[i]^{(s_i s_i^z)}_{ (j_{i-1} j_{i-1}^z \alpha_{i-1}) ; (j_{i} j_{i}^z \alpha_{i})} = \braket{j_{i-1} j_{i-1}^z s_i s_i^z \mid j_i j_i^z} T[i]^{(s_i)}_{ (j_{i-1} \alpha_{i-1}) ; (j_{i} \alpha_{i})}. \label{CGfromMPS}
\end{equation}
$A[i]^{(s_i)}$ can therefore be regarded as an irreducible tensor operator with spin $s_i$.

\subsection{Irreducible tensor operators} \label{mcculloch-irred-tensor-op-subsec}
An extra performance gain can be achieved if the operators in the Hamiltonian are irreducible tensor operators of the imposed symmetry group. For spin systems, the following operators are an example:
\begin{equation}
\left( \hat{S}_{-1}^1 , \hat{S}_{0}^1, \hat{S}_{1}^1 \right) = \left( \frac{\hat{S}_x - i \hat{S}_y}{\sqrt{2}} , \hat{S}_z, - \frac{\hat{S}_x + i \hat{S}_y}{\sqrt{2}} \right).
\end{equation}
Due to the Wigner-Eckart theorem
\begin{equation}
\braket{s_1 s_1^z \mid \hat{S}^1_m \mid s_2 s_2^z} = \braket{s_1 \mid\mid \hat{S}^1 \mid\mid s_2} \braket{s_2 s_2^z 1 m \mid s_1 s_1^z}, \label{WEforOP}
\end{equation}
renormalized operators can be obtained in reduced form by recoupling the irreducible tensor operators and the reduced renormalized basis states. Formally this boils down to contracting the common multiplets of the Clebsch-Gordan coefficients in Eqs. \eqref{CGfromMPS} and \eqref{WEforOP}. The tensor product of irreducible tensor operators can also be obtained by working solely with reduced quantities \cite{1742-5468-2007-10-P10014}, see section \ref{CheMPS2-renorm-op-subsection}.

\subsection{Singlet-embedding}
For the coupling to spin $S$ in Eq. \eqref{reducedWfn}, all spin symmetry sectors $j_A$ and $j_B$ which comply with Eq. \eqref{triangleRelation} have to be taken into account. This strategy to form a spin-$S$ wavefunction is hence less efficient for larger values of $S$. One way to circumvent the large summation, is by adding a noninteracting site at the right end of the one-dimensional lattice, with spin $S$ \cite{0295-5075-57-6-852}. At the position of the current micro-iteration, one can then simply recouple to a singlet state. Sharma and Chan called this the singlet-embedding strategy \cite{sharma:124121}. In \textsc{CheMPS2}, the singlet-embedding will arise naturally, see section \ref{subsec-chemps2-symm-imposing}.

\subsection{Advantages}
Eq. \eqref{FullWfnButRedCoeffTensor} allows to explicitly target a specific symmetry sector of the Hamiltonian. The wavefunction is then always an exact eigenstate of $\hat{S}^2$, irrespective of the virtual dimension $D$. A singlet-triplet gap can then for example be obtained by two ground state calculations, instead of several excited state calculations. For the latter, spin mixing can occur, because working in the $S^z=0$ symmetry sector does not imply anything about $S$. Explicit measurement of $\hat{S}^2$, and its evolution with $D$, should then be used to discern the spin $S$.

Another advantage is the memory reduction. $A[i]$ contains $(2s_i+1)D^2$ variables. Due to the Clebsch-Gordan coefficients in Eq. \eqref{CGfromMPS}, it becomes block-sparse. Whenever a Clebsch-Gordan coefficient is zero, the corresponding MPS tensor block does not need to be allocated. In addition, the symmetry block $(j_{i-1} , j_{i})$ in $A[i]$ is represented in reduced form in $T[i]$. $D(j_i)$ reduced renormalized basis states correspond in fact to $(2j_i+1)D(j_i)$ individual renormalized basis states. Next to block-sparsity, Eq. \eqref{CGfromMPS} hence also encompasses information compression. The block-sparsity and the compression result in faster contractions over common indices. Next to a memory advantage, there is hence also an advantage in computational time.

\section{Tensors in CheMPS2} \label{sec-CheMPS2}
\subsection{Introduction}
\textsc{CheMPS2} exploits $\mathsf{SU(2)}$ spin symmetry, $\mathsf{U(1)}$ particle-number symmetry, and the abelian point group symmetries $\mathsf{P}$ with real-valued character tables:
\begin{equation}
\mathsf{P} \in \{ C_1, C_i, C_2, C_s, D_2, C_{2v}, C_{2h}, D_{2h} \}. \label{PointGroupsInCheMPS2}
\end{equation}
$C_1$ is the trivial point group which contains only the identity operation. Because these abelian groups $\mathsf{P}$ all have real-valued character tables, the direct product of any irrep $I_j$ with itself gives the trivial irrep $I_0$: \begin{equation}
\forall I_j: ~ I_j \otimes I_j = I_0.
\end{equation}
The physical basis states of orbital $k$ correspond to the following symmetry eigenstates:
\begin{eqnarray}
\ket{-} & \rightarrow & \ket{s=0; s^z=0; N=0; I=I_0} \\
\ket{\uparrow} & \rightarrow & \ket{s=\frac{1}{2}; s^z=\frac{1}{2}; N=1; I=I_k} \\
\ket{\downarrow} & \rightarrow & \ket{s=\frac{1}{2}; s^z=-\frac{1}{2}; N=1; I=I_k} \\
\ket{\uparrow\downarrow} & \rightarrow & \ket{s=0; s^z=0; N=2; I=I_0}.
\end{eqnarray}
The virtual basis states are also labeled by the quantum numbers of $\mathsf{SU(2)} \otimes \mathsf{U(1)} \otimes \mathsf{P}$:
\begin{equation}
\ket{\alpha} \rightarrow \ket{j j^z N I \alpha}.
\end{equation}
The equivalent of Eq. \eqref{CGfromMPS} is then
\begin{equation}
A[i]_{(j_L j_L^z N_L I_L \alpha_L) ; (j_R j_R^z N_R I_R \alpha_R)}^{(s s^z N I)} = \braket{j_L j_L^z s s^z \mid j_R j_R^z} \delta_{N_L + N, N_R} \delta_{I_L \otimes I, I_R} T[i]^{(s N I)}_{(j_L N_L I_L \alpha_L);(j_R N_R I_R \alpha_R)}. \label{CheMPS2_WE_MPS}
\end{equation}
The $\mathsf{SU(2)}$, $\mathsf{U(1)}$, and $\mathsf{P}$ symmetries are locally imposed by their Clebsch-Gordan coefficients. These express nothing else than resp. local allowed spin recoupling, local particle number conservation, and local point group symmetry conservation. The index $\alpha$ keeps track of the number of reduced renormalized basis states with symmetry $(j, N, I)$. This equation again encompasses block-sparsity and information compression.

\subsection{Imposing symmetry} \label{subsec-chemps2-symm-imposing}
\begin{figure}[h!]
\centering
\includegraphics[width=0.60\textwidth]{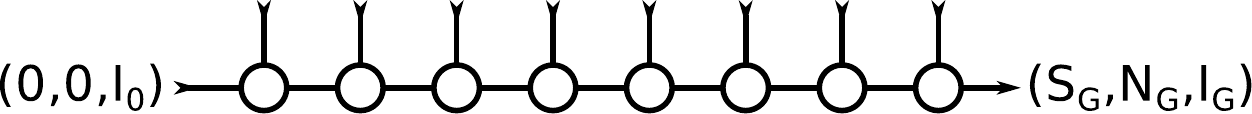} 
\caption{\label{ImposingSymmetry-plot} Imposing $\mathsf{SU(2)}$, $\mathsf{U(1)}$, and $\mathsf{P}$ symmetry.}
\end{figure}
The desired global symmetry $(S_G,N_G,I_G)$ can be imposed with the singlet-embedding strategy, see Fig. \ref{ImposingSymmetry-plot}. Assume that the MPS is part of a larger DMRG chain, to which it is connected on its left and right ends. On the left end, there is only one irrep $(j_L,N_L,I_L) = (0,0,I_0)$ in the virtual bond, which has virtual dimension 1. On the right end, there is also only one irrep $(j_R,N_R,I_R) = (S_G,N_G,I_G)$ in the virtual bond, which also has \textit{reduced} virtual dimension 1. Eq. \eqref{CheMPS2_WE_MPS} and Fig. \ref{ImposingSymmetry-plot} imply that the addition of an extra orbital to the left renormalized basis is repeated from symmetry sector $(0,0,I_0)$ at boundary 0 to symmetry sector $(S_G,N_G,I_G)$ at boundary $L$.

Towards the middle of this embedded MPS chain, the reduced virtual dimension has to grow exponentially for the MPS to represent a general FCI state. This growth can be calculated recursively from the left as
\begin{eqnarray}
D_L(i, j, N, I) & = & D_L(i-1, j, N, I) + D_L(i-1, j-\frac{1}{2}, N-1, I \otimes I_{i}) \nonumber \\
                                & + & D_L(i-1, j+\frac{1}{2}, N-1, I \otimes I_i) + D_L(i-1, j, N-2, I). \label{FCIdimFromLeft}
\end{eqnarray}
Indices $i-1$ and $i$ denote the virtual bond. The constraint above can then be formulated as $D_L(i=0, j, N, I) = \delta_{(j,N,I),(0,0,I_0)}$. The interpretation of this growth equation is quite straightforward. States of symmetry $(j, N, I)$ at boundary $i$ are constructed as certain products of renormalized basis states at boundary $i-1$ and physical basis states at site $i$:
\begin{eqnarray}
\left| \ket{jNI}\right. \otimes \left|\ket{00I_0}\right. & \rightarrow & \left|\ket{jNI}\right. ,\\
\mid\hspace{-0.1cm}\ket{(j - \frac{1}{2}) (N-1)(I \otimes I_i)} \otimes \mid\hspace{-0.1cm}\ket{\frac{1}{2} 1 I_i} & \rightarrow & \left|\ket{jNI}\right. \oplus \left| \ket{(j-1) N I} \right. , \\
\mid\hspace{-0.1cm}\ket{(j + \frac{1}{2}) (N-1)(I \otimes I_i)} \otimes \mid\hspace{-0.1cm}\ket{\frac{1}{2} 1 I_i} & \rightarrow & \left|\ket{jNI}\right. \oplus \left| \ket{(j+1) N I} \right. , \\
\left|\ket{j(N-2)I}\right. \otimes \left|\ket{0 2 I_0}\right. & \rightarrow & \left|\ket{jNI}\right. .
\end{eqnarray}
Common sense is assumed, i.e. $j \geq 0$ etc. Alternatively, the growth can be calculated recursively from the right as
\begin{eqnarray}
D_R(i, j, N, I) & = & D_R(i+1, j, N, I) + D_R(i+1, j-\frac{1}{2}, N+1, I \otimes I_{i+1}) \nonumber \\
                                & + & D_R(i+1, j+\frac{1}{2}, N+1, I \otimes I_{i+1}) + D_R(i+1, j, N+2, I), \label{FCIdimFromRight}
\end{eqnarray}
with $D_R(i=L, j, N, I) = \delta_{(j,N,I),(S_G,N_G,I_G)}$. The FCI reduced virtual dimensions are then
\begin{equation}
D_{\text{FCI}}(i,j,N,I) = \min \left( D_L(i, j, N, I), D_R(i, j, N, I)\right). \label{FCIdimFromBoth}
\end{equation}
To make the MPS ansatz in Eq. \eqref{CheMPS2_WE_MPS} of practical use, either the total reduced virtual dimension per bond, or the reduced virtual dimension per symmetry sector, has to be truncated. The former strategy is used in \textsc{CheMPS2} \cite{2013arXiv1312.2415W}, and the latter in its one-site DMRG predecessor \textsc{CheMPS} \cite{woutersJCP1}.

\begin{figure}
\centering
\includegraphics[width=0.70\textwidth]{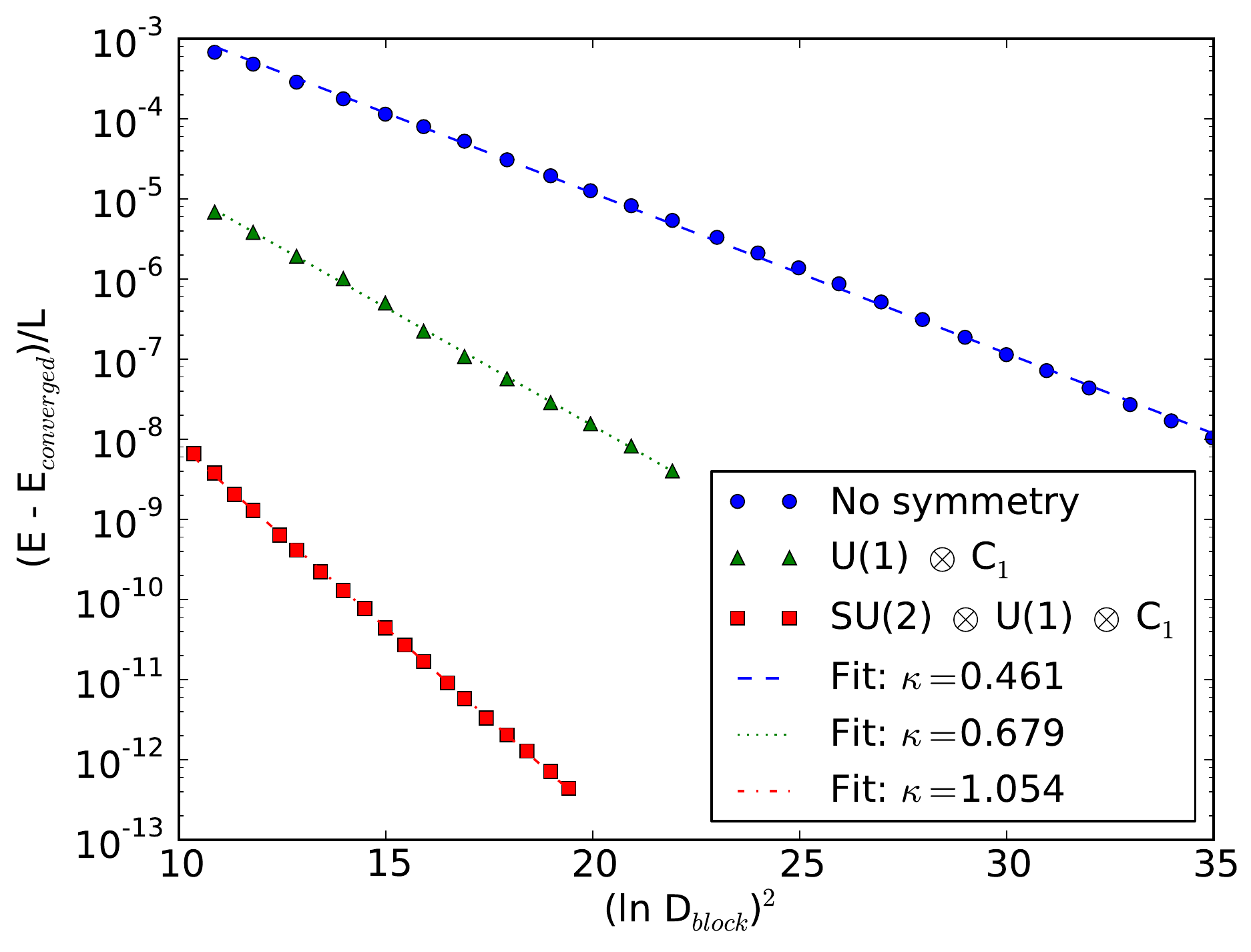} 
\caption{\label{ConvergenceHubbardSymmBlock-plot} Convergence of the one-dimensional Hubbard model with open boundary conditions, $L=36$ sites, $N=22$ electrons, $U=6$, in the $S=0$ spin singlet state. The convergence scheme \eqref{Eextrapol2} is tested for a DMRG code without any imposed symmetries, for a DMRG code with imposed particle number, and for \textsc{CheMPS}. $\kappa$ is the parameter $C_4$ of Eq. \eqref{Eextrapol2}, and $D_{block}$ denotes the number of renormalized basis states per symmetry sector. For \textsc{CheMPS}, these are the reduced ones.}
\end{figure}

\begin{figure}
\centering
\includegraphics[width=0.70\textwidth]{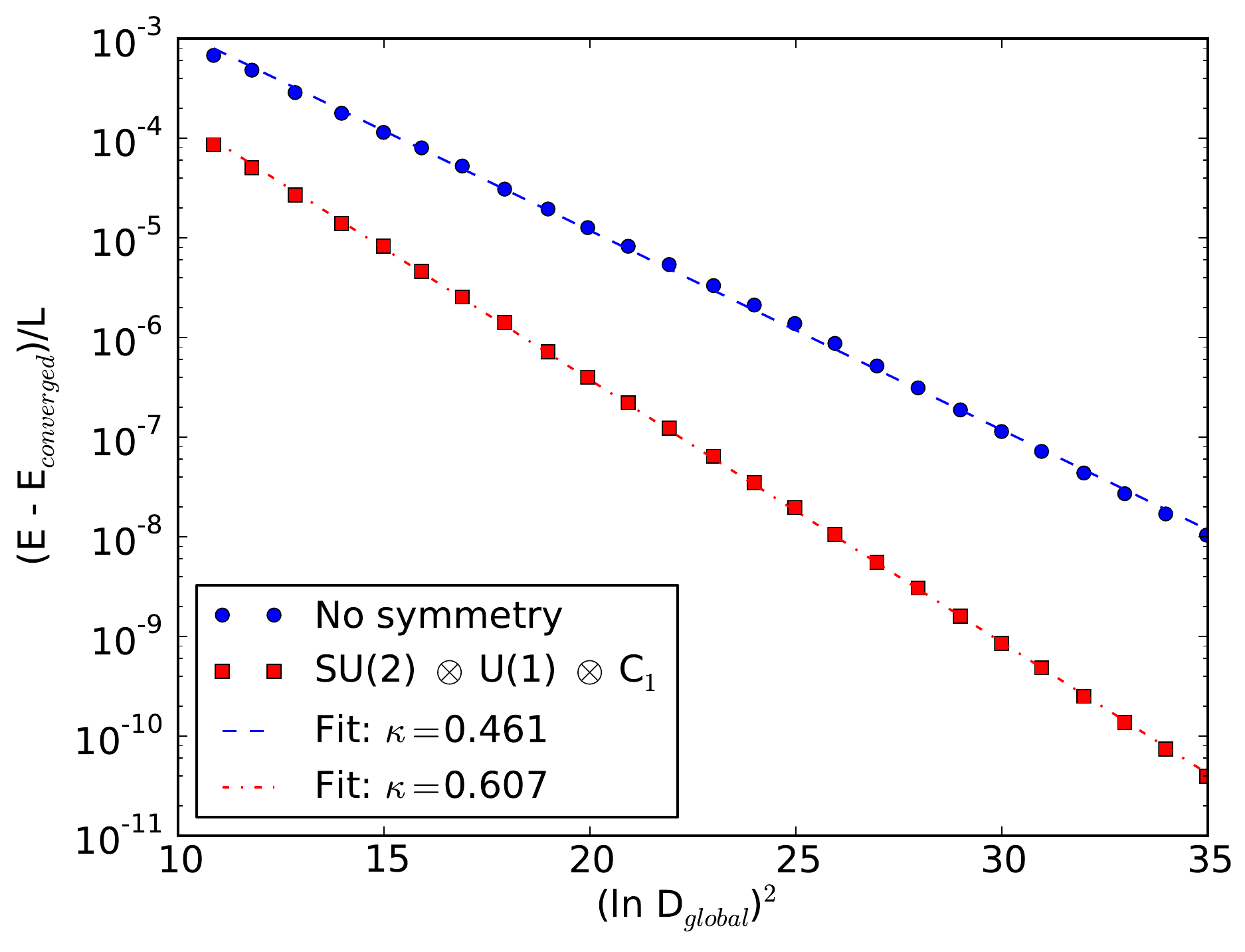} 
\caption{\label{ConvergenceHubbardSymmetry-plot} Convergence of the one-dimensional Hubbard model with open boundary conditions, $L=36$ sites, $N=22$ electrons, $U=6$, in the $S=0$ spin singlet state. The convergence scheme \eqref{Eextrapol2} is tested for a DMRG code without any imposed symmetries and for \textsc{CheMPS2}. $\kappa$ is the parameter $C_4$ of Eq. \eqref{Eextrapol2}, and $D_{global}$ denotes the total number of renormalized basis states at each virtual bond. For \textsc{CheMPS2}, these are the reduced ones.}
\end{figure}

The extrapolation scheme \eqref{Eextrapol2} is shown for the one-dimensional Hubbard model \cite{Hubbard26111963} with open boundary conditions
\begin{equation}
\hat{H} = - \sum\limits_{i=1}^{L-1} \sum\limits_{\sigma} \left( \hat{a}_{i\sigma}^{\dagger} \hat{a}_{i+1 \sigma} + \hat{a}_{i+1 \sigma}^{\dagger} \hat{a}_{i \sigma} \right) + U \sum\limits_{i=1}^{L} \hat{a}_{i\uparrow}^{\dagger} \hat{a}_{i\uparrow} \hat{a}_{i\downarrow}^{\dagger} \hat{a}_{i\downarrow}
\end{equation}
in Figs.~\ref{ConvergenceHubbardSymmBlock-plot} and \ref{ConvergenceHubbardSymmetry-plot}.
In the former, $D$ denotes the number of reduced renormalized basis states per \textit{symmetry block}. In the latter, $D$ denotes the \textit{total} number of reduced renormalized basis states at each bond. The extrapolation scheme seems to hold for the different symmetry constraints and the two truncation strategies. The comparison in Fig. \ref{ConvergenceHubbardSymmBlock-plot} is of course dubious, as the number of symmetry sectors in the middle of the MPS chain is $\mathcal{O}(L)$ for $\mathsf{U(1)}$ symmetry and $\mathcal{O}(L^2)$ for $\mathsf{SU(2)} \otimes \mathsf{U(1)}$ symmetry.

In Fig. \ref{ConvergenceHubbardSymmetry-plot}, when the total number of reduced renormalized basis states at a virtual bond is used, the curves for the calculations without symmetry and the calculations with $\mathsf{U(1)}$ symmetry will be (more or less) on top of each other. The abelian $\mathsf{U(1)}$ symmetry only results in block-sparsity, not in information compression. On the other hand, the calculation with $\mathsf{SU(2)} \otimes \mathsf{U(1)}$ symmetry will converge faster with $D$ due to the Wigner-Eckart theorem and the corresponding information compression, as can be observed from Fig. \ref{ConvergenceHubbardSymmetry-plot}.

\subsection{Canonical forms}
In the remainder of this text, the following abbreviations will often be used:
\begin{alignat}{4}
\Box_L & \rightarrow & \left( j_L j_L^z N_L I_L \right), \qquad \Box_R & \rightarrow & \left( j_R j_R^z N_R I_R \right), \\
\Box_{\text{phys}} & \rightarrow & \left( ss^z N I \right), \qquad \Box_{\text{phys}}^n & \rightarrow & \left( s_ns^z_n N_n I_n \right), \\
\bigtriangledown_L & \rightarrow & \left( j_L N_L I_L \right), \qquad \bigtriangledown_R & \rightarrow & \left( j_R N_R I_R \right), \\
\bigtriangledown_{\text{phys}} & \rightarrow & \left( s N I \right), \qquad \bigtriangledown_{\text{phys}}^n & \rightarrow & \left( s_n N_n I_n \right).
\end{alignat}
Squares ($\Box$) hence denote an $\mathsf{SU(2)} \otimes \mathsf{U(1)} \otimes \mathsf{P}$ symmetry sector \textit{with spin projection}, while triangles ($\bigtriangledown$) denote a \textit{reduced} symmetry sector. $A[i]$ is left-normalized if
\begin{eqnarray}
& \sum\limits_{\Box_{\text{phys}} \Box_L \alpha_L} \left(A[i]^{\Box_{\text{phys}}}\right)^{\dagger}_{(\Box_R \alpha_R);(\Box_L \alpha_L)}  A[i]^{\Box_{\text{phys}}}_{(\Box_L \alpha_L);(\widetilde{\Box}_R \widetilde{\alpha}_R)} \nonumber \\
= & \delta_{\Box_R, \widetilde{\Box}_R} \sum\limits_{\bigtriangledown_{\text{phys}} \bigtriangledown_L \alpha_L } \left(T[i]^{\bigtriangledown_{\text{phys}}}\right)^{\dagger}_{(\bigtriangledown_R \alpha_R);(\bigtriangledown_L \alpha_L)}  T[i]^{\bigtriangledown_{\text{phys}}}_{(\bigtriangledown_L \alpha_L);(\bigtriangledown_R \widetilde{\alpha}_R)} = \delta_{\Box_R, \widetilde{\Box}_R} \delta_{\alpha_R,\widetilde{\alpha}_R}.
\end{eqnarray}
The $\delta_{\Box_R, \widetilde{\Box}_R}$ is only due to the Clebsch-Gordan coefficients of the $\mathsf{SU(2)} \otimes \mathsf{U(1)} \otimes \mathsf{P}$ symmetry. Left-normalization can therefore be performed with a QR-decomposition on $T[i]$ per right symmetry sector. $A[i]$ is right-normalized if
\begin{eqnarray}
& \sum\limits_{\Box_{\text{phys}} \Box_R \alpha_R}  A[i]^{\Box_{\text{phys}}}_{(\widetilde{\Box}_L \widetilde{\alpha}_L);(\Box_R \alpha_R)} \left(A[i]^{\Box_{\text{phys}}}\right)^{\dagger}_{(\Box_R \alpha_R);(\Box_L \alpha_L)} \nonumber \\
= & \delta_{\Box_L, \widetilde{\Box}_L}  \sum\limits_{\bigtriangledown_{\text{phys}} \bigtriangledown_R \alpha_R} \frac{2j_R+1}{2j_L+1} T[i]^{\bigtriangledown_{\text{phys}}}_{(\bigtriangledown_L \widetilde{\alpha}_L);(\bigtriangledown_R \alpha_R)} \left(T[i]^{\bigtriangledown_{\text{phys}}}\right)^{\dagger}_{(\bigtriangledown_R \alpha_R);(\bigtriangledown_L \alpha_L)} = \delta_{\Box_L, \widetilde{\Box}_L} \delta_{\alpha_L,\widetilde{\alpha}_L}.
\end{eqnarray}
The $\delta_{\Box_L, \widetilde{\Box}_L}$ is again only due to the Clebsch-Gordan coefficients. Right-normalization can be obtained by performing the LQ-decomposition
\begin{equation}
\sqrt{\frac{2j_R+1}{2j_L+1}} T[i]^{\bigtriangledown_{\text{phys}}}_{(\bigtriangledown_L \alpha_L);(\bigtriangledown_R \alpha_R)} = \sum\limits_{\widetilde{\alpha}_L} L_{(\bigtriangledown_L \alpha_L) ; (\bigtriangledown_L \widetilde{\alpha}_L)} Q[i]^{\bigtriangledown_{\text{phys}}}_{(\bigtriangledown_L \widetilde{\alpha}_L);(\bigtriangledown_R \alpha_R)}
\end{equation}
per left symmetry sector. The quantity
\begin{equation}
\sqrt{\frac{2j_L+1}{2j_R+1}} Q[i]^{\bigtriangledown_{\text{phys}}}_{(\bigtriangledown_L \alpha_L);(\bigtriangledown_R \alpha_R)}
\end{equation}
is then the reduced part of the right-normalized MPS tensor $A[i]$.

\subsection{The reduced two-site object}
Section \ref{subsec-micro-iteraions} can be reformulated with the reduced MPS tensors $T[i]$ from Eq. \eqref{CheMPS2_WE_MPS} and a reduced two-site object $S[i]$:
\begin{eqnarray}
& S[i]^{j(s_1 s_2) N_1 N_2 I_1 I_2}_{\bigtriangledown_L \alpha_L ; \bigtriangledown_R \alpha_R} = \delta_{N_L+N_1+N_2,N_R} \delta_{I_L \otimes I_1 \otimes I_2, I_R} \sqrt{2j+1} (-1)^{j_L+j_R+s_1+s_2} \sum\limits_{j_M \alpha_M} \sqrt{2j_M+1} \nonumber \\
& \left\{ \begin{array}{ccc} j_L & j_R & j \\ s_2 & s_1 & j_M \end{array} \right\} T[i]^{\bigtriangledown_{\text{phys}}^1 }_{\bigtriangledown_L \alpha_L ; j_M (N_L+N_1) (I_L \otimes I_1) \alpha_M} T[i+1]^{\bigtriangledown_{\text{phys}}^2 }_{j_M (N_L+N_1) (I_L \otimes I_1) \alpha_M ; \bigtriangledown_R \alpha_R}. \label{TTtoS}
\end{eqnarray}
Eq. \eqref{TTtoS} is the analogue of Eq. \eqref{two-site-object-not-reduced}. The Lagrangian can be written in terms of $S[i]$, the effective Hamiltonian equation can be solved, and after convergence, Eq. (\ref{TTtoS}) can be backtransformed:
\begin{eqnarray}
& (TT)[i]^{\bigtriangledown_{\text{phys}}^1 ; \bigtriangledown_{\text{phys}}^2 ; j_M}_{\bigtriangledown_L \alpha_L ; \bigtriangledown_R \alpha_R} = \delta_{N_L+N_1+N_2,N_R} \delta_{I_L \otimes I_1 \otimes I_2, I_R} \sqrt{2j_M+1} (-1)^{j_L+j_R+s_1+s_2} \nonumber \\
& \sum\limits_j \sqrt{2j+1} \left\{ \begin{array}{ccc} j_L & j_R & j \\ s_2 & s_1 & j_M \end{array} \right\} S[i]^{j(s_1 s_2) N_1 N_2 I_1 I_2}_{\bigtriangledown_L \alpha_L ; \bigtriangledown_R \alpha_R}. \label{StoTT}
\end{eqnarray}
$(TT)[i]$ can be decomposed per middle symmetry sector $\bigtriangledown_M = (j_M, N_L + N_1, I_L \otimes I_1 )$:
\begin{eqnarray}
& \sqrt{\frac{2j_M+1}{2j_R+1}} \left( \sqrt{\frac{2j_R+1}{2j_M+1}} (TT)[i]^{\bigtriangledown_{\text{phys}}^1 ; \bigtriangledown_{\text{phys}}^2 ; j_M}_{\bigtriangledown_L \alpha_L ; \bigtriangledown_R \alpha_R} \right) \nonumber \\
& = \sqrt{\frac{2j_M+1}{2j_R+1}} \left( \sum\limits_{\alpha_M} U[i]^{\bigtriangledown_M}_{(\bigtriangledown_L \alpha_L \bigtriangledown_{\text{phys}}^1);\alpha_M} \lambda[i]^{\bigtriangledown_M}_{\alpha_M} V[i]^{\bigtriangledown_M}_{\alpha_M;(\bigtriangledown_R \alpha_R \bigtriangledown_{\text{phys}}^2)} \right) \nonumber \\
& = \sum\limits_{\alpha_M} U[i]^{\bigtriangledown_{\text{phys}}^1}_{(\bigtriangledown_L \alpha_L); (\bigtriangledown_M \alpha_M)} \lambda[i]_{\bigtriangledown_M \alpha_M} \left( \sqrt{\frac{2j_M+1}{2j_R+1}} V[i]^{\bigtriangledown_{\text{phys}}^2}_{(\bigtriangledown_M \alpha_M);(\bigtriangledown_R \alpha_R)} \right).
\end{eqnarray}
$U[i]$ is the reduced part of a left-normalized MPS site tensor, and the bracketed term is the reduced part of a right-normalized MPS site tensor. The reduced Schmidt numbers $\lambda[i]$ are related to the individual Schmidt numbers $\kappa[i]$ of Eq. \eqref{SVDofBsolution} by
\begin{equation}
\kappa[i]_{\Box_M \alpha_M} = \frac{\lambda[i]_{\bigtriangledown_M \alpha_M}}{\sqrt{ \sum\limits_{\bigtriangledown_Q \alpha_Q} (2j_Q+1) \lambda[i]_{\bigtriangledown_Q \alpha_Q}^2 }}.
\end{equation}
In the spin-adapted DMRG algorithm, the $D$ largest reduced Schmidt numbers $\lambda[i]$ are kept in the truncation step.

\subsection{(Complementary) reduced renormalized operators} \label{CheMPS2-renorm-op-subsection}
Due to the abelian point group symmetry $\mathsf{P}$, the matrix elements $h_{ij;kl}$ of the Hamiltonian \eqref{QC-ham-2} are only nonzero if $I_i \otimes I_j = I_k \otimes I_l$. If $\mathsf{P}$ is nontrivial, this considerably reduces the number of terms in the construction of the complementary renormalized operators, and in the multiplication of the effective Hamiltonian with a trial vector.

To calculate (complementary) renormalized operators, a specific ordering of the second-quantized operators $\hat{a}_{a\alpha}^{\dagger} \hat{a}_{b\beta}^{\dagger} \hat{a}_{d\delta} \hat{a}_{c\gamma}$ is initially assumed, $a \leq b$ and $c \leq d$, to keep track of the fermion signs due to the anticommutation relations. If $a = b$ or $c = d$, $\alpha = -\beta = -\delta = \gamma$ is assumed in addition. If another ordering is needed, it can be easily deduced.

Suppose the renormalized operator $\hat{a}_{c \gamma}$ is needed for the current micro-iteration at sites $(i,i+1)$ with $c<i$. All MPS site tensors to the left of site $i$ are left-normalized. To calculate the desired renormalized operator, it is hence sufficient to start at site $c$:
\begin{eqnarray}
& \braket{\Box_R \alpha_R \mid \hat{a}_{c \gamma} \mid \widetilde{\Box}_R \widetilde{\alpha}_R}_c = \sum\limits_{\Box_L \alpha_L \Box_{\text{phys}} \widetilde{\Box}_{\text{phys}}}
\left(A[c]^{\Box_{\text{phys}}}\right)^{\dagger}_{(\Box_R \alpha_R); (\Box_L \alpha_L)} \nonumber\\
& \braket{\Box_{\text{phys}} \mid \hat{a}_{c \gamma} \mid \widetilde{\Box}_{\text{phys}}}
A[c]^{\widetilde{\Box}_{\text{phys}}}_{(\Box_L \alpha_L);(\widetilde{\Box}_R \widetilde{\alpha}_R)} (-1)^{\delta_{\widetilde{N},2}\delta_{\gamma,\uparrow}} ,\label{OperatorExample} 
\end{eqnarray}
and renormalize this operator stepwise up to virtual boundary $i-1$:
\begin{eqnarray}
& \braket{\Box_R \alpha_R \mid \hat{a}_{c \gamma} \mid \widetilde{\Box}_R \widetilde{\alpha}_R}_{l+1} = \sum\limits_{\Box_L \alpha_L \widetilde{\Box}_L \widetilde{\alpha}_L \Box_{\text{phys}}}
\left(A[l+1]^{\Box_{\text{phys}}}\right)^{\dagger}_{(\Box_R \alpha_R); (\Box_L \alpha_L)} \nonumber \\
& \braket{\Box_L \alpha_L \mid \hat{a}_{c \gamma} \mid \widetilde{\Box}_L \widetilde{\alpha}_L}_l
A[l+1]^{\Box_{\text{phys}}}_{(\widetilde{\Box}_L \widetilde{\alpha}_L);(\widetilde{\Box}_R \widetilde{\alpha}_R)} (-1)^{\delta_{N,1}}. \label{OperatorPropagationExample}
\end{eqnarray}
In Eqs. \eqref{OperatorExample} and \eqref{OperatorPropagationExample} the minus signs of the Jordan-Wigner transformation \cite{JordanWignerTfo} are written explicitly.  They have their origin in the chosen orbital ordering in the occupation number representation in Eq. \eqref{occupation-number-representation-spin-orbs}. If $\gamma = \uparrow$ and orbital $c$ is doubly occupied, three second-quantized operators still have to anticommute with $\hat{a}^{\dagger}_{c \downarrow}$. They anticommute in addition with all $\hat{a}^{\dagger}_{l \tau}$, with $\min(a,b,d)>l>c$ and $\tau \in \{ \uparrow, \downarrow \}$, to propagate to their position in the ket $\ket{n_1 ... n_L}$.

With the reduced MPS ansatz in Eq. \eqref{CheMPS2_WE_MPS}, this renormalized operator becomes:
\begin{equation}
\braket{\Box_R \alpha_R \mid \hat{a}_{c \gamma} \mid \widetilde{\Box}_R \widetilde{\alpha}_R}_l = \delta_{N_R + 1, \widetilde{N}_R} \delta_{I_R \otimes I_c, \widetilde{I}_R} \braket{j_R j_R^z \frac{1}{2} \gamma \mid \widetilde{j}_R \widetilde{j}_R^z} \braket{\bigtriangledown_R \alpha_R \mid\mid \hat{L}^{\frac{1}{2}}_c \mid\mid \widetilde{\bigtriangledown}_R \widetilde{\alpha}_R}_l \label{reducedPartOfAnnihilatorLeft}
\end{equation}
with
\begin{eqnarray}
& \braket{\bigtriangledown_R \alpha_R \mid\mid \hat{L}^{\frac{1}{2}}_c \mid\mid \widetilde{\bigtriangledown}_R \widetilde{\alpha}_R}_c = \delta_{N_R + 1, \widetilde{N}_R} \delta_{I_R \otimes I_c, \widetilde{I}_R} \delta_{\mid j_R - \widetilde{j}_R \mid, \frac{1}{2}} \nonumber \\
& \left( \sum\limits_{\alpha_L} \left(T[c]^{ (0 0 I_0)  }\right)^{\dagger}_{(\bigtriangledown_R \alpha_R); (\bigtriangledown_R \alpha_L)} T[c]^{(\frac{1}{2} 1 I_c)}_{(\bigtriangledown_R \alpha_L);(\widetilde{\bigtriangledown}_R \widetilde{\alpha}_R)} \qquad + \qquad (-1)^{\widetilde{j}_R-j_R+\frac{1}{2}} \sqrt{\frac{2j_R+1}{2\widetilde{j}_R+1}} \times \right. \nonumber\\
& \left. \sum\limits_{\alpha_L} \left(T[c]^{ (\frac{1}{2} 1 I_c)  }\right)^{\dagger}_{(\bigtriangledown_R \alpha_R); (\widetilde{j}_R (N_R-1) \widetilde{I}_R \alpha_L)} T[c]^{(0 2 I_0)}_{(\widetilde{j}_R (N_R-1) \widetilde{I}_R \alpha_L);(\widetilde{\bigtriangledown}_R \widetilde{\alpha}_R)} \right)
\end{eqnarray}
and for $l > c$:
\begin{eqnarray}
& \braket{\bigtriangledown_R \alpha_R \mid\mid \hat{L}^{\frac{1}{2}}_c \mid\mid \widetilde{\bigtriangledown}_R \widetilde{\alpha}_R}_l = \delta_{N_R + 1, \widetilde{N}_R} \delta_{I_R \otimes I_c, \widetilde{I}_R} \delta_{\mid j_R - \widetilde{j}_R \mid, \frac{1}{2}} \sum\limits_{\bigtriangledown_{\text{phys}} j_L \widetilde{j}_L \alpha_L \widetilde{\alpha}_L } (-1)^{\widetilde{j}_L + j_R + \frac{1}{2} - s} \nonumber \\
& \sqrt{(2\widetilde{j}_L+1)(2j_R+1)} \left(T[l]^{ \bigtriangledown_{\text{phys}} }\right)^{\dagger}_{(\bigtriangledown_R \alpha_R); (j_L (N_R-N) (I_R \otimes I) \alpha_L)} T[l]^{ \bigtriangledown_{\text{phys}} }_{(\widetilde{j}_L (\widetilde{N}_R-N)(\widetilde{I}_R \otimes I) \widetilde{\alpha}_L );(\widetilde{\bigtriangledown}_R \widetilde{\alpha}_R)}  \nonumber \\
& \left\{ \begin{array}{ccc} j_R & \widetilde{j}_R & \frac{1}{2} \\ \widetilde{j}_L & j_L & s \end{array} \right\} \braket{j_L (N_R-N) (I_R \otimes I) \alpha_L \mid\mid \hat{L}^{\frac{1}{2}}_c \mid\mid \widetilde{j}_L (\widetilde{N}_R-N)(\widetilde{I}_R \otimes I) \widetilde{\alpha}_L}_{l-1}.
\end{eqnarray}
Note that the Jordan-Wigner transformation is incorporated in these equations. The renormalized operator $\hat{a}_{c\gamma}^{\dagger}$ can be obtained by hermitian conjugation:
\begin{equation}
\braket{\Box_R \alpha_R \mid \hat{a}^{\dagger}_{c \gamma} \mid \widetilde{\Box}_R \widetilde{\alpha}_R}_l = \delta_{N_R - 1, \widetilde{N}_R} \delta_{I_R, \widetilde{I}_R \otimes I_k} \braket{ \widetilde{j}_R \widetilde{j}_R^z \frac{1}{2} \gamma \mid j_R j_R^z } \braket{ \widetilde{\bigtriangledown}_R \widetilde{\alpha}_R \mid\mid \hat{L}^{\frac{1}{2}}_c \mid\mid \bigtriangledown_R \alpha_R }_l^{\dagger} \label{reducedPartOfCreaLeft}.
\end{equation}
The reduced $L$-tensor in Eqs. \eqref{reducedPartOfAnnihilatorLeft} and \eqref{reducedPartOfCreaLeft} is a spin-$\frac{1}{2}$ object, because the operators
\begin{eqnarray}
\hat{b}^{\dagger}_{c \gamma} & = & \hat{a}^{\dagger}_{c \gamma} \label{creaannih1--}\\
\hat{b}_{c \gamma} & = & (-1)^{\frac{1}{2}-\gamma}\hat{a}_{c -\gamma} \label{creaannih2--}
\end{eqnarray}
for orbital $c$ correspond to resp. the $(s=\frac{1}{2}, s^z=\gamma, N=1, I_c)$ row of irrep $(s=\frac{1}{2}, N=1, I_c)$ and the $(s=\frac{1}{2}, s^z=\gamma, N=-1, I_c)$ row of irrep $(s=\frac{1}{2}, N=-1, I_c)$ \cite{BookDimitri}. $\hat{b}^{\dagger}$ and $\hat{b}$ are hence both doublet irreducible tensor operators. As described in section \ref{mcculloch-irred-tensor-op-subsec}, this fact permits exploitation of the Wigner-Eckart theorem for operators and (complementary) renormalized operators. Contracting terms of the type \eqref{CheMPS2_WE_MPS} and \eqref{creaannih1--}-\eqref{creaannih2--} can be done by implicitly summing over the common multiplets and recoupling the local, virtual and operator spins. As is shown by Eqs. \eqref{reducedPartOfAnnihilatorLeft} and \eqref{reducedPartOfCreaLeft}, (complementary) renormalized operators then formally consist of terms containing Clebsch-Gordan coefficients and reduced tensors. In the actual implementation of \textsc{CheMPS2}, only the reduced tensors need to be calculated, and Wigner 3-j symbols or Clebsch-Gordan coefficients are never used. \textsc{CheMPS2} uses the GNU Scientific Library to extract Wigner 6-j and 9-j symbols for the recoupling.

To give an example of a tensor product of irreducible tensor operators, consider the renormalized operator $\hat{a}_{a\alpha}^{\dagger} \hat{a}_{c\gamma}$ with $c<a<i$. When $\hat{a}_{a\alpha}^{\dagger}$ acts on site $a$, an extra minus sign should be included due to the Jordan-Wigner transformation if $\alpha=\downarrow$ and if the site already contains an electron with spin projection $\uparrow$, because the remaining three second-quantized operators $\hat{a}_{a\downarrow}^{\dagger} \hat{a}_{b \beta}^{\dagger} \hat{a}_{d \delta}$ then still have to anticommute with $\hat{a}^{\dagger}_{a\uparrow}$:
\begin{eqnarray}
& \braket{\Box_R \alpha_R \mid \hat{a}_{a \alpha}^{\dagger} \hat{a}_{c \gamma} \mid \widetilde{\Box}_R \widetilde{\alpha}_R}_a = \sum\limits_{\Box_L \alpha_L \widetilde{\Box}_L \widetilde{\alpha}_L \Box_{\text{phys}} \widetilde{\Box}_{\text{phys}}}
\left(A[a]^{\Box_{\text{phys}}}\right)^{\dagger}_{(\Box_R \alpha_R); (\Box_L \alpha_L)} \nonumber\\
& \braket{\Box_L \alpha_L \mid \hat{a}_{c \gamma} \mid \widetilde{\Box}_L \widetilde{\alpha}_L}_{a-1} \braket{\Box_{\text{phys}} \mid \hat{a}^{\dagger}_{a \alpha} \mid \widetilde{\Box}_{\text{phys}}}
A[a]^{\widetilde{\Box}_{\text{phys}}}_{(\widetilde{\Box}_L \widetilde{\alpha}_L);(\widetilde{\Box}_R \widetilde{\alpha}_R)} (-1)^{\delta_{\widetilde{N},1}\delta_{\alpha,\downarrow}}.\label{OperatorExample2} 
\end{eqnarray}
This operator can then be renormalized stepwise up to virtual boundary $i-1$:
\begin{eqnarray}
& \braket{\Box_R \alpha_R \mid \hat{a}_{a \alpha}^{\dagger} \hat{a}_{c \gamma} \mid \widetilde{\Box}_R \widetilde{\alpha}_R}_{l+1} = \sum\limits_{\Box_L \alpha_L \widetilde{\Box}_L \widetilde{\alpha}_L \Box_{\text{phys}} }
\left(A[l+1]^{\Box_{\text{phys}}}\right)^{\dagger}_{(\Box_R \alpha_R); (\Box_L \alpha_L)} \nonumber\\
& \braket{\Box_L \alpha_L \mid \hat{a}_{a \alpha}^{\dagger} \hat{a}_{c \gamma} \mid \widetilde{\Box}_L \widetilde{\alpha}_L}_{l} 
A[l+1]^{\Box_{\text{phys}}}_{(\widetilde{\Box}_L \widetilde{\alpha}_L);(\widetilde{\Box}_R \widetilde{\alpha}_R)}.\label{OperatorPropagationExample2} 
\end{eqnarray}
No fermion signs arise in Eq. \eqref{OperatorPropagationExample2} due to the Jordan-Wigner transformation, because $\hat{a}^{\dagger}_{b\beta} \hat{a}_{d\delta}$ has to be anticommuted, which contains an even number of second-quantized operators. With Eqs. \eqref{CheMPS2_WE_MPS} and \eqref{reducedPartOfAnnihilatorLeft}, this renormalized operator becomes
\begin{eqnarray}
& \braket{\Box_R \alpha_R \mid \hat{a}_{a \alpha}^{\dagger} \hat{a}_{c \gamma} \mid \widetilde{\Box}_R \widetilde{\alpha}_R}_{l} = \delta_{N_R,\widetilde{N}_R} \delta_{I_R \otimes I_a \otimes I_c, \widetilde{I}_R} (-1)^{\frac{1}{2}-\alpha} \nonumber \\
& \left( \braket{\frac{1}{2} \gamma \frac{1}{2} -\alpha \mid 00} \delta_{j_R,\widetilde{j}_R} \delta_{j_R^z,\widetilde{j}_R^z} \braket{\bigtriangledown_R\alpha_R \mid \hat{F}^0_{c,a} \mid \widetilde{\bigtriangledown}_R \widetilde{\alpha}_R}_l \right. \nonumber \\
& + \left. \braket{\frac{1}{2} \gamma \frac{1}{2} -\alpha \mid 1 (\gamma - \alpha)} \braket{j_R j_R^z 1 (\gamma - \alpha) \mid \widetilde{j}_R \widetilde{j}_R^z } \braket{\bigtriangledown_R\alpha_R \mid \hat{F}^1_{c,a} \mid \widetilde{\bigtriangledown}_R \widetilde{\alpha}_R}_l  \right) \label{F-tensors-reduced}
\end{eqnarray}
with
\begin{eqnarray}
& \braket{\bigtriangledown_R\alpha_R \mid\mid \hat{F}^x_{c,a} \mid\mid \widetilde{\bigtriangledown}_R \widetilde{\alpha}_R}_a = \delta_{N_R,\widetilde{N}_R} \delta_{I_R \otimes I_a \otimes I_c, \widetilde{I}_R} \left( \sum\limits_{j_L \alpha_L \widetilde{\alpha}_L} \sqrt{(2x+1)(2j_R+1)} \times \right. \nonumber \\
& (-1)^{j_L+\widetilde{j}_R + \frac{1}{2} + x} \left\{ \begin{array}{ccc} \frac{1}{2} & \frac{1}{2} & x \\ j_R & \widetilde{j}_R & j_L \end{array} \right\} \left( T[a]^{(\frac{1}{2} 1 I_a)} \right)^{\dagger}_{(\bigtriangledown_R \alpha_R) ; (j_L (N_R-1) (I_R\otimes I_a) \alpha_L)} \times \nonumber \\
& \braket{j_L(N_R-1) (I_R\otimes I_a) \alpha_L \mid\mid \hat{L}^{\frac{1}{2}}_c \mid\mid \widetilde{\bigtriangledown}_R \widetilde{\alpha}_L}_{a-1}  T[a]^{00I_0}_{(\widetilde{\bigtriangledown}_R \widetilde{\alpha}_L);(\widetilde{\bigtriangledown}_R \widetilde{\alpha}_R)} \nonumber \\
& + \qquad \sum\limits_{\widetilde{j}_L \alpha_L \widetilde{\alpha}_L} \sqrt{(2x+1)(2\widetilde{j}_L+1)} (-1)^{j_R+\widetilde{j}_R + 1} \left\{ \begin{array}{ccc} \frac{1}{2} & \frac{1}{2} & x \\ j_R & \widetilde{j}_R & \widetilde{j}_L \end{array} \right\} \times \nonumber  \\
& \left( T[a]^{(0 2 I_0)} \right)^{\dagger}_{(\bigtriangledown_R \alpha_R) ; (j_R (N_R-2) I_R \alpha_L)} \braket{j_R (N_R-2) I_R \alpha_L \mid\mid \hat{L}^{\frac{1}{2}}_c \mid\mid \widetilde{j}_L (N_R-1) (I_R \otimes I_c) \widetilde{\alpha}_L}_{a-1} \times \nonumber \\
& \left. T[a]^{\frac{1}{2} 1 I_a}_{(\widetilde{j}_L (N_R-1) (I_R \otimes I_c) \widetilde{\alpha}_L);(\widetilde{\bigtriangledown}_R \widetilde{\alpha}_R)} \right)
\end{eqnarray}
and for $l>a$:
\begin{eqnarray}
& \braket{\bigtriangledown_R \alpha_R \mid\mid \hat{F}^x_{c,a} \mid\mid \widetilde{\bigtriangledown}_R \widetilde{\alpha}_R}_l = \delta_{N_R,\widetilde{N}_R} \delta_{I_R \otimes I_a \otimes I_c, \widetilde{I}_R} \sum\limits_{\bigtriangledown_{\text{phys}} j_L \widetilde{j}_L \alpha_L \widetilde{\alpha}_L } \sqrt{(2\widetilde{j}_L+1)(2j_R+1)} \nonumber \\
& (-1)^{j_R + \widetilde{j}_L + s + x} \left\{ \begin{array}{ccc} j_L & \widetilde{j}_L & x \\ \widetilde{j}_R & j_R & s \end{array} \right\} \left(T[l]^{ \bigtriangledown_{\text{phys}} }\right)^{\dagger}_{(\bigtriangledown_R \alpha_R); (j_L (N_R-N) (I_R \otimes I) \alpha_L)} \nonumber \\
& \braket{j_L (N_R-N) (I_R \otimes I) \alpha_L \mid\mid \hat{F}^x_{c,a} \mid\mid \widetilde{j}_L (N_R-N)(\widetilde{I}_R \otimes I) \widetilde{\alpha}_L}_{l-1} \nonumber \\
& T[l]^{ \bigtriangledown_{\text{phys}} }_{(\widetilde{j}_L (N_R-N)(\widetilde{I}_R \otimes I) \widetilde{\alpha}_L );(\widetilde{\bigtriangledown}_R \widetilde{\alpha}_R)}. 
\end{eqnarray}
In these equations, $x$ can be 0 or 1. The tensor product of two spin-$\frac{1}{2}$ irreducible tensor operators hence decomposes into the sum of a spin-$0$ irreducible tensor operator and a spin-$1$ irreducible tensor operator, in accordance with $\mathsf{SU(2)}$ representation theory: $\frac{1}{2} \otimes \frac{1}{2} \approx 0 \oplus 1$. The renormalized operator $\hat{a}_{c \gamma}^{\dagger} \hat{a}_{a \alpha}$ (with $c<a<i$) can be obtained by hermitian conjugation:
\begin{eqnarray}
& \braket{\Box_R \alpha_R \mid \hat{a}_{c \gamma}^{\dagger} \hat{a}_{a \alpha} \mid \widetilde{\Box}_R \widetilde{\alpha}_R}_{l} = \delta_{N_R,\widetilde{N}_R} \delta_{I_R \otimes I_a \otimes I_c, \widetilde{I}_R} (-1)^{\frac{1}{2}-\alpha} \nonumber \\
& \left( \braket{\frac{1}{2} \gamma \frac{1}{2} -\alpha \mid 00} \delta_{j_R,\widetilde{j}_R} \delta_{j_R^z,\widetilde{j}_R^z} \braket{\widetilde{\bigtriangledown}_R \widetilde{\alpha}_R \mid \hat{F}^0_{c,a} \mid \bigtriangledown_R\alpha_R}^{\dagger}_l \right. \nonumber \\
& + \left. \braket{\frac{1}{2} \gamma \frac{1}{2} -\alpha \mid 1 (\gamma - \alpha)} \braket{\widetilde{j}_R \widetilde{j}_R^z 1 (\gamma - \alpha) \mid j_R j_R^z} \braket{ \widetilde{\bigtriangledown}_R \widetilde{\alpha}_R \mid \hat{F}^1_{c,a} \mid \bigtriangledown_R\alpha_R}_l^{\dagger}  \right).
\end{eqnarray}
It is hence sufficient to restrict the calculation of the $\hat{F}^x_{c,a}$-tensors to $c \leq a$.

The main concepts to calculate renormalized operators were addressed in this section. These can be used to generate all the required (complementary) renormalized operators. For the complementary renormalized operator of three second-quantized operators, one can sum over one spin projection as either $\alpha = \gamma$ and $\beta = \delta$, or $\alpha = \delta$ and $\beta = \gamma$, which results in a spin-$\frac{1}{2}$ irreducible tensor operator.

\section{Program structure of CheMPS2}
\textsc{CheMPS2} can be obtained from its public git repository \cite{CheMPS2github}. The file \texttt{README.md} contains information about the installation, the included tests, and the extraction of the comments in Doxygen format. In this section, a short introduction to the program structure is given. The focus lies on the topics relevant to users. The file \texttt{CheMPS2/include/Options.h} contains the user-specifiable options.

\subsection{The Hamiltonian}
\textsc{CheMPS2} requires an orthonormal single-particle basis, and two-body matrix elements with eightfold permutation symmetry which do not break $\mathsf{SU(2)}$ total electronic spin. There are two ways to create and fill a \texttt{Hamiltonian} object in \textsc{CheMPS2}.

It can be created by specifying the number of orbitals $L$ in the DMRG active space, the abelian point group $\mathsf{P}$ of the molecule at hand, and an array containing the point group irreps $I_i$ for each orbital. The class \texttt{Irreps} contains the symmetry labeling conventions: integers are used to label the point groups $\mathsf{P}$ and their irreps $I_i$. Users can generate matrix elements with their preferred molecular electronic structure program. The functions \texttt{setEconst}, \texttt{setTmat}, and \texttt{setVmat} then allow to fill the \texttt{Hamiltonian} elementwise. Note that for $(ij | \hat{V} | kl) = V_{ijkl}$ the physics notation is assumed, see Eq. \eqref{physics-notation-two-body-matrix-elements}.

\textsc{Psi4} \cite{WCMS:WCMS93} can be used as well to generate molecular orbital matrix elements. Two plugins can be found in the folder \texttt{mointegrals}, with corresponding instructions in \texttt{README.md}. One plugin allows to print matrix elements as text during a \textsc{Psi4} calculation, in a format which the \texttt{Hamiltonian} object is able to read at creation. The other plugin creates a \texttt{Hamiltonian} object during a \textsc{Psi4} calculation, fills it with the molecular orbital matrix elements, and stores it to disk in binary format. The latter option requires linking of the \textsc{CheMPS2} library to the \textsc{Psi4} plugin, but allows for reduced storage requirements.

\subsection{The desired corner of the Hilbert space}
The \texttt{Problem} object contains the \texttt{Hamiltonian} and the $\mathsf{SU(2)} \otimes \mathsf{U(1)} \otimes \mathsf{P}$ symmetry sector to which the calculations are restricted. All $\mathsf{SU(2)}$ spin symmetry sectors $S$ are denoted in \textsc{CheMPS2} by their integer counterparts $2S$. The \texttt{Hamiltonian} and the symmetry sector completely determine a FCI calculation. In order to do DMRG instead of FCI, a convergence scheme for the subsequent sweeps should be set up.

\subsection{The convergence scheme} \label{ConvergenceSchemeSection}
The \texttt{ConvergenceScheme} object is divided into a number of consecutive instructions. Each instruction contains four parameters: the number of reduced renormalized basis states $D$ which should be kept, an energy threshold $E_{\text{conv}}$ for convergence, the maximum number of sweeps $N_{\text{max}}$, and the noise prefactor $\gamma_{\text{noise}}$.

The parameters $\gamma_{\text{noise}}$ and $D$ are relevant for the micro-iterations. Just before the decomposition of the reduced two-site object $S[i]$, noise is added to it. This noise is bounded in magnitude by $0.5 \gamma_{\text{noise}} w^{\text{disc}}_D$, where $w^{\text{disc}}_D$ is the maximum discarded weight obtained during the previous left or right sweep. After decomposition of the reduced two-site object $S[i]$, its reduced Schmidt spectrum $\lambda[i]$ is truncated to the largest $D$ numbers.

The parameters $E_{\text{conv}}$ and $N_{\text{max}}$ are relevant for the macro-iterations. If after one macro-iteration (left plus right sweep), the energy difference is smaller than $E_{\text{conv}}$, the sweeping stops and the next instruction is performed. If energy convergence is not reached after $N_{\text{max}}$ macro-iterations, the current instruction ends as well.

\subsection{DMRG}
Creation of a \texttt{DMRG} object requires a \texttt{Hamiltonian}, a \texttt{Problem}, and a \texttt{ConvergenceScheme}. The \texttt{DMRG} object creates, in turn, a \texttt{SyBookkeeper}. Based on Eqs. \eqref{FCIdimFromLeft}, \eqref{FCIdimFromRight}, and \eqref{FCIdimFromBoth}, the \texttt{SyBookkeeper} calculates the FCI reduced virtual dimensions of each symmetry sector at each virtual bond. The same object keeps track of the MPS reduced virtual dimensions during the DMRG sweeps. To start, the reduced virtual dimension $D_{\text{trunc}}$ of the first instruction of the \texttt{ConvergenceScheme} is distributed over the symmetry sectors as follows:
\begin{equation}
 D^{\text{ini}}_{\text{MPS}}(i,j,N,I) = \min \left( \lceil \frac{D_{\text{FCI}}(i,j,N,I) D_{\text{trunc}} }{ \sum\limits_{jNI} D_{\text{FCI}}(i,j,N,I)} \rceil , D_{\text{FCI}}(i,j,N,I) \right).
\end{equation}
This implies that if $D_{\text{FCI}}(i,j,N,I) \neq 0$, $D^{\text{ini}}_{\text{MPS}}(i,j,N,I)$ will be nonzero as well. The \texttt{DMRG} object then creates an MPS with virtual dimensions $D^{\text{ini}}_{\text{MPS}}(i,j,N,I)$, and fills it with noise. The \texttt{DMRG} object is also responsible for creating, storing and loading the (complementary) reduced renormalized operators. The function \texttt{Solve} performs the instructions of the \texttt{ConvergenceScheme}.

\texttt{Solve} relies heavily on two classes: \texttt{Sobject} and \texttt{Heff}. The former is responsible for constructing and decomposing the reduced two-site object $S[i]$. The latter performs the reduced effective Hamiltonian multiplication $\mathbf{H}[i]^{\text{eff}}_{\text{red}} \mathbf{S}[i]$, based on the (complementary) reduced renormalized operators. \texttt{Heff} contains our own implementation of the Davidson algorithm \cite{Davidson197587} to obtain the ground state of $\mathbf{H}[i]^{\text{eff}}_{\text{red}}$. After \texttt{Solve} has performed all the instructions of the \texttt{ConvergenceScheme}, it returns the minimal variational energy encountered during all the performed micro-iterations.

With the function \texttt{calc2DM}, the reduced 2-RDMs $\Gamma^A$ and $\Gamma^B$ are calculated:
\begin{eqnarray}
\Gamma_{(i \sigma) (j \tau) ; (k \sigma) (l \tau)} & = & \braket{ \hat{a}^{\dagger}_{i \sigma} \hat{a}^{\dagger}_{j \tau}  \hat{a}_{l \tau} \hat{a}_{k \sigma}}, \\
\Gamma^{A}_{ij ; kl} & = & \sum\limits_{\sigma \tau} \Gamma_{(i \sigma) (j \tau) ; (k \sigma) (l \tau)}, \\
\Gamma^{B}_{ij ; kl} & = & \sum\limits_{\sigma \tau} (-1)^{\sigma - \tau} \Gamma_{(i \sigma) (j \tau) ; (k \sigma) (l \tau)}.
\end{eqnarray}
$\Gamma^A$ can be used to calculate the energy, the particle number $N$, and the reduced 1-RDM:
\begin{eqnarray}
E & = & E_{0} + \frac{1}{2} \sum\limits_{ijkl} h_{ij ; kl} \Gamma^{A}_{ij ; kl}, \\
N(N-1) & = & \sum\limits_{ij} \Gamma^{A}_{ij ; ij}, \\
\sum\limits_{\sigma} \braket{ \hat{a}^{\dagger}_{i \sigma} \hat{a}_{k \sigma}} & = & \frac{1}{N-1} \sum\limits_j \Gamma^{A}_{ij ; kj}.
\end{eqnarray}
$\Gamma^A$ is also needed to calculate analytic nuclear gradients, as well as the gradient and the Hessian for DMRG-SCF. $\Gamma^B$ is important for certain types of spin-spin correlation functions. The strategy of Zgid and Nooijen \cite{zgid:144115} is used to obtain the reduced 2-RDMs $\Gamma^A$ and $\Gamma^B$ efficiently.

A sweep is performed, in which only the canonical form of the MPS is varied, but not the wavefunction represented by it. At each sweep step, one site $i$ is considered. All sites to the left of $i$ are left-normalized, and all sites to the right of $i$ are right-normalized. At each sweep step, only certain subsets $(x,y,z)$ of elements of $\Gamma^{A,B}$ are calculated, meaning $x$ orbital indices are smaller than $i$, $y$ indices are equal to $i$, and $z$ indices are larger than $i$. The following subsets are considered: (1,1,2), (1,2,1), (1,3,0), (0,2,2), (0,3,1), and (0,4,0). Note that these are all variations of (1,1,2), in which the index to the left of $i$, and the indices to the right of $i$, are also allowed to become equal to $i$. With this strategy, all elements of $\Gamma^{A,B}$ can be calculated with the reduced renormalized operators needed to perform the reduced effective Hamiltonian multiplication.

OpenMP parallelization is used in the \texttt{DMRG} object to speed up contractions involving tensors with a sparse block structure, for example the action of the reduced effective Hamiltonian on a particular guess, and the construction of the (often similar) (complementary) reduced renormalized operators in between two micro-iterations.

\subsection{State-specific excited states}

The \texttt{DMRG} object also contains a state-specific excited-state algorithm. After the ground state $\ket{\Psi_0}$ has been determined, the desired number of excited states can be set once with the function \texttt{activateExcitations}. Before \texttt{Solve} is called to find the next new excitation $\ket{\Psi_{m}}$, the function \texttt{newExcitation} should be called with the parameter $\eta_{m}$. This pushes back the current MPS which represents $\ket{\Psi_{m-1}}$, and sets the Hamiltonian to
\begin{equation}
 \hat{H}_m = \hat{H}_0 + \sum\limits_{k={0}}^{m-1} \eta_{k+1} \ket{\Psi_k} \bra{\Psi_k}.
\end{equation}
The state-specific excited-state DMRG algorithm hence projects out all lower-lying states in the given $\mathsf{SU(2)} \otimes \mathsf{U(1)} \otimes \mathsf{P}$ symmetry sector.

\subsection{DMRG-SCF}
A state-specific DMRG-SCF algorithm is implemented in the class \texttt{CASSCF}. Its creation requires a \texttt{Hamiltonian} object. The number of occupied, active, and virtual orbitals per point group irrep should be given with the function \texttt{setupStart} before calling the SCF routine.

The CASSCF routine which is implemented is the augmented Hessian \cite{Lengsfield, 10.1021.j100247a015} Newton-Raphson method from Ref. \cite{Roos3}, with exact Hessian. It can be called with the function \texttt{doCASSCFnewtonraphson}, which requires a \texttt{ConvergenceScheme}, the targeted symmetry sector, and the targeted root for the state-specific algorithm. When the gradient for orbital rotations reaches a predefined threshold, the routine returns the converged DMRG-SCF energy.

\section{Non-abelian spatial symmetries}
\subsection{Point groups}
\textsc{CheMPS2} can only deal with the abelian point groups \eqref{PointGroupsInCheMPS2} thus far. Sharma and Chan have recently augmented \textsc{Block} to deal with non-abelian point group symmetry as well \cite{1.4867383}.

The orbitals which form a complete basis for one of the point group irreps should then be combined to one DMRG lattice site. Consider for example $\mathsf{D_{\infty h}}$, the molecular point group of centrosymmetric linear molecules, which includes the homonuclear dimers.  The irreps of this point group are characterized by three quantum numbers: the (magnitude of the) angular momentum projection along the internuclear axis $l^z = \braket{\hat{L}_z}$, the parity under spatial inversion $u/g$, and for $l^z=0$, the parity under $\sigma_v$ reflection. The character table is shown in Tab. \ref{DinfhCharTable}. 

\begin{table}[h!]
\centering
\caption{\label{DinfhCharTable} The character table of $\mathsf{D_{\infty h}}$ \cite{BookCornwell}.}
\begin{tabular}{|c|cccccc|cc|}
\hline
$\mathsf{D_{\infty h}}$ & $E$ & $2C_{\infty}$ & $\infty \sigma_v$ & $i$ & $2 S_{\infty}$ & $\infty C'_2$ & linear & quadratic \\
\hline  
$\Sigma_g^+$  &  1    & 1              & 1     &  1    & 1               & 1     &              & $x^2 + y^2$, $z^2$ \\
$\Sigma_g^-$  &  1    & 1              & -1    &  1    & 1               & -1    &              &                    \\
$\Pi_g$       &  2    & $2\cos(\phi)$  & 0     &  2    & $-2\cos(\phi)$  & 0     &              & $(x \pm i y)z$     \\
$\Delta_g$    &  2    & $2\cos(2\phi)$ & 0     &  2    & $2\cos(2\phi)$  & 0     &              & $(x \pm i y)^2$    \\
$\Phi_g$      &  2    & $2\cos(3\phi)$ & 0     &  2    & $-2\cos(3\phi)$ & 0     &              &                    \\
$...$         & $...$ & $...$          & $...$ & $...$ & $...$           & $...$ &              &                    \\
$\Sigma_u^+$  &  1    & 1              & 1     & -1    & -1              & -1    & $z$          &                    \\
$\Sigma_u^-$  &  1    & 1              & -1    & -1    & -1              & 1     &              &                    \\
$\Pi_u$       &  2    & $2\cos(\phi)$  & 0     & -2    & $2\cos(\phi)$   & 0     & $x \pm i y$  &                    \\
$\Delta_u$    &  2    & $2\cos(2\phi)$ & 0     & -2    & $-2\cos(2\phi)$ & 0     &              &                    \\
$\Phi_u$      &  2    & $2\cos(3\phi)$ & 0     & -2    & $2\cos(3\phi)$  & 0     &              &                    \\
$...$         & $...$ & $...$          & $...$ & $...$ & $...$           & $...$ &              &                    \\
\hline
\end{tabular}
\end{table}

Consider for example a corresponding pair of bonding $\pi$-orbitals of a homonuclear dimer: $(\pi_x,\pi_y)$ . A rotation over $\frac{\pi}{2}$, with the internuclear axis as rotation axis, then transforms these orbitals into each other. The linear combinations
\begin{eqnarray}
\pi_{-1} & = &   \pi_x - i \pi_y \\
\pi_1    & = & - \pi_x - i \pi_y
\end{eqnarray}
have angular momentum projection $-1$ and $1$, respectively. They form a basis for the two-dimensional irrep $\Pi_u$. The local Hilbert space of the corresponding DMRG lattice site consists of 16 states and can be made symmetry-adapted as follows:
\begin{align}
&\ket{-}                     &&\rightarrow &&|s = 0; &&s^z = 0; &&N = 0; &&I = \Sigma_g^{+}; &&l^z = 0\rangle  \\
&\ket{\pi^{\uparrow}_1}      &&\rightarrow &&|s = \frac{1}{2}; &&s^z = \frac{1}{2};  &&N = 1; &&I = \Pi_u; &&l^z = 1\rangle  \\
&\ket{\pi^{\uparrow}_{-1}}   &&\rightarrow &&|s = \frac{1}{2}; &&s^z = \frac{1}{2};  &&N = 1; &&I = \Pi_u; &&l^z = -1\rangle \\
&\ket{\pi^{\downarrow}_1}    &&\rightarrow &&|s = \frac{1}{2}; &&s^z = -\frac{1}{2}; &&N = 1; &&I = \Pi_u; &&l^z = 1\rangle  \\
&\ket{\pi^{\downarrow}_{-1}} &&\rightarrow &&|s = \frac{1}{2}; &&s^z = -\frac{1}{2}; &&N = 1; &&I = \Pi_u; &&l^z = -1\rangle \\
& \ket{\pi^{\uparrow}_1\pi^{\downarrow}_1}       &&\rightarrow &&|s = 0; &&s^z = 0; &&N = 2; &&I = \Delta_g; &&l^z = 2\rangle \\
& \ket{\pi^{\uparrow}_{-1}\pi^{\downarrow}_{-1}} &&\rightarrow &&|s = 0; &&s^z = 0; &&N = 2; &&I = \Delta_g; &&l^z = -2\rangle\\
& \ket{\pi^{\uparrow}_{-1}\pi^{\uparrow}_{1}} &&\rightarrow &&|s = 1; &&s^z = 1; &&N = 2; &&I = \Sigma_g^{-}; &&l^z = 0\rangle\\
& \ket{\pi^{\downarrow}_{-1}\pi^{\downarrow}_{1}} &&\rightarrow &&|s = 1; &&s^z = -1; &&N = 2; &&I = \Sigma_g^{-}; &&l^z = 0\rangle\\
& \left( \ket{\pi^{\uparrow}_{-1}\pi^{\downarrow}_{1}} , \ket{\pi^{\downarrow}_{-1}\pi^{\uparrow}_{1}} \right) &&\rightarrow &&|s = 1; &&s^z = 0; &&N = 2; &&I = \Sigma_g^{-}; &&l^z = 0\rangle\\
& \left( \ket{\pi^{\uparrow}_{-1}\pi^{\downarrow}_{1}} , \ket{\pi^{\downarrow}_{-1}\pi^{\uparrow}_{1}} \right) &&\rightarrow &&|s = 0; &&s^z = 0; &&N = 2; &&I = \Sigma_g^{+}; &&l^z = 0\rangle\\
& \ket{\pi^{\uparrow}_{-1} \pi^{\downarrow}_{-1} \pi^{\uparrow}_{1}}   &&\rightarrow &&|s = \frac{1}{2}; &&s^z = \frac{1}{2};  &&N = 3; &&I = \Pi_u; &&l^z = -1\rangle  \\
& \ket{\pi^{\uparrow}_{-1} \pi^{\downarrow}_{-1} \pi^{\downarrow}_{1}} &&\rightarrow &&|s = \frac{1}{2}; &&s^z = -\frac{1}{2};  &&N = 3; &&I = \Pi_u; &&l^z = -1\rangle  \\
& \ket{\pi^{\uparrow}_{-1} \pi^{\uparrow}_{1} \pi^{\downarrow}_{1}}    &&\rightarrow &&|s = \frac{1}{2}; &&s^z = \frac{1}{2};  &&N = 3; &&I = \Pi_u; &&l^z = 1\rangle  \\
& \ket{\pi^{\downarrow}_{-1} \pi^{\uparrow}_{1} \pi^{\downarrow}_{1}}  &&\rightarrow &&|s = \frac{1}{2}; &&s^z = -\frac{1}{2};  &&N = 3; &&I = \Pi_u; &&l^z = 1\rangle  \\
& \ket{\pi^{\uparrow}_{-1} \pi^{\downarrow}_{-1} \pi^{\uparrow}_{1} \pi^{\downarrow}_{1}}  &&\rightarrow &&|s = 0; &&s^z = 0;  &&N = 4; &&I = \Sigma_g^+; &&l^z = 0\rangle .
\end{align}
Because $\pi_{\pm 1}$ are basisfunctions of $\Pi_u$, the states with even particle number are \textit{gerade} and the ones with odd particle number \textit{ungerade}. The angular momentum projection $l^z$ is an additive quantum number, and can be obtained from the $\pi_{\pm 1}$ orbital fillings. The parity under $\sigma_v$ reflection of $\|s=0/1 ; N=2; \Sigma_g^{+/-}\rangle$ can be obtained by considering the spatial and spin part of the two-electron wavefunctions. The spin part of a singlet (triplet) state is antisymmetric (symmetric) with respect to particle interchange, and the spatial part hence has to be symmetric (antisymmetric):
\begin{eqnarray}
\pi_1(\vec{r}_A) \pi_{-1}(\vec{r}_B) + \pi_1(\vec{r}_B) \pi_{-1}(\vec{r}_A) & \propto & \pi_x(\vec{r}_A) \pi_{x}(\vec{r}_B) + \pi_y(\vec{r}_A) \pi_{y}(\vec{r}_B) \rightarrow \Sigma_g^+ \\
\pi_1(\vec{r}_A) \pi_{-1}(\vec{r}_B) - \pi_1(\vec{r}_B) \pi_{-1}(\vec{r}_A) & \propto & \pi_x(\vec{r}_A) \pi_{y}(\vec{r}_B) - \pi_y(\vec{r}_A) \pi_{x}(\vec{r}_B) \rightarrow \Sigma_g^-.
\end{eqnarray}
The reduced local basis hence consists of 7 multiplets $\|s;N;I\rangle$:
$\|0;           0; \Sigma_g^{+}\rangle$,
$\|\frac{1}{2}; 1; \Pi_u\rangle$,
$\|0;           2; \Delta_g\rangle$,
$\|1;           2; \Sigma_g^{-}\rangle$,
$\|0;           2; \Sigma_g^{+}\rangle$,
$\|\frac{1}{2}; 3; \Pi_u\rangle$,
$\|0;           4; \Sigma_g^+\rangle$.
This example can be extended to all irreps of all molecular point groups. The equivalent of Eq. \eqref{CGfromMPS} then becomes:
\begin{eqnarray}
& A[i]_{(j_L j_L^z N_L I_L I_L^q \alpha_L) ; (j_R j_R^z N_R I_R I_R^q \alpha_R)}^{(s s^z N I I^q)} \nonumber \\
& = \braket{j_L j_L^z s s^z \mid j_R j_R^z} \delta_{N_L + N, N_R} \braket{I_L I_L^q I I^q \mid I_R I_R^q} T[i]^{(s N I)}_{(j_L N_L I_L \alpha_L);(j_R N_R I_R \alpha_R)}.
\end{eqnarray}

\subsection{Space groups}
Two-dimensional lattice systems are also often studied with DMRG \cite{2D-DMRG-citation}. Typically, the lattice is considered to have periodic boundary conditions in one or two of the spatial directions, as one is actually interested in the thermodynamic limit. The studied lattices are then resp. the cylinder or the torus. Extrapolations of properties measured in systems with increasing size are then used to gain insight in the thermodynamic limit \cite{scalapino}. Because DMRG only works well for one-dimensional systems, rather large virtual dimensions are needed to obtain accurate numerical results.

One way to reduce the virtual dimension requirement is to exploit the non-Abelian symmetries of the Hamiltonian. Consider for example the two-dimensional Hubbard model on a $L \times L$ torus, with L even. For half-filling, instead of using the $\mathsf{SU(2)} \otimes \mathsf{U(1)}$ spin and particle-number symmetries, one can instead exploit the $\mathsf{SO(4)} \approx \mathsf{SU(2)} \otimes \mathsf{SU(2)} / \mathsf{Z_2}$ spin and particle-hole symmetry \cite{0295-5075-57-6-852, 101142S0217984990000933}. In addition, one can augment the abelian translational symmetry with $\mathsf{C_{4v}}$ to the full $\mathsf{p4mm}$ space group of the lattice. This group consists of all possible combinations of the symmetry elements $\{C_4, \sigma_h, T_x\}$: the rotation over $\frac{\pi}{2}$, the reflection with the $x$-axis as mirror, and the translation over one lattice constant in the $x$-direction.

\begin{figure}[h!]
\centering
\includegraphics[width=0.54\textwidth]{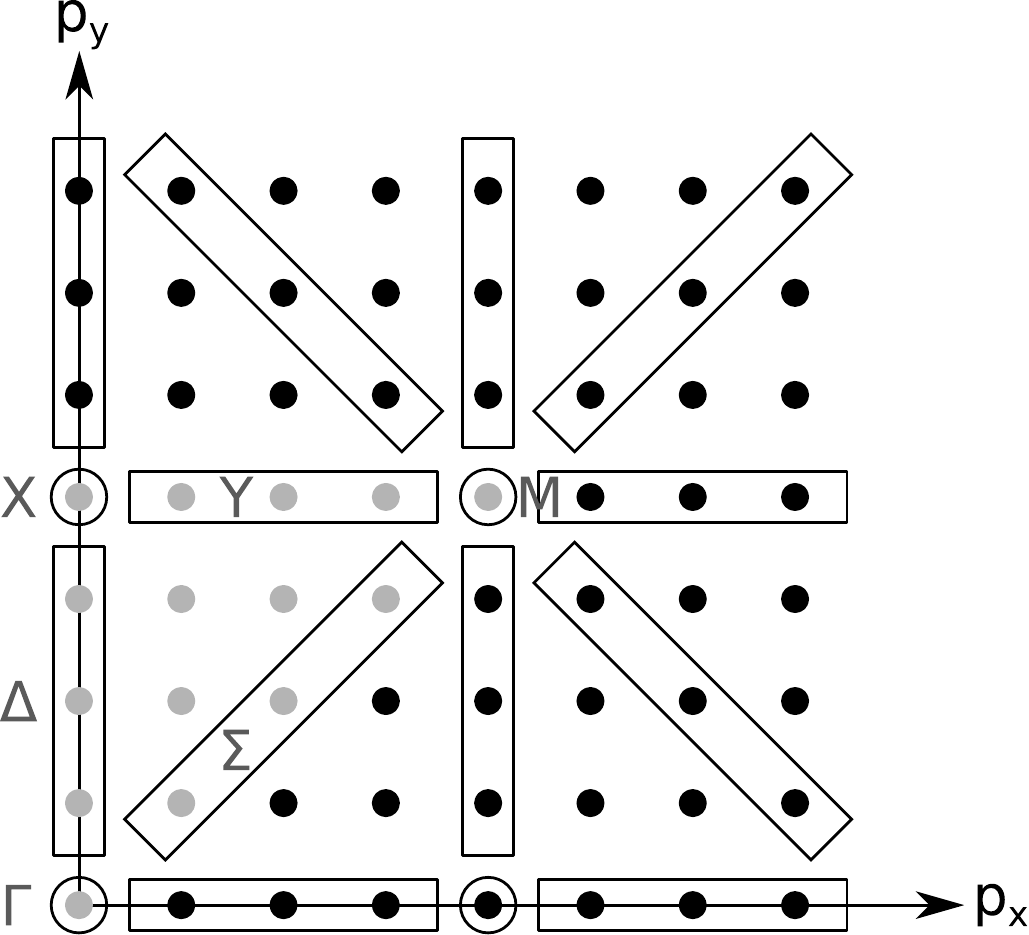}
\caption{\label{p4mm-plot} $\mathsf{p4mm}$ space group symmetry of the $L \times L$ torus.}
\end{figure}

Consider the lattice momentum vectors $\vec{k}_{\vec{p}} = \frac{2\pi}{L}(p_x,p_y)$ in Fig. \ref{p4mm-plot}. The basis functions of the irrep to which $\vec{k}_{\vec{p}}$ belongs, can be found by constructing its star \cite{BookCornwell}. This star is obtained by acting with the elements of the $\mathsf{C_{4v}}$ subgroup of $\mathsf{p4mm}$ on $\vec{k}_{\vec{p}}$ in momentum space, and by projecting the result back into the first Brillouin zone. From the grey momentum vectors in Fig. \ref{p4mm-plot}, all momentum vectors can be obtained by constructing the corresponding stars. These momentum vectors can hence be used to label all irreps. Tab. \ref{p4mm-table} gives an overview of the resulting irreps, their number, and their dimensions. The number of required lattice sites is hence reduced by a factor 8 in the leading order. This of course requires to group the eight single-particle basis functions of the bulk irreps to one DMRG lattice site. The instructions in, for example, chapter 9 of \citet{BookCornwell} allow to construct the Wigner $n$j symbols of the space groups.

\begin{table}
\centering
\caption{\label{p4mm-table} Overview of the $\mathsf{p4mm}$ irreps.}
\begin{tabular}{|c|ccc|}
\hline
name & \# irreps & irrep dimension & \# k-vectors \\
\hline
$\Gamma$ & 1 & 1 & 1\\
$\Delta$ & $\frac{L-2}{2}$ & 4 & $2L-4$\\
$X$ & 1 & 2 & 2\\
$Y$ & $\frac{L-2}{2}$ & 4 & $2L-4$\\
$M$ & 1 & 1 & 1\\
$\Sigma$ & $\frac{L-2}{2}$ & 4 & $2L-4$\\
Bulk & $\frac{(L-2)(L-4)}{8}$ & 8 & $L^2 - 6L + 8$\\
\hline
Total & $\frac{(L+2)(L+4)}{8}$ & - & $L^2$\\
\hline
\end{tabular}
\end{table}

\chapter{Longitudinal static response properties of hydrogen chains} \label{HCHAIN-chapter}
\begin{chapquote}{Harlan J. Ellison, 1985}
Apart from hydrogen, the most common thing in the universe is stupidity.
\end{chapquote}

\section{Introduction}
DMRG works extremely well for noncritical one-dimensional systems. The underlying MPS ansatz then complies with the area law for the entanglement entropy (see section \ref{entanglement-section}). Hydrogen chains have been studied extensively with QC-DMRG \cite{Chan2002, hachmann:144101, zgid:144115, zgid:144116, QUA:QUA23173, woutersJCP1, nakatani:134113, ma:224105}. Although the Coulomb interaction is nonlocal, the virtual dimension does not have to grow with chain length to maintain a constant accuracy in the insulating regime \cite{hachmann:144101}.

As the nuclear separation grows, the system exhibits a large amount of static correlation. The atoms can then be considered independent, and all possible spin states are degenerate. Hydrogen chains and lattices are therefore often used as benchmark systems to assess new MR methods in quantum chemistry \cite{PhysRevB.50.14791, Chan2002, HchainExample9, ExampleHchain8, ExampleHchain7, ExampleHchain6, PhysRevB.84.245117, ExampleHchains5, ExampleHchain4, ExampleHchain3, ExampleHchain2, ExampleHchain1}.

When the bond length in an equidistant hydrogen chain decreases, the system goes through a metal-insulator transition (MIT) \cite{PhysRevB.84.245117, QUA:QUA23047}. The initially local electrons become delocalized, and at the transition point the electrons are highly correlated. Response properties, such as the static longitudinal dipole (hyper)polarizability, which are extensive quantities in the insulator regime, diverge in the metallic regime.

The equidistant hydrogen chain cannot exist due to the Peierls instability \cite{Peierls}: it is unstable with respect to dimerization. The equidistant and dimerized hydrogen chains are toy models to mimic the effect of bond length and bond length alternation on the electron delocalization, electron correlation, and electronic response properties. A realization of a delocalized one-dimensional system with bond length alternation is the conjugated $\pi$-system of all-trans polyenes, another system which has been extensively studied with QC-DMRG \cite{Chan2002, chan:204101, hachmann:144101, ghosh:144117, yanai:024105, saitow:044118, NaokiLRTpaper}.

\section{Longitudinal static response properties}
Consider an external static electric field $\vec{F}$, changing the electronic Hamiltonian $\hat{H}_0$ to
\begin{equation}
\hat{H} = \hat{H}_0 + \vec{F} \cdot \hat{\vec{r}}.
\end{equation}
For variational wavefunctions, such as the MPS in DMRG, the electronic dipole moment can be calculated as
\begin{eqnarray}
\vec{\mu} & = & - \braket{ \Psi_0 \mid \hat{\vec{r}} \mid \Psi_0 } = - {\nabla}_{\vec{F}} \braket{ \Psi_0 \mid \hat{H}_0 + \vec{F} \cdot \hat{\vec{r}} \mid \Psi_0 } \nonumber \\
& = & - \lim\limits_{\vec{F} \rightarrow \vec{0}} ~ {\nabla}_{\vec{F}} \braket{ \Psi(\vec{F}) \mid \hat{H}_0 + \vec{F} \cdot \hat{\vec{r}} \mid \Psi(\vec{F}) } = - \lim\limits_{\vec{F} \rightarrow \vec{0}} ~ {\nabla}_{\vec{F}} E(\vec{F}),
\end{eqnarray}
due to the Hellmann-Feynman theorem \cite{HellmannFeynmanTheorem}. The dipole (hyper)polarizability is the (higher order) response of the dipole moment to a change in the electric field. The static polarizability tensor is for example:
\begin{equation}
\alpha_{ u v } = \lim\limits_{\vec{F} \rightarrow \vec{0}} ~ \frac{ \partial \mu_u }{ \partial F_v } = - \lim\limits_{\vec{F} \rightarrow \vec{0}} ~ \frac{\partial^2 E(\vec{F})}{\partial F_u \partial F_v} .
\end{equation}

In Ref. \cite{woutersJCP1}, we have studied linear centrosymmetric chains with $\mathsf{D_{\infty h}}$ symmetry. For convenience, the $z$-axis is chosen along the chain, and the center of mass coincides with the origin. The quantities of interest are the \textit{longitudinal} dipole (hyper)polarizabilities, i.e. along the $z$-axis. The external static electric field is then
\begin{equation}
\vec{F} = F \vec{1}_z.
\end{equation}
For centrosymmetric systems, all odd derivatives of the energy $E(F)$ with respect to $F$ vanish at $F=0$ due to the inversion symmetry: $E(F) = E(-F)$. The longitudinal static polarizability and second hyperpolarizability are
\begin{eqnarray}
\alpha_{zz} & = & - \lim\limits_{F \rightarrow 0} ~ \frac{\partial^2 E(F)}{\partial F^2}, \label{PolIdea}\\
\gamma_{zzzz} & = & - \lim\limits_{F \rightarrow 0} ~ \frac{\partial^4 E(F)}{\partial F^4}. \label{HyperPolIdea}
\end{eqnarray}
Both quantities are studied for chains of increasing length $L$. A small electric field $\delta F$ can cause an elementary excitation in the chain. If these excitations are localized, i.e. have a finite size, the response properties $\eqref{PolIdea}$ and $\eqref{HyperPolIdea}$ eventually have to saturate, i.e. become extensive quantities in the system size $L$:
\begin{eqnarray}
\lim\limits_{L \rightarrow \infty} \frac{\alpha_{zz}(L)}{L} & = & \text{constant}, \label{polLimit} \\
\lim\limits_{L \rightarrow \infty} \frac{\gamma_{zzzz}(L)}{L} & = & \text{another constant}. \label{hyperpolLimit}
\end{eqnarray}
In the insulating regime, this is the case. In the metallic regime, the response properties grow faster than linear, because the elementary excitations due to a small electric field $\delta F$ do not have a finite size. The metallic regime and the MIT will be discussed in greater detail in section \ref{MITsection}. In Ref. \cite{woutersJCP1}, we have mainly focussed on the insulating regime, and especially on obtaining numerical results for the limits in Eqs. \eqref{polLimit} and \eqref{hyperpolLimit}.

The (hyper)polarizabilities can be obtained in several ways. An analytic response theory can be set up, which considers the Rayleigh-Schr\"odinger perturbation expansion for the electric field $F$ in the manifold of the ansatz wavefunction $\ket{\Psi(\mathbf{A}(F))}$:
\begin{eqnarray}
\left( \hat{H}_0 + F z \right) \left( \ket{\Psi_0} + F \ket{\Psi_1} + ... \right) & = & \left( E_0 + F E_1 + ... \right) \left( \ket{\Psi_0} + F \ket{\Psi_1} + ... \right), \\
\left( \hat{H}_0 - E_0 \right) \ket{\Psi_1} & = & \left( E_1 - z  \right) \ket{\Psi_0} \qquad \text{with} \braket{\Psi_1 \mid \Psi_0} = 0, \\
\alpha_{zz} & = & -2 \Re \braket{\Psi_1 \mid z \mid \Psi_0}. 
\end{eqnarray}
For HF theory this yields the coupled-perturbed HF equations \cite{CPHF,CPHF2,CPHF3}. For DMRG the linear response theory has been derived as well \cite{dorando:184111}. For the latter, convergence problems were perceived for the polarizability calculations, and we have therefore opted to use another method.

A second method is the sum-over-states (SOS) expression \cite{PhysRev.46.618}. Instead of solving $\ket{\Psi_1}$ in the tangent space of $\ket{\Psi_0}$, the former can be written as a linear combination over many excited states:
\begin{equation}
\alpha_{zz} = 2 \sum\limits_{k \neq 0} \frac{\braket{\Psi_0 \mid z \mid \Psi_k}\braket{\Psi_k \mid z \mid \Psi_0}}{E_k - E_0}. \label{SOSexpressionIntro}
\end{equation}
For certain ground states $\ket{\Psi_0}$ an operator $\hat{O}$ can be constructed so that \cite{DalgarnoHydrogenTrick}
\begin{equation}
\left[ \hat{H}, \hat{O} \right] \ket{\Psi_0} = z \ket{\Psi_0}.
\end{equation}
This allows to remove the denominator in Eq. \eqref{SOSexpressionIntro}:
\begin{equation}
\alpha_{zz} = 2 \left( \braket{\Psi_0 \mid z \hat{O} \mid \Psi_0} - \braket{\Psi_0 \mid z \mid \Psi_0}\braket{\Psi_0 \mid \hat{O} \mid \Psi_0} \right).
\end{equation}
The operator relation
\begin{equation}
\left[ \hat{H}, \hat{O} \right] = z \label{bareCommutatorTrickEquationAndAllThat}
\end{equation}
has no general solution if $\hat{H}$ is not a one-body operator. For a $k$-body operator $\hat{H}$ and an $n$-body operator $\hat{O}$, their commutator is a $(k+n-1)$-body operator. For $M$ orbitals Eq. \eqref{bareCommutatorTrickEquationAndAllThat} yields $M^{2(k+n-1)}$ equations for the $M^{2n}$ parameters in $\hat{O}$, which implies that a solution is only guaranteed for one-body Hamiltonians. For general ground states $\ket{\Psi_0}$, the SOS expression \eqref{SOSexpressionIntro} hence requires to calculate all excited states. It is therefore also not preferred in conjunction with DMRG.

A third method is to calculate Eqs. \eqref{PolIdea} and \eqref{HyperPolIdea} by using a set of small finite fields, the finite-field method \cite{JCC:JCC540110110}. This method requires to calculate the ground states of a few Hamiltonians differing only in the one-body matrix elements, and was the method adopted in our study [Ref. \cite{woutersJCP1}]:

\newpage
\hspace{-\parindent}{\large\color{blue}{\textbf{Longitudinal static optical properties of hydrogen chains: Finite field extrapolations of matrix product state calculations} \cite{woutersJCP1}}}

\vspace{0.2cm}

Sebastian Wouters,$^{1}$ Peter A. Limacher,$^{2}$ Dimitri Van Neck,$^{1}$ and Paul W. Ayers$^{2}$

{\footnotesize $^{1}$\textit{Center for Molecular Modeling, Ghent University, Ghent, Belgium}}

{\footnotesize $^{2}$\textit{Department of Chemistry, McMaster University, Hamilton, Ontario, Canada}}

\vspace{0.5cm}

\parbox{0.90\textwidth}{
We have implemented the sweep algorithm for the variational optimization of $\mathsf{SU(2)} \otimes \mathsf{U(1)}$ (spin and particle number) invariant matrix product states (MPS) for general spin and particle number invariant fermionic Hamiltonians. This class includes non-relativistic quantum chemical systems within the Born-Oppenheimer approximation. High-accuracy \textit{ab initio} finite field results of the longitudinal static polarizabilities and second hyperpolarizabilities of one-dimensional hydrogen chains are presented. This allows to assess the performance of other quantum chemical methods. For small basis sets, MPS calculations in the saturation regime of the optical response properties can be performed. These results are extrapolated to the thermodynamic limit.
}

\vspace{0.7cm}

\hspace{-\parindent}\textbf{I. INTRODUCTION}
\vspace{0.3cm}

Non-linear optical (NLO) properties of materials are of interest to experiment, theory, and industry. They account for a wide variety of phenomena such as frequency doubling, optical control of the refractive index, and phase conjugation \cite{nlobook}. Especially the NLO properties of linearly conjugated organic polymer chains have moved to the center of attention and many theoretical studies have been published about the interplay of molecular structure, electron delocalization, and NLO properties \cite{marder93, da94, chemrevpersoons, tykwinski98, champagne99, kirtman00, perpete08, borini09, champagne11}.
An important question in many of these studies is the suitability and accuracy of different quantum chemical (QC) methods \cite{sekino08, l09}, henceforth called levels of theory (LOT). Conventional density functional theory was found to dramatically overestimate NLO properties of long molecular chains \cite{kirtman98, kirtman99}. Newly developed approaches were presented to mitigate but not fully resolve the problem \cite{oep03, sekino05}. In the meantime also certain irregularities between Hartree-Fock (HF) and post-HF methods were noticed, calling into question the importance and the influence of electron correlation on NLO properties \cite{toto95, li08, 2011JChPh.135a4111L}. It is therefore desirable to obtain the NLO properties of the fully correlated problem, i.e., at exact diagonalization (ED) accuracy. Linear chains of hydrogen are ideal test systems for assessing the quality of different LOTs \cite{PhysRevA.52.178, PhysRevA.52.1039, sekino07, QUA:QUA22177}.

A recently developed class of variational ansatzes, the tensor network states (TNS), yield compact and accurate approximations of low-lying eigenstates based on the topological properties of the Hamiltonian. The matrix product state (MPS) is the natural TNS for one-dimensional holographic geometries \cite{TNSoverview}. Conversely, it can be shown that every quantum many-body state can be rewritten as an MPS \cite{Schollwock201196}. This allows the MPS to be used as a variational ansatz for any quantum system. The optimal MPS can be found implicitly by means of the density matrix renormalization group (DMRG) or explicitly by variationally optimizing the MPS \cite{Schollwock201196, PhysRevB.55.2164}. Several groups have implemented the DMRG algorithm for \textit{ab initio} QC calculations \cite{WhiteQCDMRG, QUA:QUA1, mitrushenkov:6815, Chan2002, PhysRevB.67.125114, PhysRevB.68.195116, PhysRevB.70.205118, moritz:024107, Rissler2006519, moritz:244109, zgid:014107, kurashige:234114, PhysRevB.81.235129}. For quasi-one-dimensional chemical systems such as hydrogen chains \cite{hachmann:144101}, the MPS gives an efficient description. The mutual screening of electrons and nuclei results in an effectively local electromagnetic interaction, which explains why DMRG works well for these systems \cite{QUA:QUA1, kurashige:234114}. For systems that do not have a one-dimensional holographic geometry, the MPS is not always efficient, as can be seen by the virtual dimensions required to obtain near-ED accuracy \cite{WhiteQCDMRG, kurashige:234114}. A better choice and ordering of the single particle basis can resolve the problem partly \cite{WhiteQCDMRG, Chan2002, PhysRevB.68.195116, PhysRevB.70.205118, moritz:244109, PhysRevB.82.205105, PhysRevA.83.012508, moritz:034103}. Other TNSs such as the tree TNS (Refs. \cite{PhysRevB.82.205105} and \cite{PhysRevA.83.012508}) or different ansatzes such as correlator product states \cite{1367-2630-11-8-083026, PhysRevB.80.245116} (e.g., the complete graph TNS (Ref. \cite{1367-2630-12-10-103008})) can further improve the descriptions of such systems. It has even been suggested to use correlator product states with auxiliary indices \cite{PhysRevB.80.245116}. This leads us back to White's original proposal \cite{WhiteQCDMRG} to combine several orbitals into a single local degree of freedom in QC DMRG.

Together with an efficient TNS, the use of symmetry can make the description of eigenstates even more compact. Structuring the virtual bonds according to the irreducible representations of the applied symmetry groups introduces a sparse block structure in the tensors. For non-Abelian symmetry groups, the Wigner-Eckart theorem permits working with reduced tensors \cite{PhysRevA.82.050301, 1367-2630-12-3-033029, 0295-5075-57-6-852, 1742-5468-2007-10-P10014}.

In this paper, we use the $\mathsf{SU(2)} \otimes \mathsf{U(1)}$ invariant MPS to study the longitudinal static dipole polarizability and second hyperpolarizability of one-dimensional hydrogen chains by means of finite field extrapolations. The MPS algorithm enables us to study longer chains than with ED but not at the expense of decreasing accuracy. For small basis sets, this allows us to obtain high-accuracy data in the saturation regime of the optical response properties. The results obtained with our MPS algorithm let us assess the performance of standard QC methods. When possible, these results are extrapolated to infinite chain length to obtain quantitative results in the thermodynamic (TD) limit. Different basis sets are compared.

Related work, studying both the static and dynamic polarizabilities and second hyperpolarizabilities of conjugated $\pi$-systems, includes the analytic response theory for \textit{ab initio} QC DMRG (Ref. \cite{dorando:184111}) and the correction vector DMRG algorithm for Pariser-Parr-Pople Hamiltonians \cite{Ramasesha1,Ramasesha2}. Accurate TD limit data of the static optical response properties of hydrogen chains can also be obtained with diffusion Monte Carlo, using the modern theory of polarization \cite{PhysRevLett.95.207602, Umari2}.

The MPS ansatz is briefly addressed in Sec. II, where the variational optimization of the MPS for \textit{ab initio} QC Hamiltonians, imposing $\mathsf{SU(2)} \otimes \mathsf{U(1)}$ spin and particle number symmetry, and our implementation are also discussed. The finite field method is outlined in Sec. III. Section IV deals with the optical properties of several spin states of an equally spaced hydrogen chain, where the spacing controls the amount of static correlation. A chain of H$_2$ constituents is studied in Sec.~V: the influence of intermolecular distance (and hence the amount of electron delocalization), LOT, and basis set on the optical properties are determined. When possible, the MPS results are extrapolated to the TD limit. Section VI contains the conclusions.

\newpage

\hspace{-\parindent}\textbf{II. THE MPS ALGORITHM}
\vspace{0.3cm}

As there are already excellent works on the variational optimization of an MPS \cite{Schollwock201196}, on the implementation of DMRG for \textit{ab initio} QC calculations \cite{WhiteQCDMRG, QUA:QUA1, mitrushenkov:6815, Chan2002, PhysRevB.67.125114, PhysRevB.68.195116, PhysRevB.70.205118, moritz:024107, Rissler2006519, moritz:244109, zgid:014107, kurashige:234114, PhysRevB.81.235129, chan:3172}, and on the use of non-Abelian symmetries in TNSs \cite{PhysRevA.82.050301, 1367-2630-12-3-033029, 0295-5075-57-6-852, 1742-5468-2007-10-P10014}, we choose to focus only on how these principal concepts contribute to our algorithm.

\vspace{0.7cm}
\hspace{-\parindent}\textbf{A. DMRG and MPS}
\vspace{0.3cm}

In non-relativistic \textit{ab initio} QC, the positions of the nuclei are fixed in the Born-Oppenheimer approximation and a basis set is chosen as the orbital degrees of freedom. Because we study one-dimensional systems in this work, L\"owdin transformed Gaussian basis sets are used as they preserve locality well \cite{hachmann:144101, PhysRev.105.102}. Consider a state with $L$ orbitals and 4 possible occupations $i$ per orbital
\begin{equation}
\ket{\Psi} = \sum\limits_{\{i_1 ... i_L\}} c_{i_1 ... i_L} \ket{i_1 ... i_L} .
\end{equation}
This state can always be rewritten as an MPS \cite{Schollwock201196},
\begin{equation}
\ket{\Psi} = \sum\limits_{\{i_1 ... i_L\}} \sum\limits_{\{k_1 ... k_{L-1}\}} M^{i_1}_{k_1} M^{i_2}_{k_1 k_2} ... M^{i_L}_{k_{L-1}} \ket{i_1 ... i_L} , \label{MPSnotationeq}
\end{equation}
which associates to every orbital 4 matrices $M^{i}_{k_L k_R}$ or a single three-index tensor. The index $i$ is called the local index and represents the occupation. The indices $k_L$ and $k_R$ are called virtual indices. The dimension $D$ of the virtual indices needs to increase exponentially towards the middle of the MPS chain for Eq. \eqref{MPSnotationeq} to represent the full Hilbert space. In calculations, the virtual dimension $D$ is truncated and the MPS represents only a part of the full Hilbert space. The tensors in the MPS chain are iteratively optimized, one at a time, in the sweep algorithm \cite{Schollwock201196}. This method is strictly variational. For arbitrarily large systems with a one-dimensional holographic geometry, the ED solution can be approximated to any desired accuracy by an MPS with a finite $D$ \cite{TNSoverview}. Note that Eq. \eqref{MPSnotationeq} represents a multideterminantal wavefunction and is hence able to capture static correlation \cite{hachmann:144101}.

There are two versions of the DMRG algorithm: single-site and two-site DMRG. Their names refer to the number of neighbouring orbitals that are free at a local optimization step. The variational optimization of an MPS corresponds to (but is not equal to) single-site DMRG. In the MPS algorithm, the renormalization transformations and subsequent decimations of the DMRG algorithm are incorporated in the MPS ansatz itself. Fixed points of both DMRG algorithms can be written as MPSs \cite{PhysRevB.55.2164}. Single-site DMRG is also strictly variational, while two-site DMRG is not \cite{Chan2002,zgid:144115}.

In certain cases, the two-site DMRG algorithm and the variational optimization of the corresponding MPS both lead to the same result. This is often the case for systems that have one-dimensional holographic geometries and for which the MPS is the natural TNS, while for other systems the two-site DMRG algorithm can outperform the single-site variational optimization of an MPS as it provides more degrees of freedom for each local diagonalization step \cite{Schollwock201196, PhysRevB.67.125114}. In both DMRG algorithms, adding perturbative corrections or noise to the reduced density matrix helps to reach the true ground state within the subspace of the full Hilbert space spanned by the MPS, as they help to reintroduce lost quantum numbers in the reduced basis \cite{Schollwock201196, WhiteQCDMRG, Chan2002, kurashige:234114, zgid:144115, PhysRevB.72.180403}. Another way to achieve this, is to explicitly keep states with a certain symmetry in the reduced basis \cite{chan:3172}.

\begin{figure}
\centering
\includegraphics[width=0.70\textwidth]{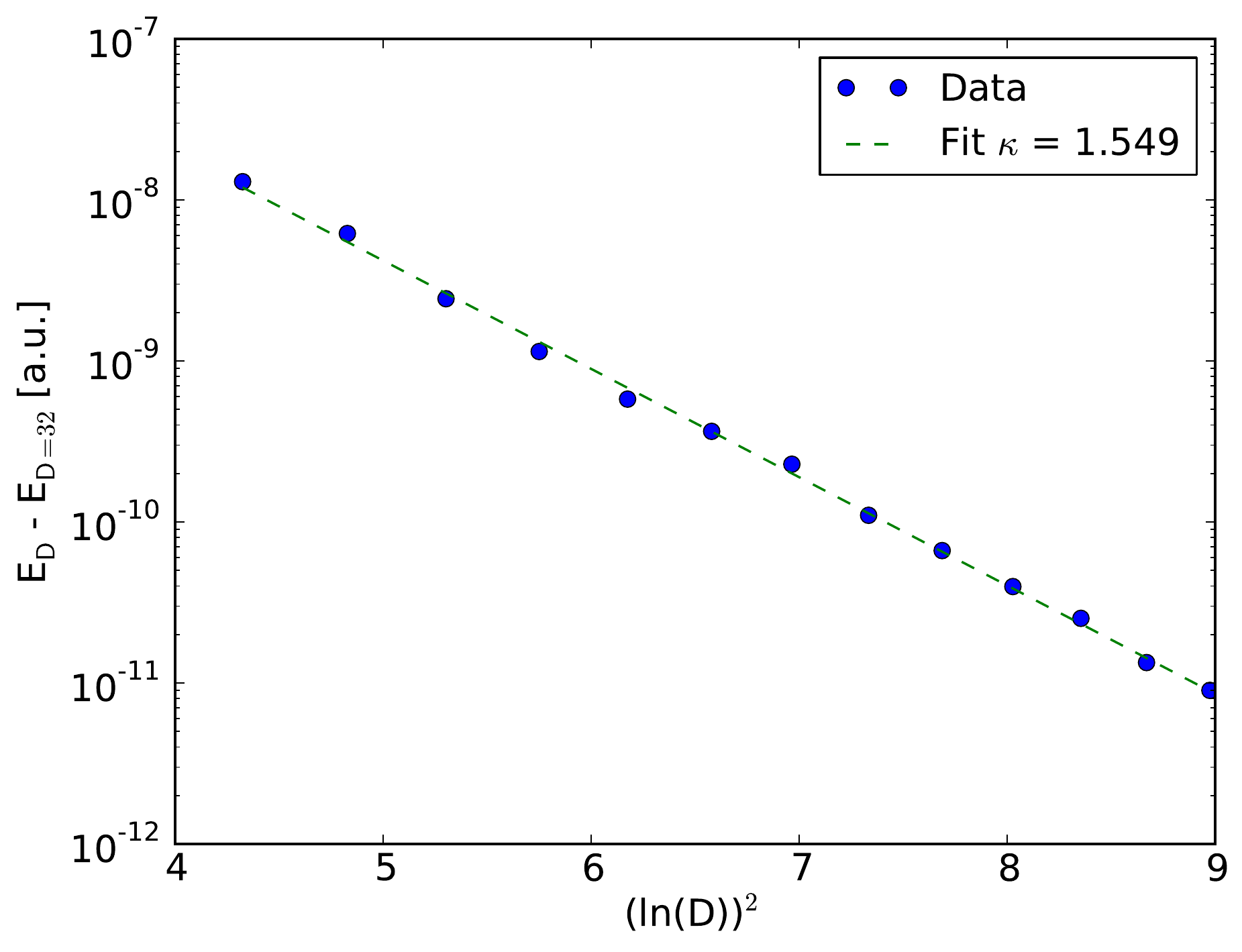}
\caption{\label{ConvergencePlotPaperJCP} The ground state of a hydrogen chain of 36 atoms, with an alternate atom spacing of 2/3 a.u. (see Sec. V A), in the L\"owdin transformed STO-6G basis, is approximated by several MPSs with increasing virtual dimension. The scaling of the ground state energy with $D$, the virtual dimension \textit{per symmetry sector} (see Sec. II D), follows Eq. \eqref{Chanscaling}. The rightmost data point corresponds to $D = 20$ and $E_{D = 32}$ is used as an approximation to the exact
result.}
\end{figure}

For the systems in our study, the holographic geometry is one-dimensional and hence the MPS ansatz is a good choice. This is confirmed by the rapid convergence of the ground state energy obtained with an MPS with increasing virtual dimension. Chan \textit{et al.} \cite{Chan2002, AyersRDMcalues} have proposed a relation for this convergence,
\begin{equation}
\ln(E_{D} - E_{\text{exact}} ) = a - \kappa (\ln(D))^2 . \label{Chanscaling}
\end{equation}
Here, $a$ and $\kappa$ are fitting parameters, $E_{\text{exact}}$ is the ED result, and $E_{D}$ the energy when an MPS with virtual dimension $D$ is used. Equation \eqref{Chanscaling} is illustrated in Fig. \ref{ConvergencePlotPaperJCP}.

\newpage

\hspace{-\parindent}\textbf{B. General two-body Hamiltonians}
\vspace{0.3cm}

In second quantization, the Hamiltonian can be written as \cite{BookDimitri}
\begin{equation}
\hat{H}_0 = \sum\limits_{i,j,\sigma} (i | \hat{T} | j) \hat{a}^{\dagger}_{i \sigma} \hat{a}_{j \sigma} \\
+ \frac{1}{2} \sum\limits_{i,j,k,l, \sigma, \tau} (i j | \hat{V} | k l) \hat{a}^{\dagger}_{i \sigma} \hat{a}^{\dagger}_{j \tau} \hat{a}_{l \tau} \hat{a}_{k \sigma} , \label{hamiltonian1}
\end{equation}
where the Latin letters denote orbitals and the Greek letters spin projections. Global spin and global particle number are conserved by this Hamiltonian. The matrix elements are calculated based on the work of Obara and Saika \cite{ObaraSaika}.

For the local optimization procedure, partial Hamiltonian  terms such as $a_{i \in \text{left} \sigma}^{\dagger} a_{j \in \text{left} \tau}^{\dagger}$ need to be stored in memory. We have taken all previous considerations in the literature into account to store as few of them as possible \cite{Chan2002, kurashige:234114, chan:3172}. These include multiplying creators/annihilators with two-body matrix elements and contracting common indices to form complementary operators, exploiting the Hermitian symmetry of matrix elements as well as exploiting the creator/annihilator swap symmetry due to the fermion anticommutation relations. Further storage reduction is possible by exploiting global symmetry.

\vspace{0.7cm}
\hspace{-\parindent}\textbf{C. Global symmetries}
\vspace{0.3cm}

Using the global symmetries of the Hamiltonian has many advantages, including the ability to explicitly scan only the desired symmetry sector of the total Hilbert space, and an improvement of computational performance. This improvement consists of a reduction in both central processing unit (CPU) time (by reducing the number of sweeps) and memory usage (the tensors adopt a sparse block structure and the required virtual dimensions are smaller; this further decreases the CPU time) \cite{0295-5075-57-6-852}. The main disadvantage is the increasing complexity of the algorithm: i.e., analytic work done beforehand and overhead in the resulting program. However, this needs to be done only once, and in many cases it does not outweigh the benefits.

We have implemented global spin and particle number symmetry. The $\mathsf{U(1)}$ particle number symmetry is an Abelian symmetry and is therefore represented by an additive quantum number \cite{1742-5468-2007-10-P10014}. Its implementation in \textit{ab initio} QC DMRG calculations is well known \cite{PhysRevB.68.195116, kurashige:234114}. The $\mathsf{SU(2)}$ spin symmetry is a non-Abelian symmetry and requires recoupling \cite{1742-5468-2007-10-P10014}.

Global symmetry can be imposed by requiring that the three-index tensors $M^i_{k_L k_R}$ in the MPS chain are irreducible tensor operators of the imposed symmetry group \cite{PhysRevA.82.050301, 1367-2630-12-3-033029, 0295-5075-57-6-852, 1742-5468-2007-10-P10014}. The local and virtual bases are represented in states with the correct symmetry, i.e., spin $s$ or $j$, spin projection $s^z$ or $j^z$, and particle number $N$. The local states $i = \ket{-}$, $\ket{\uparrow}$, $\ket{\downarrow}$ or $\ket{\uparrow\downarrow}$ then correspond to resp.~$i = \ket{s = 0; s^z = 0; N = 0}$, $\ket{\frac{1}{2}\frac{1}{2}1}$, $\ket{\frac{1}{2} -\frac{1}{2}1}$ and $\ket{002}$. Due to the Wigner-Eckart theorem, each irreducible tensor operator decomposes into a structural part and a degeneracy part $T$,
\begin{equation}
M^i_{k_Lk_R} = M^{(s s^z N)}_{(j_L j_L^z N_L \alpha_L)(j_R j_R^z N_R \alpha_R)} = \braket{j_L j_L^z s s^z | j_R j_R^z} \delta_{N_L + N, N_R} T^{(s N)}_{(j_L N_L \alpha_L)(j_R N_R \alpha_R)} . \label{tensordecomp}
\end{equation}
The $\mathsf{SU(2)}$ symmetry is imposed by the Clebsch-Gordan coefficient and the $\mathsf{U(1)}$ symmetry by the particle conserving Kronecker delta. The indices $\alpha_L$ and $\alpha_R$ are used to keep track of the number of times an irreducible representation occurs at a virtual bond. If the virtual dimension of a symmetry sector is $D(j_L N_L) = \text{size}(\alpha_L)$, this would correspond to a dimension of $(2j_L + 1)D(j_L N_L)$ in a non-symmetry adapted MPS \cite{0295-5075-57-6-852}. Global symmetry can be imposed by requiring that the left virtual index of the leftmost tensor in the MPS chain consists of one irreducible representation corresponding to $(j_L , N_L) = (0, 0)$, while the right virtual index of the rightmost tensor consists of one irreducible representation corresponding to $(j_R N_R) = (SN)$, the desired global spin, and particle number.

The operators
\begin{eqnarray}
\hat{b}^{\dagger}_{m} & = & \hat{a}^{\dagger}_{m} , \label{creaannih1}\\
\hat{b}_{m} & = & (-1)^{\frac{1}{2}-m}\hat{a}_{-m} , \label{creaannih2}
\end{eqnarray}
transform as irreducible tensor operators with spin $\frac{1}{2}$ under $\mathsf{SU(2)}$, with $m$ the spin projection \cite{BookDimitri}. Because these operators are part of a doublet, it is possible to exploit the Wigner-Eckart theorem also for operators and complementary operators, and to develop a code without any spin projections or Clebsch-Gordan coefficients. Contracting terms of the types of Eqs. \eqref{tensordecomp}-\eqref{creaannih2} can be done by implicitly summing over the common multiplets and recoupling the local, virtual, and operator spins. Examples are given in the Appendix. Operators and complementary operators then formally consist of terms containing Clebsch-Gordan coefficients, particle conserving Kronecker deltas, and reduced tensors. In our code, however, only the reduced tensors need to be calculated and stored. To the best of our knowledge, the global $\mathsf{SU(2)}$ symmetry has been implemented only once in \textit{ab initio} QC DMRG calculations \cite{zgid:014107}. In this algorithm \cite{zgid:014107}, no use is made of the Wigner-Eckart theorem to work with reduced tensors, as is often proposed \cite{PhysRevA.82.050301, 1367-2630-12-3-033029, 0295-5075-57-6-852, 1742-5468-2007-10-P10014}.

\vspace{0.7cm}
\hspace{-\parindent}\textbf{D. Implementation}
\vspace{0.3cm}

We have implemented the sweep algorithm \cite{Schollwock201196} to variationally optimize an $\mathsf{SU(2)} \otimes \mathsf{U(1)}$ invariant MPS in \texttt{C++}. Matrix operations are handled by LAPACK and BLAS. Wigner 6-j symbols are calculated by the GNU scientific library. For the local optimization of the degeneracy part of an MPS tensor, we have chosen the Lanczos method, implemented in ARPACK. Where possible, the code is parallellized on a single node with OpenMP. No multinode parallellization was needed for the results in this paper. 

The virtual dimension is truncated per symmetry sector: if the virtual dimension $D(j_L N_L)$ of a symmetry sector $(j_L N_L)$ required to represent the full Hilbert space exceeds a predefined threshold $D$, it is set to $D$. For the results presented in this paper, $D$ is chosen large enough so that no relative energy error is larger than $10^{-11}$,
\begin{equation}
\frac{E_{D}-E_{\text{exact}}}{E_{\text{exact}}} < 10^{-11} . \label{10minus11convergence}
\end{equation}
Specific choices for $D$ are mentioned when the applications are introduced. All tensors are stored in the minimum amount of memory required. Convergence is reached when both the energy and the wavefunction meet the following criteria:
\begin{eqnarray}
\mid E_n - E_{n-1} \mid & < & \epsilon_1 , \\
1 - \mid \braket{\text{MPS}_n \mid \text{MPS}_{n-1}} \mid & < & \epsilon_2 ,
\end{eqnarray}
where $n$ is the sweep number and $\epsilon_1 = \epsilon_2 = 10^{-13}$ for the calculations presented in this paper. At the start of the algorithm, the MPS is filled with noise, but during the sweeps no noise or perturbative corrections were added. For more complex chemical systems, the orbital choice, the orbital ordering, and the initial guess play an important role for the convergence and even for the qualitative properties of the solution \cite{WhiteQCDMRG, Chan2002, PhysRevB.68.195116, moritz:244109, PhysRevB.82.205105, PhysRevA.83.012508, moritz:034103}. The holographic geometry of such systems is often far from one-dimensional. In DMRG calculations, basis states with a certain symmetry are sometimes explicitly kept in the reduced basis to avoid losing quantum numbers \cite{chan:3172}. Note that the division of the virtual bonds in symmetry sectors $(j_L N_L)$ boils down to the same thing.

If there are $N$ electrons in the system, with $N \leq L$, the number of $\mathsf{SU(2)} \otimes \mathsf{U(1)}$ symmetry sectors in the middle of the chain is $\mathcal{O}(N^2)$. In that case, we obtain for our algorithm a scaling per sweep of $\mathcal{O}(D^3 L^3 N^2 + D^2 L^4 N^2)$ in time and $\mathcal{O}(D^2 L^2 N^2)$ in memory \cite{Chan2002}. For $N \geq L$, $N$ should be replaced by $(2L - N)$. Note that both the number of sweeps to reach convergence and the virtual dimension $D$ to reach a certain accuracy are smaller when global symmetry is imposed \cite{0295-5075-57-6-852}. Hachmann \textit{et al.} \cite{hachmann:144101} present a method that makes use of the numerical negligibility of certain two-body matrix elements to obtain an algorithm that scales per sweep as $\mathcal{O}(D^3 L^2)$ in time and $\mathcal{O}(D^2 L)$ in memory. When applying global $\mathsf{SU(2)} \otimes \mathsf{U(1)}$ symmetry, these order estimates have to be multiplied with $\mathcal{O}(N^2)$ when $N \leq L$ or $\mathcal{O}((2L - N )^2)$ when $N \geq L$. The efficiency gain when neglecting these matrix elements comes with the cost of losing the variational character of the algorithm, because the Hamiltonian is altered. However, the error is under control. In the current version of our program, this quadratically scaling algorithm is not yet used, but we plan to implement it in the future.

\vspace{0.7cm}
\hspace{-\parindent}\textbf{III. THE FINITE FIELD METHOD}
\vspace{0.3cm}

When a homogeneous electric field $\vec{F}$ is applied, the electrons acquire a potential energy that depends on their position \cite{JCC:JCC540110110}. The total Hamiltonian of the system becomes (atomic units)
\begin{equation}
\hat{H} = \hat{H}_0 + \vec{F}.\vec{r} . \label{hami2CH4}
\end{equation}
This total Hamiltonian still conserves global spin and global particle number.

The static polarizability $\alpha_{ij}$ and second hyperpolarizability $\gamma_{ijkl}$ tensors are resp.~the first and third order derivatives of the electric dipole moment $\vec{\mu}$ with respect to the applied field $\vec{F}$, in the limit of an infinitesimal field
\begin{eqnarray} \allowdisplaybreaks
\alpha_{ij} & = & \left(\frac{\partial \mu_i(\vec{F})}{\partial F_j}\right)_{\vec{F} \rightarrow \vec{0}} ,\\
\gamma_{ijkl} & = & \left(\frac{\partial^3 \mu_i(\vec{F})}{\partial F_j \partial F_k \partial F_l}\right)_{\vec{F} \rightarrow \vec{0}} .
\end{eqnarray}
All subscripts denote Cartesian components. Because the electric dipole moment $\vec{\mu}$ is minus the derivative of the total energy $E$ with respect to an applied electric field $F$, $\alpha_{ij}$ and $\gamma_{ijkl}$ can also be obtained from
\begin{eqnarray}
\alpha_{ij} & = & - \left(\frac{\partial^2 E(\vec{F})}{\partial F_i \partial F_j}\right)_{\vec{F} \rightarrow \vec{0}} ,\\
\gamma_{ijkl} & = & - \left(\frac{\partial^4 E(\vec{F})}{\partial F_i \partial F_j \partial F_k \partial F_l}\right)_{\vec{F} \rightarrow \vec{0}} .
\end{eqnarray}
The energy $E(\vec{F})$ has to be evaluated with a wavefunction optimized for Eq. \eqref{hami2CH4}. For molecules extending mainly in one spatial dimension (assume this to be the z-direction), the main contribution to these tensors comes from the longitudinal components $\alpha_{zz}$ and $\gamma_{zzzz}$ . The hydrogen chains under study are in addition centrosymmetric. When the origin of the Cartesian coordinate system is chosen in the center of the chain, $E(\vec{F}) = E(-\vec{F})$ and the static longitudinal components of both quantities can be obtained with the following minimal finite difference formulae, where $\vec{F} = F \hat{z}$:
\begin{eqnarray}
\alpha_{zz}(F) & = & \left(\frac{2 E(0) - 2 E(F)}{F^2}\right)_{F \rightarrow 0} , \label{FD1Ch4}\\
\gamma_{zzzz}(F) & = & \left(\frac{-6 E(0) + 8 E(F) - 2 E(2F)}{F^4}\right)_{F \rightarrow 0} . \label{FD2Ch4}
\end{eqnarray}

\begin{figure}
 \centering
 \includegraphics[width=0.70\textwidth]{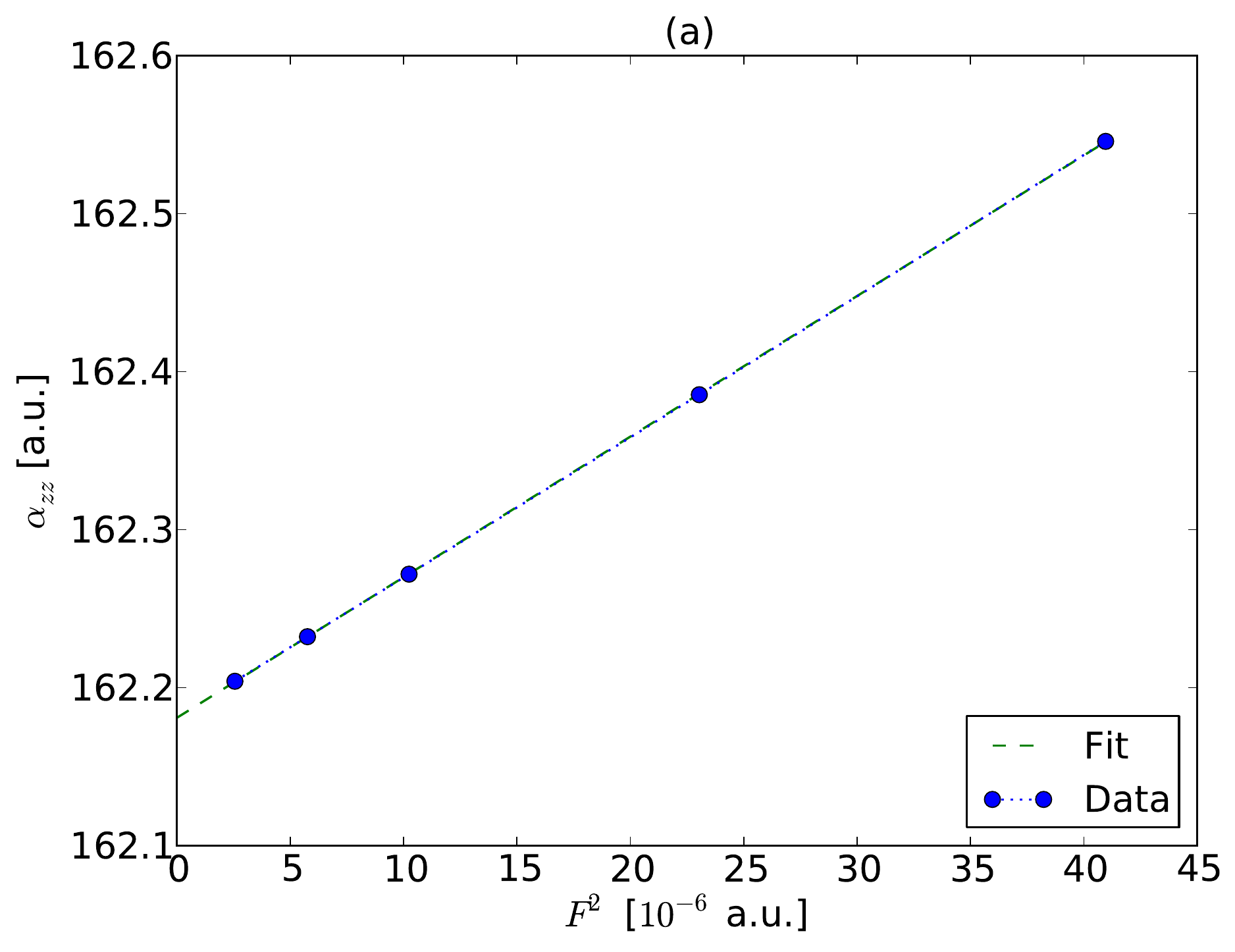}
 \includegraphics[width=0.70\textwidth]{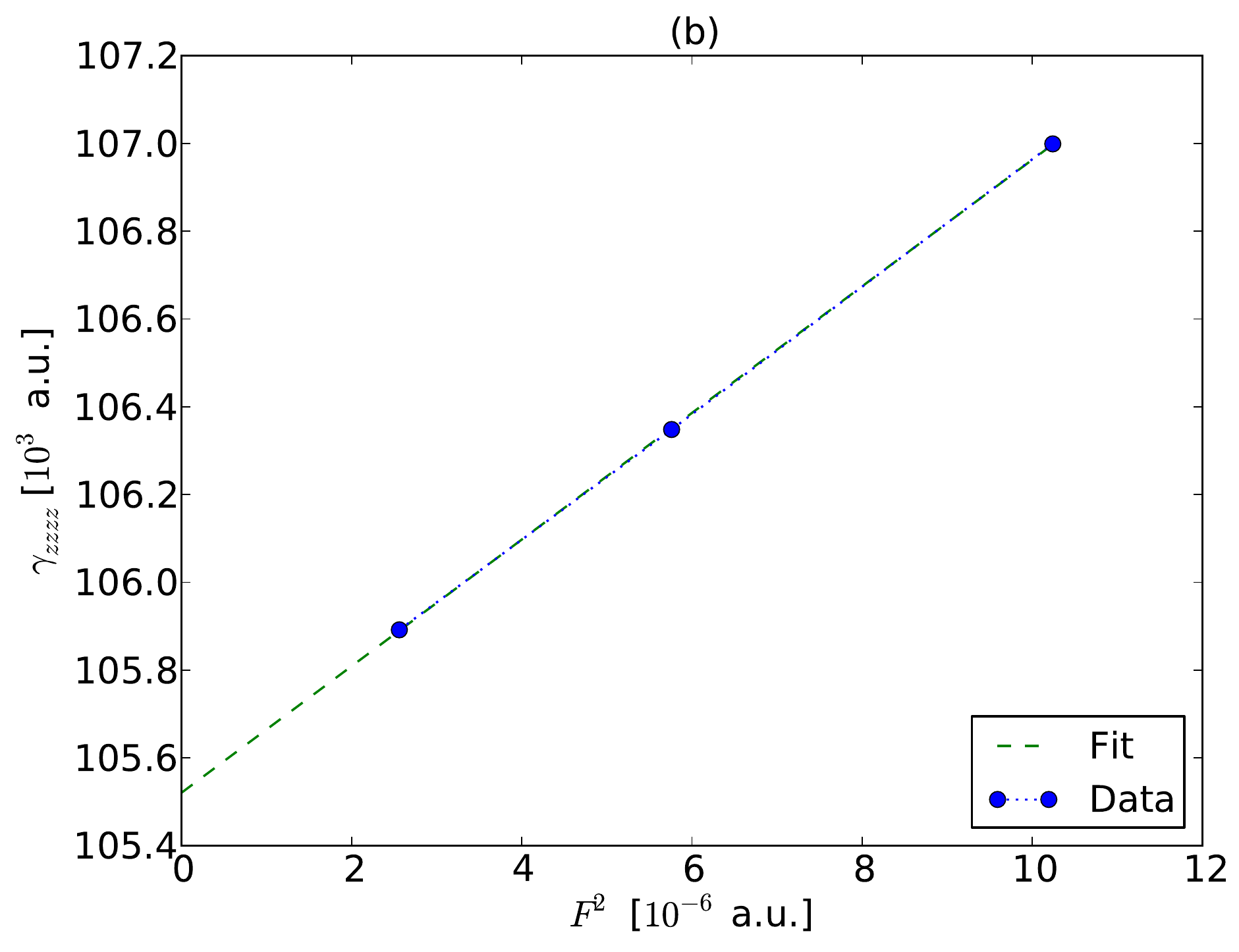}
 \caption{\label{extrapolplotJCP136} Finite field extrapolations of the static longitudinal polarizability (a) and second hyperpolarizability (b) for MPS calculations of a hydrogen chain of 36 atoms, with an alternate atom spacing of 2/3 a.u.~(see Sec.~V A), in the L\"owdin transformed STO-6G basis. The extrapolations are done with a least-squares fit to Eq. \eqref{extrapolequationCh4}.}
\end{figure}

The use of a finite field is explicitly incorporated in the notation: $\alpha_{zz}(F)$ and $\gamma_{zzzz}(F)$. We calculate both quantities for different values of $F$ and make a least-squares extrapolation to $F = 0$ according to
\begin{equation}
q(F) = q(0) + cF^2 \label{extrapolequationCh4} ,
\end{equation}
where $q$ can be $\alpha_{zz}$ or $\gamma_{zzzz}$. Values of $q(0)$ and $c$ are obtained by the fit. The procedure is illustrated in Fig. \ref{extrapolplotJCP136}.

The values of $F$ are chosen with care. If they are too large, higher order effects come into play and higher order terms have to be added to Eq. \eqref{extrapolequationCh4}. In that case, more calculations are required as more points $q(F)$ are needed to fit all parameters. Because the eigenstate energies $E_{\text{exact}}$ are approximated with MPS energies $E_D$ up to a certain accuracy, the energy differences in the numerators of Eqs. \eqref{FD1Ch4} and \eqref{FD2Ch4} have a constant error. If the field values become smaller, this absolute error for the energy differences is multiplied by increasing values of $F^{-2}$ or $F^{-4}$ and the absolute error of $\alpha_{zz}(F)$ and $\gamma_{zzzz}(F)$ becomes larger. The rms deviation of the quantities from the fit (as in Fig. \ref{extrapolplotJCP136}) will then be larger.

\newpage
\hspace{-\parindent}\textbf{IV. EQUALLY SPACED HYDROGEN CHAIN}
\vspace{0.3cm}

Our MPS program was tested for many small systems and the results were compared with ED, confirming the correctness of our implementation. Both for this application and the next one, all presented MPS data are converged according to Eq. \eqref{10minus11convergence}.

\vspace{0.7cm}
\hspace{-\parindent}\textbf{A. Introduction}
\vspace{0.3cm}

As a benchmark calculation, illustrating the possibilities of the program, the energy, as well as the static longitudinal polarizability and second hyperpolarizability of a hydrogen chain with 20 atoms are studied for different interatomic distances. The interatomic distance $R$ is defined by
\begin{equation}
\vcenter{\hbox{\setlength{\unitlength}{1cm}
\begin{picture}(6.7,1)
\put(0.1,0.375){H}
\put(0.4,0.5){\line(1,0){0.5}}
\put(1.0,0.375){H}
\put(1.3,0.5){\line(1,0){0.5}}
\put(1.9,0.375){H}
\put(2.2,0.5){\line(1,0){0.5}}
\put(2.8,0.375){H}
\put(3.1,0.5){\line(1,0){0.5}}
\put(3.7,0.375){H}
\put(4.0,0.5){\line(1,0){0.5}}
\put(4.6,0.375){H}
\put(4.9,0.5){\line(1,0){0.5}}
\put(5.5,0.375){H}
\put(5.8,0.5){\line(1,0){0.5}}
\put(6.4,0.375){H}
\put(0.5,0.6){$R$}
\put(1.4,0.6){$R$}
\put(2.3,0.6){$R$}
\put(3.2,0.6){$R$}
\put(4.1,0.6){$R$}
\put(5.0,0.6){$R$}
\put(5.9,0.6){$R$}
\end{picture}}} . \label{JCP136equallySpacedFormula}
\end{equation}

\begin{figure}
 \centering
 \includegraphics[width=0.70\textwidth]{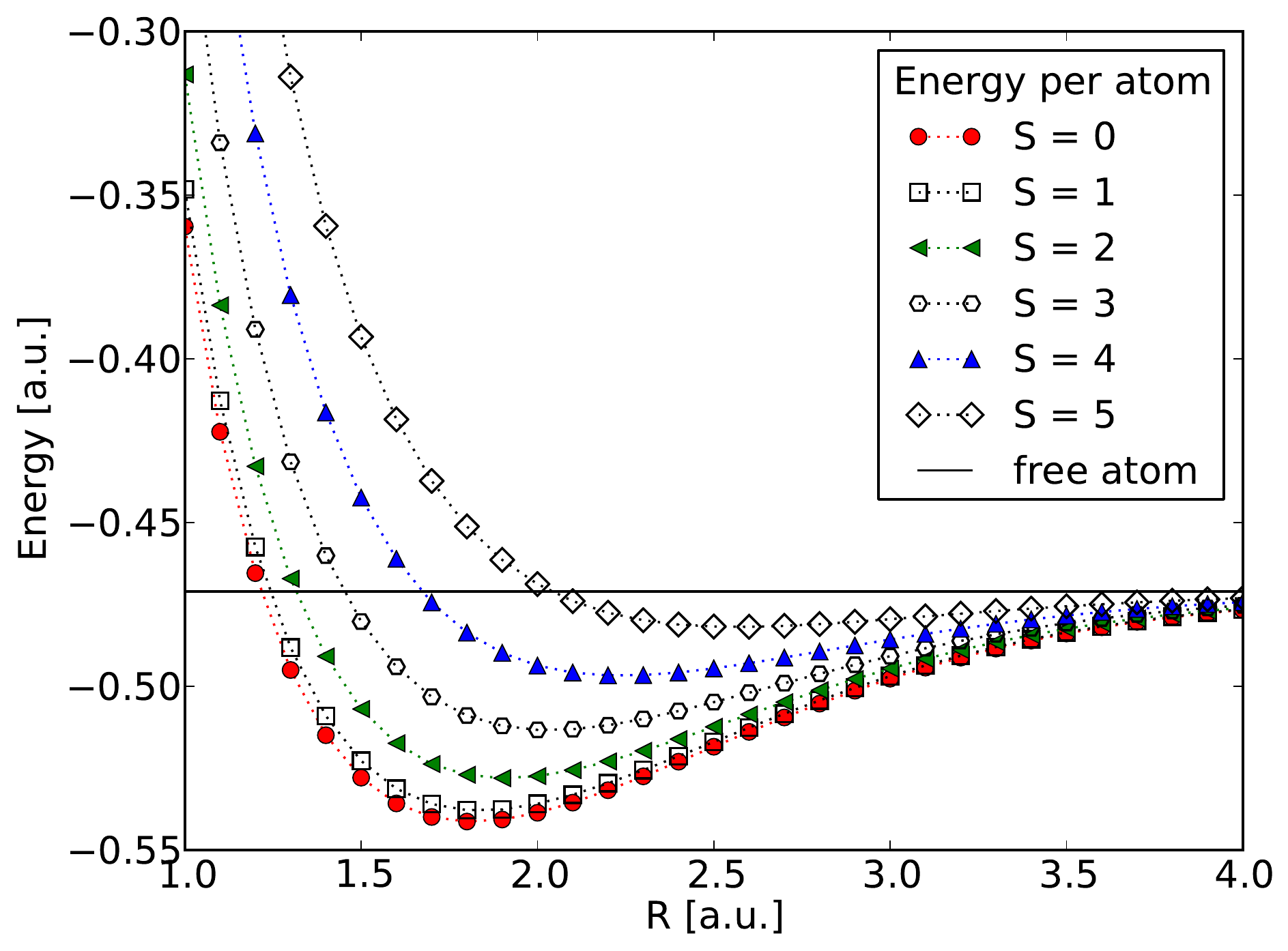}
 \caption{\label{MITenergyJCP136} The ground state energy per atom for an equally spaced hydrogen chain of 20 atoms is shown for 6 different spin states.}
\end{figure}

The study is performed for the ground states in 6 different spin symmetry sectors $S = 0,1,...,5$. The virtual dimension per symmetry sector was truncated to $D = 64$ for all the results in this section. The energies were determined for 6 field values $F = 0$, 0.0008, 0.0012, 0.0016, 0.0024, and 0.0032 a.u., yielding 5 points for the $\alpha_{zz}$ extrapolation and 3 points for the $\gamma_{zzzz}$ extrapolation. The minimal basis set STO-6G (Ref. \cite{PopleIniPaper}) was used as single-particle degrees of freedom.
\newpage
\hspace{-\parindent}\textbf{B. Results and discussion}
\vspace{0.3cm}

As is already well known, the MPS ansatz is able to capture static correlation and hence gives correct potential energy surfaces (PES) whereas HF based methods break down for large interatomic distances \cite{hachmann:144101}. The energy per atom as a function of interatomic distance is shown for the 6 spin states in Fig. \ref{MITenergyJCP136}. The energy rises with increasing spin. In the limit of large $R$, all PESs converge in accordance with the noninteracting atom picture.

In the range of R values shown, the equally spaced hydrogen chain is known to make a metal-insulator transition. The transition point is marked by diverging response properties in the TD limit. An earlier ED study has shown that $\alpha_{zz} N^{-2}$ in function of interatomic distance $R$, with $N$ the number of atoms, converges to a limiting curve in the TD limit \cite{QUA:QUA23047}.

\begin{figure}
 \centering
 \includegraphics[width=0.70\textwidth]{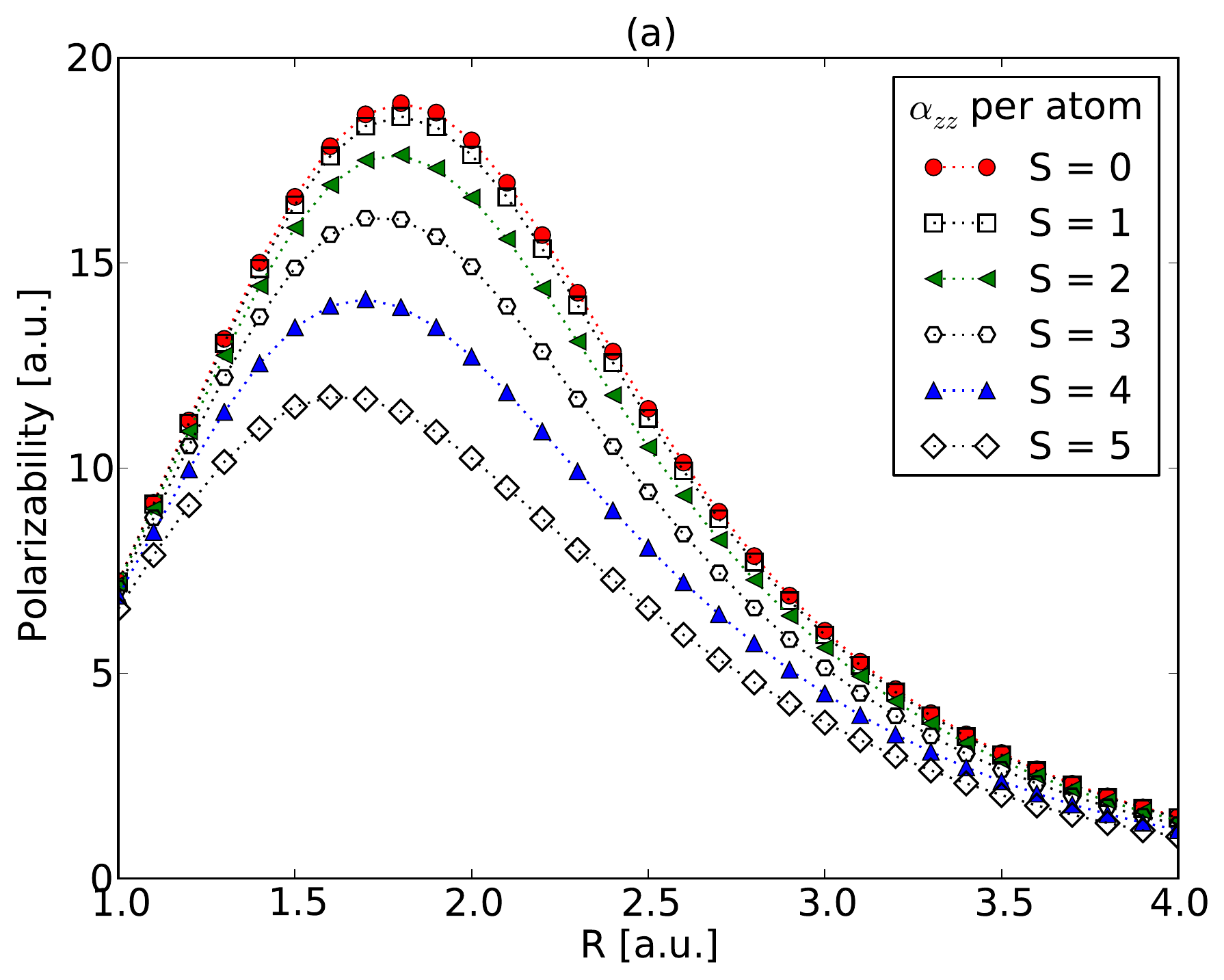}
 \includegraphics[width=0.70\textwidth]{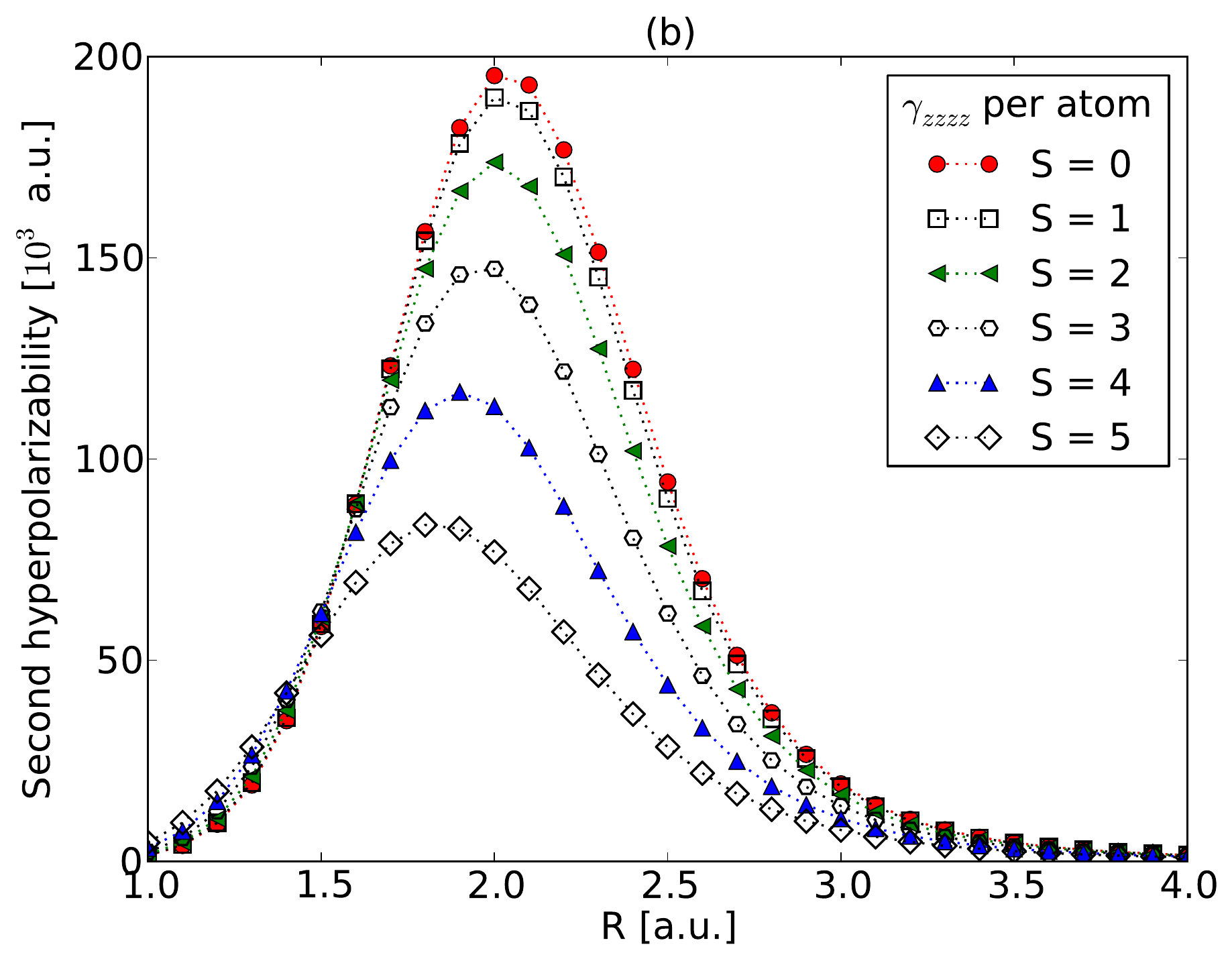}
 \caption{\label{MITresponsesJCP136} The polarizabilities (a) and second hyperpolarizabilities (b) per atom for an equally spaced hydrogen chain of 20 atoms are shown for 6 different spin ground states.}
\end{figure}

The spin dependence of the optical response properties is shown in Fig. \ref{MITresponsesJCP136}. For increasing spin, both the polarizability and second hyperpolarizability peaks decrease and shift towards smaller values of $R$. The peaks of the polarizability also occur at slightly smaller values of $R$ than the corresponding peaks of the second hyperpolarizability. Both responses vanish in the limit of large $R$ as a minimal basis set is used \cite{QUA:QUA23047}.

An alternative method to determine the polarizability and second hyperpolarizability is the sum over states (SOS) perturbation expansion \cite{chemrevpersoons}. Note that the dipole moment in the SOS expression commutes with spin operators. Different spin states can hence be treated separately. Two counteracting effects occur in this expression. The number of terms in the summation rapidly decreases with increasing spin because fewer high-spin configurations can be built with $N$ electrons in $L$ orbitals. The magnitude of the terms is expected to be larger for higher spin states due to the smaller energy differences in the denominator. Both effects combined result in properties of the same order of magnitude for the different spin states treated in this paper. The diminishing peak can then be attributed to the smaller number of possible spin configurations. Note that this is only a heuristic argument, as we have not performed any calculations related to the SOS expression. 

\vspace{0.7cm}
\hspace{-\parindent}\textbf{V. A CHAIN OF H$_2$ MOLECULES}
\vspace{0.3cm}

In Sec.~IV, we have studied a system with changing static correlation. Here, we look at a system where the static correlation remains roughly the same but where the electron delocalization changes. 

\vspace{0.7cm}
\hspace{-\parindent}\textbf{A. Introduction}
\vspace{0.3cm}

In this section, the optical properties of hydrogen chains with different intra- and intermolecular distances are studied
\begin{equation}
\vcenter{\hbox{\setlength{\unitlength}{1cm}
\begin{picture}(6.7,1)
\put(0.1,0.375){H}
\put(0.4,0.5){\line(1,0){0.5}}
\put(1.0,0.375){H}
\put(1.3,0.45){......}
\put(1.9,0.375){H}
\put(2.2,0.5){\line(1,0){0.5}}
\put(2.8,0.375){H}
\put(3.1,0.45){......}
\put(3.7,0.375){H}
\put(4.0,0.5){\line(1,0){0.5}}
\put(4.6,0.375){H}
\put(4.9,0.45){......}
\put(5.5,0.375){H}
\put(5.8,0.5){\line(1,0){0.5}}
\put(6.4,0.375){H}
\put(0.5,0.6){$R_f$}
\put(1.4,0.6){$R$}
\put(2.3,0.6){$R_f$}
\put(3.2,0.6){$R$}
\put(4.1,0.6){$R_f$}
\put(5.0,0.6){$R$}
\put(5.9,0.6){$R_f$}
\end{picture}}} . \label{JCP136jetajetaDimerizedHchain}
\end{equation}

The intramolecular distance is kept fixed at $R_f$ = 2 a.u., whereas the intermolecular distance $R$ can be 2.5, 3 or 4 a.u., in analogy with previous studies \cite{PhysRevA.52.178, PhysRevA.52.1039, QUA:QUA22177}. In the following, an H$_2$ constituent will be called a molecule even if $R_f$ is far from the H$_2$ equilibrium distance. With decreasing $R$, the system changes from a collection of separated H$_2$ molecules to a chain where the electrons are delocalized \cite{Umari2}, whereas the static correlation remains similar due to the constant bond length $R_f$ of the H$_2$ molecule.

\begin{table}
\centering
\caption{\label{dentabelFFapplic2} Values of $F$ per intermolecular distance $R$.}
\begin{tabular}{c c c c c c c c c}
  \hline
  \hline
  $R$ (a.u.) & $\quad$ & \multicolumn{7}{c}{$F$ ($10^{-3}$ a.u.)} \\
  \hline
  2.5 & & 0.0$~$ & 0.8$~$ & 1.2$~$ & 1.6$~$ & 2.4$~$ & 3.2$~$ & \\
  3.0 & & 0.0$~$ & 1.6$~$ & 2.4$~$ & 3.2$~$ & 4.8$~$ & 6.4$~$ & \\
  4.0 & & 0.0$~$ & 1.6$~$ & 1.8$~$ & 2.0$~$ & 3.2$~$ & 3.6$~$ & 4.0\\
  \hline
  \hline
\end{tabular}
\end{table}

Only the absolute ground state ($S = 0$) was targeted, but for different chain lengths, LOT, and basis sets. All calculations for the basis sets STO-6G, 6-31G \cite{BasisSet2JCP136}, and 6-31G(d,p) (Ref. \cite{BasisSet3JCP136}) were performed with a virtual dimension per symmetry sector $D$ of resp. 32, 64, and 120, independent of chain length and $R$. The fields for which ground state calculations were performed are shown in Table \ref{dentabelFFapplic2}. They depend on the intermolecular distance $R$, but are independent of chain length, basis set, and LOT. The LOTs that were studied are MPS, HF, second order M{\o}ller-Plesset perturbation theory (MP2), coupled cluster with singles and doubles (CCSD) and coupled cluster with singles and doubles and perturbative triples (CCSD(T)). The HF, MP2, CCSD, and CCSD(T) calculations were performed with the molecular electronic structure program DALTON \cite{dalton}.

\vspace{0.7cm}
\hspace{-\parindent}\textbf{B. Results and discussion}
\vspace{0.3cm}

\begin{table}
\caption{\label{allH8} All polarizability and second hyperpolarizability data for $\text{H}_8$.}
\centering
\begin{tabular}{c c c r r r r r}
  \hline
  \hline
  Quantity & R (a.u.) & Basis set & HF & MP2 & CCSD & CCSD(T) & MPS \\
  \hline
  $\alpha_{zz}$ (a.u.) & 2.5 & STO-6G & 63.93 & 53.77 & 41.61 & 42.26 & 42.47\\
  & 2.5 & 6-31G & 105.38 & 96.68 & 80.20 & 81.34 & 81.78\\
  & 2.5 & 6-31G(d,p) & 106.03 & 102.48 & 91.61 & 92.75 & 93.12\\
  & 3.0 & STO-6G & 43.63 & 36.67 & 29.73 & 30.00 & 30.10\\
  & 3.0 & 6-31G & 80.75 & 73.16 & 61.80 & 62.40& 62.66 \\
  & 3.0 & 6-31G(d,p) & 80.44 & 76.20 & 68.73 & 69.31 & 69.50\\
  & 4.0 & STO-6G & 29.26 & 25.21 & 21.20 & 21.27 & 21.31\\
  & 4.0 & 6-31G & 61.77 & 55.46 & 47.62 & 47.84 & 47.97\\
  & 4.0 & 6-31G(d,p) & 60.90 & 56.49 & 51.52 & 51.70 & 51.77\\
  \hline
  $\gamma_{zzzz}$ ($10^3$ a.u.) & 2.5 & STO-6G & 33.00 & 36.96 & 24.36 & 24.78 & 25.30\\
  & 2.5 & 6-31G & 79.02 & 104.36 & 89.03 & 90.56 & 91.72\\
  & 2.5 & 6-31G(d,p) & 74.40 & 97.37 & 90.28 & 93.57 & 94.87\\
  & 3.0 & STO-6G & 15.50 & 14.33 & 9.90 & 10.20 & 10.30 \\
  & 3.0 & 6-31G & 48.98 & 58.89 & 47.53 & 48.80 & 49.34\\
  & 3.0 & 6-31G(d,p) & 47.25 & 58.10 & 49.78 & 51.98 & 52.62\\
  & 4.0 & STO-6G & 3.17 & 2.66 & 2.41 & 2.44 & 2.44\\
  & 4.0 & 6-31G & 17.63 & 19.75 & 17.60 & 17.85 & 17.92\\
  & 4.0 & 6-31G(d,p) & 17.38 & 19.53 & 17.42 & 17.88 & 18.00\\
  \hline
  \hline
\end{tabular}
\end{table}

\begin{figure}
 \centering
 \includegraphics[width=0.70\textwidth]{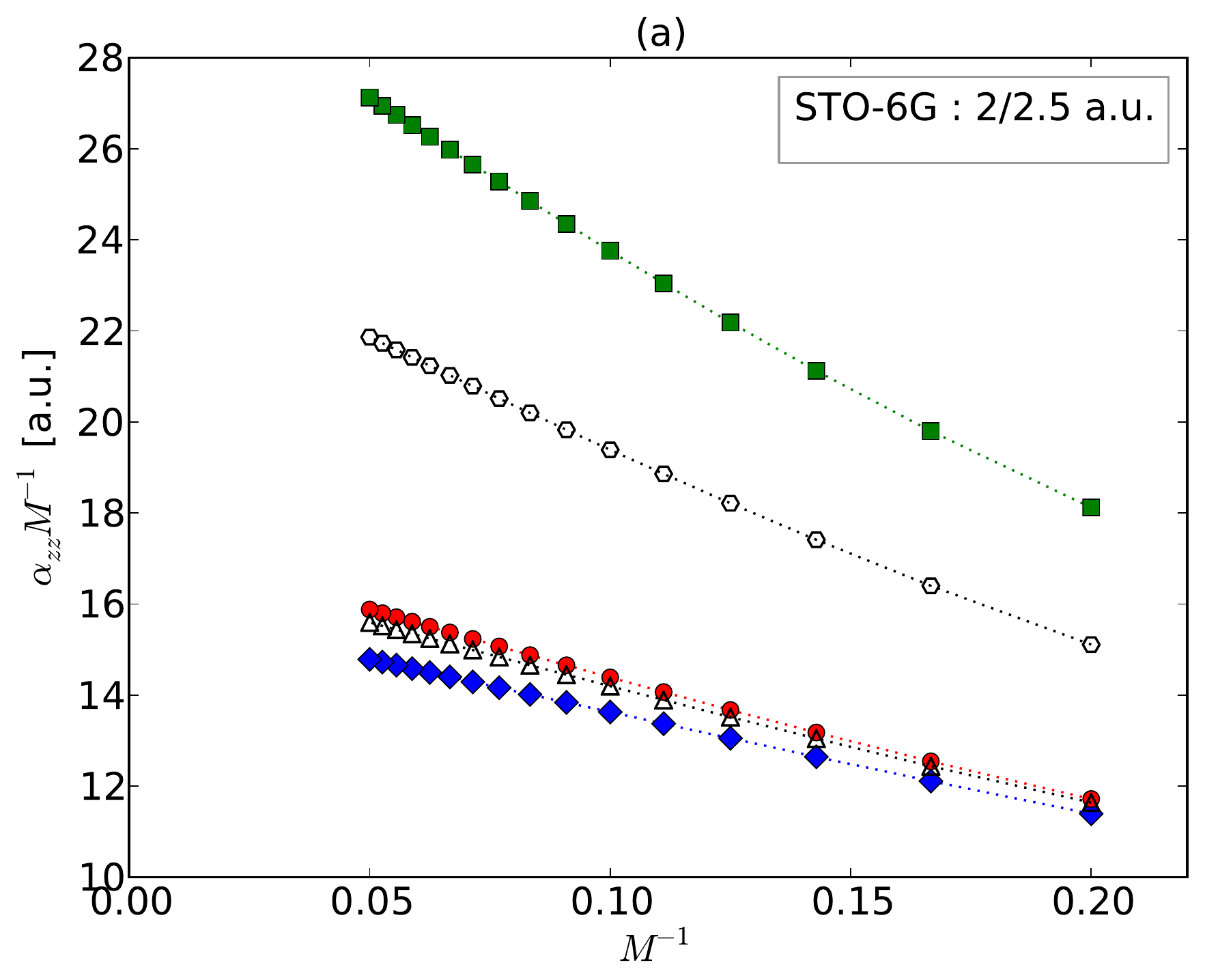}
 \includegraphics[width=0.70\textwidth]{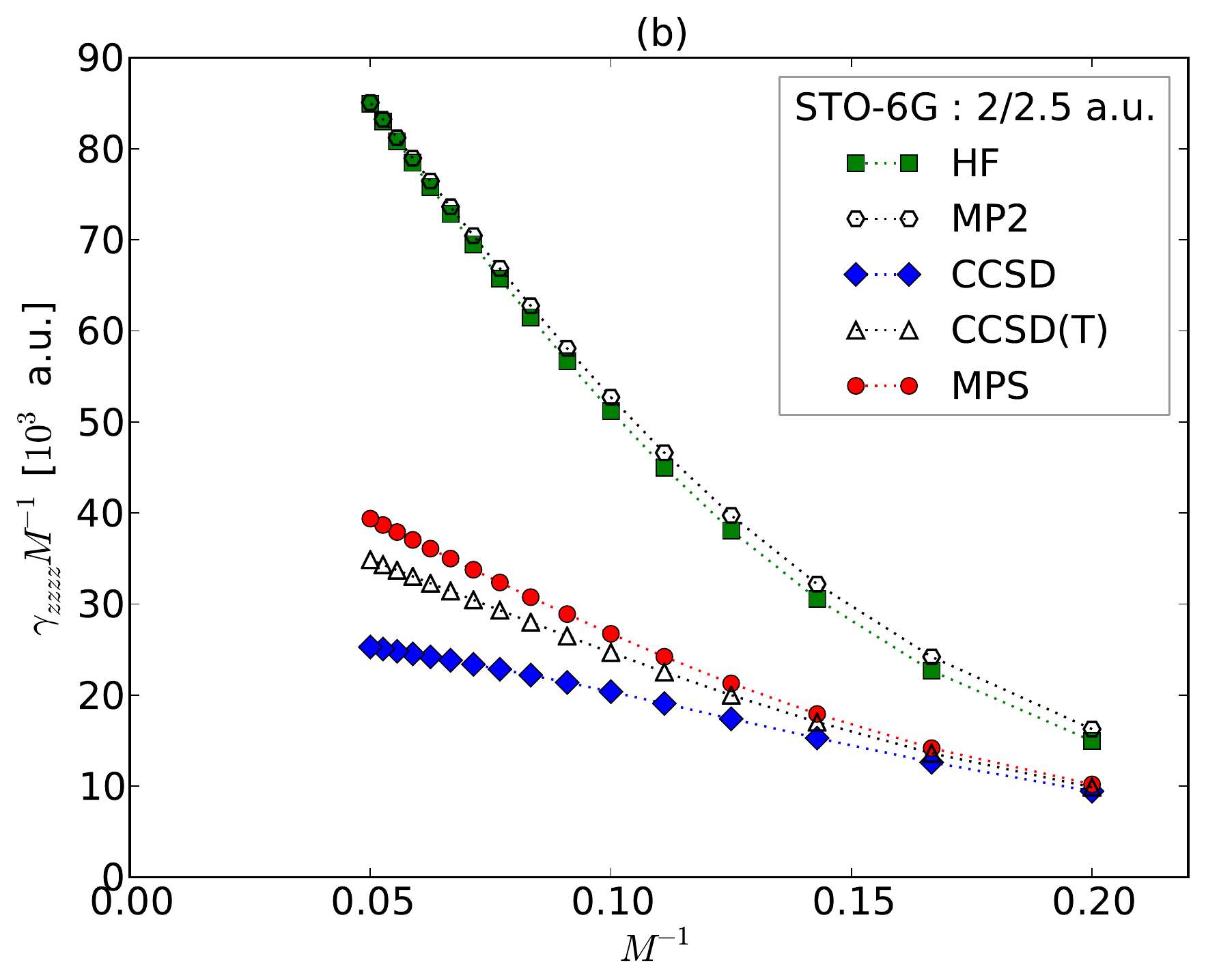}
 \caption{\label{Fig5JCP136-ab} Polarizabilities (a) and second hyperpolarizabilities (b) of hydrogen chains with intramolecular distance 2 a.u. and intermolecular distance 2.5 a.u., calculated with several LOTs in the L\"owdin transformed STO-6G basis.}
\end{figure}

\begin{figure}
 \centering
 \includegraphics[width=0.70\textwidth]{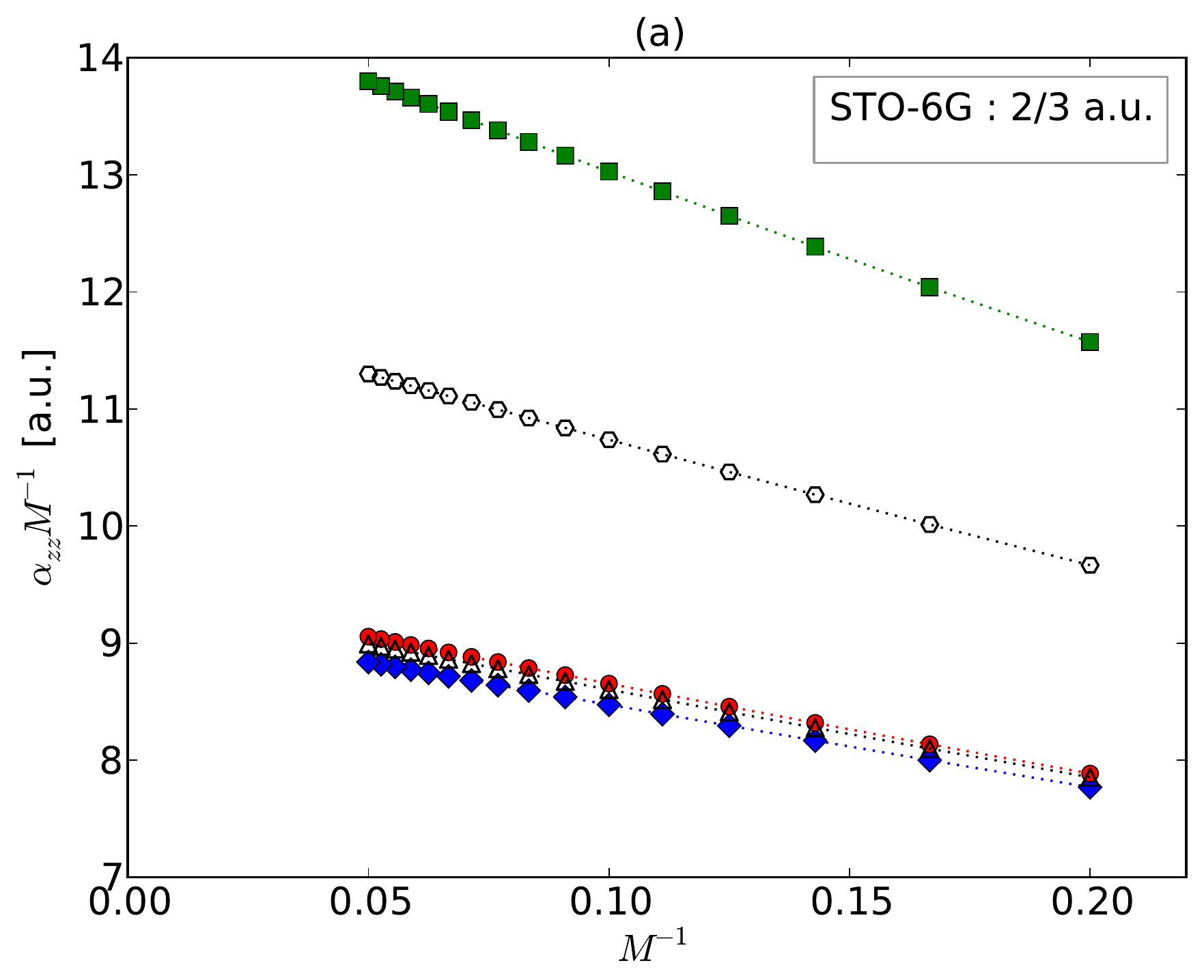}
 \includegraphics[width=0.70\textwidth]{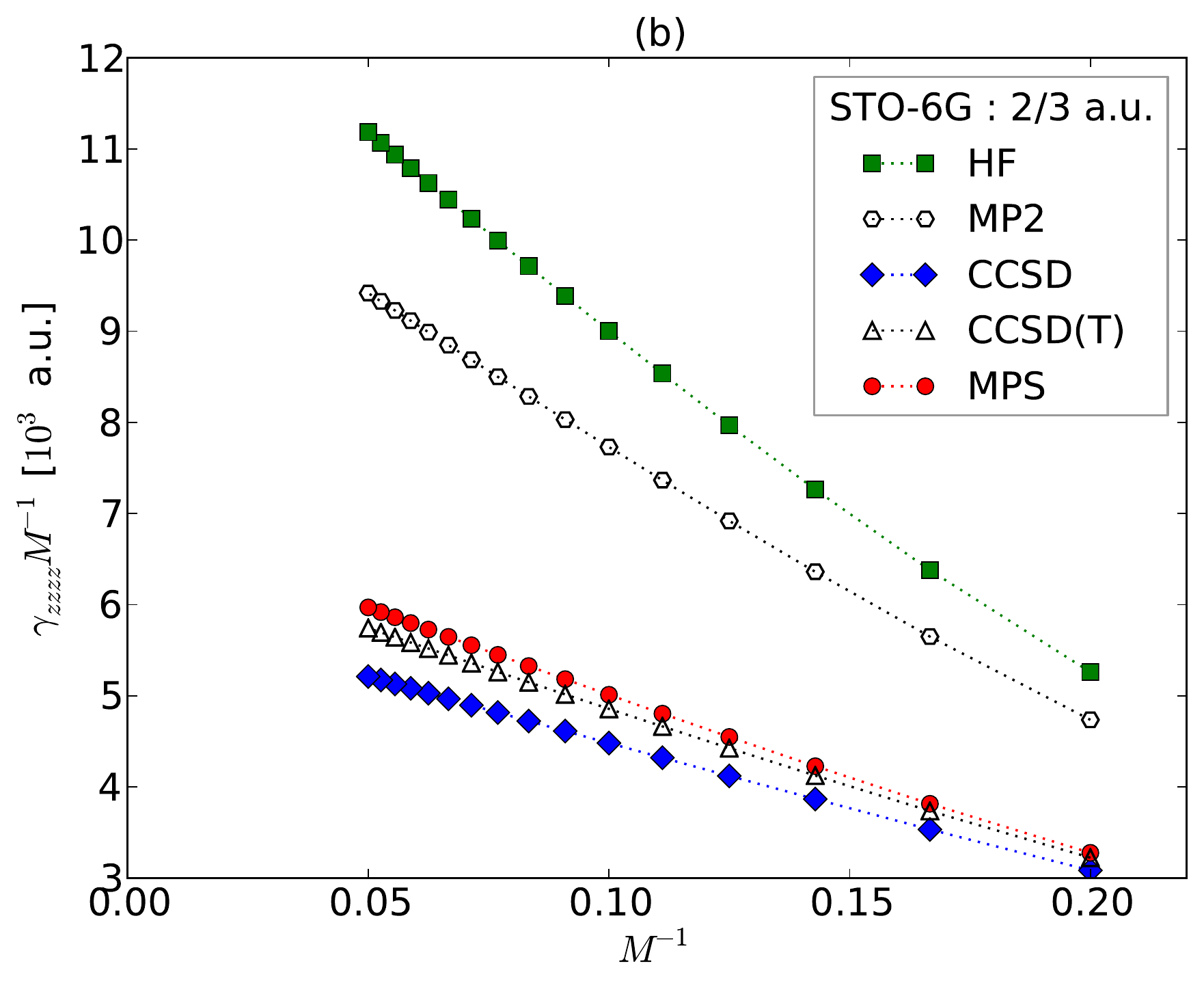}
 \caption{\label{Fig5JCP136-cd} Polarizabilities (a) and second hyperpolarizabilities (b) of hydrogen chains with intramolecular distance 2 a.u. and intermolecular distance 3.0 a.u., calculated with several LOTs in the L\"owdin transformed STO-6G basis.}
\end{figure}

\begin{figure}
 \centering
 \includegraphics[width=0.70\textwidth]{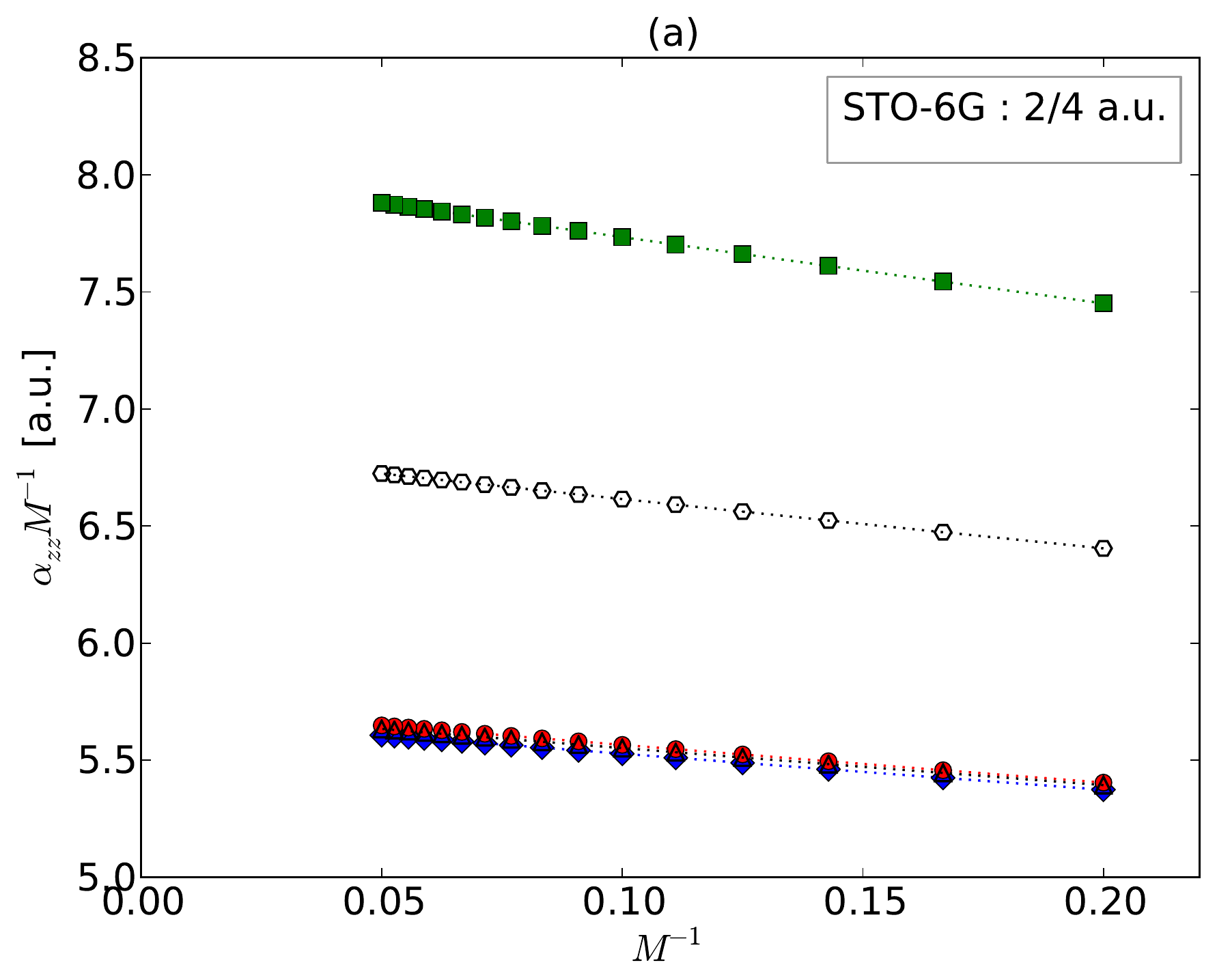}
 \includegraphics[width=0.70\textwidth]{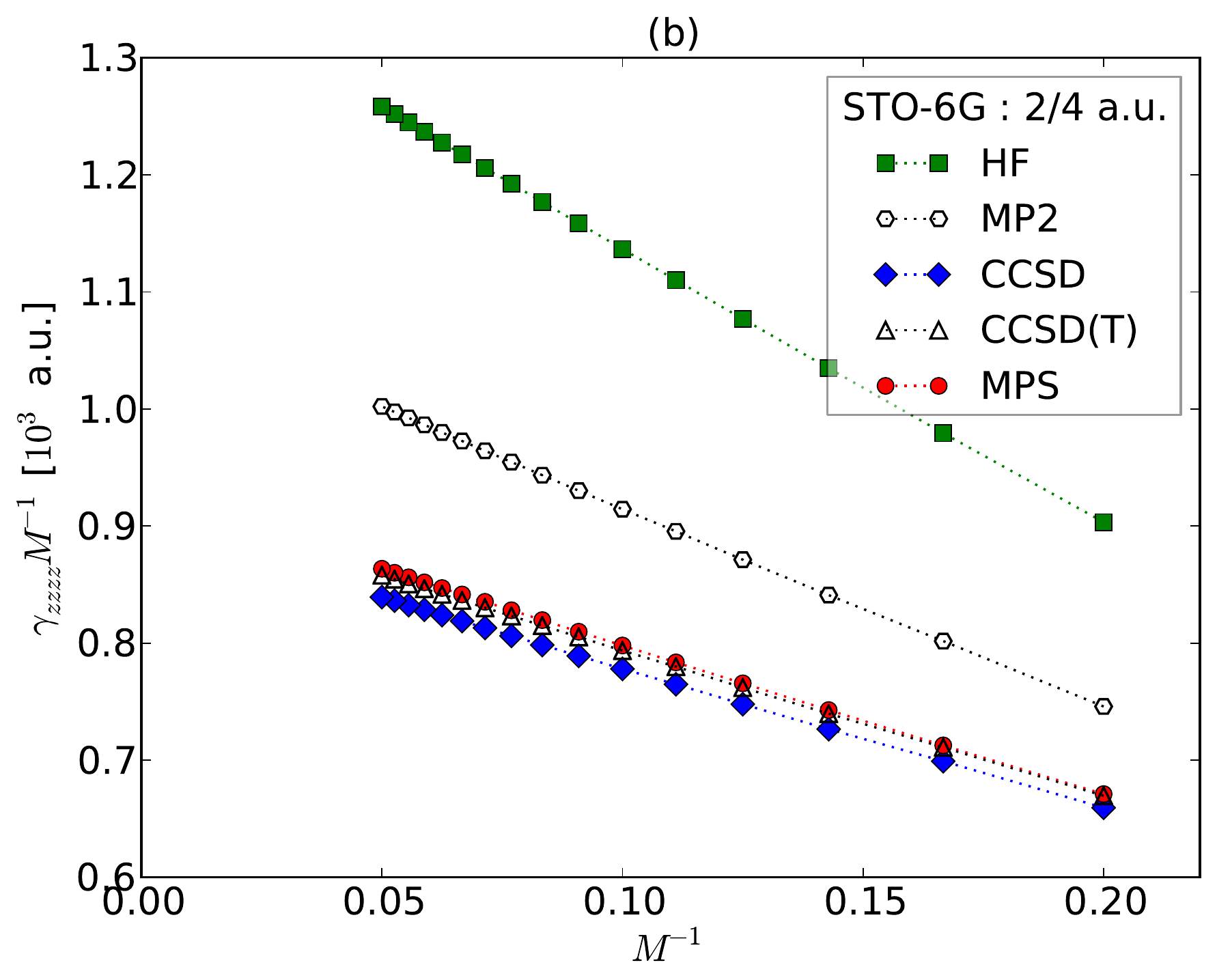}
 \caption{\label{Fig5JCP136-ef} Polarizabilities (a) and second hyperpolarizabilities (b) of hydrogen chains with intramolecular distance 2 a.u. and intermolecular distance 4.0 a.u., calculated with several LOTs in the L\"owdin transformed STO-6G basis.}
\end{figure}

\begin{figure}
 \centering
 \includegraphics[width=0.70\textwidth]{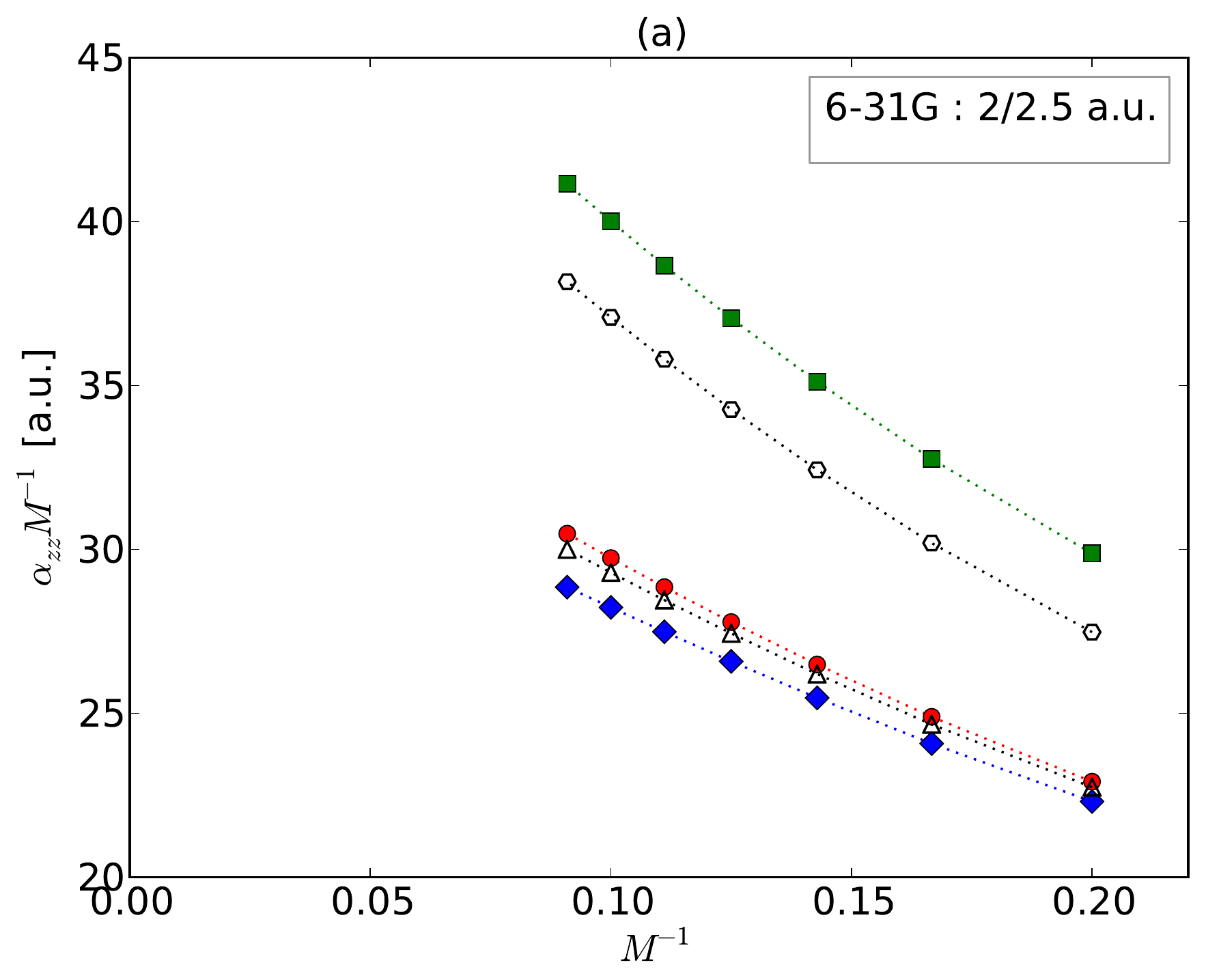}
 \includegraphics[width=0.70\textwidth]{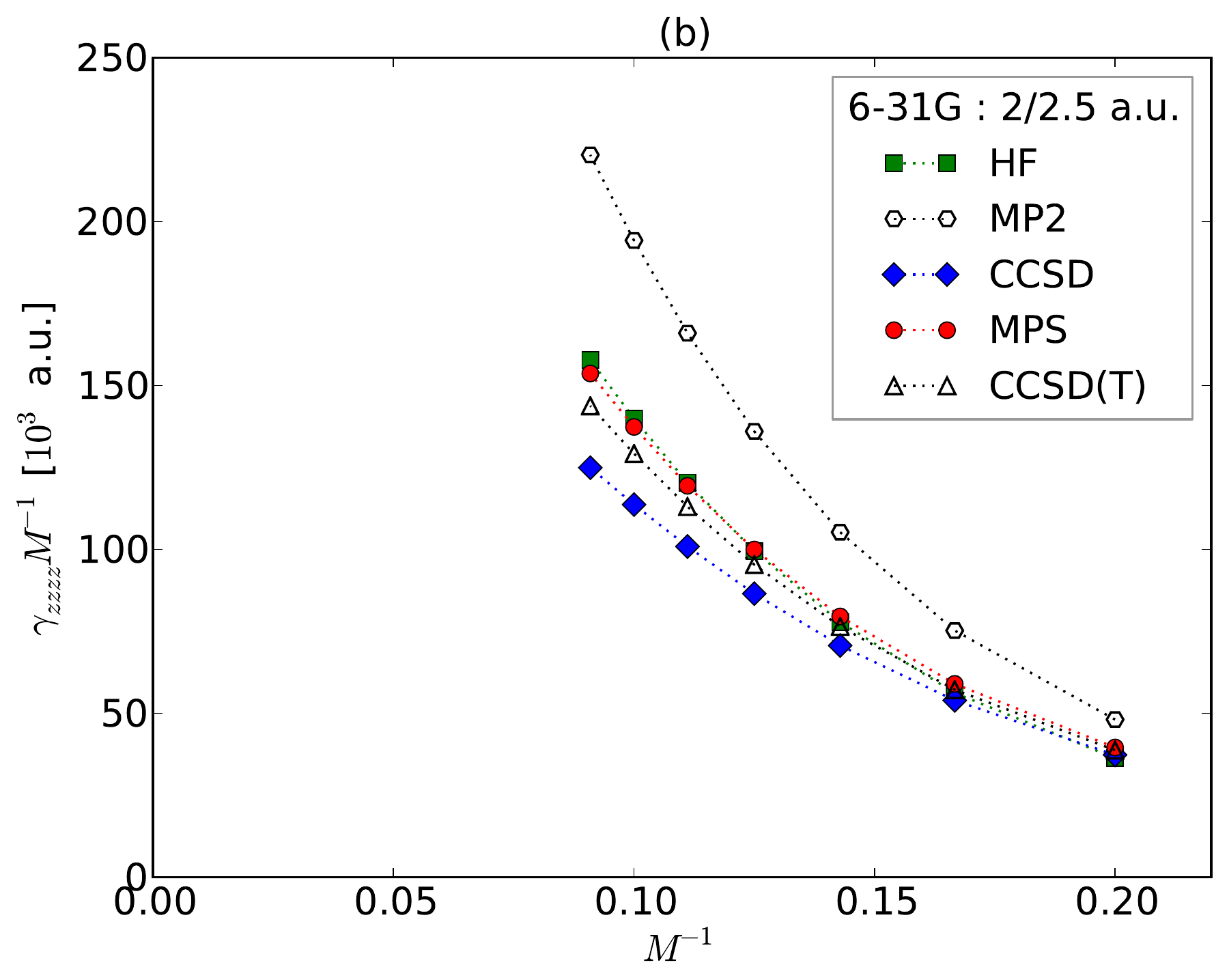}
 \caption{\label{Fig6JCP136-ab} Polarizabilities (a) and second hyperpolarizabilities (b) of hydrogen chains with intramolecular distance 2 a.u. and intermolecular distance 2.5 a.u., calculated with several LOTs in the L\"owdin transformed 6-31G basis.}
\end{figure}

\begin{figure}
 \centering
 \includegraphics[width=0.70\textwidth]{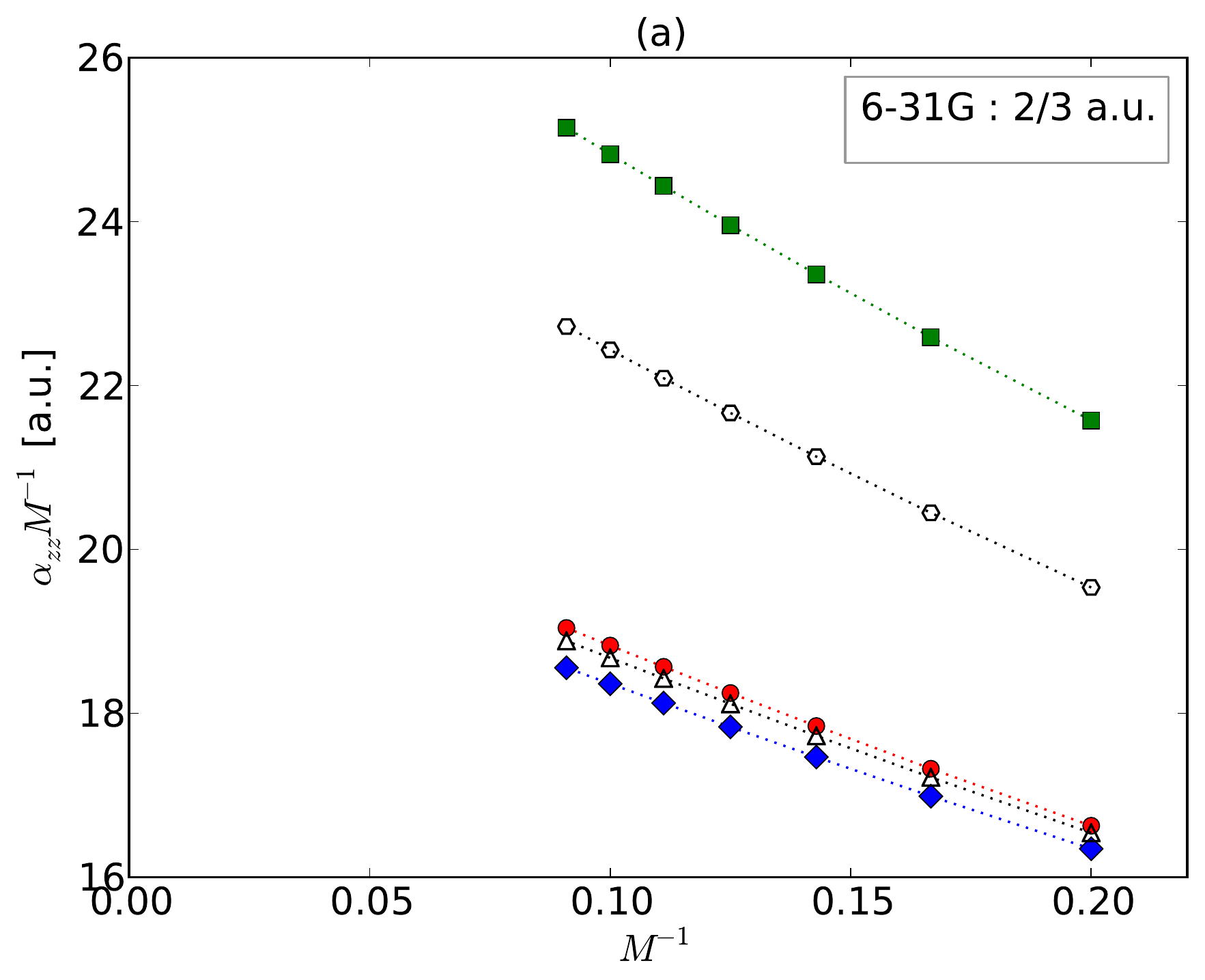}
 \includegraphics[width=0.70\textwidth]{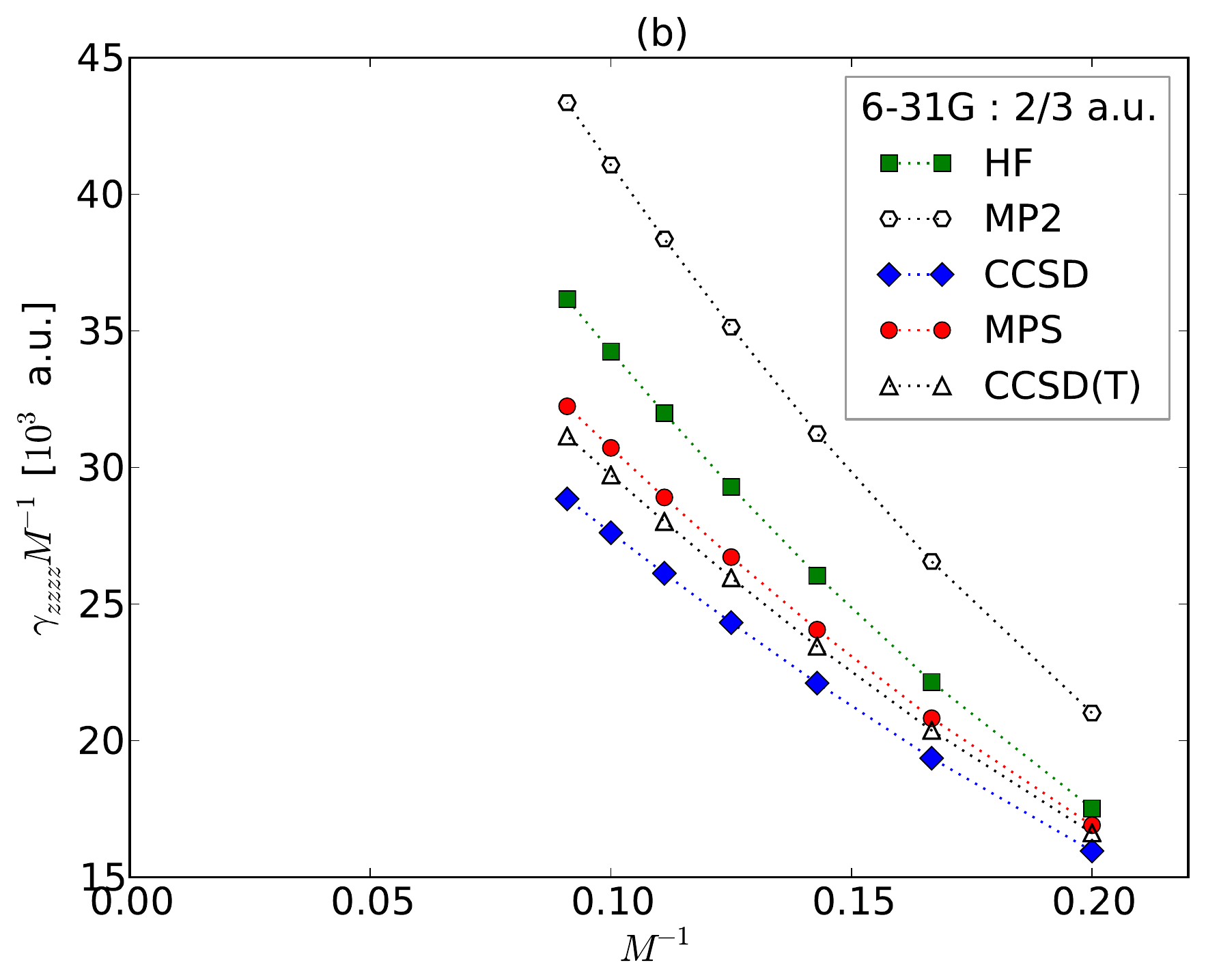}
 \caption{\label{Fig6JCP136-cd} Polarizabilities (a) and second hyperpolarizabilities (b) of hydrogen chains with intramolecular distance 2 a.u. and intermolecular distance 3.0 a.u., calculated with several LOTs in the L\"owdin transformed 6-31G basis.}
\end{figure}

\begin{figure}
 \centering
 \includegraphics[width=0.70\textwidth]{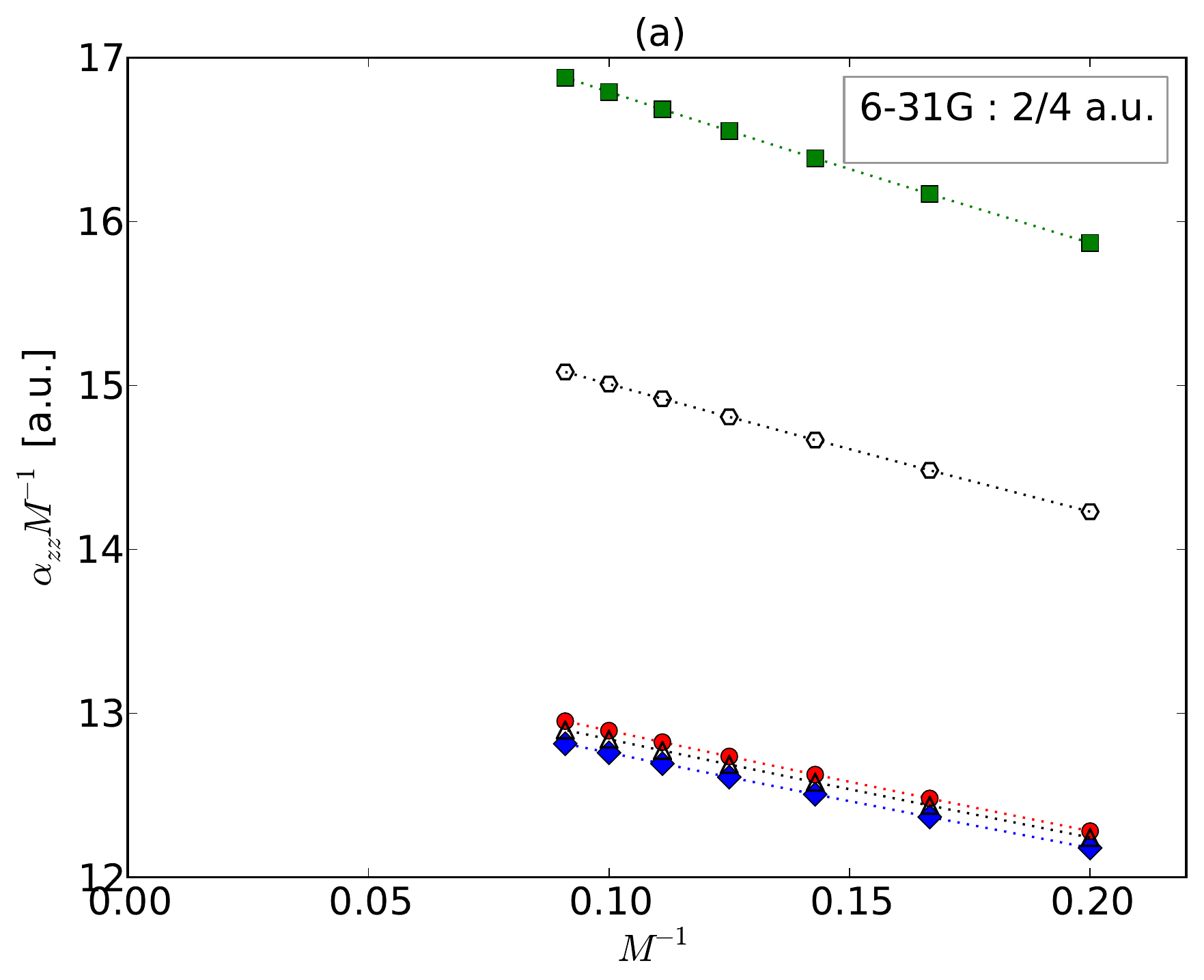}
 \includegraphics[width=0.70\textwidth]{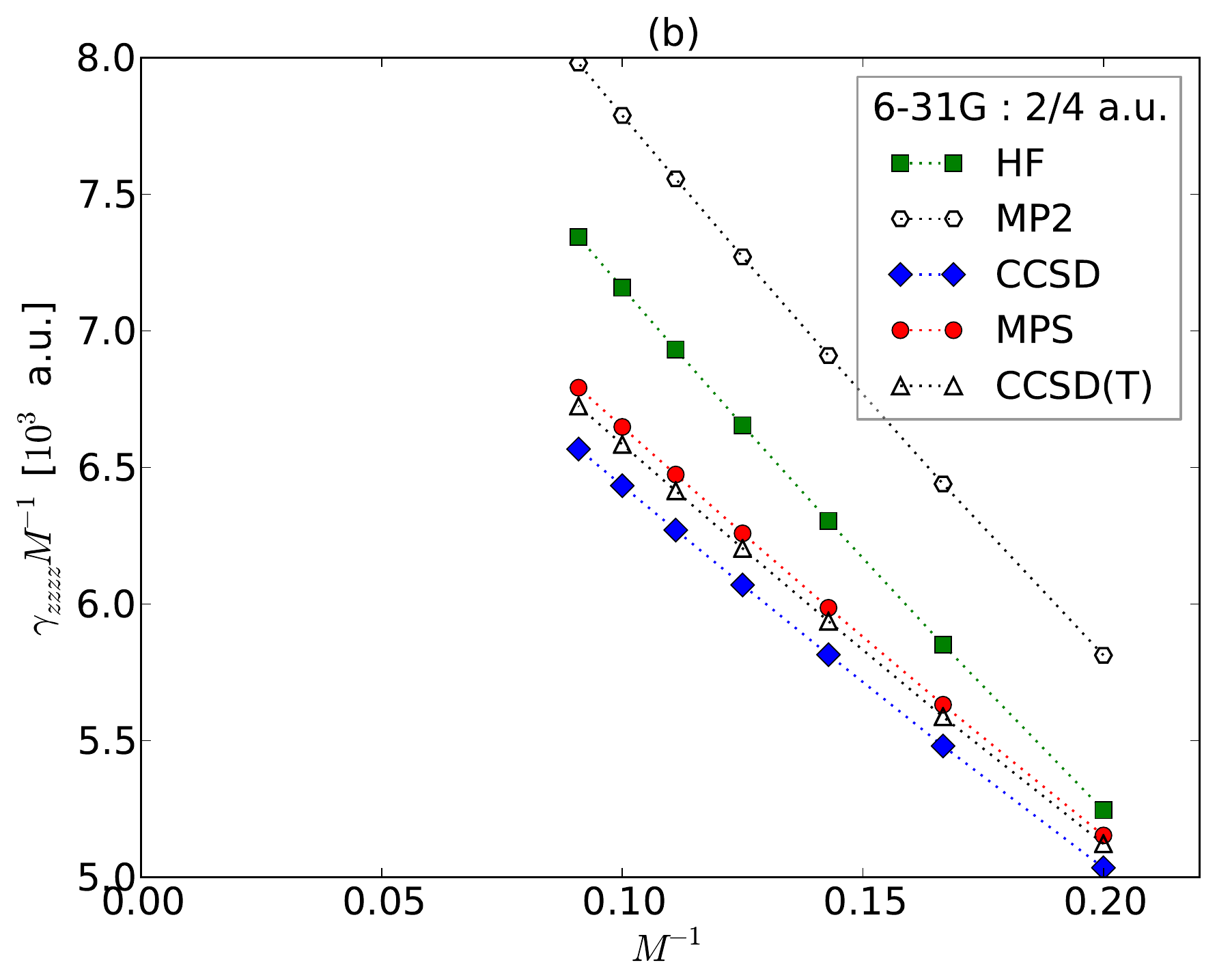}
 \caption{\label{Fig6JCP136-ef} Polarizabilities (a) and second hyperpolarizabilities (b) of hydrogen chains with intramolecular distance 2 a.u. and intermolecular distance 4.0 a.u., calculated with several LOTs in the L\"owdin transformed 6-31G basis.}
\end{figure}

For the basis sets STO-6G and 6-31G, $\alpha_{zz}$ and $\gamma_{zzzz}$ were calculated for an increasing number of H$_2$ units $M$. The values per molecule, $\alpha_{zz} M^{-1}$ and $\gamma_{zzzz} M^{-1}$, are presented for STO-6G in Figs. \ref{Fig5JCP136-ab}, \ref{Fig5JCP136-cd} and \ref{Fig5JCP136-ef} and for 6-31G in Figs. \ref{Fig6JCP136-ab}, \ref{Fig6JCP136-cd} and \ref{Fig6JCP136-ef} for the different intermolecular distances and LOTs. For the 6-31G(d,p) basis set, the largest chain was H$_8$. All H$_8$ data are shown in Table \ref{allH8}.

From Table \ref{allH8}, it can be observed that for corresponding intermolecular distances and LOTs, the STO-6G polarizability and second hyperpolarizability values are significantly lower than the values obtained with the 6-31G and 6-31G(d,p) basis sets. The possible movement of electrons in a minimal basis set is of course restricted. The 6-31G and 6-31G(d,p) results are also much closer to each other than to the minimal basis set results, in agreement with Champagne \textit{et al.} \cite{PhysRevA.52.178, PhysRevA.52.1039}.

For the polarizability of long chains, a clear order exists for the LOTs, which is the same for the three intermolecular distances and the STO-6G and 6-31G basis sets,
\begin{equation}
\alpha_{zz}^{\text{HF}} > \alpha_{zz}^{\text{MP2}} > \alpha_{zz}^{\text{MPS}} > \alpha_{zz}^{\text{CCSD(T)}} > \alpha_{zz}^{\text{CCSD}} . \label{polorderCh4}
\end{equation}
This order is in agreement with previous work \cite{PhysRevA.52.1039}, which looks at small basis sets. For larger basis sets, it was found that the HF polarizability tends to drop below the MP2 values for decreasing values of the intermolecular distance $R$ (increasing electron delocalization) \cite{QUA:QUA22177}. There is also a clear order in the deviation between the polarizability obtained with a certain LOT and the MPS result
\begin{equation}
\Delta \alpha_{zz}^{\text{HF}} > \Delta \alpha_{zz}^{\text{MP2}} > \Delta \alpha_{zz}^{\text{CCSD}} > \Delta \alpha_{zz}^{\text{CCSD(T)}} . \label{hyperpolorderCh4}
\end{equation}

For the second hyperpolarizability of long chains, a clear order exists for all LOTs except HF. Again this order is the same for the three intermolecular distances and the STO-6G and 6-31G basis sets, and equals the one in Eq. \eqref{polorderCh4} when $\alpha_{zz}^{\text{HF}}$ is excluded. The HF second hyperpolarizability tends to drop below the MP2 values for decreasing values of the intermolecular distance $R$ (increasing electron delocalization) and for increasing basis sets. For even larger basis sets, the HF values drop below the CCSD values, but the order of the other methods is also left unchanged \cite{QUA:QUA22177}. It is intriguing that for the second hyperpolarizability, the mean-field (HF) results have no fixed position relative to the other correlated methods. This shows that the approximate treatment of electron correlation by MP2 or CCSD and CCSD(T) does not lead to a smooth transition from mean-field theory towards ED. Instead, the final value of $\gamma_{zzzz}$ is the result of a delicate balance of positive and negative contributions from the various excited determinants that are summed up with different weights. This fluctuating nature of electron correlation on NLO properties was also observed in linearly $\pi$ conjugated chains \cite{2011JChPh.135a4111L}. For the deviations, the same order as in Eq. \eqref{hyperpolorderCh4} is found, when $\Delta\alpha_{zz}^{\text{HF}}$ is excluded.

\begin{figure}
 \centering
 \includegraphics[width=0.70\textwidth]{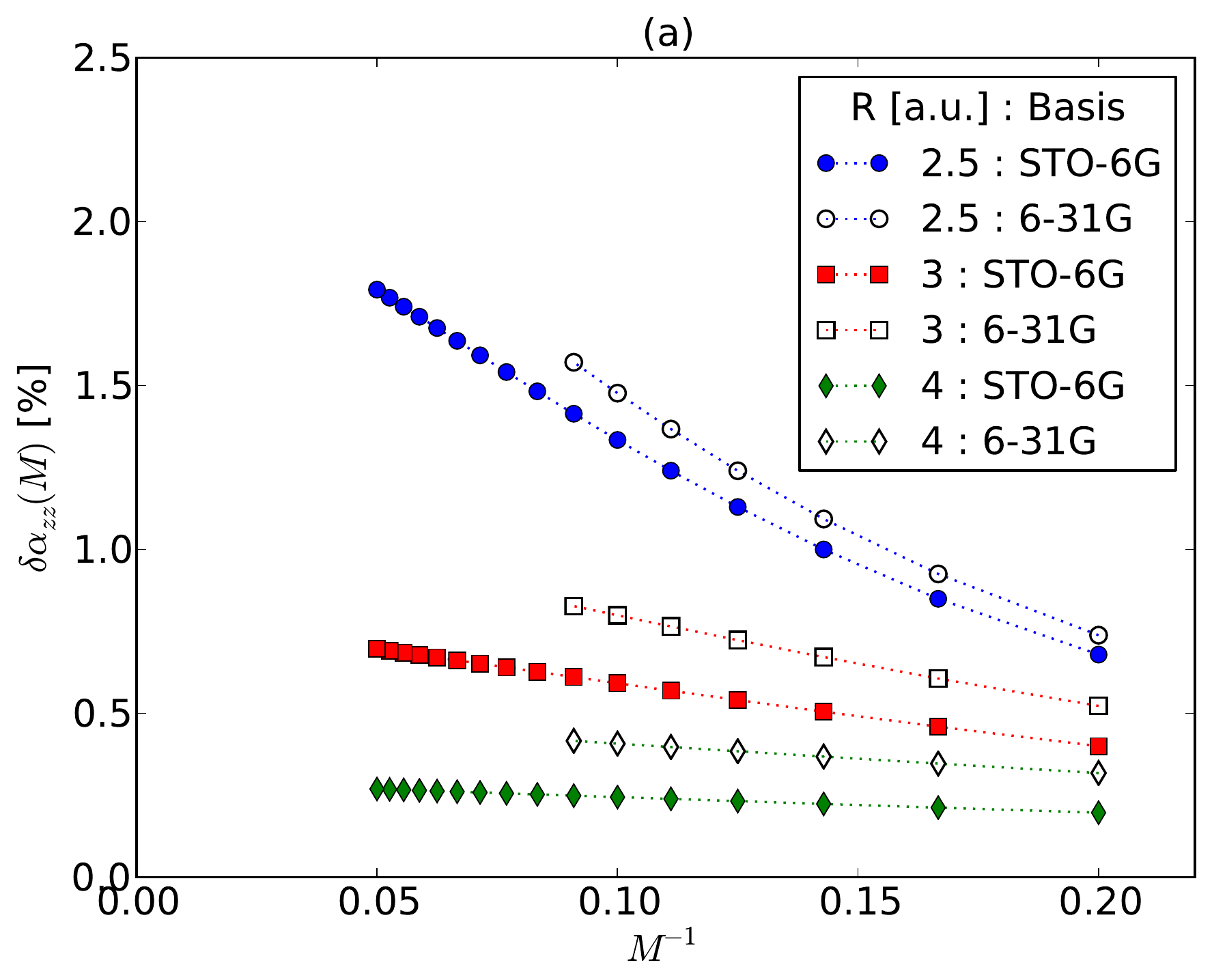}
 \includegraphics[width=0.70\textwidth]{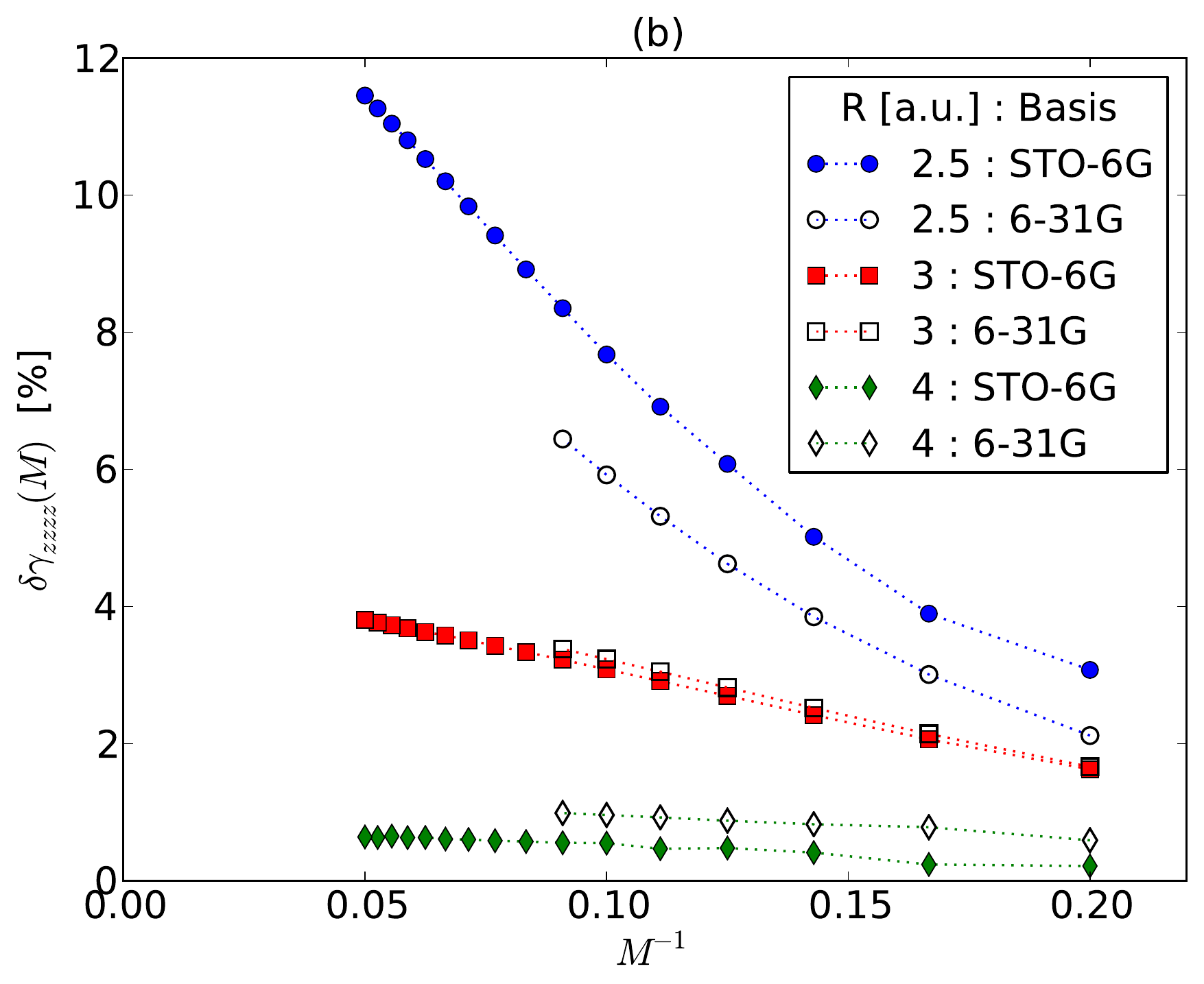}
 \caption{\label{relativeDeviationJCP136} The relative deviation (see Eq. \eqref{reldeveqCh4}) of the CCSD(T) polarizability (a) and second hyperpolarizability (b) to the MPS values for the data in Figs. \ref{Fig5JCP136-ab} to \ref{Fig6JCP136-ef}.}
\end{figure}

CCSD(T) is often used as the benchmark method to test the performance of LOTs for linear and non-linear optical properties \cite{QUA:QUA22177}. Of the four HF based LOTs we have tested, CCSD(T) indeed consistently gives the best results. To check the performance of CCSD(T) for the data in Figs. \ref{Fig5JCP136-ab} to \ref{Fig6JCP136-ef}, the relative deviation
\begin{equation}
 \delta q (M) = \frac{q^{\text{MPS}}(M) - q^{\text{CCSD(T)}}(M)}{q^{\text{MPS}}(M)} \label{reldeveqCh4}
\end{equation}
is defined. $q$ can again be $\alpha_{zz}$ or $\gamma_{zzzz}$. This relative deviation is shown in Fig. \ref{relativeDeviationJCP136}. The deviation is larger for the second hyperpolarizability than for the polarizability. For both parameters, the deviation increases with decreasing intermolecular distance (increasing electron delocalization). For chains with small intermolecular distance (delocalized electrons), the deviation also rapidly increases with the number of molecules. Note that the $\gamma_{zzzz}^{\text{CCSD(T)}} (M = 20)$ result for the intermolecular distance $R = 2.5$ a.u.~and the STO-6G basis set already deviates by 12\% from the exact result and a simple extrapolation to the TD limit shows that this deviation can become as large as 15\%. The breakdown of the CCSD(T) method can be understood by the following heuristic argument in terms of elementary optical excitations. For large intermolecular distances, the electrons are localized in H$_2$ molecules and the maximum number of electrons involved in an elementary excitation is 2. These effects can be captured by the CCSD(T) method. For small intermolecular distances, the electrons are delocalized over the chain and a larger number of electrons are involved in elementary excitations. This number also increases with chain length. CCSD(T) cannot adequately capture this effect and the CCSD(T) results start to deviate from the exact ones.

\begin{figure}
\centering
\includegraphics[width=0.70\textwidth]{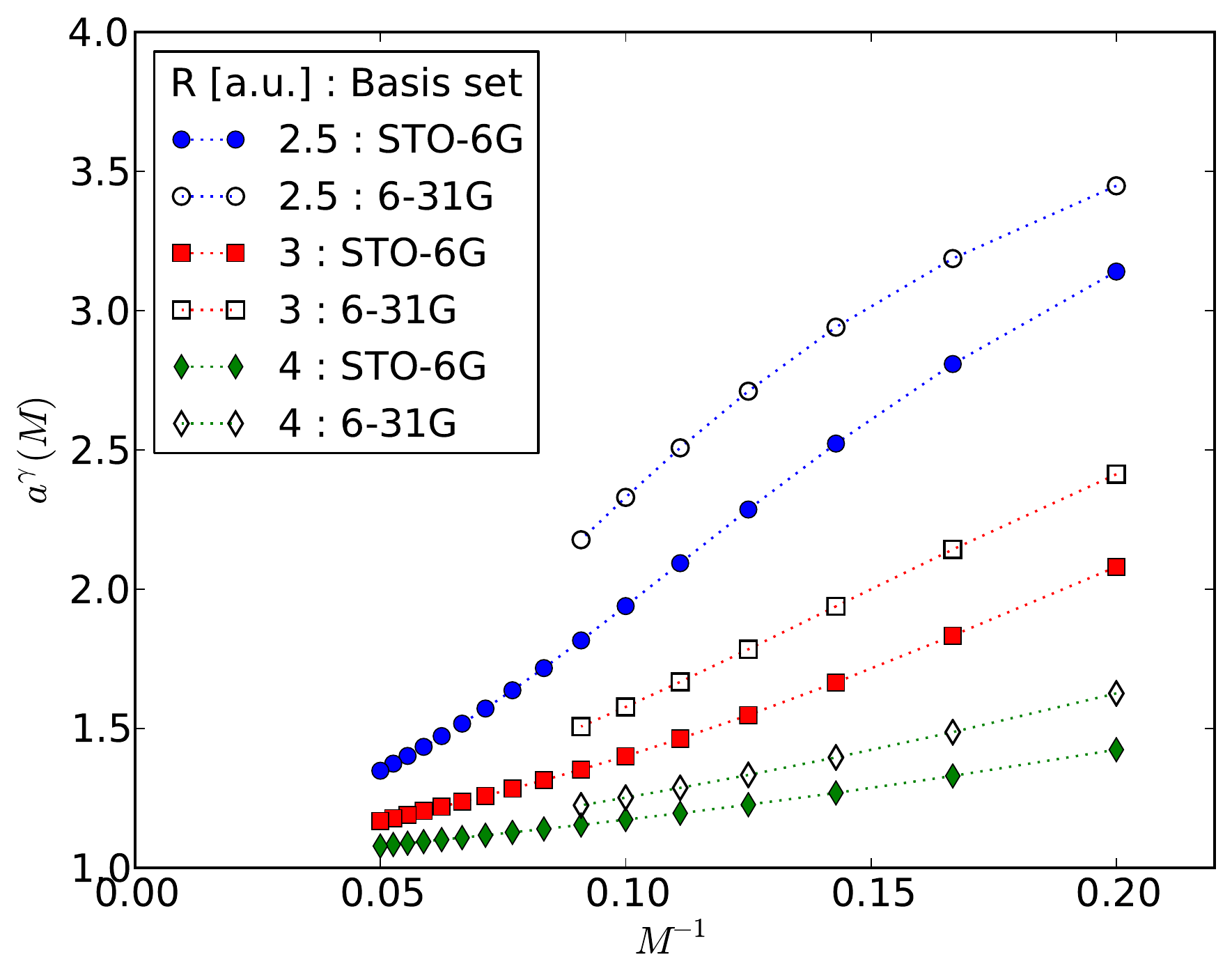}
\caption{\label{powergammaCh4} The power approximation of Eq. \eqref{powerapproxeqCh4}, applied to the MPS calculations for the STO-6G and 6-31G basis sets.}
\end{figure}

For the second hyperpolarizability, the scaling
\begin{equation}
 \gamma(M) \propto M^{a(M)} \label{JCP136theScalingRelationForTheSecondHyperpol}
\end{equation}
is often proposed \cite{chemrevpersoons}. The power $a(M)$ depends weakly on the number of molecules $M$. Its initially constant value drops eventually towards one in the TD limit. This can be explained in terms of a delocalized optical excitation, with a typical length scale. With increasing lengths, the possibility for such excitations opens up. When the chain can contain the delocalized excitations completely, the power tends to 1 and it is said that the system is in the saturation regime \cite{chemrevpersoons}. As can be seen in Figs. \ref{Fig5JCP136-ab} to \ref{Fig6JCP136-ef}, the saturation regime indeed sets in later when the intermolecular distance is smaller (electron delocalization larger). This can be confirmed by the following approximation to $a(M)$:
\begin{equation}
a^{\gamma}(M) = \frac{\ln{ (\gamma_{zzzz}(M) )} - \ln{(\gamma_{zzzz}(M-1))} }{\ln{(M)} - \ln{(M-1})} , \label{powerapproxeqCh4}
\end{equation}
which is shown in Fig. \ref{powergammaCh4} for the MPS calculations. From this figure, two extra conclusions can be made. The power for $R = 2.5$ a.u. and the 6-31G basis set is still above 2 for the chain lengths studied. Accurate extrapolations of the second hyperpolarizability to the TD limit are therefore not possible for this data set. The estimated powers are larger for the 6-31G basis than for the STO-6G basis, a result of the increased number of possibilities for optical excitations in 6-31G, but the effect of electron delocalization predominates.

\begin{figure}
 \centering
 \includegraphics[width=0.70\textwidth]{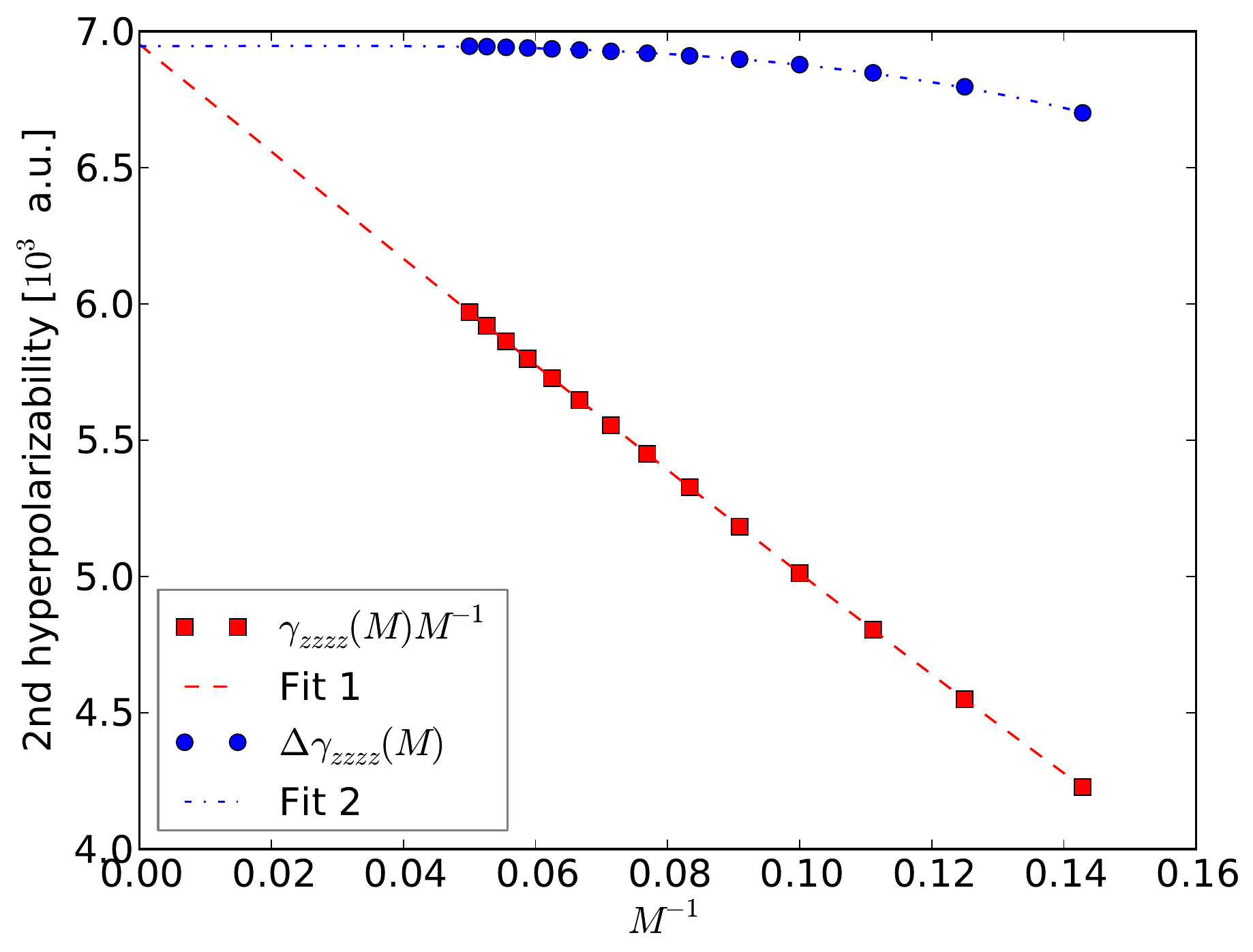}
 \caption{\label{extrapolexamplefigCh4} Extrapolation of the second hyperpolarizability per $\text{H}_2$ unit for the configuration with $R$ = 3.0 a.u. in the STO-6G basis. The extrapolation schemes in Eqs. \eqref{propfitCh4} and \eqref{propfit2Ch4} were used to obtain resp.~Fits 1 and 2.}
\end{figure}

From the data in Figs. \ref{Fig5JCP136-ab} to \ref{Fig6JCP136-ef}, values for $\alpha_{zz}^{\text{MPS}} M^{-1}$ and $\gamma_{zzzz}^{\text{MPS}} M^{-1}$ in the TD limit can be extrapolated. A scaling relation of the form
\begin{equation}
\frac{q(M)}{M} = a_0 + \frac{a_1}{M} + \frac{a_2}{M^2} + \frac{a_3}{M^3} \label{propfitCh4}
\end{equation}
is assumed, where $q$ can again be $\alpha_{zz}$ or $\gamma_{zzzz}$ and the $a_n$ are obtained from a least-squares fit. The parameter $a_0$ then corresponds to the desired TD limit value. From Eq. \eqref{propfitCh4}, the following equation can be derived:
\begin{equation}
\Delta q(M) = q(M) - q(M-1) = a_0 + \frac{b_2}{M^2} + \frac{b_3}{M^3} + \mathcal{O}(M^{-4}) . \label{propfit2Ch4}
\end{equation}
To check the extrapolations, a least-squares fit of Eq. \eqref{propfit2Ch4} to $\Delta q(M)$ is performed too. In both extrapolation schemes, the cut-off value for $M$ was 5 for the polarizability and 7 for the second hyperpolarizability. An example is shown in Fig. \ref{extrapolexamplefigCh4}. All obtained data are presented in Table \ref{extrapoltableCh4}. Except for the second hyperpolarizability for $R = 2.5$ a.u. and the 6-31G basis, the results of both extrapolation schemes are within 1\% relative deviation.

\begin{table}
\centering
\caption{\label{extrapoltableCh4} Extrapolated values for the polarizability and second hyperpolarizability per $\text{H}_2$ unit in the TD limit.}
\begin{tabular}{ c c c r r }
  \hline\hline
  Quantity & Basis set & R (a.u.) & Eq. \eqref{propfitCh4} & Eq. \eqref{propfit2Ch4}\\
  \hline
  $\alpha_{zz}$ (a.u.) & STO-6G & 2.5 & 17.41 & 17.41 \\
  & STO-6G & 3.0 & 9.464 & 9.462  \\
  & STO-6G & 4.0 & 5.733 & 5.733 \\
  & 6-31G & 2.5 & 39.00 & 39.20 \\
  & 6-31G & 3.0 & 21.27 & 21.27 \\
  & 6-31G & 4.0 & 13.55 & 13.55\\
  \hline
  $\gamma_{zzzz}$ ($10^3$ a.u.) & STO-6G & 2.5 & 52.64 & 52.74\\
  & STO-6G & 3.0 & 6.953 & 6.945\\
  & STO-6G & 4.0 & 0.9303 & 0.9301 \\
  & 6-31G & 2.5 & 410.8$^{(a)}$ & 424.0$^{(a)}$\\
  & 6-31G & 3.0 & 48.52 & 48.45 \\
  & 6-31G & 4.0 & 8.275 & 8.269 \\
  \hline
  \hline
\end{tabular}

{\small (a) These extrapolated values lie far apart and have to be treated with care as the powers $a^{\gamma}(M)$ for the largest chain lengths studied are still rather large.}
\end{table}

\vspace{0.7cm}
\hspace{-\parindent}\textbf{VI. CONCLUSIONS}
\vspace{0.3cm}

There is a lot of interest in the optical properties of chemical systems extended in one spatial dimension. The MPS ansatz works well for quasi-one-dimensional non-critical systems and yields highly accurate results. It can hence be used to study the optical properties of one-dimensional systems. We have implemented the sweep algorithm for the variational optimization of $\mathsf{SU(2)} \otimes \mathsf{U(1)}$ invariant MPSs to study the static longitudinal polarizability and second hyperpolarizability of hydrogen chains by means of finite field extrapolations.

As a first application, the optical response properties of an equally spaced hydrogen chain were studied for the ground states in different spin symmetry sectors. It is well known that HF based methods break down in the limit of large interatomic distances, whereas an MPS can capture the relevant static correlation needed to obtain accurate energy results. It was shown that accurate optical response properties can also be obtained with the MPS ansatz. The peaks of the polarizability and second hyperpolarizability decrease with increasing spin and shift towards smaller interatomic distances. Arguments based on an SOS expansion can be invoked to explain which terms contribute to these optical response properties.

CCSD(T) is often used as a reference method for the calculation of optical response properties. For roughly constant static correlation, avoiding the expected breakdown of HF based methods, the deviation of the optical properties calculated with CCSD(T) and the quasi-exact MPS method was studied. For increasing electron delocalization, the deviation becomes larger. For a large electron delocalization, the deviation rapidly increases with increasing chain length. The increasing deviation was explained in terms of delocalized optical excitations, which CCSD(T) cannot accurately capture. For small basis sets, the MPS algorithm gives accurate optical response properties in the saturation regime. These results were extrapolated to the TD limit.

In the future, we aim to implement the quadratically scaling algorithm of Hachmann \textit{et al.} \cite{hachmann:144101} and try to find a better choice of virtual dimension truncation to extend the range of our algorithm. We also aim to extend our algorithm to find excited states, allowing a study of the dominant terms in the SOS expression.

The MPS algorithm is hence a promising method to assess the performance of other QC methods for quasi-one-dimensional chemical systems. It allows to maintain ED accuracy for larger system sizes, e.g., to obtain accurate results of optical response properties in the saturation regime.

\vspace{0.7cm}
\hspace{-\parindent}\textbf{ACKNOWLEDGEMENTS}
\vspace{0.3cm}

This research was supported by the Research Foundation Flanders (S.W.), the Swiss National Science Foundation fellowship PBEZP2-134449 (P.A.L.), and NSERC (P.W.A.). The authors acknowledge a generous allocation of computer time granted by the Stevin Supercomputer Infrastructure at Ghent University, funded by Ghent University, the Hercules Foundation and the Flemish Government Department of EWI. Additional computing resources were granted by SHARCNET, a partner consortium in the Compute Canada national HPC platform.

\vspace{0.7cm}
\hspace{-\parindent}\textbf{APPENDIX: REDUCED TENSORS}
\vspace{0.3cm}

Note that during a sweep, we work with left normalized tensors in the left part and right normalized tensors in the right part. Consider the following partial contraction in the graphical notation \cite{Schollwock201196}:
\begin{equation}
\vcenter{\hbox{\scriptsize{
\setlength{\unitlength}{1cm}
\begin{picture}(2.6,2)
\put(0.4,1.8){\circle{0.4}}
\put(0.6,1.8){\line(1,0){0.75}}
\put(1.4,1.7){$j_R j_R^z N_R \alpha_R$}
\put(0.28,1.7){\scriptsize{M}}
\put(0.4,1.3){\line(0,1){0.3}}
\put(0.1,0.7){\line(1,0){0.6}}
\put(0.1,1.3){\line(1,0){0.6}}
\put(0.1,0.7){\line(0,1){0.6}}
\put(0.7,0.7){\line(0,1){0.6}}
\put(0.2,0.9){$a_m$}
\put(0.4,0.4){\line(0,1){0.3}}
\put(0.4,0.2){\circle{0.4}}
\put(0.6,0.2){\line(1,0){0.75}}
\put(0.28,0.1){M}
\put(1.40,0.1){$\widetilde{j}_R \widetilde{j}_R^z \widetilde{N}_R \widetilde{\alpha}_R$}
\put(0.2,1.0){\oval(1.0,1.6)[l]}
\end{picture}
}}} . \label{example1-appJCP136}
\end{equation}

With Eq. \eqref{tensordecomp}, it is easy to show that Eq. \eqref{example1-appJCP136} can be written as
\begin{equation}
\delta_{\widetilde{N}_R,N_R+1} \braket{j_R j_R^z \frac{1}{2} m \mid \widetilde{j}_R \widetilde{j}_R^z} \vcenter{\hbox{\scriptsize{
\setlength{\unitlength}{1cm}
\begin{picture}(2.0,2.4)
\put(0.2,0.9){\line(1,0){0.4}}
\put(0.2,1.5){\line(1,0){0.4}}
\put(0.2,0.9){\line(0,1){0.6}}
\put(0.6,0.9){\line(0,1){0.6}}
\put(0.3,1.1){$\Lambda$}
\put(0.4,1.5){\line(0,1){0.5}}
\put(0.4,0.4){\line(0,1){0.5}}
\put(0.4,2.0){\line(1,0){0.75}}
\put(0.4,0.4){\line(1,0){0.75}}
\put(1.2,1.9){$j_R N_R \alpha_R$}
\put(1.2,0.3){$\widetilde{j}_R (N_R + 1) \widetilde{\alpha}_R$}
\end{picture}
}}} \label{againWE-JCP136}
\vspace{0.01\textwidth}
\end{equation}
with
\begin{eqnarray}
\vcenter{\hbox{\scriptsize{
\setlength{\unitlength}{1cm}
\begin{picture}(2.8,2.4)
\put(0.2,0.9){\line(1,0){0.4}}
\put(0.2,1.5){\line(1,0){0.4}}
\put(0.2,0.9){\line(0,1){0.6}}
\put(0.6,0.9){\line(0,1){0.6}}
\put(0.3,1.1){$\Lambda$}
\put(0.4,1.5){\line(0,1){0.5}}
\put(0.4,0.4){\line(0,1){0.5}}
\put(0.4,2.0){\line(1,0){0.75}}
\put(0.4,0.4){\line(1,0){0.75}}
\put(1.2,1.9){$j_R N_R \alpha_R$}
\put(1.2,0.3){$\widetilde{j}_R (N_R + 1) \widetilde{\alpha}_R$}
\end{picture}
}}}
& = & \sum\limits_{\alpha_L} \vcenter{\hbox{\scriptsize{
\setlength{\unitlength}{1cm}
\begin{picture}(3.7,2)
\put(1.1,1.8){\circle{0.4}}
\put(1.3,1.8){\line(1,0){0.75}}
\put(2.1,1.7){$j_R N_R \alpha_R$}
\put(1.0,1.7){T}
\put(1.1,1.2){\line(0,1){0.4}}
\put(1.2,1.3){$0 0$}
\put(1.1,0.4){\line(0,1){0.4}}
\put(1.2,0.5){$\frac{1}{2} 1$}
\put(1.1,0.2){\circle{0.4}}
\put(1.3,0.2){\line(1,0){0.75}}
\put(1.0,0.1){T}
\put(2.1,0.1){$\widetilde{j}_R (N_R+1) \widetilde{\alpha}_R$}
\put(0.9,1.0){\oval(1.0,1.6)[l]}
\put(0.0,0.4){\rotatebox{90}{$j_R N_R \alpha_L$}}
\end{picture}
}}} \nonumber \\
& + & (-1)^{\widetilde{j}_R - j_R + \frac{1}{2}} \sqrt{\frac{2 j_R + 1}{2 \widetilde{j}_R + 1}}\sum\limits_{\alpha_L} \vcenter{\hbox{\scriptsize{
\setlength{\unitlength}{1cm}
\begin{picture}(3.7,2)
\put(1.1,1.8){\circle{0.4}}
\put(1.3,1.8){\line(1,0){0.75}}
\put(2.1,1.7){$j_R N_R \alpha_R$}
\put(1.0,1.7){T}
\put(1.1,1.2){\line(0,1){0.4}}
\put(1.2,1.3){$\frac{1}{2} 1$}
\put(1.1,0.4){\line(0,1){0.4}}
\put(1.2,0.5){$0 2$}
\put(1.1,0.2){\circle{0.4}}
\put(1.3,0.2){\line(1,0){0.75}}
\put(1.0,0.1){T}
\put(2.1,0.1){$\widetilde{j}_R (N_R+1) \widetilde{\alpha}_R$}
\put(0.9,1.0){\oval(1.0,1.6)[l]}
\put(0.0,0.0){\rotatebox{90}{$\widetilde{j}_R (N_R-1) \alpha_L$}}
\end{picture}
}}} .
\end{eqnarray}
Equation \eqref{example1-appJCP136} can hence be decomposed into a structural part (Clebsch-Gordan coefficient and particle conserving Kronecker delta) and a degeneracy part (the reduced $\Lambda$ tensor with spin $\frac{1}{2}$), as is shown in Eq. \eqref{againWE-JCP136}. As a second example, consider the partial contraction
\begin{align} \allowdisplaybreaks
& \vcenter{\hbox{\scriptsize{
\setlength{\unitlength}{1cm}
\begin{picture}(4.0,2)
\put(0.4,1.8){\circle{0.4}}
\put(1.6,1.8){\line(1,0){0.75}}
\put(2.4,1.7){$j_R j_R^z N_R \alpha_R$}
\put(0.28,1.7){\scriptsize{M}}
\put(0.4,1.3){\line(0,1){0.3}}
\put(0.1,0.7){\line(1,0){0.6}}
\put(0.1,1.3){\line(1,0){0.6}}
\put(0.1,0.7){\line(0,1){0.6}}
\put(0.7,0.7){\line(0,1){0.6}}
\put(0.19,0.9){$a_{m_1}$}
\put(0.4,0.4){\line(0,1){0.3}}
\put(0.4,0.2){\circle{0.4}}
\put(0.6,1.8){\line(1,0){0.6}}
\put(0.6,0.2){\line(1,0){0.6}}
\put(1.4,1.8){\circle{0.4}}
\put(1.4,0.2){\circle{0.4}}
\put(1.1,0.7){\line(0,1){0.6}}
\put(1.7,0.7){\line(0,1){0.6}}
\put(1.1,0.7){\line(1,0){0.6}}
\put(1.1,1.3){\line(1,0){0.6}}
\put(1.4,1.3){\line(0,1){0.3}}
\put(1.19,0.9){$a_{m_2}^{\dagger}$}
\put(1.4,0.4){\line(0,1){0.3}}
\put(1.28,1.7){\scriptsize{M}}
\put(1.28,0.1){\scriptsize{M}}
\put(1.6,0.2){\line(1,0){0.75}}
\put(0.28,0.1){M}
\put(2.40,0.1){$\widetilde{j}_R \widetilde{j}_R^z \widetilde{N}_R \widetilde{\alpha}_R$}
\put(0.2,1.0){\oval(1.0,1.6)[l]}
\end{picture}
}}} \hspace{-0.7cm} = \delta_{N_R,\widetilde{N}_R} (-1)^{\frac{1}{2}-m_2} \hspace{-0.1cm}\left(\hspace{-0.1cm} \braket{\frac{1}{2} m_1 \frac{1}{2} -m_2 \mid 0 0} \braket{j_R j_R^z 0 0 \mid \widetilde{j}_R \widetilde{j}_R^z}\hspace{-0.2cm} \vcenter{\hbox{\scriptsize{
\setlength{\unitlength}{1cm}
\begin{picture}(2.0,2.4)
\put(0.1,0.9){\line(1,0){0.6}}
\put(0.1,1.5){\line(1,0){0.6}}
\put(0.1,0.9){\line(0,1){0.6}}
\put(0.7,0.9){\line(0,1){0.6}}
\put(0.2,1.1){$F^0$}
\put(0.4,1.5){\line(0,1){0.5}}
\put(0.4,0.4){\line(0,1){0.5}}
\put(0.4,2.0){\line(1,0){0.75}}
\put(0.4,0.4){\line(1,0){0.75}}
\put(1.2,1.9){$j_R N_R \alpha_R$}
\put(1.2,0.3){$j_R N_R \widetilde{\alpha}_R$}
\end{picture}
}}} \right. \nonumber \\ 
& + \left. \braket{\frac{1}{2} m_1 \frac{1}{2} -m_2 \mid 1 (m_1 - m_2)} \braket{j_R j_R^z 1 (m_1-m_2) \mid \widetilde{j}_R \widetilde{j}_R^z} \vcenter{\hbox{\scriptsize{
\setlength{\unitlength}{1cm}
\begin{picture}(2.0,2.4)
\put(0.1,0.9){\line(1,0){0.6}}
\put(0.1,1.5){\line(1,0){0.6}}
\put(0.1,0.9){\line(0,1){0.6}}
\put(0.7,0.9){\line(0,1){0.6}}
\put(0.2,1.1){$F^1$}
\put(0.4,1.5){\line(0,1){0.5}}
\put(0.4,0.4){\line(0,1){0.5}}
\put(0.4,2.0){\line(1,0){0.75}}
\put(0.4,0.4){\line(1,0){0.75}}
\put(1.2,1.9){$j_R N_R \alpha_R$}
\put(1.2,0.3){$\widetilde{j}_R N_R \widetilde{\alpha}_R$}
\end{picture}
}}} \quad \right)
\end{align}
with
\begin{align}
\vcenter{\hbox{\scriptsize{
\setlength{\unitlength}{1cm}
\begin{picture}(2.8,2.4)
\put(0.1,0.9){\line(1,0){0.6}}
\put(0.1,1.5){\line(1,0){0.6}}
\put(0.1,0.9){\line(0,1){0.6}}
\put(0.7,0.9){\line(0,1){0.6}}
\put(0.2,1.1){$F^0$}
\put(0.4,1.5){\line(0,1){0.5}}
\put(0.4,0.4){\line(0,1){0.5}}
\put(0.4,2.0){\line(1,0){0.75}}
\put(0.4,0.4){\line(1,0){0.75}}
\put(1.2,1.9){$j_R N_R \alpha_R$}
\put(1.2,0.3){$j_R N_R \widetilde{\alpha}_R$}
\end{picture}
}}}
= & \sum\limits_{j_L \alpha_L \widetilde{\alpha}_L} \frac{1}{\sqrt{2}} \vcenter{\hbox{\scriptsize{
\setlength{\unitlength}{1cm}
\begin{picture}(4.4,2.4)
\put(0.6,0.9){\line(1,0){0.4}}
\put(0.6,1.5){\line(1,0){0.4}}
\put(0.6,0.9){\line(0,1){0.6}}
\put(1.0,0.9){\line(0,1){0.6}}
\put(0.7,1.1){$\Lambda$}
\put(0.8,1.5){\line(0,1){0.5}}
\put(0.8,0.4){\line(0,1){0.5}}
\put(0.8,2.0){\line(1,0){1.0}}
\put(0.8,0.4){\line(1,0){1.0}}
\put(2.0,2.0){\circle{0.4}}
\put(2.0,0.4){\circle{0.4}}
\put(1.9,1.9){T}
\put(1.9,0.3){T}
\put(2.0,1.4){\line(0,1){0.4}}
\put(2.0,0.6){\line(0,1){0.4}}
\put(2.1,1.5){$\frac{1}{2} 1$}
\put(2.1,0.7){$0 0$}
\put(2.2,0.4){\line(1,0){0.4}}
\put(2.2,2.0){\line(1,0){0.4}}
\put(2.7,1.9){$j_R N_R \alpha_R$}
\put(2.7,0.3){$j_R N_R \widetilde{\alpha}_R$}
\put(0.0,2.1){$j_L (N_R-1) \alpha_L$}
\put(0.8,0.1){$j_R N_R \widetilde{\alpha}_L$}
\end{picture}
}}} \nonumber \\
+ & \sum\limits_{\widetilde{j}_L \alpha_L \widetilde{\alpha}_L} \frac{1}{\sqrt{2}} \sqrt{\frac{2 \widetilde{j}_L+1}{2 j_R+1}} (-1)^{j_R - \widetilde{j}_L + \frac{1}{2}} \vcenter{\hbox{\scriptsize{
\setlength{\unitlength}{1cm}
\begin{picture}(4.4,2.4)
\put(0.6,0.9){\line(1,0){0.4}}
\put(0.6,1.5){\line(1,0){0.4}}
\put(0.6,0.9){\line(0,1){0.6}}
\put(1.0,0.9){\line(0,1){0.6}}
\put(0.7,1.1){$\Lambda$}
\put(0.8,1.5){\line(0,1){0.5}}
\put(0.8,0.4){\line(0,1){0.5}}
\put(0.8,2.0){\line(1,0){1.0}}
\put(0.8,0.4){\line(1,0){1.0}}
\put(2.0,2.0){\circle{0.4}}
\put(2.0,0.4){\circle{0.4}}
\put(1.9,1.9){T}
\put(1.9,0.3){T}
\put(2.0,1.4){\line(0,1){0.4}}
\put(2.0,0.6){\line(0,1){0.4}}
\put(2.1,1.5){$0 2$}
\put(2.1,0.7){$\frac{1}{2} 1$}
\put(2.2,0.4){\line(1,0){0.4}}
\put(2.2,2.0){\line(1,0){0.4}}
\put(2.7,1.9){$j_R N_R \alpha_R$}
\put(2.7,0.3){$j_R N_R \widetilde{\alpha}_R$}
\put(0.0,2.1){$j_R (N_R-2) \alpha_L$}
\put(0.0,0.1){$\widetilde{j}_L (N_R - 1) \widetilde{\alpha}_L$}
\end{picture}
}}}
\end{align}
and
\begin{align}
\vcenter{\hbox{\scriptsize{
\setlength{\unitlength}{1cm}
\begin{picture}(2.8,2.4)
\put(0.1,0.9){\line(1,0){0.6}}
\put(0.1,1.5){\line(1,0){0.6}}
\put(0.1,0.9){\line(0,1){0.6}}
\put(0.7,0.9){\line(0,1){0.6}}
\put(0.2,1.1){$F^1$}
\put(0.4,1.5){\line(0,1){0.5}}
\put(0.4,0.4){\line(0,1){0.5}}
\put(0.4,2.0){\line(1,0){0.75}}
\put(0.4,0.4){\line(1,0){0.75}}
\put(1.2,1.9){$j_R N_R \alpha_R$}
\put(1.2,0.3){$\widetilde{j}_R N_R \widetilde{\alpha}_R$}
\end{picture}
}}} \hspace{-0.7cm}
= & \sum\limits_{j_L \alpha_L \widetilde{\alpha}_L} \sqrt{3(2 j_R + 1)} (-1)^{\widetilde{j}_R + j_L + \frac{3}{2}} \left\{ \begin{array}{ccc} \frac{1}{2} & \frac{1}{2} & 1 \\ j_R & \widetilde{j}_R & j_L \end{array} \right\} \hspace{-0.7cm} \vcenter{\hbox{\scriptsize{
\setlength{\unitlength}{1cm}
\begin{picture}(4.4,2.4)
\put(0.6,0.9){\line(1,0){0.4}}
\put(0.6,1.5){\line(1,0){0.4}}
\put(0.6,0.9){\line(0,1){0.6}}
\put(1.0,0.9){\line(0,1){0.6}}
\put(0.7,1.1){$\Lambda$}
\put(0.8,1.5){\line(0,1){0.5}}
\put(0.8,0.4){\line(0,1){0.5}}
\put(0.8,2.0){\line(1,0){1.0}}
\put(0.8,0.4){\line(1,0){1.0}}
\put(2.0,2.0){\circle{0.4}}
\put(2.0,0.4){\circle{0.4}}
\put(1.9,1.9){T}
\put(1.9,0.3){T}
\put(2.0,1.4){\line(0,1){0.4}}
\put(2.0,0.6){\line(0,1){0.4}}
\put(2.1,1.5){$\frac{1}{2} 1$}
\put(2.1,0.7){$0 0$}
\put(2.2,0.4){\line(1,0){0.4}}
\put(2.2,2.0){\line(1,0){0.4}}
\put(2.7,1.9){$j_R N_R \alpha_R$}
\put(2.7,0.3){$\widetilde{j}_R N_R \widetilde{\alpha}_R$}
\put(0.0,2.1){$j_L (N_R-1) \alpha_L$}
\put(0.8,0.1){$\widetilde{j}_R N_R \widetilde{\alpha}_L$}
\end{picture}
}}} \hspace{-0.7cm} \nonumber \\
+ & \sum\limits_{\widetilde{j}_L \alpha_L \widetilde{\alpha}_L} \sqrt{3(2 \widetilde{j}_L + 1)} (-1)^{\widetilde{j}_R + j_R + 1} \left\{ \begin{array}{ccc} \frac{1}{2} & \frac{1}{2} & 1 \\ j_R & \widetilde{j}_R & \widetilde{j}_L \end{array} \right\} \hspace{-0.7cm} \vcenter{\hbox{\scriptsize{
\setlength{\unitlength}{1cm}
\begin{picture}(4.4,2.4)
\put(0.6,0.9){\line(1,0){0.4}}
\put(0.6,1.5){\line(1,0){0.4}}
\put(0.6,0.9){\line(0,1){0.6}}
\put(1.0,0.9){\line(0,1){0.6}}
\put(0.7,1.1){$\Lambda$}
\put(0.8,1.5){\line(0,1){0.5}}
\put(0.8,0.4){\line(0,1){0.5}}
\put(0.8,2.0){\line(1,0){1.0}}
\put(0.8,0.4){\line(1,0){1.0}}
\put(2.0,2.0){\circle{0.4}}
\put(2.0,0.4){\circle{0.4}}
\put(1.9,1.9){T}
\put(1.9,0.3){T}
\put(2.0,1.4){\line(0,1){0.4}}
\put(2.0,0.6){\line(0,1){0.4}}
\put(2.1,1.5){$0 2$}
\put(2.1,0.7){$\frac{1}{2} 1$}
\put(2.2,0.4){\line(1,0){0.4}}
\put(2.2,2.0){\line(1,0){0.4}}
\put(2.7,1.9){$j_R N_R \alpha_R$}
\put(2.7,0.3){$\widetilde{j}_R N_R \widetilde{\alpha}_R$}
\put(0.0,2.1){$j_R (N_R-2) \alpha_L$}
\put(0.0,0.1){$\widetilde{j}_L (N_R - 1) \widetilde{\alpha}_L$}
\end{picture}
}}}\hspace{-0.7cm},
\end{align}
where the curly brackets denote Wigner 6-j symbols. The second example can hence also be decomposed in terms containing a structural part and a degeneracy part. The reduced tensors corresponding to the direct product of two spin $\frac{1}{2}$ operators are a spin 0 tensor ($F^0$ in the example) and a spin 1 tensor ($F^1$ in the example).

{\color{blue}{
\noindent\makebox[\linewidth]{\rule{\textwidth}{0.4pt}}

\vspace{-0.40cm}

\noindent\makebox[\linewidth]{\rule{\textwidth}{0.4pt}}
}}
\newpage

\section{The metal-insulator transition} \label{MITsection}
The equally spaced hydrogen chain \eqref{JCP136equallySpacedFormula} is studied in this section, using the minimal basis set STO-6G. This model has one half-filled conduction band, which suggests that the chain is conducting for all interatomic distances $R$. However, at large interatomic distance, the hydrogen chain consists of isolated atoms and is therefore an insulator. Because the insulating behaviour cannot be explained by band theory (a single-particle theory), the hydrogen chain is a Mott insulator \cite{Gebhard, PhysRevB.84.245117}. For decreasing interatomic distance $R$, the equally spaced hydrogen chain goes through an MIT \cite{PhysRevB.84.245117, QUA:QUA23047}. Three properties then simultaneously change \cite{kudinov, PhysRevB.62.1666, Resta2006}:
\begin{enumerate}
\item The static dipole polarizability \textit{per electron}
\begin{equation}
  \lim\limits_{L \rightarrow \infty} \frac{\alpha_{zz}(L)}{L} \label{polPerEl}
\end{equation}
is infinite in a conductor (metal) and finite in an insulator.
\item The excitation gap
\begin{equation}
\Delta E = \lim\limits_{L \rightarrow \infty} \left( E_1(L) - E_0(L) \right)
\end{equation}
is closed (zero) in a conductor and open (nonzero) in an insulator.
\item The fluctuation of the dipole moment \textit{per electron}
\begin{equation}
\lambda_{zz} = \lim\limits_{L \rightarrow \infty} \frac{1}{L} \left( \braket{ \Psi_0 \mid z z \mid \Psi_0} - \braket{ \Psi_0 \mid z \mid \Psi_0}\braket{ \Psi_0 \mid z \mid \Psi_0} \right),
\end{equation}
is infinite in a conductor and finite in an insulator.
\end{enumerate}
A static electric field induces a current in a conductor (infinite electron displacement), while the electrons in an insulator are only displaced over a finite distance, which explains the behaviour of the static polarizability. An infinite response (displacement) can only occur if the corresponding energy cost is zero, i.e. when the excitation gap is closed. The behaviour of the dipole moment fluctuation is less intuitive. The following relation can be proven \cite{PhysRevB.62.1666}:
\begin{equation}
\lambda_{zz} \propto \int\limits_0^{\infty} \frac{d\omega}{\omega} \Re \sigma(\omega),
\end{equation}
where $\sigma(\omega)$ is the conductivity at frequency $\omega$. Because the real part of the static conductivity is nonzero for conductors and zero for insulators, the dipole moment fluctation is infinite for conductors and finite for insulators.

\begin{figure}
 \centering
 \includegraphics[width=0.70\textwidth]{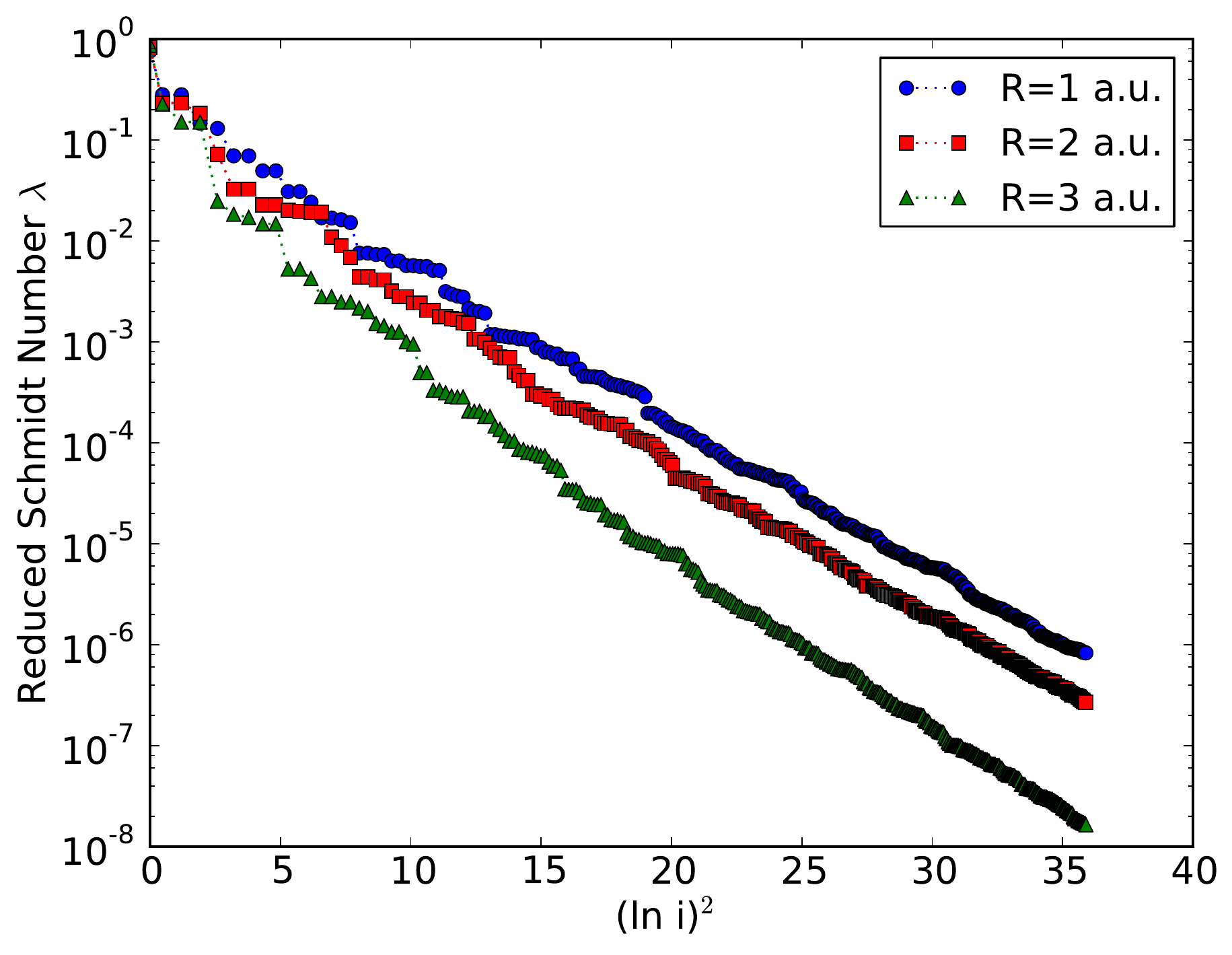}
 \caption{\label{MITschmidtfigure} 
Ground-state calculations for equally spaced hydrogen chains with 36 atoms were performed, for three interatomic distances $R$.
The L\"owdin transformed STO-6G basis was used as single-particle degrees of freedom. The reduced Schmidt spectra at the central MPS bond are shown. The index $i$ counts the ordered reduced Schmidt numbers. They decay according to Eq. \eqref{schmidtspectrumdecayrelationcroot}. As the system becomes more metallic, a larger virtual dimension is required.}
\end{figure}

DMRG works well for gapped one-dimensional systems. The excitation gap of metallic hydrogen chains only closes in the TD limit. For finite-size systems, a larger virtual dimension is required as the system becomes more metallic. This is illustrated in Fig. \ref{MITschmidtfigure}.

\begin{figure}
 \centering
 \includegraphics[width=0.70\textwidth]{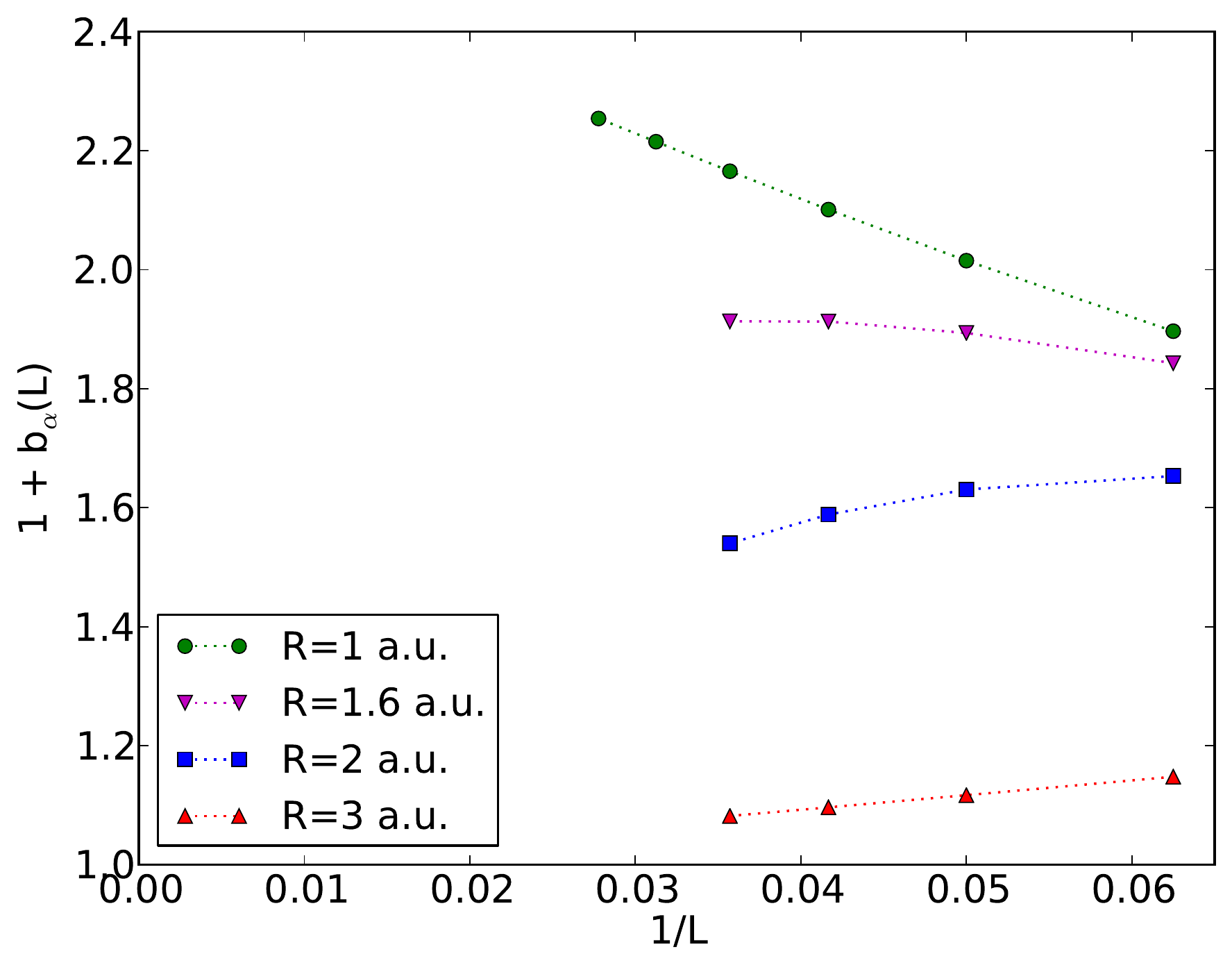}
 \includegraphics[width=0.70\textwidth]{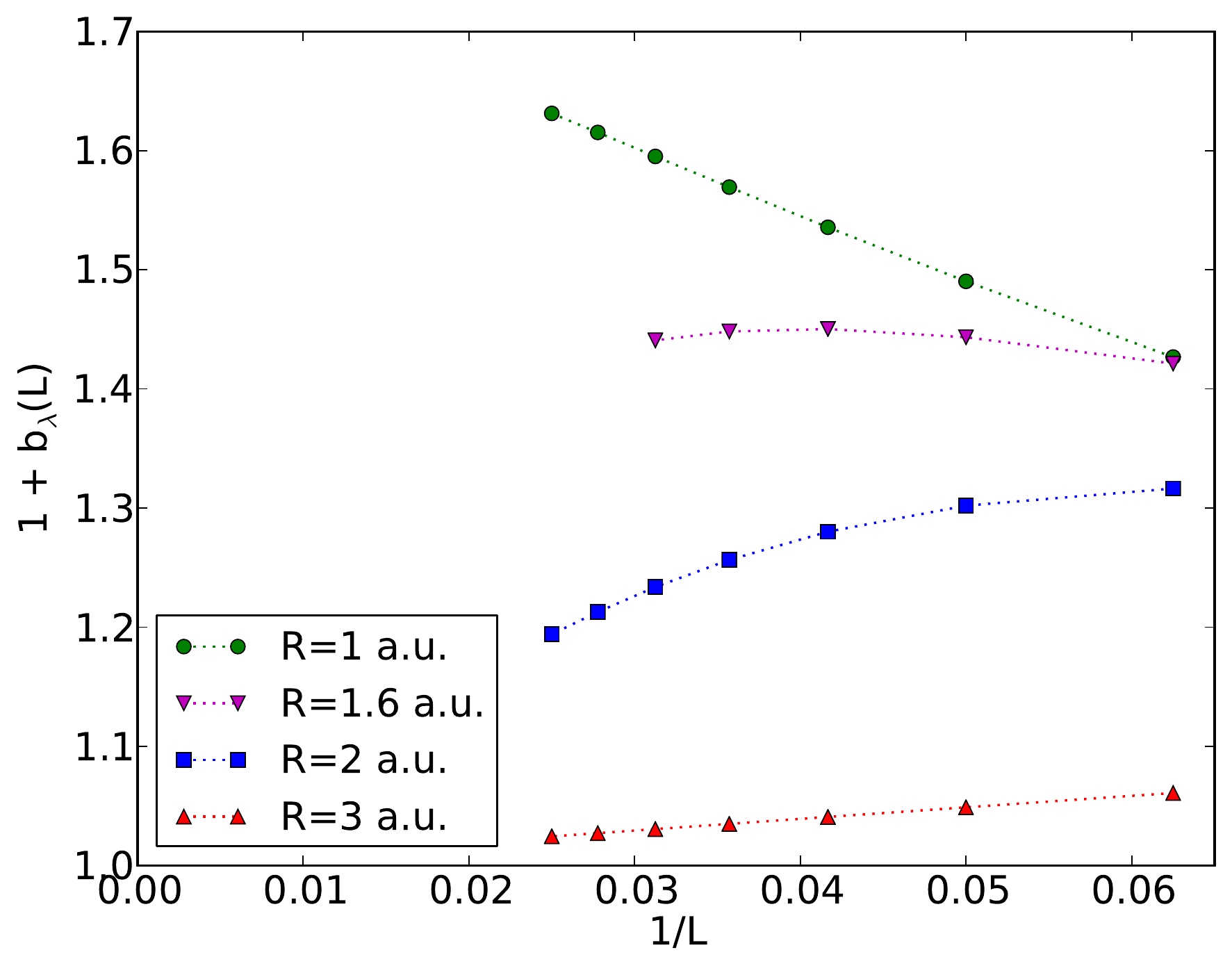}
 \caption{\label{MITpolandloclength} Extrapolation of $b_{\alpha,\lambda}(L)$ to infinite chain length ($\frac{1}{L} \rightarrow 0$). The L\"owdin transformed STO-6G basis was used as single-particle degrees of freedom. The chosen values of $L$ are multiples of 4. The \textit{reduced} virtual dimension per $\mathsf{SU(2)} \otimes \mathsf{U(1)}$ \textit{symmetry sector} was truncated to 96. The finite fields $F = 0.0000$, 0.0008, 0.0012, and 0.0016 a.u.~were used to extrapolate $\alpha_{zz}(L,F)$ to zero field. The matrix elements required to calculate $L \lambda_{zz}(L)$ are discussed in Ref.~\cite{QUA:QUA23047}.}
\end{figure}

Suppose the dipole polarizability and fluctuation \textit{per electron} scale as:
\begin{eqnarray}
\frac{\alpha_{zz}(L)}{L} & = & a_{\alpha} L^{b_{\alpha}} + ... \\
\lambda_{zz}(L) & = & a_{\lambda} L^{b_{\lambda}} + ...
\end{eqnarray}
then their increments \textit{for the entire hydrogen chain} will be
\begin{eqnarray}
\Delta \alpha_{zz}(L) & = & \frac{\alpha_{zz}(L) - \alpha_{zz}(L - 4)}{4} = a_{\alpha} (b_{\alpha} + 1) L^{b_{\alpha}} + ... \\
\Delta (L \lambda_{zz}(L)) & = & \frac{L \lambda_{zz}(L) - (L-4) \lambda_{zz}(L-4)}{4} = a_{\lambda} (b_{\lambda} + 1) L^{b_{\lambda}} + ...
\end{eqnarray}
This allows to estimate $b_{\alpha}$ and $b_{\lambda}$ as
\begin{eqnarray}
b_{\alpha}(L) +1 =  \frac{ L \Delta (\alpha_{zz}(L)) }{\alpha_{zz}(L)}, \\
b_{\lambda}(L) +1 =  \frac{\Delta (L \lambda_{zz}(L)) }{\lambda_{zz}(L)}.
\end{eqnarray}
Extrapolation to infinite chain length then yields the desired exponents. This is illustrated in Fig. \ref{MITpolandloclength}. For $R=3$ a.u. the exponents $b_{\alpha,\lambda}$ immediately tend to zero. For this interatomic distance, the hydrogen chain consists of separated atoms. For $R=1.6$ and $2$ a.u. the possibility for elementary excitations (with a finite length scale) first opens up, and when the chain is sufficiently long, the saturation regime sets in: $b_{\alpha,\lambda}(L) \rightarrow 0$. For $R=1$ a.u. the system appears to be a metal, and a least-squares fit yields $b_{\alpha} \approx 1.60$ and $b_{\lambda} \approx 0.79$.

\chapter{Low-lying bond dissociation curves of the carbon dimer} \label{C2-chapter}
\begin{chapquote}{Peifeng Su, 2010}
What emerges from all of these high level studies is the extreme difficulty in calculating the ground state and the low-lying excited states of C$_2$ in a meaningful way, owing to the multireference character of the wave functions and the near degeneracies which change very rapidly as a function of the C-C distance.
\end{chapquote}

\section{Introduction}
The carbon dimer is a challenging system for molecular electronic structure methods. The ground state has significant MR character \cite{10.1021.ct100577v}. The low-lying states are quasi-degenerate, and many crossings and avoided crossings occur between the low-lying states of this homonuclear dimer \cite{BoggioPasqua2000159, AbramsC2, Varandas, TheoChemAcc2014}. The core correlation and core-valence correlation are also important \cite{CoreCorrelationInC2, 10.1080.00268976.2011.564593, CoreCorrelation2011JCP, TheoChemAcc2014}.

Traditionally, bonding is interpreted in terms of localized electrons. A covalent bond between two atoms is formed by a shared singlet pair of electrons. An ionic bond arises due to the electrostatic stabilization between two oppositely charged species. In $\pi$-conjugated systems such as benzene, the energy stabilization can only be explained by a delocalized electron picture: the resonance between Kekul\'e structures lowers the energy. In one-dimensional $\pi$-conjugated polyenes, this resonance induces ionic contributions, in which the ends of the polyene are oppositely charged. This charge-shift bonding is also of importance in homonuclear diatomics such as F$_2$, C$_2$, Cl$_2$, and Br$_2$ \cite{naturechemChargeShift, 10.1021.ct100577v}. The FCI solution then contains both covalent and charge-shift (ionic) Slater determinants with significant weights:
\begin{equation}
\ket{\text{FCI}(\text{F}_2)} \approx C_0 \ket{\text{F} - \text{F}} + C_1 \ket{\text{F}^- ~ \text{F}^+} + C_2 \ket{\text{F}^+ ~ \text{F}^-}.
\end{equation}

\begin{figure}
\centering
\includegraphics[width=0.40\textwidth]{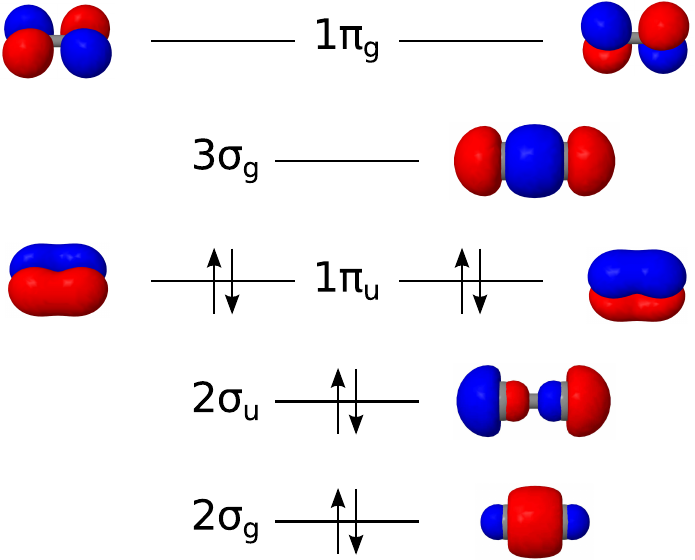}
\caption{\label{MOdiagramFig} Molecular orbital diagram for the carbon dimer at $R$ = 2.4 a.u. The molecular orbitals were obtained with \textsc{Psi4} at the RHF/cc-pVDZ level of theory.}
\end{figure}

In the last four years, the bond order of the carbon dimer has come under debate \cite{10.1021.ct100577v, naturechemFourthBond, 10.1021.ct400867h}. In organic chemistry, two carbon atoms can have a single (H$_3$C$-$CH$_3$, ethane, [$\sigma$]), a double (H$_2$C$=$CH$_2$, ethylene, [$\sigma+\pi$]), or a triple (HC$\equiv$CH, acetylene, [$\sigma+2\pi$]) bond. With increasing bond order, the bond length decreases. The bond length of the carbon dimer is in between the corresponding bond lengths of ethylene and acetylene. A valence bond study \cite{10.1021.ct100577v} has shown that the ground state of C$_2$ indeed has important contributions of double [$\sigma + \pi$] and triple [$\sigma + 2\pi$] bonded Slater determinants, while the double [$2\pi$] bonded contributions are small. The molecular orbital diagram in Fig. \ref{MOdiagramFig} predicts that the Slater determinant with largest weight is
\begin{equation}
\ket{(\text{core}) 2\sigma_g^2 2\sigma_u^2 1\pi_u^4}. \label{MOdiagram}
\end{equation}
This determinant represents a double [$2\pi$] bond. Based on the shapes of the NOs, it can be argued that Eq. \eqref{MOdiagram} effectively represents a triple [$\sigma + 2\pi$] bond: the NO $2\sigma_u$ is only weakly antibonding, while the NO $2\sigma_g$ is strongly bonding \cite{10.1021.ct100577v}. The determinant
\begin{equation}
\ket{(\text{core}) 2\sigma_g^2 1\pi_u^4 3\sigma_g^2}
\end{equation}
also contributes significantly to the triple [$\sigma + 2\pi$] bond. In the double [$\sigma + \pi$] and triple [$\sigma + 2\pi$] bonds, the $\pi$-bond is of the charge-shift type, while the $\sigma$-bond is covalent \cite{10.1021.ct100577v}.

\begin{figure}
\centering
\includegraphics[width=0.70\textwidth]{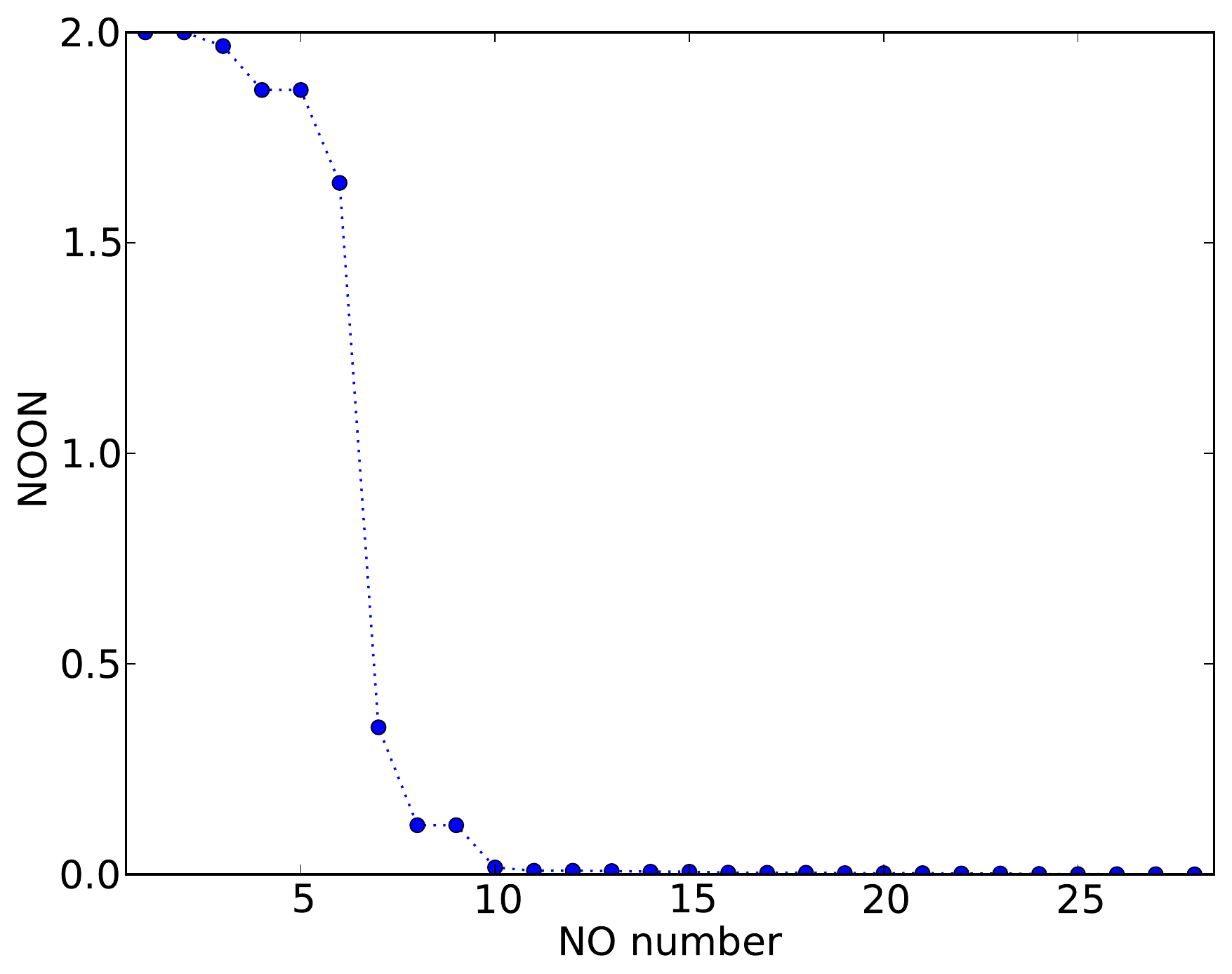}
\caption{\label{NOONspectrumPlot} NOON spectrum for the $X^1\Sigma_g^+$ state of the carbon dimer at $R$ = 2.4 a.u. The calculations were performed with \textsc{CheMPS2} at the DMRG(28o,12e,D$_{\mathsf{SU(2)}}=2500$)/cc-pVDZ level of theory.}
\end{figure}

From the NO occupation number (NOON) spectrum in Fig. \ref{NOONspectrumPlot} it can be observed that, while the MR character is significant, the carbon dimer is not a real diradical. Remember that the carbon dimer has important contributions of both double [$\sigma + \pi$] and triple [$\sigma + 2\pi$] bonded Slater determinants. In the former, the valence electrons are all paired (:C=C:), but contributions of this type only carry 20-25\% of the total weight \cite{10.1021.ct100577v}.

The authors of Ref. \cite{naturechemFourthBond} suggested that the carbon dimer has a fourth bond, which suppresses the diradical character further. Six of the eight valence electrons then participate in the triple [$\sigma + 2\pi$] bond ($\cdot \text{C} \equiv \text{C} \cdot$). The fourth bond is formed by the singlet pairing of the two remaining valence electrons, which reside in the outward pointing orbitals. The breaking of this singlet pair corresponds to the transition $X^1\Sigma_g^+ \rightarrow c^3\Sigma_u^+$, which allows to estimate the strength of the suggested fourth bond: $14~\text{kcal}~\text{mol}^{-1}$ \cite{naturechemFourthBond}. As this bond is stronger than a hydrogen bond (about $1~\text{to}~7~\text{kcal}~\text{mol}^{-1}$), the authors nominate it as the fourth bond in the carbon dimer.

A recent valence bond study \cite{10.1021.ct400867h} has shown that the interpretation of the fourth bond as the singlet pairing of the two remaining valence electrons in the triple [$\sigma + 2\pi$] bonded configuration is wrong. They have found that a more accurate description of the dimer is obtained by an antiferromagnetic coupling of the electrons. Two valence electrons are coupled to a singlet in the $\sigma$-bond. The remaining three valence electrons of each carbon atom first couple locally (per carbon atom) to a spin quartet. The two local spin quartets then couple to a global spin singlet ($X^1\Sigma_g^+$).

\section{Low-lying bond dissociation curves}

The debate on the bond order of the carbon dimer is still not settled. The only thing which has generally been agreed upon is that only MR methods provide a \textit{meaningful way} to calculate the bond dissociation curves of the low-lying states of this dimer. We have used C$_2$ as a benchmark test for \textsc{CheMPS2} [Ref. \cite{2013arXiv1312.2415W}]:

\newpage
\hspace{-\parindent}{\large\color{blue}{\textbf{\textsc{CheMPS2}: A free open-source spin-adapted implementation of the density matrix renormalization group for ab initio quantum chemistry} \cite{PhysRevB.88.075122}}}

\vspace{0.2cm}

Sebastian Wouters,$^{a}$ Ward Poelmans,$^{a}$ Paul W. Ayers,$^{b}$ and Dimitri Van Neck$^{a}$

{\footnotesize $^{a}$\textit{Center for Molecular Modelling, Ghent University, Technologiepark 903, 9052 Zwijnaarde, Belgium}}

{\footnotesize $^{b}$\textit{Department of Chemistry, McMaster University, Hamilton, Ontario L8S 4M1, Canada}}

\vspace{0.5cm}

\parbox{0.90\textwidth}{
The density matrix renormalization group (DMRG) has become an indispensable numerical tool to find exact eigenstates of finite-size quantum systems with strong correlation. In the fields of condensed matter, nuclear structure and molecular electronic structure, it has significantly extended the system sizes that can be handled compared to full configuration interaction, without losing numerical accuracy. For quantum chemistry (QC), the most efficient implementations of DMRG require the incorporation of particle number, spin and point group symmetries in the underlying matrix product state (MPS) ansatz, as well as the use of so-called complementary operators. The symmetries introduce a sparse block structure in the MPS ansatz and in the intermediary contracted tensors. If a symmetry is non-abelian, the Wigner-Eckart theorem allows to factorize a tensor into a Clebsch-Gordan coefficient and a reduced tensor. In addition, the fermion signs have to be carefully tracked. Because of these challenges, implementing DMRG efficiently for QC is not straightforward. Efficient and freely available implementations are therefore highly desired. In this work we present \textsc{CheMPS2}, our free open-source spin-adapted implementation of DMRG for ab initio QC. Around \textsc{CheMPS2}, we have implemented the augmented Hessian Newton-Raphson complete active space self-consistent field method, with exact Hessian. The bond dissociation curves of the 12 lowest states of the carbon dimer were obtained at the DMRG(28 orbitals, 12 electrons, D$_{ \mathsf{ SU(2) }}$=2500)/cc-pVDZ level of theory. The contribution of $1s$ core correlation to the $X^1\Sigma_g^+$ bond dissociation curve of the carbon dimer was estimated by comparing energies at the DMRG(36o, 12e, D$_{\mathsf{SU(2)}}$=2500)/cc-pCVDZ and DMRG-SCF(34o, 8e, D$_{\mathsf{SU(2)}}$=2500)/cc-pCVDZ levels of theory.
}

\vspace{0.7cm}
\hspace{-\parindent}\textbf{Program Summary}
\vspace{0.3cm}
\def \CPCspacing {-0.35cm}

\noindent{\em Program title:} \textsc{CheMPS2}                                  \\

\vspace{\CPCspacing}
\noindent{\em Catalogue identifier:} AESE\_v1\_0                         \\

\vspace{\CPCspacing}
\noindent{\em Program summary URL:} \url{http://cpc.cs.qub.ac.uk/summaries/AESE\_v1\_0.html} \\

\vspace{\CPCspacing}
\noindent{\em Program obtainable from:} CPC Program Library, Queen's University, Belfast, N. Ireland \\

\vspace{\CPCspacing}
\noindent{\em Licensing provisions:} GNU General Public License, version 2 \\

\vspace{\CPCspacing}
\noindent{\em No. of lines in distributed program, including test data, etc.:} 131472 \\

\vspace{\CPCspacing}
\noindent{\em No. of bytes in distributed program, including test data, etc.:} 1645700 \\

\vspace{\CPCspacing}
\noindent{\em Distribution format:} tar.gz \\

\vspace{\CPCspacing}
\noindent{\em Programming language:} \texttt{C++}.                             \\

\vspace{\CPCspacing}
\noindent{\em Computer:} x86-64.                                        \\

\vspace{\CPCspacing}
\noindent{\em Operating system:} Scientific Linux 6.0.                  \\

\vspace{\CPCspacing}
\noindent{\em RAM:} 10 MB - 64 GB                                        \\

\vspace{\CPCspacing}
\noindent{\em Classification:} 16.1. \\

\vspace{\CPCspacing}
\noindent{\em External routines:} Basic Linear Algebra Subprograms (BLAS), Linear Algebra Package (LAPACK), GNU Scientific Library (GSL), and Hierarchical Data Format Release 5 (HDF5) \\

\vspace{\CPCspacing}
\noindent{\em Nature of problem:}\\
The many-body Hilbert space grows exponentially with the number of single-particle states. Exact diagonalization solvers can therefore only handle small systems, of up to 18 electrons in 18 orbitals. Interesting active spaces are often significantly larger. \\

\vspace{\CPCspacing}
\noindent{\em Solution method:}\\
The density matrix renormalization group allows the extension of the size of active spaces, for which numerically exact solutions can be found, to about 40 electrons in 40 orbitals. In addition, it provides a rigorous variational upper bound to energies, as it has an underlying wavefunction ansatz, the matrix product state. \\

\vspace{\CPCspacing}
\noindent{\em Restrictions:}\\
Our implementation of the density matrix renormalization group is spin-adapted. This means that targeted eigenstates in the active space are exact eigenstates of the total electronic spin operator. Hamiltonians which break this symmetry (a magnetic field term for example) cannot be handled by our code. As electron repulsion integrals in Gaussian basis sets have eightfold permutation symmetry, we have used this property in our code. \\

\vspace{\CPCspacing}
\noindent{\em Unusual features:}\\
The nature of the matrix product state ansatz allows for exact spin coupling. In \textsc{CheMPS2}, the total electronic spin is imposed (not just the spin projection), in addition to the particle-number and abelian point-group symmetries. \\

\vspace{\CPCspacing}
\noindent{\em Running time:}\\
The running time depends on the size of the targeted active space, the number of desired eigenstates, their symmetry, the density of states, the individual orbital symmetries, the orbital ordering, the desired level of convergence, and the chosen convergence scheme. To converge a single point of one of the dissociation curves of the carbon dimer ($D_{\infty h} \rightarrow D_{2h}$ symmetry) in the cc-pVDZ basis (28 orbitals; their ordering is described in Section 5.3) with 2500 reduced renormalized basis states (see the convergence scheme in Section 5.4; the variational energy then lies 0.1 $mE_h$ above the fully converged result) takes about 8 h on a single node with a dual-socket octa-core Intel Xeon Sandy Bridge (E5-2670) (16 cores at 2.6 GHz), and requires 6 GB of RAM.

\vspace{0.7cm}
\hspace{-\parindent}\textbf{1. Introduction}
\vspace{0.3cm}

Conventional molecular electronic structure methods such as density functional theory, Hartree-Fock theory, and coupled cluster theory start with the assumption that a single Slater determinant (SD) provides a qualitatively good description of the molecule at hand \cite{helgaker2}. While this assumption is valid for some molecules near equilibrium geometry, the static correlation which arises in other molecules, as well as for geometries far from equilibrium, requires the use of multireference (MR) methods. These provide a qualitative description which is equivalent to multiple SDs, thereby resolving the static correlation. One of these MR methods is the exact diagonalization of the many-body Hamiltonian in the full Hilbert space, also known as full configuration interaction (FCI) in quantum chemistry (QC). Because the many-body Hilbert space grows exponentially with the number of single-particle states, only small systems, of up to 18 electrons in 18 orbitals, can be treated by FCI. In 1999, the density matrix renormalization group (DMRG) was introduced in QC \cite{WhiteQCDMRG}. This MR method allows to extend the system sizes for which numerically exact solutions can be found to about 40 electrons in 40 orbitals, depending on the nature of the system.

DMRG originated in 1992 in the field of condensed matter \cite{PhysRevLett.69.2863, PhysRevB.48.10345}. Although it was originally introduced as a renormalization group flow for increasing many-body Hilbert spaces, in 1995 it was realized that DMRG can be reformulated as the variational optimization of a particular wavefunction ansatz, the matrix product state (MPS) \cite{PhysRevLett.75.3537,PhysRevB.55.2164}. This not only provided the theoretical validation that an energy obtained with DMRG is always an upper bound to the exact eigenvalue, but also shed light on DMRG from a quantum information perspective. Non-critical quantum mechanical ground states are believed to obey the so-called area law for the entanglement entropy \cite{1742-5468-2007-08-P08024}. This implies that quantum correlation is local in such a ground state. For one-dimensional systems, the boundary of a line segment consists of two points, and the entanglement entropy is a constant, independent of system length. This is the reason why DMRG works extremely well for one-dimensional non-critical systems. Quantum information theory also induced the development of other so-called tensor network states (TNS), which capture the entanglement entropy well in higher dimensional and/or critical systems \cite{PEPS-arxiv,PhysRevLett.99.220405}. There even exists a continuous MPS ansatz for quantum fields \cite{PhysRevLett.104.190405}.

Although the active orbital space of most molecular systems is far from one-dimensional, DMRG has been very useful for ab initio QC \cite{WhiteQCDMRG, QUA:QUA1, mitrushenkov:6815, Chan2002, PhysRevB.67.125114, chan:8551, DMRG_LiF, mitrushenkov:4148, PhysRevB.68.195116, chan:3172, chan:6110, PhysRevB.70.205118, moritz:024107, chan:204101, moritz:184105, moritz:034103, hachmann:144101, Rissler2006519, moritz:244109, dorando:084109, hachmann:134309, marti:014104, zgid:014107, zgid:144115, zgid:144116, ghosh:144117, ChanB805292C, ChanQUA:QUA22099, dorando:184111,kurashige:234114,yanai:024105, neuscamman:024106, RichardsonControlReiher, PhysRevB.81.235129, mizukami:091101, PhysRevA.83.012508, boguslawski:224101, kurashige:094104, QUA:QUA23173, sharma:124121, woutersJCP1, JCTCspindens, C2CP23767A, JPCLentanglement, JCTCgrapheneNano, JCTCbondForm, naturechem, ma:224105, saitow:044118, Spiropyran, C3CP53975J, Knecht-4c-DMRG}. The variational upper bound to the true eigenvalue, obtained with DMRG, can be systematically improved by increasing the so-called bond or virtual dimension of the MPS ansatz. This provides a way to check the convergence of DMRG calculations.

In ab initio QC methods which use FCI, the FCI solver can be replaced by DMRG. Ab initio DMRG allows for an efficient extraction of the reduced two-body density matrix (2-RDM) \cite{zgid:144115}. The 2-RDM of the active space is required in the complete active space self-consistent field (CASSCF) method to compute the gradient and the Hessian. It is therefore natural to introduce a CASSCF variant with DMRG as active space solver, DMRG-SCF \cite{zgid:144116}. This allows one to describe static correlation in large active spaces. To add dynamic correlation as well, three DMRG-based methods have been introduced. (a) With a little more effort, the 3-RDM and contracted 4-RDMs can be extracted from DMRG as well. These are required to apply second order perturbation theory to a CASSCF wavefunction, called CASPT2. The DMRG variant is DMRG-CASPT2 \cite{kurashige:094104}. (b) Based on a CASSCF wavefunction, a configuration interaction expansion can be introduced, called MRCI. Recently, an approximate DMRG-MRCI variant was proposed \cite{saitow:044118}. (c) Yet another way is to perform a canonical transformation (CT) on top of an MR wavefunction. When an MPS is used as MR wavefunction, the method is called DMRG-CT \cite{yanai:024105}.

In addition to ground states, DMRG can also find excited states. By projecting out lower lying eigenstates, or by targeting a specific energy \cite{dorando:084109}, the DMRG algorithm solves for a particular excited state. In these state-specific algorithms, the whole renormalized basis is used to represent one single eigenstate. In state-averaged DMRG, several eigenstates are targeted at once. Their RDMs are weighted and summed to perform the DMRG renormalization step \cite{HallbergBook}. The renormalized basis then represents several eigenstates at once.

DMRG linear response theory (DMRG-LRT) can be used as well to find excited states. Once the ground state has been found, the MPS tangent vectors to this optimized point can be used as an (incomplete) variational basis to approximate excited states \cite{dorando:184111, PhysRevB.85.035130, PhysRevB.85.100408, PhysRevB.88.075122, PhysRevB.88.075133, NaokiLRTpaper}. As the tangent vectors to an optimized SD yield the configuration interaction with singles (CIS), also called the Tamm-Dancoff approximation (TDA), for Hartree-Fock theory \cite{helgaker2}, the same names are used for DMRG: DMRG-CIS or DMRG-TDA. By linearizing the time-dependent variational principle for matrix product states \cite{PhysRevLett.107.070601}, the DMRG random phase approximation (DMRG-RPA) is found \cite{PhysRevB.88.075122, PhysRevB.88.075133, NaokiLRTpaper}, again in complete analogy with RPA for Hartree-Fock theory. The variational optimization in an (incomplete) basis of MPS tangent vectors can be extended to higher-order tangent spaces as well. DMRG-CISD, or DMRG configuration interaction with singles and doubles, is a variational approximation to target both ground and excited states in the space spanned by the MPS reference and its single and double tangent spaces \cite{PhysRevB.88.075122}.

In ab initio QC, two other TNSs have been employed as well: the tree TNS \cite{PhysRevB.82.205105, nakatani:134113} and the complete-graph TNS \cite{1367-2630-12-10-103008}. While they require a smaller virtual dimension to achieve the same accuracy, their optimization algorithms are less efficient, and as a result an MPS is currently still the preferred choice for ab initio QC.

In Section 2, the DMRG algorithm is briefly introduced, and remarks specific to ab initio QC are discussed. In Section 3, the implementation of particle number, spin, and abelian point group symmetries is presented. An overview of the structure of \textsc{CheMPS2} is given in Section 4. Results on the low-lying states of the carbon dimer are presented in Section 5. A summary is given in Section 6. Atomic units are used in this work: $E_h = 4.35974434(19) \times 10^{-18}$ J and $a_0 = 5.2917721092(17) \times 10^{-11}$ m \cite{RevModPhys.84.1527}.

\newpage
\hspace{-\parindent}\textbf{2. DMRG for ab initio quantum chemistry}
\vspace{0.3cm}

\hspace{-\parindent}\textbf{2.1 The MPS ansatz}
\vspace{0.3cm}

DMRG can be formulated as the variational optimization of an MPS. The MPS ansatz with open boundary conditions is given by
\begin{equation}
\ket{\Psi} = \sum\limits_{\{ n_k \}, \{ \alpha_j \}} A[1]^{n_1}_{\alpha_1} A[2]^{n_2}_{\alpha_1;\alpha_2} ... A[L-1]^{n_{L-1}}_{\alpha_{L-2};\alpha_{L-1}} A[L]^{n_L}_{\alpha_{L-1}} \ket{n_1 n_2 ... n_L}
\end{equation}
where $n_k$ denotes the occupancy of orbital $k$ ($\ket{-}$, $\ket{\uparrow}$, $\ket{\downarrow}$, or $\ket{\uparrow\downarrow}$) and the $\{\alpha_j\}$ are the so-called bond or virtual indices. With increasing dimension $D$ of these virtual indices, a larger part of the Hilbert space can be reached. Note that it is of no use to make virtual dimension $D_j$ larger than min$(4^j,4^{L-j})$, the minimum of the sizes of the partial Hilbert spaces spanned by resp. the first $j$ and the last $L-j$ orbitals.

\vspace{0.7cm}
\hspace{-\parindent}\textbf{2.2 Canonical forms}
\vspace{0.3cm}

The wavefunction $\ket{\Psi}$ does not uniquely define the ansatz, in analogy with a Slater determinant. For the latter, a rotation in the occupied orbital space alone, or a rotation in the virtual orbital space alone, does not change the physical wavefunction. Only occupied-virtual rotations change the wavefunction. In an MPS, there is gauge freedom as well. If for two neighbouring sites $i$ and $i+1$, the left MPS tensors are right-multiplied with the non-singular matrix $G$
\begin{equation}
\tilde{A}[i]^{n_i}_{\alpha_{i-1};\alpha_i} = \sum\limits_{\alpha_j} A[i]^{n_i}_{\alpha_{i-1};\alpha_j} G_{\alpha_j;\alpha_i}
\end{equation}
and the right MPS tensors are left-multiplied with the inverse of $G$
\begin{equation}
\tilde{A}[i+1]^{n_{i+1}}_{\alpha_{i};\alpha_{i+1}} = \sum\limits_{\alpha_j} G^{-1}_{\alpha_{i};\alpha_{j}} A[i+1]^{n_{i+1}}_{\alpha_{j};\alpha_{i+1}}
\end{equation}
the wavefunction does not change, i.e. $\forall n_i,n_{i+1},\alpha_{i-1},\alpha_{i+1}$:
\begin{equation}
\sum\limits_{\alpha_i} \tilde{A}[i]^{n_i}_{\alpha_{i-1};\alpha_i} \tilde{A}[i+1]^{n_{i+1}}_{\alpha_{i};\alpha_{i+1}} = \sum\limits_{\alpha_i} A[i]^{n_i}_{\alpha_{i-1};\alpha_i} A[i+1]^{n_{i+1}}_{\alpha_{i};\alpha_{i+1}}.
\end{equation}
\textsc{CheMPS2} is a two-site DMRG algorithm, were at each so-called micro-iteration two neighbouring sites are simultaneously optimized. Suppose these sites are $i$ and $i+1$. The gauge freedom of the MPS is used to bring it in a particular canonical form. For all sites to the left of $i$, the MPS tensors are left-normalized:
\begin{equation}
\sum\limits_{\alpha_{k-1}, n_k} \left(A[k]^{n_k}\right)^{\dagger}_{\alpha_k; \alpha_{k-1}} A[k]^{n_k}_{\alpha_{k-1};\beta_k} = \delta_{\alpha_k, \beta_k} \label{CPCleft-normalized}
\end{equation}
and for all sites to the right of $i+1$, the MPS tensors are right-normalized:
\begin{equation}
\sum\limits_{\alpha_{k}, n_k} A[k]^{n_k}_{\alpha_{k-1};\alpha_k} \left(A[k]^{n_k}\right)^{\dagger}_{\alpha_k; \beta_{k-1}} = \delta_{\alpha_{k-1}, \beta_{k-1}}. \label{CPCright-normalized}
\end{equation}

\vspace{0.7cm}
\hspace{-\parindent}\textbf{2.3 The effective Hamiltonian equation}
\vspace{0.3cm}

Combine the MPS tensors of the two sites under consideration into a single two-site tensor:
\begin{equation}
\sum\limits_{\alpha_i} A[i]^{n_i}_{\alpha_{i-1};\alpha_i} A[i+1]^{n_{i+1}}_{\alpha_{i};\alpha_{i+1}} = B[i]_{\alpha_{i-1};\alpha_{i+1}}^{n_i;n_{i+1}}. \label{CPCtwo-site-object-not-reduced}
\end{equation}
At the current micro-iteration of the DMRG algorithm, $\mathbf{B}[i]$ (the flattened form of the tensor $B[i]$) is used as an initial guess for the effective Hamiltonian equation. This equation is obtained by variation of the Lagrangian \cite{ChanB805292C}
\begin{equation}
\mathcal{L} = \braket{\Psi(\mathbf{B}[i]) \mid \hat{H} \mid \Psi(\mathbf{B}[i])} - \lambda \braket{\Psi(\mathbf{B}[i]) \mid \Psi(\mathbf{B}[i])}
\end{equation}
to the complex conjugate of $\mathbf{B}[i]$:
\begin{equation}
\mathbf{H}^{\text{eff}} \mathbf{B}[i] = \lambda \mathbf{B}[i]. \label{CPCeffHameq}
\end{equation}
The specific canonical choice of Eqs. \eqref{CPCleft-normalized}-\eqref{CPCright-normalized} ensured that no overlap matrix is present in this effective Hamiltonian equation. The lowest eigenvalue and corresponding eigenvector of this equation are searched. In \textsc{CheMPS2}, this is done with our implementation of Davidson's algorithm \cite{Davidson197587}. Once found, it is decomposed with a singular value decomposition:
\begin{equation}
B[i]_{ \left( \alpha_{i-1} n_i \right) ; \left( n_{i+1} \alpha_{i+1} \right)} = \sum\limits_{\beta} U[i]_{ \left( \alpha_{i-1} n_i \right) ; \beta} \kappa[i]_{\beta} V[i]_{ \beta ; \left( n_{i+1} \alpha_{i+1} \right)}.
\end{equation}
Note that $U[i]$ is hence left-normalized and $V[i]$ right-normalized. In the DMRG algorithm, the original sum over $\beta$ of dimension $\min(4D_{i-1},4D_{i+1})$ is truncated to $D_i$, thereby keeping the $D_i$ largest $\kappa[i]_{\beta}$.

\vspace{0.7cm}
\hspace{-\parindent}\textbf{2.4 Sweeping}
\vspace{0.3cm}

So far, we have looked at a micro-iteration of the DMRG algorithm. This micro-iteration happens during left or right sweeps. During a left sweep, $B[i]$ is constructed, the corresponding effective Hamiltonian equation solved, the solution $B[i]$ decomposed, the singular value spectrum truncated, $A[i]$ is set to $U[i] \times \kappa[i]$, $A[i+1]$ is set to $V[i]$, and $i$ is decreased by 1. Note that $A[i+1]$ is right-normalized for the next micro-iteration as required. This stepping to the left occurs until $i=0$, and then the sweep direction is reversed from left to right. Based on energy differences, or wavefunction overlaps, between consecutive sweeps, a convergence criterion is triggered, and the sweeping stops. One sweep is called a macro-iteration in DMRG.

\vspace{0.7cm}
\hspace{-\parindent}\textbf{2.5 Complementary operators}
\vspace{0.3cm}

The effective Hamiltonian in Eq. \eqref{CPCeffHameq} is too large to be fully constructed. Only its action on a particular guess $\mathbf{B}[i]$ is available as a function. In order to construct $\mathbf{H}^{\text{eff}} \mathbf{B}[i]$ efficiently for general quantum chemistry Hamiltonians, several tricks are used. (a) The one-body matrix elements $(i|T|k)$ are incorporated in the two-body matrix elements $(ij|V|kl)$:
\begin{equation}
(ij|h|kl) = (ij|V|kl) + \frac{1}{N-1} \left[ (i|T|k) \delta_{j,l} + (j|T|l) \delta_{i,k} \right] 
\end{equation}
where $N$ is the targeted particle number. (b) Suppose we want to optimize sites $i$ and $i+1$, and that $\ket{\alpha_{i-1}}$ are the corresponding $D_{i-1}$ left renormalized basis states. Renormalized operators such as $\braket{\alpha_{i-1} \mid \hat{a}_{k\sigma}^{\dagger} \hat{a}_{l\tau} \mid \beta_{i-1}}$ with $k$ and $l$ both smaller than $i$ are constructed and stored on disk \cite{Chan2002}. For the second quantized operators $\hat{a}^{\dagger}$ and $\hat{a}$, the Latin indices denote orbitals and the Greek indices spin projections. (c) Once three second quantized operators are on one side of $B[i]$, they are multiplied with the matrix elements $(ij|h|kl)$, and a summation is performed over the common indices to construct complementary operators \cite{PhysRevB.53.R10445}:
\begin{equation}
\sum\limits_{\sigma} \sum\limits_{k,l,m<i} \braket{\alpha_{i-1} \mid \hat{a}_{k\sigma}^{\dagger} \hat{a}^{\dagger}_{l\tau} \hat{a}_{m\sigma} \mid \beta_{i-1}} \times (kl|h|mn) \rightarrow \braket{\alpha_{i-1} \mid \hat{O}_{n\tau} \mid \beta_{i-1}}.
\end{equation}
For two, three, and four second quantized operators on one side of $B[i]$, these complementary operators are constructed. A bare (without matrix elements) renormalized operator is only constructed for one or two second quantized operators on one side of $B[i]$. (d) Hermitian conjugation
\begin{equation}
\braket{\alpha_{i-1} \mid \hat{a}_{k\sigma}^{\dagger} \hat{a}^{\dagger}_{l\tau} \mid \beta_{i-1}} = \braket{\beta_{i-1} \mid \hat{a}_{l\tau} \hat{a}_{k\sigma} \mid \alpha_{i-1}}^{\dagger}
\end{equation}
and commutation relations between the second quantized operators are also used to further limit the storage requirement for the renormalized partial Hamiltonian terms.

\vspace{0.7cm}
\hspace{-\parindent}\textbf{2.6 Convergence}
\vspace{0.3cm}

There is also a one-site DMRG algorithm, in which only one MPS site tensor is optimized at each micro-iteration, but this algorithm is more likely to get stuck in a local minimum. To help prevent the two-site DMRG algorithm from getting stuck in a local minimum, a small amount of noise can be added to the solution $B[i]$, just before it is decomposed. This way, renormalized basis states corresponding to lost symmetries (which should be there, but are not) can be reintroduced \cite{Chan2002}.

The choice of orbitals and their ordering on the one-dimensional DMRG lattice have a significant influence both on getting stuck in local minima, as well as on how fast the variational energy $E_D$ converges with increasing $D$ \cite{WhiteQCDMRG}. The optimal choice and ordering are still under debate, although two rules of thumb are widely used. Active space orbitals in elongated molecular systems (think about polyenes for example) should be localized as much as possible to respect the area law for the entanglement entropy \cite{QUA:QUA23173}. For small molecules with a high point group symmetry, it is beneficial to put bonding and anti-bonding orbitals close to each other on the one-dimensional DMRG lattice, as they are most strongly correlated \cite{ma:224105}.

One possibility to settle this ongoing debate might be to look at the so-called two-orbital mutual information $I_{p,q}$ in the future \cite{Rissler2006519}. This is a measure from quantum information theory for the amount of correlation between two orbitals, and is a two-point correlation function on the one-dimensional DMRG lattice. A cost function can be associated with this measure, e.g. $F = \sum_{p,q} I_{p,q} (p-q)^z$, which requires highly correlated orbitals to be close. Its gradient and Hessian with respect to orbital rotations can be calculated by resp. three- and four-point correlation functions on the one-dimensional DMRG lattice. These can be obtained efficiently \cite{zgid:144115}. If local minima can be avoided, this yields a set of minimally entangled orbitals and their optimal ordering, from which extra rules of thumb can be drawn.

Two extrapolation schemes exist to assess the convergence of the variational energy $E_D$ with increasing number of renormalized basis states $D$. The first is the scaling relation
\begin{equation}
\ln(E_D - E_{exact} ) = C_1 - C_2 (\ln(D))^2 
\end{equation}
proposed by Chan \cite{Chan2002,AyersRDMcalues,woutersJCP1} which is nowadays not often used. The $C_i$ are constants which are determined by the fit. The second and most widely used extrapolation scheme is based on the so-called maximal discarded weight $w^{disc}(D)$ during the last DMRG sweep for a certain value of $D$:
\begin{equation}
w^{disc}(D) = \max\limits_{i} \left\{ \sum\limits_{\beta = D+1}^{4D} \kappa[i]^2_{\beta} \right\}.
\end{equation}
It proposes a linear relation between the variational energy $E_D$ and the discarded weight $w^{disc}(D)$ \cite{PhysRevB.53.14349, Chan2002, 2013arXiv1307.1002V}:
\begin{equation}
E(D) = E_{exact} + C_1 ~ w^{disc}(D). \label{CPCextrapolSchemeEq}
\end{equation}
By increasing $D$ stepwise, $E_{exact}$ can be extrapolated.

\vspace{0.7cm}
\hspace{-\parindent}\textbf{3. Symmetry-adapted DMRG}
\vspace{0.3cm}

\hspace{-\parindent}\textbf{3.1 Introduction}
\vspace{0.3cm}

The symmetry group of the Hamiltonian can be used to label eigenstates by symmetry. To find an eigenstate with a particular symmetry, it is sufficient to restrict an optimization to the corresponding corner of the many-body Hilbert space. For DMRG, it is well understood how both abelian and non-abelian symmetries can be imposed \cite{0295-5075-57-6-852, 1742-5468-2007-10-P10014, 1367-2630-12-3-033029, PhysRevA.82.050301}. Each MPS tensor and intermediary contracted tensor decompose into a Clebsch-Gordan coefficient and a reduced tensor. The Clebsch-Gordan coefficient introduces a sparse block structure in the reduced tensor. If the symmetry group of the Hamiltonian is non-abelian, some irreducible representations (irrep) have a dimension larger than one, and then this factorization also presents an information compression, as the size of the full tensor is larger than the size of the reduced tensor. In addition to the possibility of restricting an optimization to a particular symmetry corner of the many-body Hilbert space, this sparsity and compression result in smaller requirements in disk, memory and computer time.

In \textsc{CheMPS2}, we have implemented three global symmetries for the MPS wavefunction: $\mathsf{SU(2)}$ total electronic spin, $\mathsf{U(1)}$ particle number, and abelian point group symmetry $\mathsf{P}$. As we work real-valued in \textsc{CheMPS2}, the latter are restricted to $\mathsf{P} \in \left\{C_1, C_i, C_2, C_s, D_2, C_{2v}, C_{2h}, D_{2h} \right\}$ \cite{BookCornwell}.

\vspace{0.7cm}
\hspace{-\parindent}\textbf{3.2 Reduced MPS tensors}
\vspace{0.3cm}

These global symmetries are imposed by requiring that the MPS site tensors $A[i]^{n_i}_{\alpha_{i-1};\alpha_i}$ are irreducible tensor operators of the total symmetry group \cite{PhysRevA.82.050301, 1367-2630-12-3-033029, 1742-5468-2007-10-P10014, 0295-5075-57-6-852}. The local and virtual basis states ($\ket{n_k}$ and $\ket{\alpha_j}$) then have to transform according to the rows of the irreps of this symmetry group. This is realized by rotating the basis states so that they can be represented by good spin ($s$ and $j$), spin projection ($s^z$ and $j^z$), particle number ($N$), and point group irrep ($I$) quantum numbers.

The local basis states of orbital $k$ are labelled as
\begin{eqnarray}
\ket{-} & \rightarrow & \ket{s=0; s^z=0, N=0; I=I_0} \\
\ket{\uparrow} & \rightarrow & \ket{s=\frac{1}{2}; s^z=\frac{1}{2}, N=1; I=I_k} \\
\ket{\downarrow} & \rightarrow & \ket{s=\frac{1}{2}; s^z=-\frac{1}{2}, N=1; I=I_k} \\
\ket{\uparrow\downarrow} & \rightarrow & \ket{s=0; s^z=0, N=2; I=I_0}
\end{eqnarray}
where $I_0$ and $I_k$ are resp. the trivial and orbital $k$ point group irreps. $\ket{\uparrow\downarrow}$ corresponds to $I_0$ because for the abelian point groups with real-valued character tables, $\forall I_k : I_k \otimes I_k = I_0$. In the same way, the virtual basis states are labelled as
\begin{equation}
\ket{\alpha} \rightarrow \ket{j j^z N I \alpha}
\end{equation}
where the $\alpha$ on the right-hand side allows to distinguish between separate virtual basis states which belong to the same symmetry.

Due to the Wigner-Eckart theorem, each irreducible tensor operator $A[i]$ factorizes into Clebsch-Gordan coefficients and a reduced tensor $T[i]$:
\begin{eqnarray}
& A[i]^{n_i}_{\alpha_{i-1};\alpha_i} = A[i]^{s s^z N I}_{j_L j_L^z N_L I_L \alpha_{i-1}; j_R j_R^z N_R I_R \alpha_i} \nonumber\\
& = \braket{j_L j_L^z s s^z | j_R j_R^z} \delta_{N_L + N, N_R} \delta_{I_L \otimes I, I_R} T[i]^{(s N I)}_{(j_L N_L I_L \alpha_L)(j_R N_R I_R \alpha_R)} . \quad \label{CPCtensordecomp}
\end{eqnarray}
The $\mathsf{SU(2)}$, $\mathsf{U(1)}$, and $\mathsf{P}$ symmetries are imposed by their corresponding Clebsch-Gordan coefficients, and express nothing else than resp. local allowed spin recoupling, local particle conservation, and local point group symmetry conservation. The indices $\alpha_L$ and $\alpha_R$ keep track of the number of times an irrep occurs at a virtual bond. If the virtual dimension of a symmetry sector is $D(j_L N_L I_L)$, this would correspond to a dimension of $(2 j_L + 1) D(j_L N_L I_L)$ in an MPS which is not symmetry-adapted \cite{0295-5075-57-6-852}. If a Clebsch-Gordan coefficient is zero by symmetry, the corresponding blocks in $T[i]$ do not need to be allocated, resulting in sparse block structure. If $j$ or $s$ are not spin-0, there is in addition data compression.

The desired global symmetry can be imposed on the MPS by requiring that the left virtual index of the leftmost tensor in the MPS chain consists of one irrep corresponding to $(j_L, N_L, I_L) = (0,0,I_0)$, while the right virtual index of the rightmost tensor consists of one irrep corresponding to $(j_R, N_R, I_R) = (S_G, N_G, I_G)$, the desired global spin, particle number, and point group symmetry. This corresponds to the singlet-embedding strategy of Sharma and Chan \cite{sharma:124121}.

The operators
\begin{eqnarray}
\hat{b}^{\dagger}_{k \sigma} & = & \hat{a}^{\dagger}_{k \sigma} \label{CPCcreaannih1}\\
\hat{b}_{k \sigma} & = & (-1)^{\frac{1}{2}-\sigma}\hat{a}_{k -\sigma} \label{CPCcreaannih2}
\end{eqnarray}
for orbital $k$ correspond to resp. the $(s=\frac{1}{2}, s^z=\sigma, N=1, I_k)$ row of irrep $(s=\frac{1}{2}, N=1, I_k)$ and the $(s=\frac{1}{2}, s^z=\sigma, N=-1, I_k)$ row of irrep $(s=\frac{1}{2}, N=-1, I_k)$ \cite{BookDimitri}. $\hat{b}^{\dagger}$ and $\hat{b}$ are hence both doublet irreducible tensor operators. This fact permits exploitation of the Wigner-Eckart theorem also for renormalized operators and complementary operators, and to develop a code without any spin projections or $\mathsf{SU(2)}$ Clebsch-Gordan coefficients. Contracting terms of the type \eqref{CPCtensordecomp} and \eqref{CPCcreaannih1}-\eqref{CPCcreaannih2} can be done by implicitly summing over the common multiplets and recoupling the local, virtual and operator spins. An example is given in the Appendix. Operators and complementary operators then formally consist of terms containing Clebsch-Gordan coefficients and reduced tensors. In our code, however, only the reduced tensors need to be calculated and stored. \textsc{CheMPS2} uses the GNU Scientific Library \cite{GSLcitation} to extract Wigner 6-$j$ and 9-$j$ symbols for the recoupling. No Wigner 3-$j$ symbols or Clebsch-Gordan coefficients are used in the program.

\vspace{0.7cm}
\hspace{-\parindent}\textbf{3.3 The reduced two-site object}
\vspace{0.3cm}

Section 2.3 can be reformulated with the reduced $T$-tensors from Eq. \eqref{CPCtensordecomp} and a reduced two-site object $S[i]$:
\begin{eqnarray}
& \hspace{-0.1cm} S[i]^{j(s_1 s_2) N_1 N_2 I_1 I_2}_{j_L N_L I_L \alpha_L ; j_R N_R I_R \alpha_R} = \delta_{N_L+N_1+N_2,N_R} \delta_{I_L \otimes I_1 \otimes I_2, I_R} \sqrt{2j+1} (-1)^{j_L+j_R+s_1+s_2} \sum\limits_{j_M \alpha_M} \sqrt{2j_M+1} \nonumber \\
& \hspace{-0.9cm} \times \left\{ \begin{array}{ccc} j_L & j_R & j \\ s_2 & s_1 & j_M \end{array} \right\} T[i]^{s_1 N_1 I_1}_{j_L N_L I_L \alpha_L ; j_M (N_L+N_1) (I_L \otimes I_1) \alpha_M} T[i+1]^{s_2 N_2 I_2}_{j_M (N_L+N_1) (I_L \otimes I_1) \alpha_M ; j_R N_R I_R \alpha_R}. \label{CPCTTtoS}
\end{eqnarray}
Eq. \eqref{CPCTTtoS} is the analogue of Eq. \eqref{CPCtwo-site-object-not-reduced}. The Lagrangian can be written in terms of $S[i]$, the effective Hamiltonian equation can be solved, and after convergence, Eq. \eqref{CPCTTtoS} can be backtransformed:
\begin{eqnarray}
& (TT)[i]^{s_1 N_1 I_1 ; s_2 N_2 I_2 ; j_M}_{j_L N_L I_L \alpha_L ; j_R N_R I_R \alpha_R} = \delta_{N_L+N_1+N_2,N_R} \delta_{I_L \otimes I_1 \otimes I_2, I_R} \sqrt{2j_M+1} (-1)^{j_L+j_R+s_1+s_2} \nonumber \\
& \times \sum\limits_j \left\{ \begin{array}{ccc} j_L & j_R & j \\ s_2 & s_1 & j_M \end{array} \right\} \sqrt{2j+1} S[i]^{j(s_1 s_2) N_1 N_2 I_1 I_2}_{j_L N_L I_L \alpha_L ; j_R N_R I_R \alpha_R}. \label{CPCStoTT}
\end{eqnarray}
Per group of $\left\{ j_M, N_M = N_L + N_1, I_M = I_L \otimes I_1 \right\}$, we can perform a singular value decomposition:
\begin{eqnarray}
& (TT)[i]^{s_1 N_1 I_1 ; s_2 N_2 I_2 ; j_M}_{j_L N_L I_L \alpha_L ; j_R N_R I_R \alpha_R} = \sum\limits_{\alpha_M} U[i]^{j_M N_M I_M}_{(j_L N_L I_L \alpha_L s_1 N_1 I_1);\alpha_M} \nonumber \\
& \times \lambda[i]^{j_M N_M I_M}_{\alpha_M} \left( \sqrt{\frac{2j_M+1}{2j_R+1}} V[i]^{j_M N_M I_M}_{\alpha_M;(j_R N_R I_R \alpha_R s_2 N_2 I_2)} \right).
\end{eqnarray}
After reshaping the indices to the normal form, it can be checked that $U[i]$ is the reduced part of a left-normalized MPS site tensor and that the term between brackets is the reduced part of a right-normalized MPS site tensor. The relation between $\lambda[i]$ and $\kappa[i]$ is given by
\begin{equation}
\kappa[i]_{j_M N_M I_M \alpha_M} = \frac{\lambda[i]_{j_M N_M I_M \alpha_M}}{\sqrt{ \sum\limits_{j_Q N_Q I_Q \alpha_Q} (2j_Q+1) \lambda[i]_{j_Q N_Q I_Q \alpha_Q}^2 }}.
\end{equation}
The $D_i$ largest values of $\lambda[i]$ are kept.

\vspace{0.7cm}
\hspace{-\parindent}\textbf{4. \textsc{CheMPS2} library}
\vspace{0.3cm}

\textsc{CheMPS2} can be obtained from the CPC Program Library, and from its public git repository \cite{CheMPS2github}. The source code contains comments in Doxygen format. A complete reference manual can be generated from these comments. See \texttt{README} on how to install the library and on how to generate the manual. In this section, we give an overview of the basic structure of \textsc{CheMPS2} so that new users can easily understand and alter the test runs to their own needs.

\vspace{0.7cm}
\hspace{-\parindent}\textbf{4.1 The Hamiltonian}
\vspace{0.3cm}

Most molecular electronic structure programs have the ability to print matrix elements or to save them in binary format. \textsc{CheMPS2} requires two-body matrix elements with eightfold permutation symmetry, which do not break $\mathsf{SU(2)}$ total electronic spin. A \texttt{CheMPS2::Hamiltonian} object should be created at the beginning of a calculation, and filled with the matrix elements of the problem at hand.

Users can utilize their preferred molecular electronic structure program to generate the matrix elements. The functions \texttt{setEconst}, \texttt{setTmat}, and \texttt{setVmat} then fill the \texttt{CheMPS2::Hamiltonian} object elementwise. Note that for $(ij | V | kl) = V_{ijkl}$ we have assumed the physics notation. This means that orbital $k$ at position $r_1$ (denoted by $k(r_1)$) scatters from orbital $l(r_2)$ into orbitals $i(r_1)$ and $j(r_2)$.

We have used \textsc{Psi4} \cite{WCMS:WCMS93} to generate molecular orbital matrix elements. Two plugins can be found in the folder \texttt{mointegrals}, with corresponding instructions in \texttt{README}. One plugin allows to print matrix elements as text during a \textsc{Psi4} calculation, in a format which \textsc{CheMPS2} is able to read. The other plugin creates a   \texttt{CheMPS2::Hamiltonian} object during a \textsc{Psi4} calculation, fills it with the molecular orbital matrix elements, and stores it to disk in binary format. The latter option requires linking of the \textsc{CheMPS2} library to the \textsc{Psi4} plugin, but allows for reduced storage requirements.

In the \texttt{CheMPS2::Problem} object, users can specify the symmetry sector to which the calculations are restricted. The \texttt{CheMPS2::Hamiltonian} and the desired total electronic spin, particle number, and point group symmetry then completely determine a FCI calculation. In order to do DMRG or DMRG-SCF instead of resp. FCI or CASSCF, a convergence scheme for the subsequent sweeps should be set up.

\vspace{0.7cm}
\hspace{-\parindent}\textbf{4.2 Convergence scheme}
\vspace{0.3cm}

The \texttt{CheMPS2::ConvergenceScheme} object controls the DMRG sweeps. It is divided into a number of consecutive instructions. Each instruction contains four parameters: the number of reduced renormalized basis states $D$ which should be kept, an energy threshold $E_{conv}$ for convergence, the maximum number of sweeps $N_{max}$, and the noise prefactor $\gamma_{noise}$.

The parameters $\gamma_{noise}$ and $D$ are relevant for the micro-iterations. Just before the decomposition of the reduced $S[i]$-tensor, random noise is added to it. This random noise is bounded in magnitude by $0.5 \gamma_{noise} w^{disc}(D)$, where $w^{disc}(D)$ is the maximum discarded weight obtained during the previous left- or right-sweep. After decomposition of the reduced $S[i]$-tensor, its reduced Schmidt spectrum $\lambda[i]$ is truncated to $D$.

The parameters $E_{conv}$ and $N_{max}$ are relevant for the macro-iterations. If after one macro-iteration (left- plus right-sweep), the energy difference is smaller than $E_{conv}$, the sweeping stops and the next instruction is performed. If energy convergence is not reached after $N_{max}$ macro-iterations, the current instruction ends as well.

\vspace{0.7cm}
\hspace{-\parindent}\textbf{4.3 DMRG}
\vspace{0.3cm}

Creation of a \texttt{CheMPS2::DMRG} object requires a \texttt{CheMPS2::Hamiltonian}, a \texttt{CheMPS2:: Problem}, and a \texttt{CheMPS2::ConvergenceScheme}. Each DMRG calculation starts by creating a new MPS. Its virtual dimension $D$ is obtained from the first instruction of the \texttt{CheMPS2::ConvergenceScheme} object. At each MPS bond, this virtual dimension $D$ is distributed over all possible symmetry sectors, ensuring that the dimension of a certain symmetry sector does not exceed the corresponding FCI dimension. The so-created MPS is filled with random noise.

The function \texttt{Solve} performs the instructions of the convergence scheme. Afterwards, it returns the minimal variational energy encountered during all the performed micro-iterations.

With the function \texttt{calc2DM}, the reduced 2-RDMs $\Gamma^A$ and $\Gamma^B$ are calculated:
\begin{eqnarray}
\Gamma_{(i \sigma) (j \tau) ; (k \sigma) (l \tau)} & = & \braket{ \hat{a}^{\dagger}_{i \sigma} \hat{a}^{\dagger}_{j \tau}  \hat{a}_{l \tau} \hat{a}_{k \sigma}}\\
\Gamma^{A}_{ij ; kl} & = & \sum\limits_{\sigma \tau} \Gamma_{(i \sigma) (j \tau) ; (k \sigma) (l \tau)} \\
\Gamma^{B}_{ij ; kl} & = & \sum\limits_{\sigma \tau} (-1)^{\sigma - \tau} \Gamma_{(i \sigma) (j \tau) ; (k \sigma) (l \tau)}
\end{eqnarray}
$\Gamma^A$ can be used to calculate the energy, the particle number $N$, and the 1-RDM:
\begin{eqnarray}
E & = & E_{const} + \frac{1}{2} \sum\limits_{ijkl} \Gamma^{A}_{ij ; kl} (ij | h | kl) \\
N(N-1) & = & \sum\limits_{ij} \Gamma^{A}_{ij ; ij} \\
\sum\limits_{\sigma} \braket{ \hat{a}^{\dagger}_{i \sigma} \hat{a}_{k \sigma}} & = & \frac{1}{N-1} \sum\limits_j \Gamma^{A}_{ij ; kj}
\end{eqnarray}
and is needed for the DMRG-SCF algorithm, while $\Gamma^B$ is important for spin-spin correlation functions.

The \texttt{CheMPS2::DMRG} object can also calculate excited states. After the ground state $\ket{\Psi_0}$ has been determined, the desired number of excited states can be set once with the function \texttt{activateExcitations}. Before \texttt{Solve} is called to find the next new excitation $\ket{\Psi_{m}}$, the function \texttt{newExcitation} should be called with the parameter $\eta_{m}$. This pushes back the current MPS which represents $\ket{\Psi_{m-1}}$, and sets the Hamiltonian to
\begin{equation}
 \hat{H}_m = \hat{H}_0 + \sum\limits_{k={0}}^{m-1} \eta_{k+1} \ket{\Psi_k} \bra{\Psi_k}.
\end{equation}
Our excited state DMRG algorithm is hence a state-specific algorithm, which projects out lower-lying states in the given $\mathsf{SU(2)} \otimes \mathsf{U(1)} \otimes \mathsf{P}$ symmetry sector. An example can be found in \texttt{tests/test5.cpp}.

OpenMP parallelization is used in the \texttt{CheMPS2::DMRG} object to speed up (a) contractions involving tensors with a sparse block structure, for example the action of the effective Hamiltonian on a particular guess, and (b) the construction of the (often similar) renormalized operators in between two micro-iterations.

\vspace{0.7cm}
\hspace{-\parindent}\textbf{4.4 DMRG-SCF}
\vspace{0.3cm}

A state-specific DMRG-SCF algorithm is implemented in \texttt{CheMPS2::CASSCF}. Its creation requires a \texttt{CheMPS2::Hamiltonian} object. The number of occupied, active, and virtual orbitals per point group irrep should be given with the function \texttt{setupStart} before calling the SCF routine.

The CASSCF routine which is implemented is the augmented Hessian \cite{Lengsfield} Newton-Raphson method from Ref. \cite{Roos3}, with exact Hessian. It can be called with the function \texttt{doCASSCFnewtonraphson}, which requires the targeted symmetry sector, the convergence scheme, and the targeted root for the state-specific algorithm. When the gradient for orbital rotations reaches a predefined threshold, the routine returns the converged DMRG-SCF energy. An example can be found in \texttt{tests/test6.cpp}.

\newpage
\hspace{-\parindent}\textbf{5. Carbon dimer}
\vspace{0.3cm}

\hspace{-\parindent}\textbf{5.1 Introduction}
\vspace{0.3cm}

Despite its simplicity at first sight, the carbon dimer provides a rich source of interesting physics. The bond between the two carbon atoms is of the charge-shift type \cite{10.1021.ct100577v, naturechemChargeShift}. Its strength tempts chemists to classify it as a quadruple bond \cite{PhysRev.56.778, 10.1021.j100174a058, vonRaguSchleyer19936387, WeinholdBook, naturechemFourthBond, ANIE:ANIE201208206}, and recent research indicates how this fourth bond can be interpreted \cite{10.1021.ct400867h}. The $1s$ core correlation is significant \cite{CoreCorrelationInC2, 10.1080.00268976.2011.564593}. The low-lying bond dissociation curves are quasi-degenerate, and avoided crossings occur between states with the same spin and $D_{\infty h}$ point group symmetry \cite{BoggioPasqua2000159, AbramsC2, Varandas}. This happens for example between the $X^1\Sigma_g^+$ and $B'^1 \Sigma_g^+$ states, and between the $c^3\Sigma_u^+$ and $2^3\Sigma_u^+$ states. Fortunately, relativistic effects are small \cite{Kokkin, CoreCorrelation2011JCP}.

Accurate data for the low-lying states, preferably at the FCI level of theory for a given basis set, are useful to assess the accuracy of approximate molecular electronic structure methods. The $X^1\Sigma_g^+$, $B^1 \Delta_g$, and $B'^1\Sigma_g^+$ bond dissociation curves of Ref. \cite{AbramsC2} at the frozen core FCI/6-31G* level of theory are utilized to this end in several works \cite{useOfAbrams1, useofAbrams2, useofAbrams3, useofAbrams4}.

The 12 lowest states of the carbon dimer are $X^1\Sigma_g^+$, $a^3\Pi_u$, $b^3\Sigma_g^-$, $A^1\Pi_u$, $c^3\Sigma_u^+$, $B^1 \Delta_g$, $B'^1\Sigma_g^+$, $d^3\Pi_g$, $C^1\Pi_g$, $1^1\Sigma_u^-$, $1^3\Delta_u$, and $2^3\Sigma_u^+$ \cite{BoggioPasqua2000159}. In Section 5.5, we present the bond dissociation curves of these states at the DMRG(28o, 12e, D$_{ \mathsf{ SU(2) }}$=2500)/cc-pVDZ level of theory.

To estimate the contribution of $1s$ core correlation to the $X^1\Sigma_g^+$ bond dissociation curve, we compare energies at the DMRG(28o, 12e, D$_{\mathsf{SU(2)}}$=2500)/cc-pVDZ, DMRG-SCF(26o, 8e, D$_{\mathsf{SU(2)}}$=2500)/cc-pVDZ, DMRG(36o, 12e, D$_{\mathsf{SU(2)}}$=2500)/cc-pCVDZ, and DMRG-SCF(34o, 8e, D$_{\mathsf{SU(2)}}$=2500)/cc-pCVDZ levels of theory in Section 5.6. The cc-pCVDZ basis augments the cc-pVDZ basis with extra $1s$ and $1p$ functions to treat core and core-valence correlation \cite{ccpcvdzreference}. 

For all calculations, the variational energies are converged to $0.1 mE_h$ from the extrapolated value. This implies that, for all practical purposes, we present data at the FCI/cc-pVDZ, CASSCF(26o, 8e)/cc-pVDZ, FCI/cc-pCVDZ, and CASSCF(34o, 8e)/cc-pCVDZ levels of theory.

\vspace{0.7cm}
\hspace{-\parindent}\textbf{5.2 Symmetry labelling}
\vspace{0.3cm}

Since \textsc{CheMPS2} can only handle abelian point groups, we use $D_{2h}$ point group symmetry to obtain these 12 states:
{\allowdisplaybreaks
\begin{eqnarray}
X^1\Sigma_g^+ ; B^1 \Delta_g ; B'^1\Sigma_g^+ & \rightarrow & ^1A_g \label{CPCmeuh1}\\
c^3\Sigma_u^+ ; 1^3\Delta_u ; 2^3\Sigma_u^+ &\rightarrow & ^3B_{1u} \label{CPCmeuh2}\\
C^1\Pi_g & \rightarrow & ^1B_{2g} \label{CPCfirstOfThed2hIrreps}\\
A^1\Pi_u & \rightarrow & ^1B_{2u} \label{CPCmeuh3}\\
1^1\Sigma_u^- & \rightarrow & ^1A_u \label{CPCmeuh4}\\
b^3 \Sigma_g^- & \rightarrow & ^3B_{1g} \label{CPCmeuh5}\\
d^3\Pi_g & \rightarrow & ^3B_{2g} \label{CPCmeuh6}\\
a^3\Pi_u &\rightarrow & ^3B_{2u}. \label{CPClastOfThed2hIrreps}
\end{eqnarray}}

\begin{figure}[t!]
 \centering
 \includegraphics[width=0.7\textwidth]{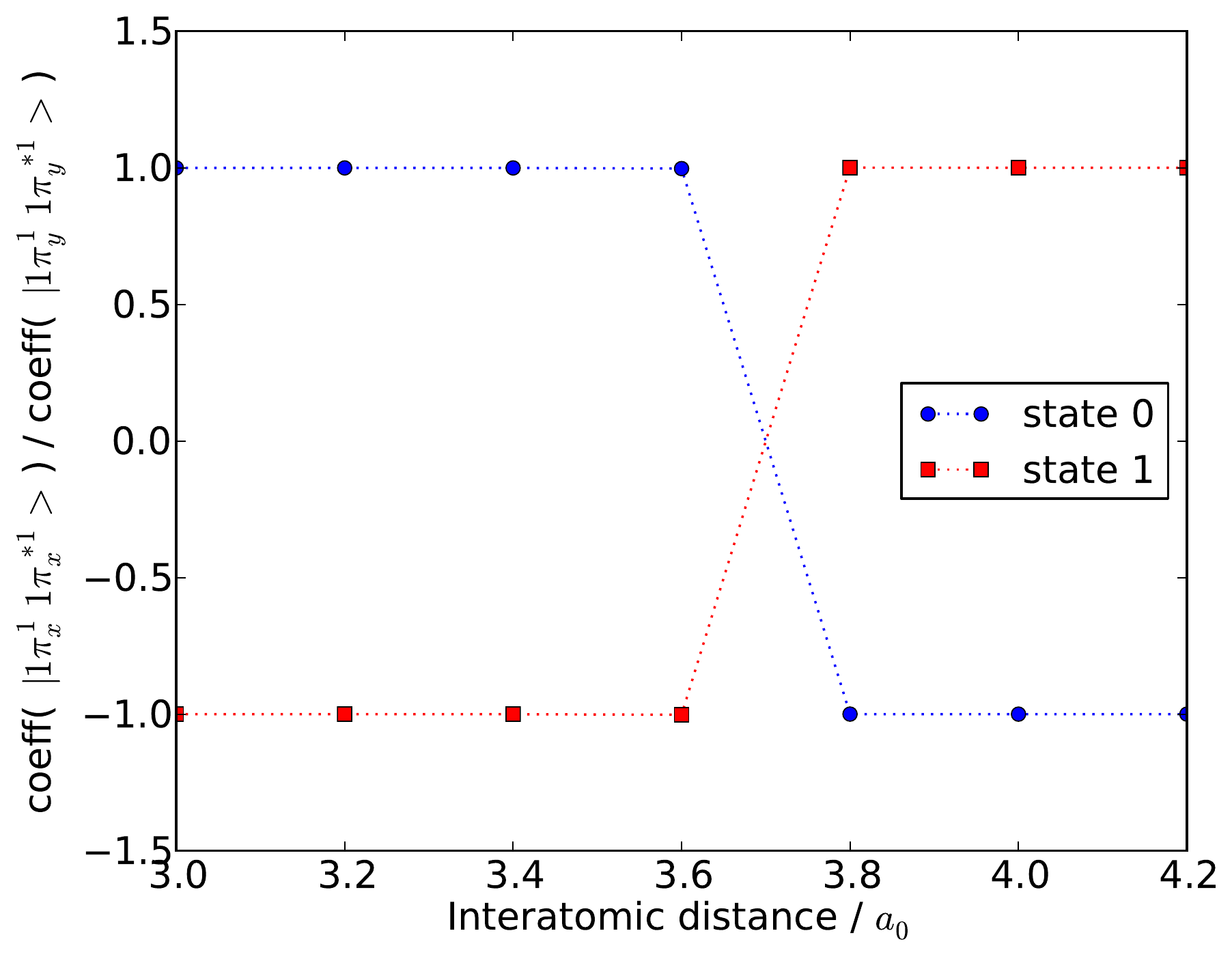}
 \caption{\label{CPCfig3B1ucoeff} For the cc-pVDZ basis, the $1^3\Delta_u$ state drops below the $c^3\Sigma_u^+$ state at an interatomic distance between 3.6 and 3.8 $a_0$. The $\ket{1\pi_x^1 1\pi_x^{*1}}$ and $\ket{1\pi_y^1 1\pi_y^{*1}}$ FCI coefficients allow to correctly label the $^3B_{1u}$ ground state (state 0) and the first excited state (state 1).}
\end{figure}

\noindent For the states (\ref{CPCfirstOfThed2hIrreps})-(\ref{CPClastOfThed2hIrreps}), we have calculated one extra state to check that no unexpected curve crossings occur. To discern the lowest three $^1A_g$ states, we have extracted the following FCI coefficients from the DMRG object \cite{AbramsC2}:
\begin{eqnarray}
\ket{1\pi_x^2} & = & \ket{1\sigma_g^2 1\sigma_u^2 2\sigma_g^2 2\sigma_u^2 \mathbf{1\pi_x^2} 3\sigma_g^2} \\
               & = & \ket{1A_g^2 1B_{1u}^2 2A_g^2 2B_{1u}^2 \mathbf{1B_{3u}^2} 3A_g^2} \\
\ket{1\pi_y^2} & = & \ket{1\sigma_g^2 1\sigma_u^2 2\sigma_g^2 2\sigma_u^2 \mathbf{1\pi_y^2} 3\sigma_g^2} \\
               & = & \ket{1A_g^2 1B_{1u}^2 2A_g^2 2B_{1u}^2 \mathbf{1B_{2u}^2} 3A_g^2}.
\end{eqnarray}
When the FCI coefficients are equal, the state has $^1\Sigma_g^+$ symmetry, and when the FCI coefficients are each other's additive inverse, the state has $^1\Delta_g$ symmetry. To discern the lowest three $^3B_{1u}$ states, we have extracted the following FCI coefficients from the DMRG object:
\begin{eqnarray}
\ket{1\pi_x^1 1\pi_x^{*1}} & = & \ket{1\sigma_g^2 1\sigma_u^2 2\sigma_g^2 2\sigma_u^2 \mathbf{1\pi_x^1} 3\sigma_g^2 \mathbf{1\pi_x^{*1}}} \\
               & = & \ket{1A_g^2 1B_{1u}^2 2A_g^2 2B_{1u}^2 \mathbf{1B_{3u}^1} 3A_g^2 \mathbf{1B_{2g}^1}} \\
\ket{1\pi_y^1 1\pi_y^{*1}} & = & \ket{1\sigma_g^2 1\sigma_u^2 2\sigma_g^2 2\sigma_u^2 \mathbf{1\pi_y^1} 3\sigma_g^2 \mathbf{1\pi_y^{*1}}} \\
               & = & \ket{1A_g^2 1B_{1u}^2 2A_g^2 2B_{1u}^2 \mathbf{1B_{2u}^1} 3A_g^2 \mathbf{1B_{3g}^1}}.
\end{eqnarray}
When the FCI coefficients are equal, the state has $^3\Sigma_u^+$ symmetry, and when the FCI coefficients are each other's additive inverse, the state has $^3\Delta_u$ symmetry. An example is shown in Fig. \ref{CPCfig3B1ucoeff}.

\vspace{0.7cm}
\hspace{-\parindent}\textbf{5.3 Irrep ordering}
\vspace{0.3cm}

\begin{figure}
 \centering
 \includegraphics[width=0.7\textwidth]{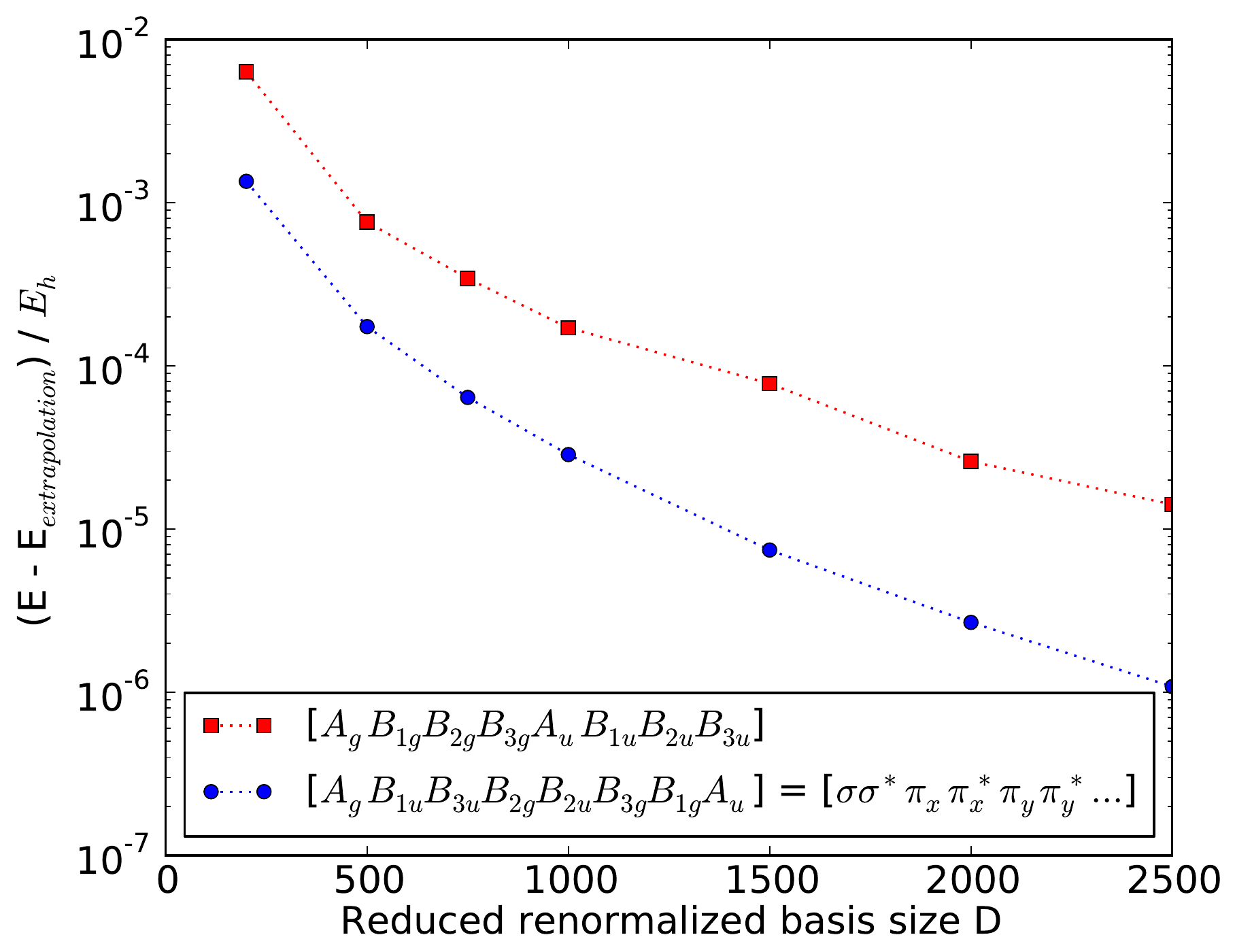}
 \caption{\label{CPCfigIrrepOrder} The orbital choice and ordering influence the convergence behaviour of DMRG. The convergence behaviour of two irrep orderings is shown for the carbon dimer with interatomic distance 2.4 $a_0$ in the cc-pVDZ basis. The extrapolated energy was obtained from the ordering where bonding and anti-bonding orbitals are grouped, with the method described in Section 5.4.}
\end{figure}

The standard $D_{2h}$ irrep order is not optimal to study the carbon dimer with DMRG. As stated in Section 2.6, it is best to group bonding and anti-bonding orbitals together on the DMRG lattice. The convergence behaviour of these two irrep orderings is shown in Fig. \ref{CPCfigIrrepOrder}. We have used the latter ordering for our calculations.
\begin{table}
\caption{\label{CPCtableC2convscheme} Convergence scheme for the carbon dimer calculations. The symbols are explained in Section 4.2.}
\begin{center}
\begin{tabular}{cccc}
\hline
$D_{\mathsf{SU(2)}}$ & $\gamma_{noise}$ & $E_{conv} / E_h$ & $N_{max}$\\
\hline
200 & 0.03 & $10^{-8}$ & 2 \\
200 & 0.00 & $10^{-8}$ & 3 \\
500 & 0.03 & $10^{-8}$ & 2 \\
500 & 0.00 & $10^{-8}$ & 5 \\
1000 & 0.03 & $10^{-8}$ & 2 \\
1000 & 0.00 & $10^{-8}$ & 5 \\
1500 & 0.03 & $10^{-8}$ & 2 \\
1500 & 0.00 & $10^{-8}$ & 5 \\
2000 & 0.03 & $10^{-8}$ & 2 \\
2000 & 0.00 & $10^{-8}$ & 5 \\
2500 & 0.03 & $10^{-8}$ & 2 \\
2500 & 0.00 & $10^{-8}$ & 12 \\
\hline
\end{tabular}
\end{center}
\end{table}

\vspace{0.7cm}
\hspace{-\parindent}\textbf{5.4 Extrapolation}
\vspace{0.3cm}

\begin{figure}
 \centering
 \includegraphics[width=0.7\textwidth]{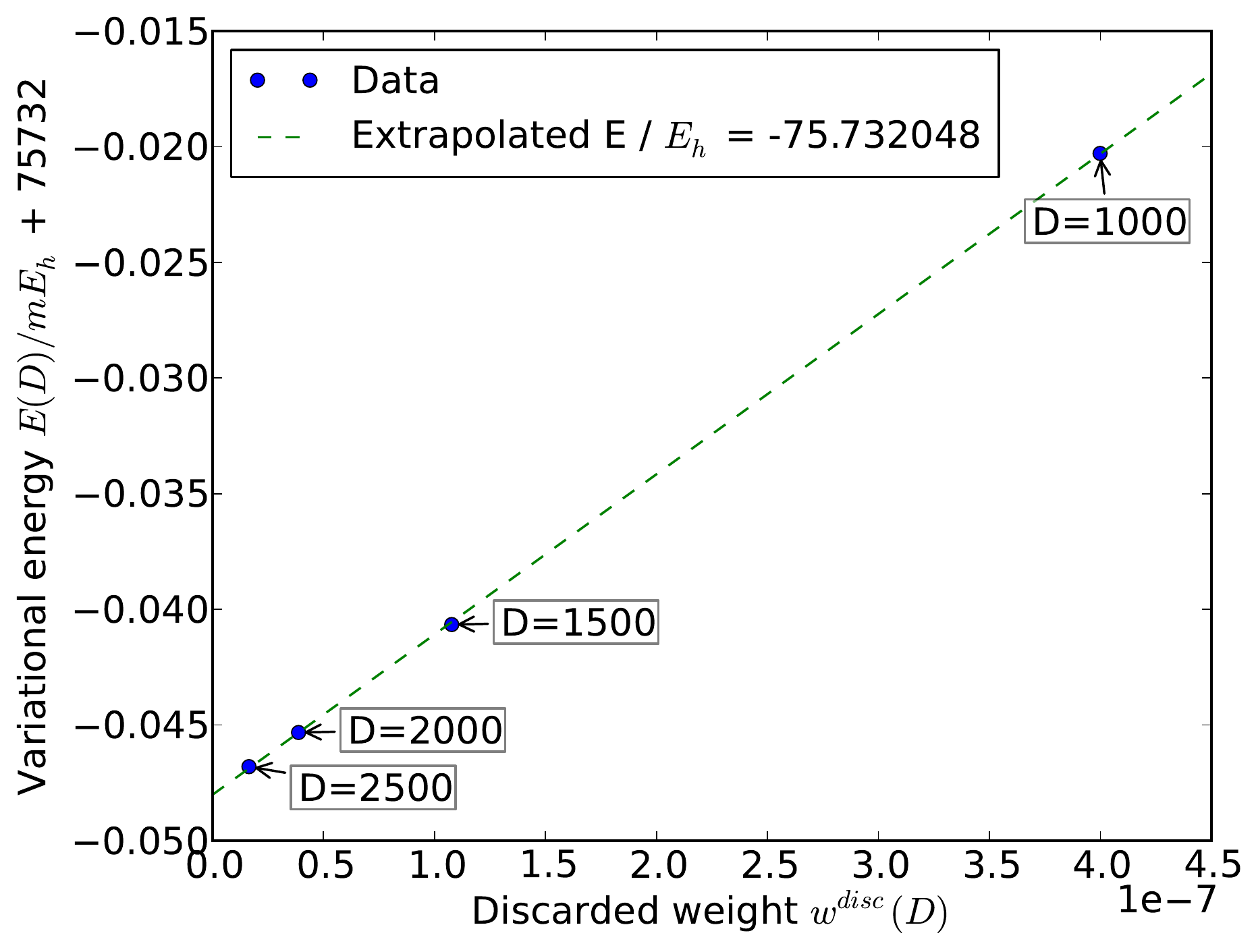}
 \caption{\label{CPCfigExtrapolationExample} The extrapolation scheme of Eq. \eqref{CPCextrapolSchemeEq} is used to obtain energies which are correct up to 0.01 $mE_h$. The example shown here is for the $X^1\Sigma_g^+$ state of the carbon dimer at an interatomic distance of 2.35 $a_0$ in the cc-pVDZ basis.}
\end{figure}

We have used the convergence scheme in Table \ref{CPCtableC2convscheme} for all the calculations of the carbon dimer. The extrapolation scheme of Eq. \eqref{CPCextrapolSchemeEq} is used to obtain energies which are correct up to 0.01 $mE_h$. An example of such an extrapolation is shown in Fig. \ref{CPCfigExtrapolationExample}. The energies shown in Sections 5.5 and 5.6 are the extrapolated values.

\begin{sidewaystable}
\caption{\label{CPClargeTableC2} Extrapolated energies for the 12 lowest states of the carbon dimer at the DMRG(28o, 12e, D$_{\mathsf{SU(2)}}$=2500)/cc-pVDZ level of theory. The energies are shifted 75 $E_h$ upwards, and are expressed in $mE_h$.}
\begin{center}
{\small
\vspace{-0.45cm}
\begin{tabular}{lrrrrrrrrrrrr}
\hline
R / $a_0$ & \multicolumn{12}{l}{(Energy + 75 $E_h$) / $mE_h$} \\
          & $X^1\Sigma_g^+$ & $a^3\Pi_u$ & $b^3\Sigma_g^-$ & $A^1\Pi_u$ & $c^3\Sigma_u^+$ & $B^1 \Delta_g$ & $B'^1\Sigma_g^+$ & $d^3\Pi_g$ & $C^1\Pi_g$ &  $1^1\Sigma_u^-$ & $1^3\Delta_u$ & $2^3\Sigma_u^+$ \\
\hline
1.8 &  -454.96  &  -357.88  &  -253.42  &  -314.72  &  -439.01  &  -207.85  &  -263.42  &  -311.47  &  -250.35  &  -4.51  &  -35.05  &  -70.74  \\
1.9 &  -562.08  &  -485.42  &  -396.48  &  -442.69  &  -541.14  &  -353.53  &  -381.99  &  -430.96  &  -368.18  &  -145.91  &  -177.61  &  -212.67  \\
2.0 &  -635.85  &  -576.98  &  -501.10  &  -534.76  &  -609.66  &  -460.60  &  -471.77  &  -514.53  &  -449.98  &  -251.39  &  -284.10  &  -318.58  \\
2.1 &  -684.30  &  -640.94  &  -576.18  &  -599.30  &  -652.70  &  -537.96  &  -538.97  &  -570.80  &  -504.41  &  -329.17  &  -362.69  &  -396.52  \\
2.2 &  -713.63  &  -683.80  &  -628.60  &  -642.81  &  -676.60  &  -592.52  &  -587.67  &  -606.46  &  -538.20  &  -385.65  &  -419.78  &  -452.75  \\
2.3 &  -728.68  &  -710.64  &  -663.75  &  -670.34  &  -686.31  &  -629.65  &  -621.40  &  -626.70  &  -556.61  &  -425.92  &  -460.33  &  -492.17  \\
2.35 &  -732.05  &  -719.33  &  -676.19  &  -679.39  &  -687.10  &  -643.04  &  -633.65  &  -632.36  &  -561.40  &  -441.36  &  -475.66  &  -506.80  \\
2.4 &  -733.18  &  -725.42  &  -685.81  &  -685.86  &  -685.73  &  -653.57  &  -643.30  &  -635.62  &  -563.85  &  -454.45  &  -488.25  &  -518.58  \\
2.5 &  -730.05  &  -731.22  &  -698.04  &  -692.42  &  -677.93  &  -667.55  &  -656.08  &  -636.43  &  -563.36  &  -477.13  &  -506.68  &  -534.93  \\
2.6 &  -721.58  &  -730.43  &  -702.98  &  -692.43  &  -665.39  &  -674.15  &  -661.94  &  -631.72  &  -558.20  &  -499.94  &  -518.21  &  -543.51  \\
2.7 &  -709.54  &  -724.91  &  -702.58  &  -687.72  &  -650.09  &  -675.32  &  -662.63  &  -623.62  &  -551.37  &  -519.32  &  -525.47  &  -546.10  \\
2.8 &  -695.37  &  -716.10  &  -698.35  &  -679.74  &  -633.70  &  -672.60  &  -659.48  &  -613.89  &  -545.69  &  -533.86  &  -532.15  &  -544.31  \\
2.9 &  -680.23  &  -705.08  &  -691.43  &  -669.58  &  -617.56  &  -667.13  &  -653.44  &  -603.99  &  -542.36  &  -544.27  &  -539.84  &  -541.16  \\
3.0 &  -665.20  &  -692.69  &  -682.70  &  -658.08  &  -602.65  &  -659.80  &  -645.08  &  -594.90  &  -540.48  &  -551.39  &  -546.21  &  -543.11  \\
3.2 &  -638.95  &  -666.17  &  -662.28  &  -633.46  &  -578.29  &  -642.09  &  -622.59  &  -579.90  &  -537.09  &  -558.59  &  -553.13  &  -549.22  \\
3.4 &  -617.95  &  -639.87  &  -640.64  &  -609.35  &  -561.37  &  -623.07  &  -597.29  &  -567.38  &  -532.80  &  -559.79  &  -554.31  &  -549.69  \\
3.6 &  -599.65  &  -615.55  &  -619.67  &  -587.68  &  -552.43  &  -604.72  &  -575.15  &  -556.01  &  -528.47  &  -557.69  &  -552.17  &  -544.05  \\
3.8 &  -583.60  &  -594.03  &  -600.33  &  -569.45  &  -547.56  &  -588.06  &  -557.98  &  -546.10  &  -525.09  &  -553.95  &  -548.30  &  -536.60  \\
4.0 &  -569.91  &  -575.68  &  -583.08  &  -555.06  &  -542.97  &  -573.57  &  -545.70  &  -538.27  &  -523.05  &  -549.57  &  -543.76  &  -531.21  \\
4.2 &  -558.63  &  -560.66  &  -568.16  &  -544.44  &  -538.59  &  -561.46  &  -537.47  &  -532.75  &  -522.22  &  -545.17  &  -539.23  &  -527.76  \\
4.4 &  -549.67  &  -548.99  &  -555.69  &  -537.12  &  -534.71  &  -551.75  &  -532.20  &  -529.22  &  -522.24  &  -541.13  &  -535.16  &  -525.67  \\
4.6 &  -542.81  &  -540.54  &  -545.74  &  -532.36  &  -531.58  &  -544.27  &  -528.91  &  -527.13  &  -522.69  &  -537.63  &  -531.81  &  -524.43  \\
4.8 &  -537.73  &  -534.90  &  -538.26  &  -529.39  &  -529.24  &  -538.70  &  -526.87  &  -525.96  &  -523.30  &  -534.74  &  -529.26  &  -523.70  \\
5.0 &  -534.05  &  -531.40  &  -533.02  &  -527.59  &  -527.64  &  -534.66  &  -525.60  &  -525.34  &  -523.89  &  -532.41  &  -527.46  &  -523.27  \\
5.2 &  -531.41  &  -529.29  &  -529.61  &  -526.50  &  -526.56  &  -531.78  &  -524.80  &  -525.01  &  -524.38  &  -530.57  &  -526.25  &  -523.03  \\
5.4 &  -529.51  &  -528.01  &  -527.51  &  -525.82  &  -525.87  &  -529.72  &  -524.29  &  -524.84  &  -524.73  &  -529.13  &  -525.48  &  -522.93  \\
5.6 &  -528.14  &  -527.19  &  -526.27  &  -525.38  &  -525.42  &  -528.23  &  -523.96  &  -524.73  &  -524.96  &  -528.00  &  -524.99  &  -522.90  \\
5.8 &  -527.13  &  -526.62  &  -525.53  &  -525.08  &  -525.10  &  -527.15  &  -523.75  &  -524.65  &  -525.08  &  -527.12  &  -524.68  &  -522.93  \\
6.0 &  -526.36  &  -526.20  &  -525.08  &  -524.87  &  -524.87  &  -526.38  &  -523.61  &  -524.58  &  -525.12  &  -526.43  &  -524.49  &  -522.99  \\
\hline
\end{tabular}}
\end{center}
\end{sidewaystable}

\newpage
\hspace{-\parindent}\textbf{5.5 Bond dissociation curves}
\vspace{0.3cm}

\begin{figure}[t!]
 \centering
 \includegraphics[width=0.7\textwidth]{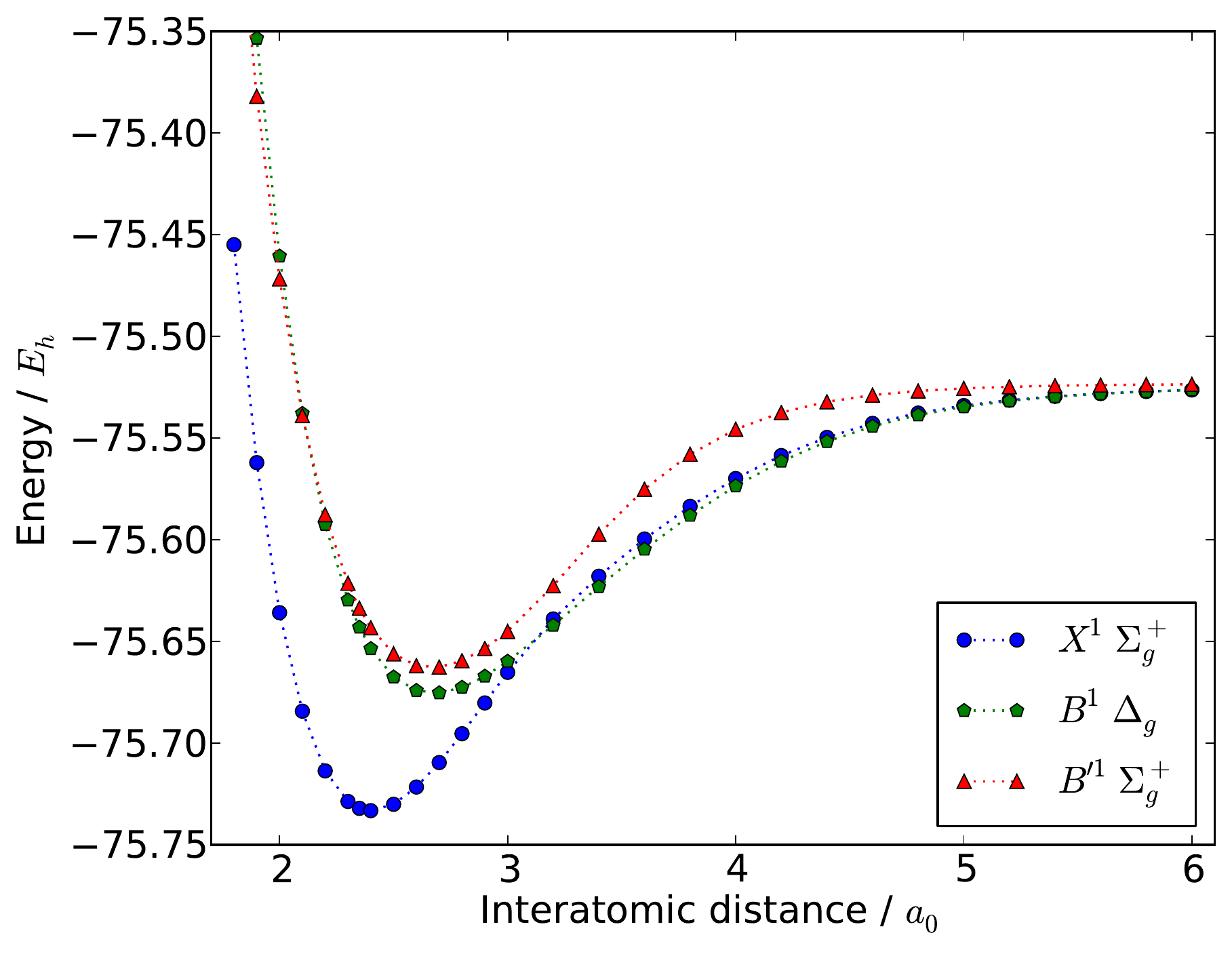}
 \caption{\label{CPC1AgStates} Bond dissociation curves for the low-lying $^1A_g$ states of the carbon dimer in the cc-pVDZ basis.}
\end{figure}
\begin{figure}[t!]
 \centering
 \includegraphics[width=0.7\textwidth]{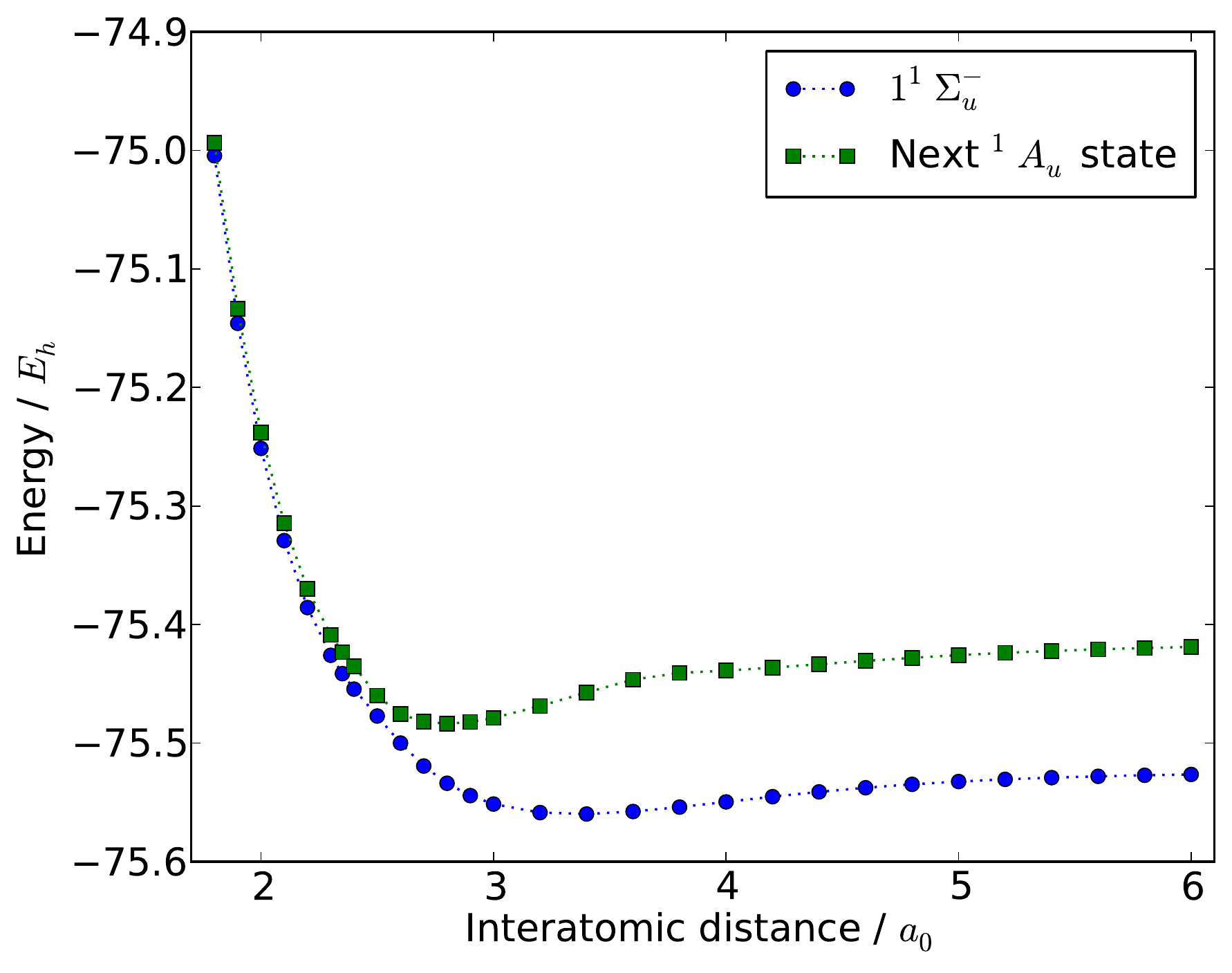}
 \caption{\label{CPC1AuStates} Bond dissociation curves for the low-lying $^1A_u$ states of the carbon dimer in the cc-pVDZ basis.}
\end{figure}
\begin{figure}[t!]
 \centering
 \includegraphics[width=0.7\textwidth]{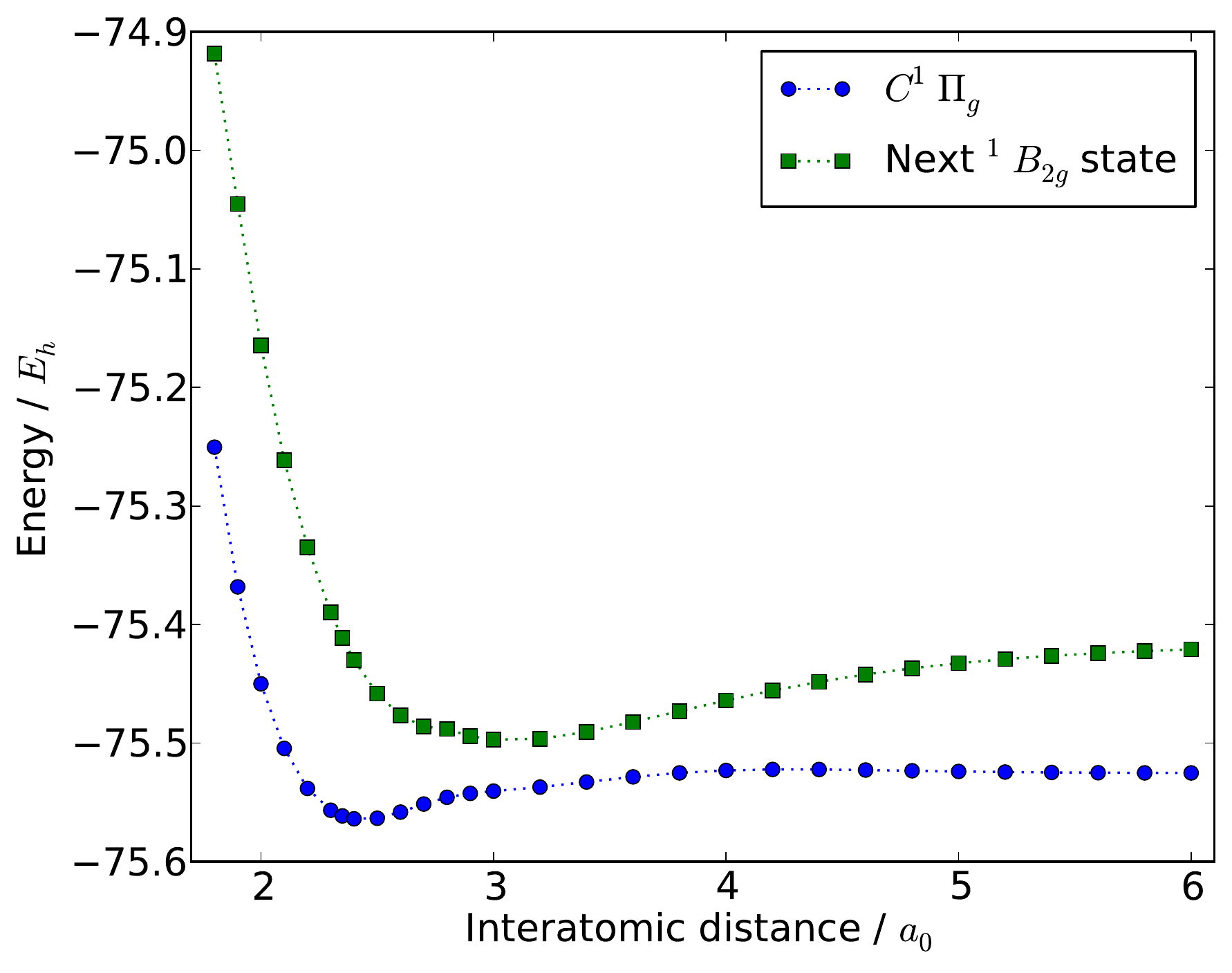}
 \caption{\label{CPC1B2gStates} Bond dissociation curves for the low-lying $^1B_{2g}$ states of the carbon dimer in the cc-pVDZ basis.}
\end{figure}
\begin{figure}[t!]
 \centering
 \includegraphics[width=0.7\textwidth]{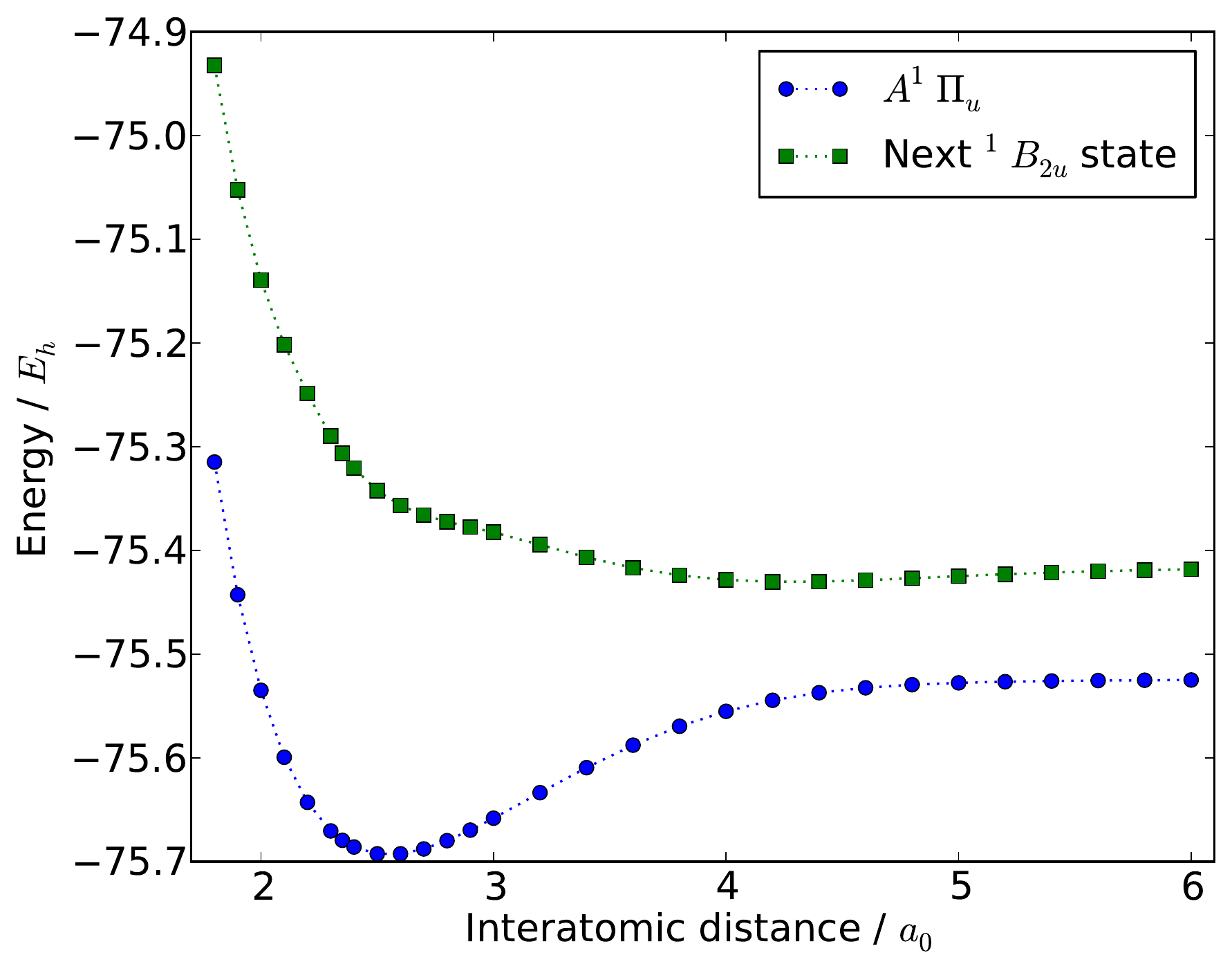}
 \caption{\label{CPC1B2uStates} Bond dissociation curves for the low-lying $^1B_{2u}$ states of the carbon dimer in the cc-pVDZ basis.}
\end{figure}
\begin{figure}[t!]
 \centering
 \includegraphics[width=0.7\textwidth]{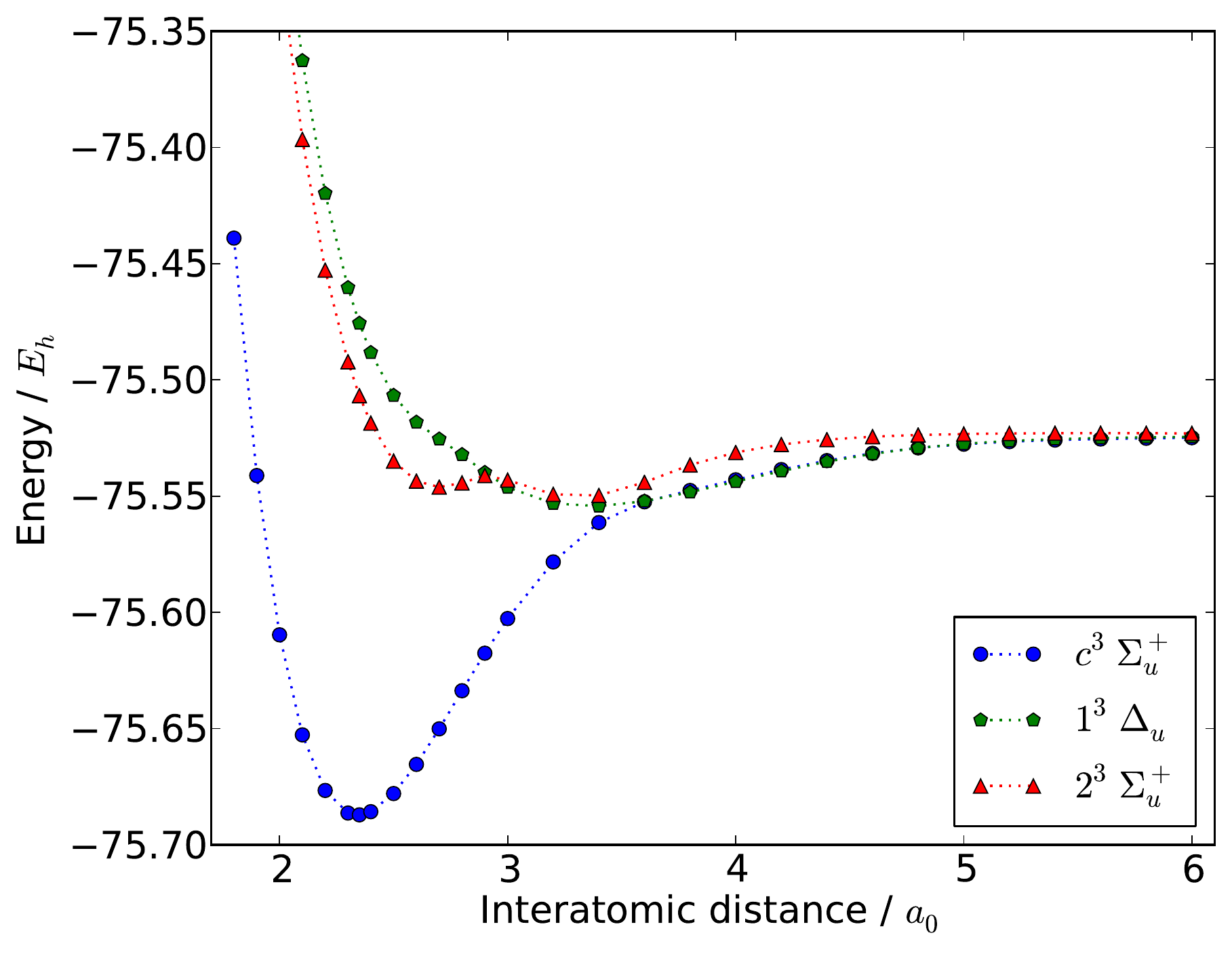}
 \caption{\label{CPC3B1uStates} Bond dissociation curves for the low-lying $^3B_{1u}$ states of the carbon dimer in the cc-pVDZ basis.}
\end{figure}
\begin{figure}[t!]
 \centering
 \includegraphics[width=0.7\textwidth]{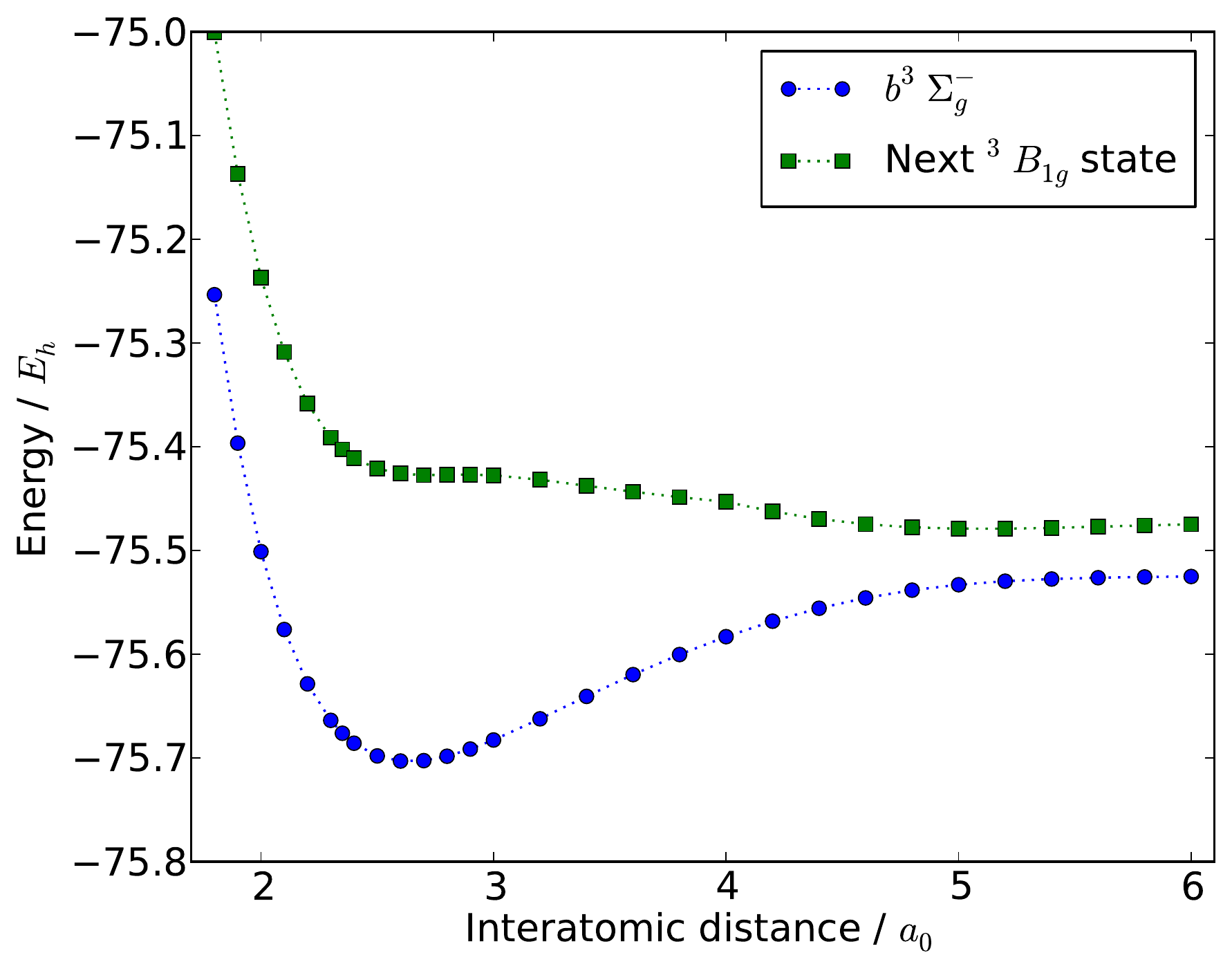}
 \caption{\label{CPC3B1gStates} Bond dissociation curves for the low-lying $^3B_{1g}$ states of the carbon dimer in the cc-pVDZ basis.}
\end{figure}
\begin{figure}[t!]
 \centering
 \includegraphics[width=0.7\textwidth]{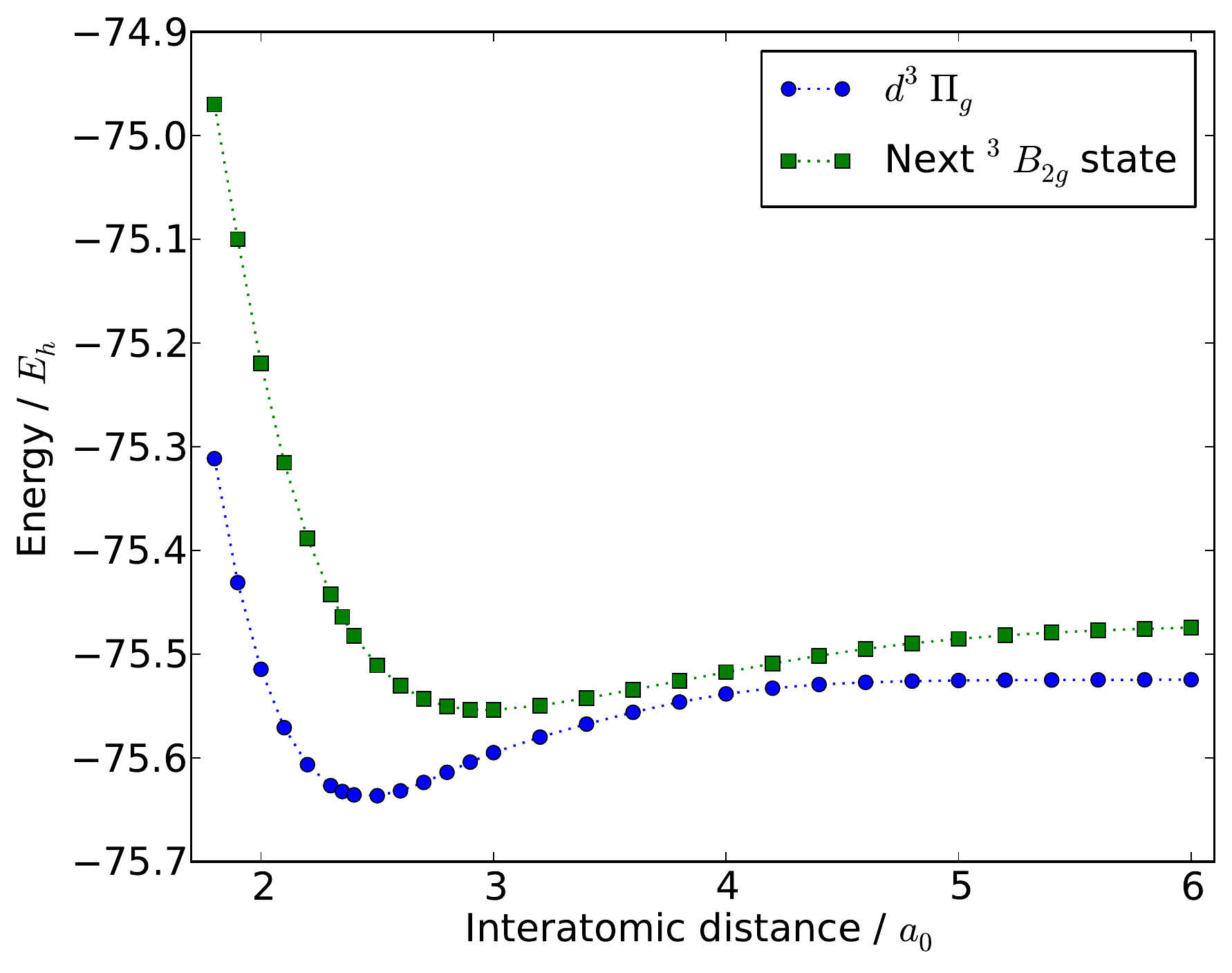}
 \caption{\label{CPC3B2gStates} Bond dissociation curves for the low-lying $^3B_{2g}$ states of the carbon dimer in the cc-pVDZ basis.}
\end{figure}
\begin{figure}[t!]
 \centering
 \includegraphics[width=0.7\textwidth]{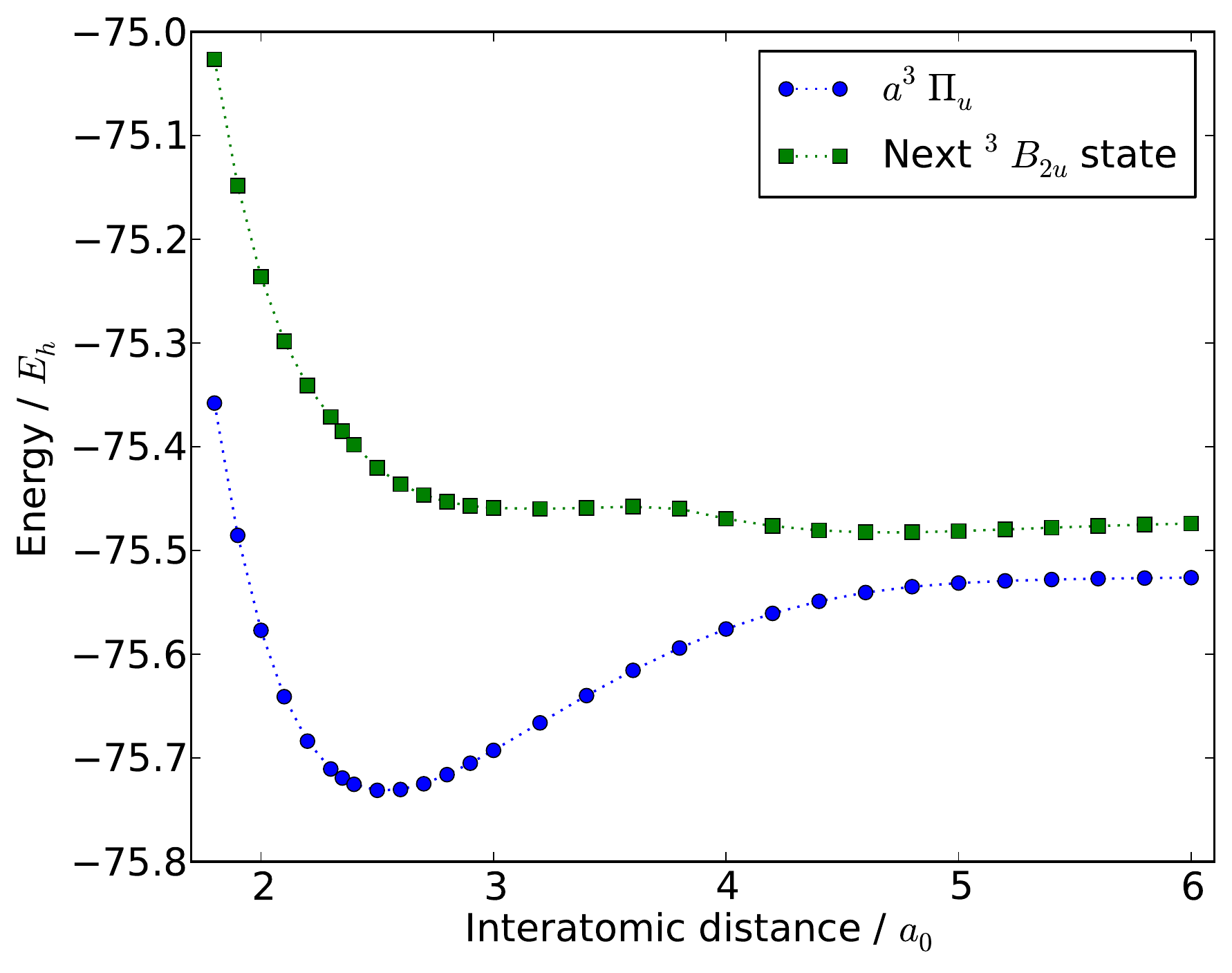}
 \caption{\label{CPC3B2uStates} Bond dissociation curves for the low-lying $^3B_{2u}$ states of the carbon dimer in the cc-pVDZ basis.}
\end{figure}

The extrapolated energies at the DMRG(28o, 12e, D$_{\mathsf{SU(2)}}$=2500)/cc-pVDZ level of theory are summarized in Table \ref{CPClargeTableC2} and are shown per targeted symmetry sector in Figs. \ref{CPC1AgStates}-\ref{CPC3B2uStates}. For the $^1A_g$ symmetry, the $B^1\Delta_g$ state drops below the $B'^1\Sigma_g^+$ state at an interatomic distance between $2 a_0$ and $2.1 a_0$, and it drops below the $X^1\Sigma_g^+$ state at an interatomic distance between $3a_0$ and $3.2 a_0$. The $X^1\Sigma_g^+$ and $B'^1\Sigma_g^+$ states have an avoided crossing. For the $^3B_{1u}$ symmetry, the $1^3\Delta_u$ state drops below the $2^3\Sigma_u^+$ state at an interatomic distance between $2.9 a_0$ and $3.0 a_0$, and it drops below the $c^3\Sigma_u^+$ state at an interatomic distance between $3.6 a_0$ and $3.8 a_0$. The $c^3\Sigma_u^+$ and $2^3\Sigma_u^+$ states have an avoided crossing. The intermediary peak of the $2^3\Sigma_u^+$ state near $2.9 a_0$ was also observed in Ref. \cite{BoggioPasqua2000159}, and is due to an avoided crossing with the $3^3\Sigma_u^+$ state. The $C^1\Pi_g$ and $d^3\Pi_g$ states also clearly show an avoided crossing with the next corresponding excited state.

\vspace{0.7cm}
\hspace{-\parindent}\textbf{5.6 Core correlation}
\vspace{0.3cm}

\begin{table}
\caption{\label{CPCCoreCorrTable} Extrapolated energies for the $X^1\Sigma_g^+$ state of the carbon dimer. (26o, 8e), (28o, 12e), (34o, 8e), and (36o, 12e) are shorthands for resp. DMRG-SCF(26o, 8e, D$_{\mathsf{SU(2)}}$=2500)/cc-pVDZ, DMRG(28o, 12e, D$_{\mathsf{SU(2)}}$=2500)/cc-pVDZ, DMRG-SCF(34o, 8e, D$_{\mathsf{SU(2)}}$=2500)/cc-pCVDZ, and DMRG(36o, 12e, D$_{\mathsf{SU(2)}}$=2500)/cc-pCVDZ. The energies are shifted 75 $E_h$ upwards, and are expressed in $mE_h$.}
\begin{center}
\begin{tabular}{lrrrr}
\hline
R / $a_0$ & \multicolumn{4}{l}{(Energy + 75 $E_h$) / $mE_h$}\\
          & (26o, 8e) & (28o, 12e) & (34o, 8e) & (36o, 12e) \\
\hline
1.8  &  -450.44  &  -454.96  &  -459.72  &  -534.24  \\
1.9  &  -557.90  &  -562.08  &  -564.84  &  -639.06  \\
2.0  &  -631.96  &  -635.85  &  -637.31  &  -711.29  \\
2.1  &  -680.64  &  -684.30  &  -684.95  &  -758.71  \\
2.2  &  -710.17  &  -713.63  &  -713.80  &  -787.37  \\
2.3  &  -725.38  &  -728.68  &  -728.57  &  -801.98  \\
2.35 &  -728.82  &  -732.05  &  -731.86  &  -805.19  \\
2.4  &  -730.02  &  -733.18  &  -732.93  &  -806.19  \\
2.5  &  -727.02  &  -730.05  &  -729.75  &  -802.89  \\
2.6  &  -718.65  &  -721.58  &  -721.28  &  -794.31  \\
2.7  &  -706.72  &  -709.54  &  -709.29  &  -782.22  \\
2.8  &  -692.64  &  -695.37  &  -695.19  &  -768.03  \\
2.9  &  -677.59  &  -680.23  &  -680.15  &  -752.93  \\
3.0  &  -662.64  &  -665.20  &  -665.25  &  -737.98  \\
3.2  &  -636.59  &  -638.95  &  -639.33  &  -711.89  \\
3.4  &  -615.74  &  -617.95  &  -618.53  &  -690.94  \\
3.6  &  -597.53  &  -599.65  &  -600.32  &  -672.66  \\
3.8  &  -581.54  &  -583.60  &  -584.32  &  -656.62  \\
4.0  &  -567.88  &  -569.91  &  -570.65  &  -642.92  \\
4.2  &  -556.62  &  -558.63  &  -559.38  &  -631.62  \\
4.4  &  -547.67  &  -549.67  &  -550.41  &  -622.64  \\
4.6  &  -540.83  &  -542.81  &  -543.54  &  -615.76  \\
4.8  &  -535.75  &  -537.73  &  -538.44  &  -610.67  \\
5.0  &  -532.08  &  -534.05  &  -534.75  &  -606.96  \\
\hline

\end{tabular}
\end{center}
\end{table}

\begin{figure}[t!]
 \centering
 \includegraphics[width=0.7\textwidth]{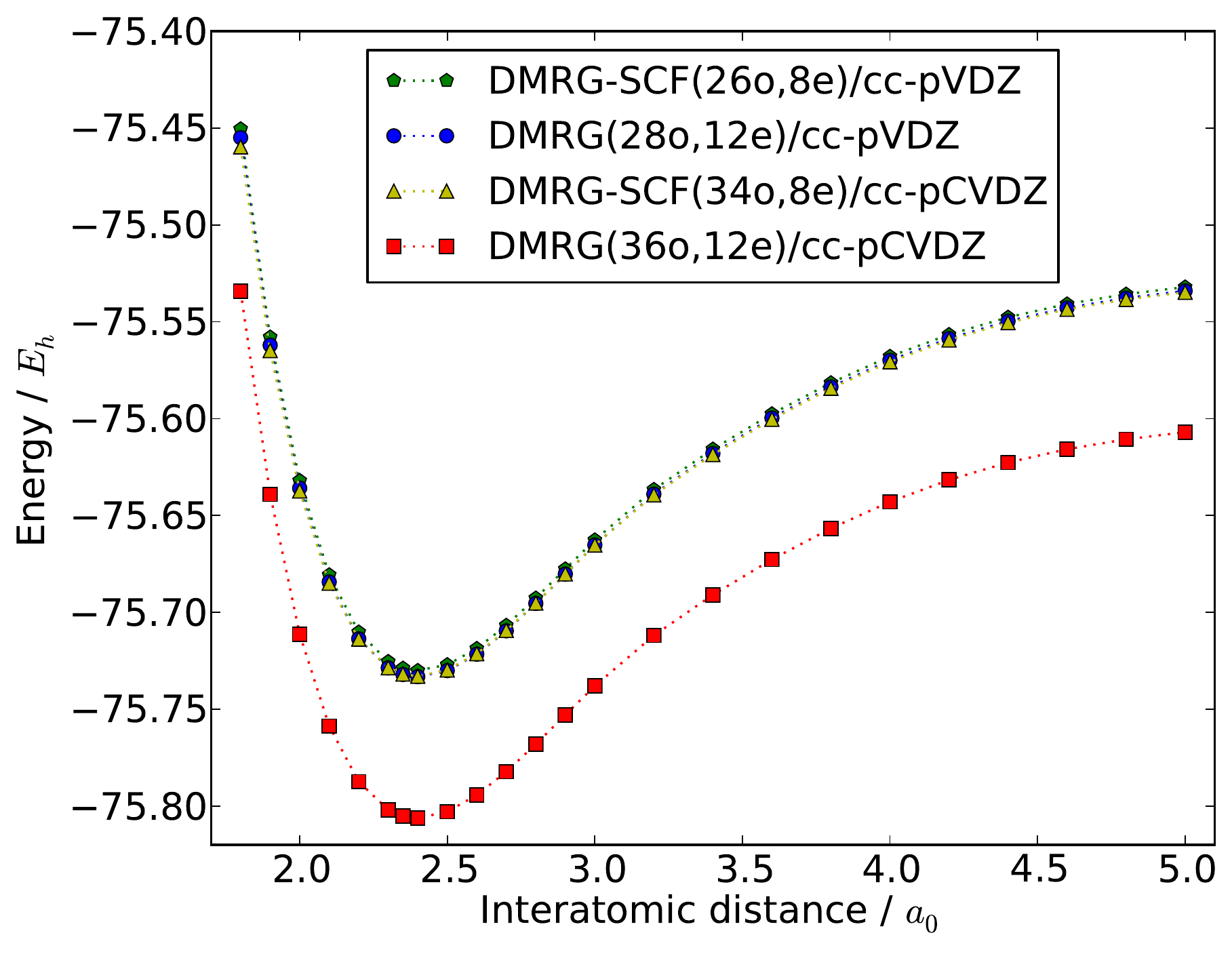}
 \caption{\label{CPCCoreCorrAbsolute} Assessment of the importance of $1s$ core correlation. This effect is captured at the DMRG(36o, 12e, D$_{\mathsf{SU(2)}}$=2500)/cc-pCVDZ level of theory.}
\end{figure}
\begin{figure}[t!]
 \centering
 \includegraphics[width=0.7\textwidth]{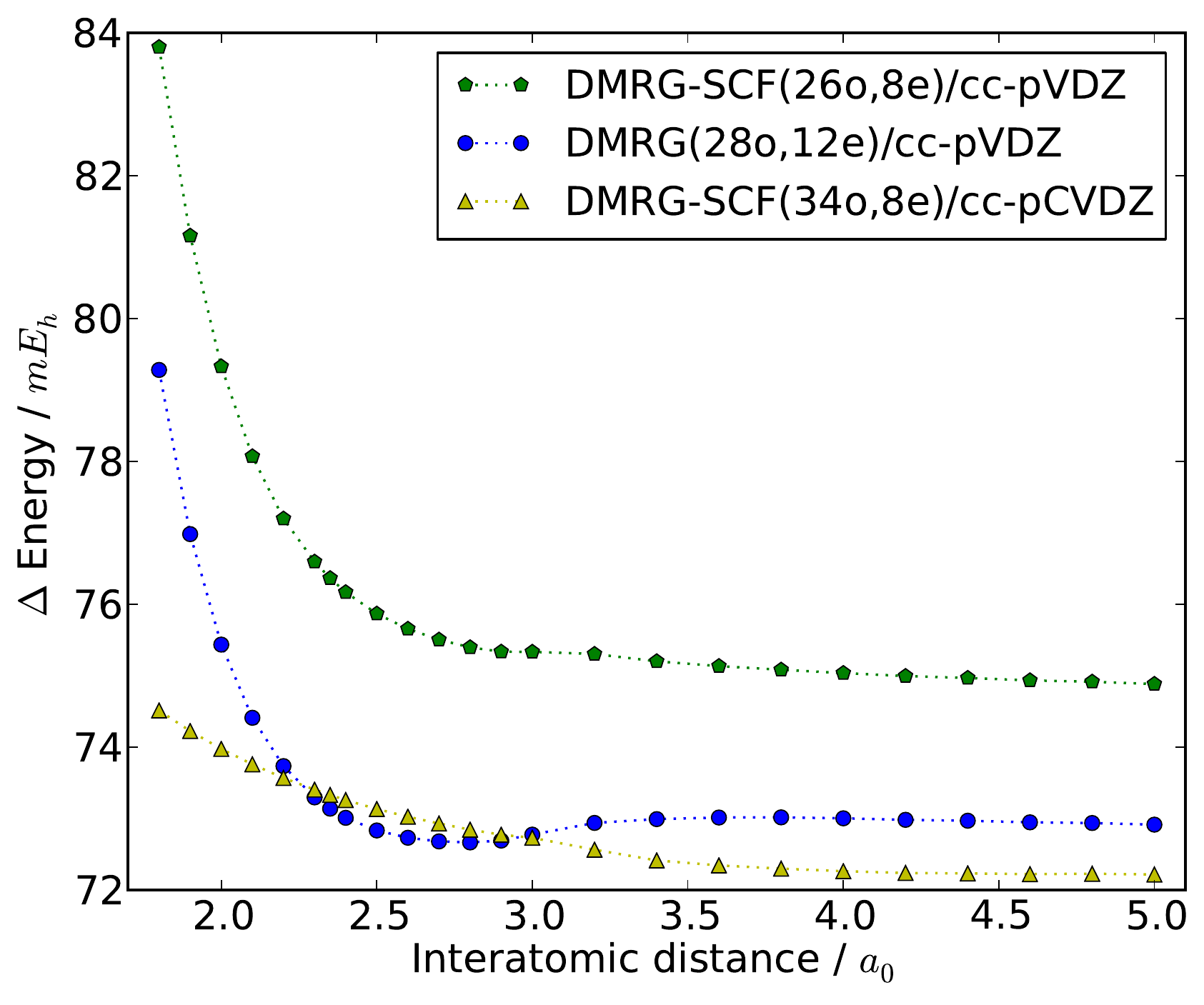}
 \caption{\label{CPCCoreCorrRelative} Assessment of the importance of $1s$ core correlation. The relative energies with respect to the DMRG(36o, 12e, D$_{\mathsf{SU(2)}}$=2500)/cc-pCVDZ calculations are shown.}
\end{figure}

The extrapolated energies at the DMRG-SCF(26o, 8e, D$_{\mathsf{SU(2)}}$=2500)/cc-pVDZ, DMRG (28o, 12e, D$_{\mathsf{SU(2)}}$=2500)/cc-pVDZ, DMRG-SCF(34o, 8e, D$_{\mathsf{SU(2)}}$=2500)/cc-pCVDZ, and DMRG(36o, 12e, D$_{\mathsf{SU(2)}}$=2500)/cc-pCVDZ levels of theory are given in Table \ref{CPCCoreCorrTable} and are shown in Fig. \ref{CPCCoreCorrAbsolute}. The relative energies with respect to the DMRG(36o, 12e, D$_{\mathsf{SU(2)}}$=2500)/cc-pCVDZ calculations are shown in Fig. \ref{CPCCoreCorrRelative}.

The $1s$ core correlation is only captured at the DMRG(36o, 12e, D$_{\mathsf{SU(2)}}$=2500)/cc-pCVDZ level of theory. Without the necessary orbital freedom, the $1s$ core correlation cannot be captured. The non-parallelity of the DMRG-SCF(34o, 8e, D$_{\mathsf{SU(2)}}$=2500)/cc-pCVDZ curve in Fig. \ref{CPCCoreCorrRelative} is of the order of 2 $mE_h$, far below the error due to basis set incompleteness.

For small interatomic distances, the cc-pCVDZ curves show a different behaviour than the cc-pVDZ curves, as can be seen in Fig. \ref{CPCCoreCorrRelative}. Extra basis set freedom is required to capture the more complicated core dynamics in the united atom limit. This can be understood as the transition from two light atoms, each with a doubly filled $1s$ orbital, to one single heavy atom, with several orbitals tightly packed around the nucleus.

\vspace{0.7cm}
\hspace{-\parindent}\textbf{6. Summary}
\vspace{0.3cm}

In Section 1, we discussed how DMRG can be useful for ab initio quantum chemistry, and we gave an overview of DMRG-related methods. These methods can be divided into two categories: DMRG can play the role of a large active space FCI solver, or it can provide an approximate MPS wavefunction, on which excitations can be built.

The DMRG algorithm was introduced in Section 2, where we discussed the use of complementary operators and how to overcome convergence difficulties. Both issues have to be addressed for DMRG to be an efficient and reliable approach for ab initio quantum chemistry.

With symmetry-adapted DMRG, a huge performance gain can be obtained both in computation time and memory. Section 3 introduced an MPS ansatz which is an exact eigenstate of the symmetry group of the Hamiltonian. The Wigner-Eckart theorem allows the introduction of a sparse block structure in this ansatz. For non-abelian groups, the Wigner-Eckart theorem also allows for data compression.

An overview of the high-level structure of \textsc{CheMPS2} is given in Section 4. The required input for the \texttt{CheMPS2::DMRG} class and its output are discussed. A DMRG-SCF algorithm was implemented in \texttt{CheMPS2::CASSCF}. Section 4 should help new users to understand the provided tests, and to alter them to their own needs.

As an application, we have calculated the 12 lowest bond dissociation curves of the carbon dimer at the DMRG(28o, 12e, D$_{\mathsf{SU(2)}}$=2500)/cc-pVDZ level of theory. In addition, we assessed the contribution of $1s$ core correlation to the $X^1\Sigma_g^+$ bond dissociation curve of the carbon dimer by comparing calculations at the DMRG(36o, 12e, D$_{\mathsf{SU(2)}}$=2500)/cc-pCVDZ and DMRG-SCF(34o, 8e, D$_{\mathsf{SU(2)}}$=2500)/cc-pCVDZ levels of theory. These results were presented in Section 5. The low-lying bond dissociation curves of the carbon dimer were resolved with \textsc{CheMPS2} to sub-$mE_h$ accuracy. The non-parallelity due to $1s$ core correlation is of the order of 2 $mE_h$ in the cc-pCVDZ basis.

In the future, we would like to incorporate the two-orbital mutual information $I_{p,q}$ \cite{Rissler2006519} in \textsc{CheMPS2}, as well as its gradient and hessian, to retrieve optimal orbitals and their corresponding ordering, as discussed in Section 2.6.

We are also working on an MPI implementation of \textsc{CheMPS2}, in which the product $\mathbf{H}^{\text{eff}} \mathbf{B}[i]$ is distributed over several processors. Each processor is then responsible for certain renormalized operators \cite{chan:3172}. Updated versions of \textsc{CheMPS2} will be provided at its public git repository \cite{CheMPS2github}.

The oxo-Mn(salen) complex \cite{Ivanic, SearsJohn} is a great challenge for molecular electronic structure methods. We are currently performing large active space DMRG-SCF calculations with \textsc{CheMPS2} to provide new insights in the relative order of the lowest singlet, triplet, and quintet states. Understanding the active space structure of this complex and several of its transition states will be of benefit for the experimentalists in our group \cite{C3CC44473B}.

\vspace{0.7cm}
\hspace{-\parindent}\textbf{Acknowledgements}
\vspace{0.3cm}

S.W. received a Ph.D. fellowship from the Research Foundation Flanders (FWO Vlaanderen). W.P. acknowledges support from a project funded by the Research Foundation Flanders (FWO Vlaanderen). P.W.A. acknowledges support from NSERC. This work was carried out using the Stevin Supercomputer Infrastructure at Ghent University, funded by Ghent University, the Hercules Foundation and the Flemish Government - department EWI. We would like to thank Veronique Van Speybroeck for providing extra resources on the Stevin Supercomputer Infrastructure, and Wim Dewitte for designing the cover of this month's issue.

\newpage
\hspace{-\parindent}\textbf{Appendix. Reduced tensors}
\vspace{0.3cm}

Note that during a sweep, we work with left-normalized tensors to the left and right-normalized tensors to the right of the current position. Consider the following renormalized partial Hamiltonian term in the graphical notation \cite{Schollwock201196}:
\begin{equation}
\vcenter{\hbox{\scriptsize{
\setlength{\unitlength}{1cm}
\begin{picture}(2.6,2)
\put(0.4,1.8){\circle{0.7}}
\put(0.75,1.8){\line(1,0){0.65}}
\put(1.4,1.7){$j_R j_R^z N_R I_R \alpha_R$}
\put(0.08,1.7){A[k]}
\put(0.4,1.30){\line(0,1){0.15}}
\put(0.1,0.7){\line(1,0){0.6}}
\put(0.1,1.3){\line(1,0){0.6}}
\put(0.1,0.7){\line(0,1){0.6}}
\put(0.7,0.7){\line(0,1){0.6}}
\put(0.2,0.9){$\hat{a}_{k \sigma}$}
\put(0.4,0.55){\line(0,1){0.15}}
\put(0.4,0.2){\circle{0.7}}
\put(0.75,0.2){\line(1,0){0.65}}
\put(0.08,0.1){A[k]}
\put(1.40,0.1){$\widetilde{j}_R \widetilde{j}_R^z \widetilde{N}_R \widetilde{I}_R \widetilde{\alpha}_R$}
\put(0.05,1.0){\oval(1.0,1.6)[l]}
\end{picture}
}}} \label{CPCexample1}
\end{equation}
With (\ref{CPCtensordecomp}), it is easy to show that (\ref{CPCexample1}) can be written as
\begin{equation}
\delta_{N_R+1,\widetilde{N}_R} \delta_{I_R \otimes I_k, \widetilde{I}_R} \braket{j_R j_R^z \frac{1}{2} \sigma \mid \widetilde{j}_R \widetilde{j}_R^z} 
\vcenter{\hbox{\scriptsize{
\setlength{\unitlength}{1cm}
\begin{picture}(2.0,2.4)
\put(0.1,0.9){\line(1,0){0.6}}
\put(0.1,1.5){\line(1,0){0.6}}
\put(0.1,0.9){\line(0,1){0.6}}
\put(0.7,0.9){\line(0,1){0.6}}
\put(0.15,1.1){L[k]}
\put(0.4,1.5){\line(0,1){0.5}}
\put(0.4,0.4){\line(0,1){0.5}}
\put(0.4,2.0){\line(1,0){0.50}}
\put(0.4,0.4){\line(1,0){0.50}}
\put(0.95,1.9){$j_R N_R I_R \alpha_R$}
\put(0.95,0.43){$\widetilde{j}_R (N_R + 1)$}
\put(0.95,0.17){$(I_R \otimes I_k) \widetilde{\alpha}_R$}
\end{picture}
}}} \label{CPCagainWE}
\vspace{0.01\textwidth}
\end{equation}
with
\begin{eqnarray}
& \vcenter{\hbox{\scriptsize{
\setlength{\unitlength}{1cm}
\begin{picture}(2.0,2.4)
\put(0.1,0.9){\line(1,0){0.6}}
\put(0.1,1.5){\line(1,0){0.6}}
\put(0.1,0.9){\line(0,1){0.6}}
\put(0.7,0.9){\line(0,1){0.6}}
\put(0.15,1.1){L[k]}
\put(0.4,1.5){\line(0,1){0.5}}
\put(0.4,0.4){\line(0,1){0.5}}
\put(0.4,2.0){\line(1,0){0.50}}
\put(0.4,0.4){\line(1,0){0.50}}
\put(0.95,1.9){$j_R N_R I_R \alpha_R$}
\put(0.95,0.43){$\widetilde{j}_R (N_R + 1)$}
\put(0.95,0.17){$(I_R \otimes I_k) \widetilde{\alpha}_R$}
\end{picture}
}}}
= \sum\limits_{\alpha_L} \vcenter{\hbox{\scriptsize{
\setlength{\unitlength}{1cm}
\begin{picture}(3.7,2)
\put(1.1,1.8){\circle{0.7}}
\put(1.45,1.8){\line(1,0){0.45}}
\put(1.95,1.7){$j_R N_R I_R \alpha_R$}
\put(0.85,1.7){T[k]}
\put(1.1,1.20){\line(0,1){0.25}}
\put(1.2,1.25){$0 0 I_0$}
\put(1.1,0.55){\line(0,1){0.25}}
\put(1.2,0.65){$\frac{1}{2} 1 I_k$}
\put(1.1,0.2){\circle{0.7}}
\put(1.45,0.2){\line(1,0){0.45}}
\put(0.85,0.1){T[k]}
\put(1.95,0.23){$\widetilde{j}_R (N_R+1)$}
\put(1.95,-0.03){$(I_R \otimes I_k) \widetilde{\alpha}_R$}
\put(0.75,1.0){\oval(1.0,1.6)[l]}
\put(-0.05,0.3){\rotatebox{90}{$j_R N_R I_R \alpha_L$}}
\end{picture}
}}} \nonumber \\
& + (-1)^{\widetilde{j}_R - j_R + \frac{1}{2}} \sqrt{\frac{2 j_R + 1}{2 \widetilde{j}_R + 1}}\sum\limits_{\alpha_L} \vcenter{\hbox{\scriptsize{
\setlength{\unitlength}{1cm}
\begin{picture}(3.7,2)
\put(1.5,1.8){\circle{0.7}}
\put(1.85,1.8){\line(1,0){0.45}}
\put(2.35,1.7){$j_R N_R I_R \alpha_R$}
\put(1.25,1.7){T[k]}
\put(1.5,1.2){\line(0,1){0.25}}
\put(1.6,1.25){$\frac{1}{2} 1 I_k$}
\put(1.5,0.55){\line(0,1){0.25}}
\put(1.6,0.55){$0 2 I_0$}
\put(1.5,0.2){\circle{0.7}}
\put(1.85,0.2){\line(1,0){0.45}}
\put(1.25,0.1){T[k]}
\put(2.35,0.23){$\widetilde{j}_R (N_R+1)$}
\put(2.35,-0.03){$(I_R \otimes I_k) \widetilde{\alpha}_R$}
\put(1.15,1.0){\oval(1.0,1.6)[l]}
\put(-0.05,0.3){\rotatebox{90}{$\widetilde{j}_R (N_R-1)$}}
\put(0.35,0.3){\rotatebox{90}{$(I_R \otimes I_k) \alpha_L$}}
\end{picture}
}}}
\end{eqnarray}
Eq. \eqref{CPCexample1} can hence be factorized into Clebsch-Gordan coefficients and a reduced spin-$\frac{1}{2}$ $L$-tensor. The $L$-tensor has spin-$\frac{1}{2}$ because $\hat{a}_{k \sigma}$ is a spin-$\frac{1}{2}$ operator.

It is shown in Ref. \cite{woutersJCP1}, that for two second quantized operators acting on different sites, the renormalized operator can be decomposed into two terms: one with a spin-0 reduced tensor and one with a spin-1 reduced tensor. This follows from $\mathsf{SU(2)}$ representation theory: $\frac{1}{2} \otimes \frac{1}{2} \approx 0 \oplus 1$.

{\color{blue}{
\noindent\makebox[\linewidth]{\rule{\textwidth}{0.4pt}}

\vspace{-0.40cm}

\noindent\makebox[\linewidth]{\rule{\textwidth}{0.4pt}}
}}

\begin{sidewaysfigure}
\centering
\includegraphics[width=0.85\textwidth]{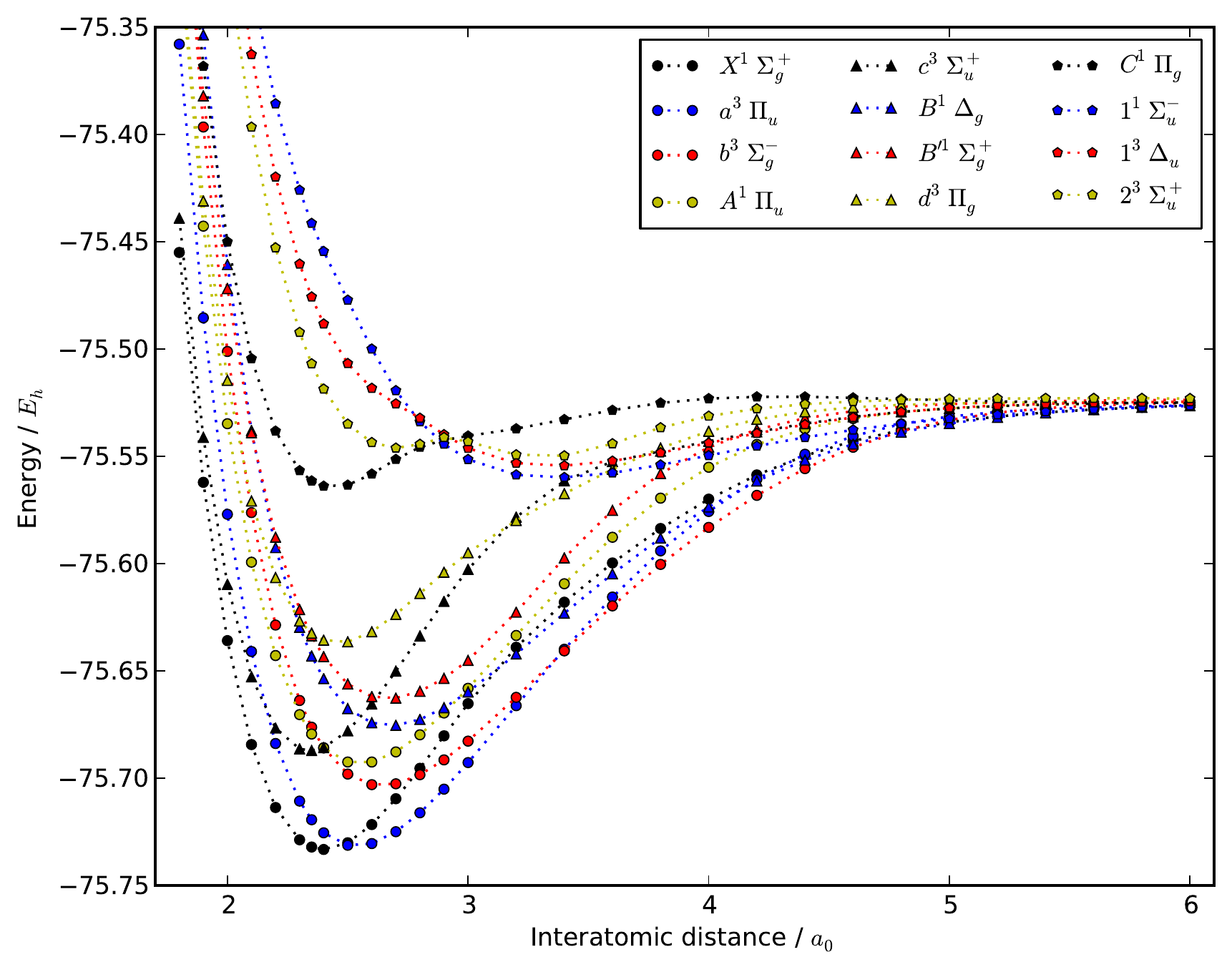}
\caption{\label{C2AllStates} All bond dissociation curves of Tab. \ref{CPClargeTableC2}.}
\end{sidewaysfigure}

\chapter{Thouless theorem for MPS and post-DMRG methods} \label{Thouless-chapter}
\begin{chapquote}{William S. Jevons, 1874}
Whoever wishes to acquire a deep acquaintance with nature must observe that there are analogies which connect whole branches of science in a parallel manner, and enable us to infer of one class of phenomena what we know of another. It has thus happened on several occasions that the discovery of an unsuspected analogy between two branches of knowledge has been the starting point for a rapid course of discovery. 
\end{chapquote}

\section{Post-DMRG as the analogon of post-HF}
DMRG is considered to be a self-consistent mean-field theory in the lattice sites (see section \ref{subsec-macro-it} or Ref. \cite{ChanB805292C}), just like HF forms a self-consistent mean-field theory for particles. On top of the optimized MPS, the zeroth order reference wavefunction, excitations and correlations can be built with post-DMRG methods. Inspiration for these post-DMRG methods has been found in their post-HF counterparts (see section \ref{DMRG-ExcitedStatesSectionInChatperTwo}).

The MPS geometry \cite{PhysRevLett.107.070601, 2012arXiv1210.7710H} played a crucial role in the development of the post-DMRG hierarchy. DMRG-LRT works in the MPS tangent space \cite{dorando:184111}. The variational optimization in this tangent space is called DMRG-CIS or DMRG-TDA \cite{PhysRevB.85.035130, PhysRevB.85.100408, PhysRevB.88.075122, PhysRevB.88.075133, NaokiLRTpaper}. Linearization of the time-dependent variational principle for MPSs \cite{PhysRevLett.107.070601} yields DMRG-RPA \cite{2011arXiv1103.2155K, PhysRevB.88.075122, PhysRevB.88.075133, NaokiLRTpaper}. The nonredundant parameterization of the entire MPS manifold is called the Thouless theorem for MPS, and it generates the DMRG CI expansion \cite{PhysRevB.88.075122}:

\newpage

\hspace{-\parindent}{\large\color{blue}{\textbf{Thouless theorem for matrix product states and subsequent post density matrix renormalization group methods} \cite{PhysRevB.88.075122}}}

\vspace{0.2cm}

Sebastian Wouters,$^{1}$ Naoki Nakatani,$^{2}$ Dimitri Van Neck,$^{1}$ and Garnet Kin-Lic Chan$^{2}$

{\footnotesize $^{1}$\textit{Center for Molecular Modeling, Ghent University, Technologiepark 903, 9052 Zwijnaarde, Belgium}}

{\footnotesize $^{2}$\textit{Department of Chemistry, Princeton University, Frick Chemistry Laboratory, Princeton, New Jersey 08544, USA}}

\vspace{0.5cm}

\parbox{0.90\textwidth}{
The similarities between Hartree-Fock (HF) theory and the density matrix renormalization group (DMRG) are explored. Both methods can be formulated as the variational optimization of a wave-function \textit{Ansatz}. Linearization of the time-dependent variational principle near a variational minimum allows to derive the random phase approximation (RPA). We show that the nonredundant parameterization of the matrix product state (MPS) tangent space [J. Haegeman, J. I. Cirac, T. J. Osborne, I. Pi\u{z}orn, H. Verschelde, and F. Verstraete, Phys. Rev. Lett. \textbf{107}, 070601 (2011)] leads to the Thouless theorem for MPS, i.e., an explicit nonredundant parameterization of the entire MPS manifold, starting from a specific MPS reference. Excitation operators are identified, which extends the analogy between HF and DMRG to the Tamm-Dancoff approximation (TDA), the configuration interaction (CI) expansion, and coupled cluster theory. For a small one-dimensional Hubbard chain, we use a CI-MPS \textit{Ansatz} with single and double excitations to improve on the ground state and to calculate low-lying excitation energies. For a symmetry-broken ground state of this model, we show that RPA-MPS allows to retrieve the Goldstone mode. We also discuss calculations of the RPA-MPS correlation energy. With the long-range quantum chemical Pariser-Parr-Pople Hamiltonian, low-lying TDA-MPS and RPA-MPS excitation energies for polyenes are obtained.
}

\vspace{0.7cm}
\hspace{-\parindent}\textbf{I. INTRODUCTION}
\vspace{0.3cm}

The standard classification of quantum ground states dates back to Landau \cite{Landau1, Landau2}. Mean-field theory is used to describe a state, and a phase transition is marked by the breaking of a symmetry. The particle-conserving mean-field theory for fermions is Hartree-Fock (HF) theory \cite{PSP.1733252, PSP.1733252bis, PhysRev.32.339, Fock1}. In HF theory, the exact ground state is approximated by a Slater determinant (SD) \cite{PhysRev.34.1293}, and the energy of a Hamiltonian is minimized within this variational \textit{Ansatz} space. To obtain excited states or a more accurate description of the ground state, post-HF (post mean-field) methods \cite{helgaker2} can be carried out such as the Tamm-Dancoff approximation (TDA) \cite{Tamm, PhysRev.78.382}, the random-phase approximation (RPA) \cite{PhysRev.92.609}, M\o{}ller-Plesset perturbation theory \cite{PhysRev.46.618}, the configuration interaction (CI) expansion \cite{PhysRev.34.1293, PhysRev.36.1121}, and coupled cluster (CC) theory \cite{Coester1958421, Coester1960477, CizekCC}.

Within the framework of second quantization \cite{Dirac01031927, Fock2ndquant, BookDimitri}, the reference SD obtains a simple product form when the canonical HF orbitals are used to construct the Fock space. Occupied-virtual (OV) excitation operators allow to connect the reference SD to post-HF wave-function \textit{Ans\"atze}. The Thouless theorem gives a nonredundant parameterization to generate \textit{all} possible SDs from any given SD reference, by means of its OV excitation operators \cite{PhysRevA.22.2362, PhysRevA.24.673, Thouless1960225, Thouless196178}.

Recently, a new way to understand the qualitative structure of quantum many-body states has appeared, whereby the state is approximated by a tensor network, i.e., a contracted product of tensors where each tensor represents a local degree of freedom. These \textit{Ans\"atze} are efficient representations of low-energy states because they capture the boundary law for the entanglement entropy. In one dimension, the tensor network is known as a matrix product state (MPS). The MPS is the wave-function \textit{Ansatz} for the density matrix renormalization group (DMRG) algorithm \cite{PhysRevLett.69.2863, PhysRevB.48.10345, PhysRevLett.75.3537, PhysRevB.55.2164, 1742-5468-2007-08-P08024, PhysRevB.78.035124}.

DMRG can capture states beyond the realm of Landau (or mean-field) theory, i.e., states with topological order \cite{WenTopoOrder, PhysRevB.83.035107, PhysRevB.83.075103, PhysRevB.83.075102, PhysRevB.84.165139, PhysRevB.86.245305}. DMRG has also been shown to be a powerful method to treat the static correlation problem in electronic structure theory \cite{WhiteQCDMRG, Chan2002, woutersJCP1, sharma:124121, chan:annurevphys}. Static correlation arises when a state consists of several significant SD contributions, which HF theory is of course unable to deal with, because a single SD does not describe the qualitative structure of the targeted state. Post-HF methods, which start from a single SD reference, have difficulty building in large static correlation \textit{a posteriori}. In these situations, DMRG has provided a new ability to access the electronic structure. The analog of static correlation for DMRG is a quantum critical system, which introduces corrections to the entanglement boundary law, which cannot be captured by DMRG.

DMRG can be interpreted as a mean-field theory in the sites, which is analogous to HF, which is a mean-field theory in the particles \cite{ChanB805292C, 2011arXiv1103.2155K}. Therefore it is natural to search for extensions to DMRG that are analogous to post-HF methods: post-DMRG methods. One example is linear response theory. Time-dependent HF theory is obtained by using an SD \textit{Ansatz} in the time-dependent variational principle (TDVP) \cite{PhysRevA.22.2362, PhysRevA.24.673, PSP:2040328, RevModPhys.44.602, Kerman1976332, Kramer}. Time-dependent DMRG (which stays within the MPS \textit{Ansatz} space) is similarly obtained by using an MPS \textit{Ansatz} in the TDVP \cite{2011arXiv1103.2155K, PhysRevLett.107.070601, 2006cond.mat.12480U, JuthoThesis}. RPA, or linear response theory for HF, is obtained by linearizing the time-dependent HF equations in the vicinity of a variational mimimum \cite{zyrianov, PhysRev.107.450, PhysRev.107.1631}. Equivalently, the RPA equations can be derived from an equation of motion (EOM) approach with excitation operators \cite{PhysRev.106.372, PhysRev.108.507, PhysRev.112.1900, PhysRev.115.786, RPAderivationGoldstone}. RPA yields a mean-field description of quasi-particle excitations. The linear response theory for DMRG was first derived by Dorando \textit{et al.} \cite{dorando:184111} and was later recast as RPA for MPS \cite{2011arXiv1103.2155K, JuthoThesis}.

In this work, we construct a more complete analog of the mean-field framework, which allows us to define a full set of post-DMRG methods. We give a nonredundant parameterization of the entire MPS manifold, starting from a specific MPS reference. This is the analog of the Thouless theorem for HF. We identify the excitation operators of the Thouless theorem. These excitation operators allow for a complete rederivation of RPA for MPS by means of the EOM, in complete analogy with HF. All other results, such as an improvement of the ground-state theory by the fluctuation-dissipation theorem, follow. With these excitation operators, we can define the analogs of other post-HF methods for MPS, such as CC and CI.

For a small one-dimensional Hubbard chain, we use a numerical CI-MPS \textit{Ansatz} with single and double excitations to improve on the ground state and to calculate low-lying excitation energies. For a symmetry-broken ground state of this model, we show that RPA-MPS allows to retrieve the Goldstone mode. We also discuss calculations of the RPA-MPS correlation energy. With the long-range quantum chemical Pariser-Parr-Pople (PPP) Hamiltonian, low-lying TDA-MPS and RPA-MPS excitation energies for polyenes are also obtained.

\vspace{0.7cm}
\hspace{-\parindent}\textbf{II. HF MEAN-FIELD THEORY}
\vspace{0.3cm}

This section provides a brief introduction to the variational principles, HF mean-field theory, the Thouless theorem, and post-HF methods. It focusses on the topics for which a DMRG analog will be constructed in this paper. For readers familiar with HF, this section can be a good guideline to understand our post-DMRG discussion.

\vspace{0.7cm}
\hspace{-\parindent}\textbf{A. Variational principles}
\vspace{0.3cm}

Because the Hilbert space increases exponentially with system size, a variational wave-function \textit{Ansatz} $\ket{\Phi(\mathbf{z})}$ with parameterization $\mathbf{z}$ is often used to make calculations feasible. In order to minimize the energy functional
\begin{equation}
E(\mathbf{z},\overline{\mathbf{z}}) = \frac{\braket{ \overline{\Phi} \mid \hat{H} \mid \Phi}}{\braket{\overline{\Phi} \mid \Phi}}
\end{equation}
to approximate ground states, the time-independent variational principle (TIVP) $\frac{\delta \mathcal{L}}{\delta \overline{z}} = 0$ can be employed, where the Lagrangian is \cite{ChanB805292C}
\begin{equation}
\mathcal{L} = \braket{\overline{\Phi} \mid \hat{H} \mid \Phi} - \lambda \left( \braket{\overline{\Phi} \mid \Phi} - 1 \right). \label{TIVPPRB88}
\end{equation}
The overline denotes complex conjugation. This yields the time-independent self-consistent field (SCF) equations. To approximate time evolution, the time-dependent variational principle (TDVP) $\frac{\delta \mathcal{S}}{\delta \overline{z}} = 0$ can be employed, where the action is \cite{PhysRevA.22.2362, PhysRevA.24.673, PSP:2040328, RevModPhys.44.602, Kerman1976332, Kramer, 2006cond.mat.12480U}
\begin{equation}
\mathcal{S} = \int\limits_{t_1}^{t_2} dt \left( \frac{i \hbar}{2} \braket{\overline{\Phi} \mid \dot{\Phi}} - \frac{i \hbar}{2} \braket{\dot{\overline{\Phi}} \mid \Phi} -  \braket{\overline{\Phi} \mid \hat{H} \mid \Phi}  \right).
\end{equation}
The dot denotes time derivation. This yields the time-dependent SCF equations.

\vspace{0.7cm}
\hspace{-\parindent}\textbf{B. The Slater determinant}
\vspace{0.3cm}

From a given single-particle basis, any other single-particle basis can be constructed by a unitary transformation: $\hat{a}^{\dagger}_j = \hat{b}_k^{\dagger} U^k_{~j}$. Second quantization is used to denote the single-particle states \cite{Dirac01031927, Fock2ndquant, BookDimitri}, and the summation convention is used for double indices. An $N$-particle SD is an antisymmetrized product of $N$ single-particle states (called occupied orbitals) \cite{PhysRev.34.1293}:
\begin{equation}
\ket{\Psi} = \hat{a}_1^{\dagger} \hat{a}_2^{\dagger} ...  \hat{a}_N^{\dagger} \ket{-}. \label{theSDeqPRB88}
\end{equation}
The variational freedom is a unitary transformation from the given single-particle basis of $L$ orbitals to another basis where the first $N$ orbitals are used to construct the SD. There is gauge freedom in the \textit{Ansatz}, as any unitary transformation that does not mix the $N$ occupied orbitals with the $L-N$ virtual orbitals, does not change the wave function (except for a global phase). An SD is therefore described by the Grassmann manifold $\mathsf{U}_L / (\mathsf{U}_{N} \times \mathsf{U}_{L-N})$, with $\mathsf{U}_k$ the unitary group of $k \times k$ unitary matrices. This manifold has dimension $2N(L-N)$, and can be parameterized by $N(L-N)$ complex numbers \cite{PhysRevA.22.2362, PhysRevA.24.673}. This will henceforth be called a complex dimension $N(L-N)$.

\vspace{0.7cm}
\hspace{-\parindent}\textbf{C. The Fock equations}
\vspace{0.3cm}

If the particles of a system interact pairwise, the Hamiltonian can always be written in second quantization as
\begin{equation}
\hat{H} = \hat{b}^{\dagger}_i \mathsf{T}^{i}_{~j} \hat{b}^j + \frac{1}{2} \hat{b}^{\dagger}_i \hat{b}^{\dagger}_j \mathsf{V}^{ij}_{~~kl} \hat{b}^l \hat{b}^k. \label{HamiltonianSecondQuantizationPRB88}
\end{equation}
The TIVP can be expressed in terms of the unitary transformation generating the occupied orbitals:
\begin{equation}
\mathcal{L} = U^{\dagger \alpha}_{\quad i} \mathsf{T}^{i}_{~j} U^{j}_{~\alpha} + \frac{1}{2} U^{\dagger \alpha}_{\quad i} U^{\dagger \beta}_{\quad j} \mathsf{V}^{ij}_{~~kl} U^{l}_{~\beta} U^{k}_{~\alpha} - \frac{1}{2} U^{\dagger \alpha}_{\quad i} U^{\dagger \beta}_{\quad j} \mathsf{V}^{ij}_{~~kl} U^{l}_{~\alpha} U^{k}_{~\beta} - \lambda_{~\alpha}^{\beta} \left(  U^{\dagger \alpha}_{\quad k} U^{k}_{~\beta} - \delta^{\alpha}_{~\beta} \right).
\end{equation}
The Greek indices denote occupied orbitals, while the Latin indices denote all single-particle basis states. Varying with respect to $U^{\dagger m}_{\quad i}$ leads to the Fock equations \cite{PSP.1733252, PSP.1733252bis, PhysRev.32.339, Fock1}:
\begin{equation}
\mathsf{F}^{i}_{~k} U^{k}_{~p} = \left( \mathsf{T}^{i}_{~k} + U^{\dagger \beta}_{\quad j} \mathsf{V}^{ij}_{~~kl} U^{l}_{~\beta} - U^{\dagger \beta}_{\quad j} \mathsf{V}^{ij}_{~~lk} U^{l}_{~\beta} \right)  U^{k}_{~p} = U^{i}_{~q} \lambda_{~p}^{q}. \label{HFequationsPRB88}
\end{equation}
The gauge can be partially fixed by requiring that the Lagrangian multiplier matrix $\lambda$ to enforce orthonormal orbitals becomes diagonal, and that the diagonal elements are sorted in ascending order. These diagonal elements are then interpreted as the single-particle energy levels \cite{helgaker2, PhysRevA.22.2362, PhysRevA.24.673}. The remaining gauge freedom is then $\mathsf{U}_1^{~\otimes L}$, i.e., the phase of each HF orbital. The lowest $N$ single-particle states are used to construct the SD.

The Fock equations are orbital-based mean-field equations. There is self-consistency because the Fock operator in Eq. \eqref{HFequationsPRB88} which determines the orbitals, also depends on the orbitals.

\vspace{0.7cm}
\hspace{-\parindent}\textbf{D. The Thouless theorem}
\vspace{0.3cm}

The Thouless theorem for HF \cite{Thouless1960225, Thouless196178} and its unitary counterpart \cite{PhysRevA.22.2362, PhysRevA.24.673} state that any $N$-electron SD can be globally parameterized, respectively, as
\begin{eqnarray}
\ket{\Psi} & \propto & \exp{\left( X^{vo} \hat{B}_{vo}^{\dagger} \right)} \ket{\Psi^0}, \label{HFThoulessPRB88}\\
\ket{\Psi} & = & \exp{\left( X^{vo} \hat{B}_{vo}^{\dagger} - \overline{X}_{vo} \hat{B}^{vo} \right)} \ket{\Psi^0} \label{counterpartThoulessPRB88}
\end{eqnarray}
with $\ket{\Psi^0}$ a random SD, with $N$ occupied ($o$) and $L-N$ virtual ($v$) orbitals. $\hat{B}_{vo}^{\dagger}$ is a shorthand for $\hat{a}_{v}^{\dagger} \hat{a}^{o}$. Note that the summation convention was used. The equality in Eq. \eqref{counterpartThoulessPRB88} holds because the exponential of an anti-Hermitian operator is unitary and hence does not change the norm.

This parameterization is of complex dimension $N(L-N)$ and is thus nonredundant. All parameters $X$ are needed to parameterize the neighbourhood of $\ket{\Psi^0}$. For all possible combinations $\ket{\Psi^0}$ and $\ket{\Psi}$, a solution $X$ can always be found. The theorem does not state that this $X$ is unique. In fact, $X$ is not unique, see, e.g., the discussion in the appendix of Rowe \textit{et al.} \cite{PhysRevA.22.2362}. The reader can think about the Lie group $\mathsf{O}(3)$, where several combinations of successive rotations along different axes can generate the same global rotation. Instead of working with the redundant parameters $U$, we can equivalently work with the nonredundant parameters $X$.

The $n$th-order variation of a wave function defines its $n$th-order tangent space. The (first-order) tangent space of this nonredundant parameterization consists of the single OV excitations $\hat{B}_{vo}^{\dagger} \ket{\Psi^0}$.

\vspace{0.7cm}
\hspace{-\parindent}\textbf{E. Time evolution}
\vspace{0.3cm}

The TDVP leads to the time-dependent SCF equations \cite{PhysRevA.22.2362, PhysRevA.24.673, PSP:2040328, RevModPhys.44.602, Kerman1976332, Kramer, RingAndSchuck}:
\begin{equation}
i \hbar \dot{U}(t)^{i}_{~ p} = F(t)^{i}_{~k} U(t)^{k}_{~ p}. \label{TDHFPRB88}
\end{equation}
The Fock operator dictates how orbitals are rotated into each other over time. Rotations within the space of occupied orbitals or within the space of virtual orbitals do not change the SD wave function as it represents a Grassmann manifold. Only the rotation of occupied and virtual orbitals into each other has physical meaning.

To obtain the rate of OV rotation determined by Eq. \eqref{TDHFPRB88} in the point $\ket{\Psi^0}$, the Thouless parameterization of a general SD can be used in the TDVP:
\begin{equation}
i \hbar \dot{X}^{vo}( ~\mathbf{X}=\mathbf{0}~ ) = \braket{\overline{\Psi^0} \mid \hat{B}^{vo} \hat{H} \mid \Psi^0}. \label{OVrotationsPRB88}
\end{equation}
The parameters $X^{vo}$ are flattened to a column $\mathbf{X}$. The same equation is obtained by inserting Eq. \eqref{counterpartThoulessPRB88} in the time-dependent Schr\"odinger equation and by projecting this equation onto $\hat{B}_{vo}^{\dagger} \ket{\Psi} = \hat{B}_{vo}^{\dagger} \ket{\Psi(\mathbf{X},\overline{\mathbf{X}})}$. The time evolution and its projection are respectively given by
\begin{eqnarray}
& i \hbar \left( \dot{X}^{wp} \frac{\partial}{\partial X^{wp}} + \dot{\overline{X}}_{wp}  \frac{\partial}{\partial \overline{X}_{wp}} \right) \ket{\Psi} = \left( \hat{H} - E_{\text{HF}} \right) \ket{\Psi}, ~ \quad\\
& i \hbar \bra{\overline{\Psi}} \hat{B}^{vo} \left( \dot{X}^{wp} \frac{\partial}{\partial X^{wp}} + \dot{\overline{X}}_{wp}  \frac{\partial}{\partial \overline{X}_{wp}} \right) \ket{\Psi} = \bra{\overline{\Psi}} \hat{B}^{vo} \left( \hat{H} - E_{\text{HF}} \right) \ket{\Psi}, \label{toLinearizePRB88}
\end{eqnarray}
where $w$ denotes virtual orbitals and $p$ occupied orbitals. Evaluation for $\mathbf{X} = \mathbf{0}$ yields Eq. \eqref{OVrotationsPRB88}.

\newpage

\hspace{-\parindent}\textbf{F. RPA}
\vspace{0.3cm}

Linearization of the TDVP near a variational minimum leads to RPA \cite{PhysRevA.22.2362, PhysRevA.24.673, zyrianov, PhysRev.107.450, PhysRev.107.1631, RingAndSchuck}. Take the variational minimum as the reference $\ket{\Psi^0}$. Expand Eq. \eqref{toLinearizePRB88} up to first-order around $\mathbf{X} = \mathbf{0}$. The zeroth order terms vanish because the expansion point is a variational minimum: $\braket{\overline{\Psi^0} \mid \hat{B}^{vo} \hat{H} \mid \Psi^0} = 0$. This is Brillouin's theorem \cite{Brillouin}. The linearized equations are
\begin{eqnarray}
& i \hbar \dot{X}^{vo} = \braket{\overline{\Psi^0} \mid \hat{B}^{wp} \hat{B}^{vo} \left(\hat{H} - E_{\text{HF}} \right) \mid \Psi^0} \overline{X}_{wp} + \braket{\overline{\Psi^0} \mid \hat{B}^{vo} \left(\hat{H} - E_{\text{HF}} \right) \hat{B}^{\dagger}_{wp} \mid \Psi^0} X^{wp} \qquad\\
& = - \braket{\overline{\Psi^0} \mid \left[ \hat{B}^{vo}, \left[ \hat{H} , \hat{B}^{wp} \right] \right] \mid \Psi^0} \overline{X}_{wp} + \braket{\overline{\Psi^0} \mid \left[ \hat{B}^{vo} , \left[ \hat{H} , \hat{B}^{\dagger}_{wp} \right] \right] \mid \Psi^0} X^{wp}.
\end{eqnarray}
Assume a harmonic motion of the form $\mathbf{X} = \mathbf{Y} e^{-i \omega t} + \overline{\mathbf{Z}} e^{i \omega t}$. This leads to the RPA equations:
\begin{equation}
\hbar \omega \left[ \begin{array}{cc} I & 0 \\ 0 & -I \end{array} \right] \left( \begin{array}{c} \mathbf{Y} \\ \mathbf{Z} \end{array} \right) = \left[ \begin{array}{cc} A & B \\ \overline{B} & \overline{A} \end{array} \right] \left( \begin{array}{c} \mathbf{Y} \\ \mathbf{Z} \end{array} \right) \label{eqRPAunitPRB88}
\end{equation}
with $A_{vo;wp} = \braket{\overline{\Psi^0} \mid \left[ \hat{B}^{vo} , \left[ \hat{H}, \hat{B}^{\dagger}_{wp} \right] \right] \mid \Psi^0}$ and $B_{vo;wp} = - \braket{\overline{\Psi^0} \mid \left[ \hat{B}^{vo} , \left[ \hat{H}, \hat{B}^{wp} \right] \right] \mid \Psi^0}$. Note that if $(\omega, \mathbf{Y}, \mathbf{Z})$ is a solution, $(-\omega, \overline{\mathbf{Z}}, \overline{\mathbf{Y}})$ is a solution too.

Consider the energy functional

\vspace{-0.1cm}
\begin{equation}
E(\mathbf{X},\overline{\mathbf{X}}) = \braket{\overline{\Psi(\mathbf{X},\overline{\mathbf{X}})} \mid \hat{H} \mid \Psi(\mathbf{X},\overline{\mathbf{X}})}
\end{equation}
and its expansion up to second order in $\mathbf{X}$:

\vspace{-0.1cm}
\begin{equation}
E^{(2)}(\mathbf{X},\overline{\mathbf{X}}) - E_{\text{HF}} = \frac{1}{2} \left( \begin{array}{c} \mathbf{X} \\ \overline{\mathbf{X}} \end{array} \right)^{\dagger} \left[ \begin{array}{cc} A & B \\ \overline{B} & \overline{A} \end{array} \right]  \left( \begin{array}{c} \mathbf{X} \\ \overline{\mathbf{X}} \end{array} \right).
\end{equation}
The RPA method searches for the harmonic modes of this potential near its minimum, akin to normal mode analysis in analytical mechanics.

In linear response theory, the RPA frequencies occur as poles in the response function. Because the exact response function for the exact ground state has the excitation energies of the Hamiltonian as poles, the RPA frequencies are interpreted as approximate excitation energies \cite{RingAndSchuck}. A second argument to interpret the RPA frequencies as excitation energies is given by the alternative derivation of RPA by means of the EOM \cite{PhysRev.106.372, PhysRev.108.507, PhysRev.112.1900, RingAndSchuck}. Assume we know the exact ground state $\ket{0}$ and the exact excitation operators, which connect the ground state to the excited states $\hat{Q}^{\dagger}_n = \ket{n}\bra{\overline{0}}$. The operator $\hat{Q}_n$ then destroys the ground state. With $\hbar \omega_n = E_n - E_0$, the excitation energy of the excited state $\ket{n}$, it is easy to derive the EOM:

\vspace{-0.1cm}
\begin{equation}
\braket{ \overline{0} \mid \left[ \delta \hat{Q}, \left[ \hat{H}, \hat{Q}^{\dagger}_n \right] \right] \mid 0 } = \hbar \omega_n \braket{ \overline{0} \mid \left[ \delta \hat{Q}, \hat{Q}^{\dagger}_n \right] \mid 0}. \label{EOMeqPRB88}
\end{equation}
For RPA, two assumptions are made: the excitation operators are approximated by $\hat{Q}_n^{\dagger} = Y^{vo}_n \hat{B}_{vo}^{\dagger} - Z^{vo}_n \hat{B}^{vo}$ and the expectation values of the commutators are calculated with the HF reference wave function. The latter approximation is called the quasiboson approximation.  Equation \eqref{eqRPAunitPRB88} is then retrieved.

As an alternative, an exact bosonic algebra can be set up:
\begin{eqnarray}
\left[\hat{B}^{vo},\hat{B}_{wp}^{\dagger}\right] & = & \delta^{v}_{w} \delta^{o}_{p}, \\
\left[\hat{B}^{vo},\hat{B}^{wp}\right] & = & 0
\end{eqnarray}
by adding higher order terms to $\hat{B}_{vo}^{\dagger} = \hat{a}^{\dagger}_v \hat{a}^o + \mathcal{O}(\hat{a}^4)$. The Hamiltonian can then be written in terms of these new operators:
\begin{equation}
\hat{H}_B = E_{\text{HF}} - \frac{\text{tr}A}{2} + \frac{1}{2} \left( \hat{\mathbf{B}}^{\dagger} \hat{\mathbf{B}} \right) \left[ \begin{array}{cc} A & B \\ \overline{B} & \overline{A} \end{array} \right] \left(  \begin{array}{c} \hat{\mathbf{B}}\\ \hat{\mathbf{B}}^{\dagger} \end{array} \right) + \mathcal{O}(\hat{B}^3).
\end{equation}
RPA coincides with neglecting all terms of $\mathcal{O}(\hat{B}^3)$ in the bosonic expansion. This leads to the RPA correlation energy and wave function \cite{Thouless1960225, Thouless196178, RingAndSchuck}:
\begin{eqnarray}
E_{\text{cRPA}} & = & - \frac{\text{tr}A}{2} + \sum\limits_{\omega_n>0} \frac{\hbar \omega_n}{2} = - \sum\limits_{n} \hbar \omega_n \sum\limits_{vo} \mid Z_{n}^{vo} \mid^2 , \label{RPAcorrEnergyEqPRB88}\quad\\
\ket{\text{RPA}} & \propto & e^{\frac{1}{2} \left( \overline{ZY^{-1}} \right)^{vo;wp} \hat{B}^{\dagger}_{vo} \hat{B}^{\dagger}_{wp}} \ket{\Phi^0}. \label{RPAwavefunctionPRB88}
\end{eqnarray}
The RPA correlation energy has contributions from the zero point energy of the harmonic oscillators with frequency $\omega_n$. The RPA wave function vanishes by the action of deexcitation operators: $\hat{Q}_n \ket{\text{RPA}} = 0$.

If the Hamiltonian has a continuous symmetry, and the exact ground state is degenerate due to this symmetry, a ground-state calculation typically breaks this symmetry. Think for example about a spin-$\frac{1}{2}$ ground state. A calculation will lead to one possibility: $\alpha \ket{s^z = \frac{1}{2}} + \beta \ket{s^z = -\frac{1}{2}}$. A  gapless bosonic degree of freedom remains, which corresponds to rotating within the spin-$\frac{1}{2}$ multiplet, called a Goldstone boson \cite{PhysRev.117.648, GoldstoneBoson}. An interesting feature of RPA is its ability to retrieve Goldstone modes. The excitation energy of a Goldstone mode is of course zero, and the mode is its own dual solution $(\omega=0, \mathbf{Y}, \mathbf{Z}) = (\omega=0, \overline{\mathbf{Z}}, \overline{\mathbf{Y}})$ \cite{BookDimitri, Thouless1960225, Thouless196178}. This implies that
\begin{equation}
\sum_{vo} \left( \overline{Y}_{vo} Y^{vo} - \overline{Z}_{vo} Z^{vo} \right) = 0. \label{GoldstoneEqPRB88}
\end{equation}

\vspace{0.7cm}
\hspace{-\parindent}\textbf{G. Post-HF methods}
\vspace{0.3cm}

With the excitation operators $\hat{B}^{\dagger}_{vo} = \hat{a}_{v}^{\dagger} \hat{a}^o$, a set of orthonormal vectors can be generated: $\ket{\text{HF}}$, $\hat{B}^{\dagger}_{vo} \ket{\text{HF}}$, $ \hat{B}^{\dagger}_{vo} \hat{B}^{\dagger}_{wp} \ket{\text{HF}}$.... They correspond to the zeroth, first and second order tangent space of the Thouless parameterization of a general SD. With the CI method, eigenstates of the Hamiltonian are approximated by working in an incomplete basis of such vectors \cite{PhysRev.34.1293, PhysRev.36.1121}. Consider, for example, the second-order expansion CISD, or CI with single and double excitations:
\begin{equation}
\ket{\text{CISD}} \propto \left( x + y^{vo}\hat{B}^{\dagger}_{vo} + \frac{1}{2} z^{vo;wp} \hat{B}^{\dagger}_{vo} \hat{B}^{\dagger}_{wp} \right) \ket{\text{HF}}.
\end{equation}
With CIS, or CI with only single excitations, the lowest energy state is again $\ket{\text{HF}}$ due to Brillouin's theorem, and the eigenstates approximated in the basis $\hat{B}^{\dagger}_{vo} \ket{\text{HF}}$ are therefore excited states. Note that this corresponds to diagonalizing the $A$ matrix of RPA in Eq. \eqref{eqRPAunitPRB88}. Historically, this method is known as TDA \cite{Tamm, PhysRev.78.382}.

The RPA wave function in Eq. \eqref{RPAwavefunctionPRB88} suggests a CC \textit{Ansatz} \cite{Coester1958421, Coester1960477, CizekCC}. Consider, for example, CCSD or CC with single and double excitations:
\begin{equation}
\ket{\text{CCSD}} \propto e^{\left( y^{vo}\hat{B}^{\dagger}_{vo} + \frac{1}{2} z^{vo;wp} \hat{B}^{\dagger}_{vo} \hat{B}^{\dagger}_{wp} \right)} \ket{\text{HF}}.
\end{equation}
An important property of \textit{Ansatz} wave functions is their size consistency, i.e., the property that for two noninteracting subsystems, the compound wave function is multiplicatively separable and the total energy additively separable. CISD is not size consistent if there are more than two electrons in the compound system, whereas CCSD is always size consistent because of the exponential \textit{Ansatz} \cite{helgaker2}.

\vspace{0.7cm}
\hspace{-\parindent}\textbf{III. THE MATRIX PRODUCT STATE}
\vspace{0.3cm}

\hspace{-\parindent}\textbf{A. The \textit{Ansatz}}
\vspace{0.3cm}

Consider the many-body Hilbert space $\ket{n_1 n_2 ... n_L}$, formed by taking the direct product of $L$ local Hilbert spaces $\ket{n_i}$. The local degrees of freedom can be, e.g., the spin projections of spins on a lattice, or the occupancies of orbitals. In the latter case, the states $\ket{n_1 n_2 ... n_L}$ form the Fock space \cite{Fock2ndquant}. An MPS can be seen as a linear combination of these vectors, where the coefficient of each vector is a product of matrices:
\begin{equation}
\ket{\Phi} = \sum\limits_{\{ n_i \}} A[1]^{n_1} A[2]^{n_2}  ... A[L]^{n_L} \ket{n_1 n_2 ... n_L}. \label{EqWhatIsAnMPSPRB88}
\end{equation}
We assume an MPS with open boundary conditions, i.e., the first matrix has row dimension 1 and the last matrix has column dimension 1. The bond dimension (virtual dimension) $D_i$ of an MPS at boundary $i$ is the column dimension of the matrices at site $i$ and the row dimension of the matrices at site $i+1$. With our assumption, $D_0=D_L=1$. The total number of complex parameters in this \textit{Ansatz} is $\text{dim} \mathbb{A} = \sum_{i=1}^{L} D_{i-1} d D_{i}$, with $d$ the size of the local Hilbert space $\ket{n_i}$. Just as in the SD, there is gauge freedom in the \textit{Ansatz}. Right multiplying the $d$ site matrices on site $i$ with the nonsingular matrix $G$ ($\tilde{A}[i]^{n_i} = A[i]^{n_i} G$), and simultaneously left multiplying the $d$ site matrices on site $i+1$ with its inverse $G^{-1}$ ($\tilde{A}[i+1]^{n_{i+1}} = G^{-1} A[i+1]^{n_{i+1}}$), does not change the wave function ($\tilde{A}[i]^{n_i} \tilde{A}[i+1]^{n_{i+1}} = A[i]^{n_i} A[i+1]^{n_{i+1}}$). A global scalar multiplication does not change the wave function either. The MPS manifold, i.e., the quotient space of the general parameterization (complex dimension $\text{dim} \mathbb{A}$) and all gauge freedom (complex dimension $\sum_{i=1}^L D_i^2$), has complex dimension $\text{dim} \mathbb{T} = \sum_{i=1}^{L} \left( dD_{i-1}-D_i \right) D_i$ \cite{PhysRevLett.107.070601, 2012arXiv1210.7710H, MPSmanifoldSchneider, Uschmajew2013133}.

\newpage

\hspace{-\parindent}\textbf{B. The SD as low bond dimension limit}
\vspace{0.3cm}

An interesting connection to HF can be made by considering an MPS where the $L$ orbitals are the HF orbitals. As each orbital occupation number is definite in an SD, an MPS with bond dimension 1 suffices to represent it. Conversely, if an MPS has bond dimension 1 and represents a state with definite particle number, each orbital has a definite occupation number. If this is not the case, two or more orbitals must be entangled (there is static correlation between them), and the bond dimension has to be larger than 1 to represent this. An MPS with bond dimension 1 and definite particle number can hence always be represented by an SD. An SD is the low bond dimension limit of an MPS, while a general full CI (FCI) solution requires an exponentially large bond dimension to be represented by an MPS \cite{Schollwock201196}.

The SD \textit{Ansatz} provides a single variational approximation to the ground state, which unfortunately fails to represent static correlation. On the contrary, the MPS \textit{Ansatz} allows to systematically improve the approximation to the ground state by increasing the bond dimension, up to the point where all static correlation is resolved \cite{WhiteQCDMRG, Chan2002, woutersJCP1, sharma:124121, chan:annurevphys}.

\vspace{0.7cm}
\hspace{-\parindent}\textbf{IV. THE DMRG EQUATIONS}
\vspace{0.3cm}

The TIVP leads to the DMRG equations \cite{ChanB805292C}. The canonical DMRG equations for site $i$ are retrieved when additional constraints are added to the Lagrangian to enforce that the site matrices to the left of site $i$ are left-normalized:
\begin{equation}
\forall j<i : \sum\limits_{n_j} (A^{n_j}[j])^{\dagger} A^{n_j}[j] = I_{D_j} , \label{left-can-indexPRB88}
\end{equation}
and that the site matrices to the right of site $i$ are right-normalized:
\begin{equation}
\forall j>i : \sum\limits_{n_j} A^{n_j}[j] (A^{n_j}[j])^{\dagger} = I_{D_{j-1}}.
\end{equation}
With $(A^{n_j}[j])_{\alpha \beta} = A[j]^{n_j \alpha \beta}$, the Lagrangian becomes
\begin{eqnarray}
& \mathcal{L} = \braket{\overline{\Phi} \mid \hat{H} \mid \Phi} - \lambda \left( \overline{A[i]_{n_i \alpha \beta}} A[i]^{n_i \alpha \beta}  - 1 \right) - \sum\limits_{j<i} \lambda[j]_{\gamma}^{~\beta} \left( \overline{A[j]_{n_j \alpha \beta}} A[j]^{n_j \alpha \gamma}  - \delta^{~\gamma}_{\beta}  \right) \nonumber \\
& - \sum\limits_{j>i} \lambda[j]_{\gamma}^{~\alpha} \left( \overline{A[j]_{n_j \alpha \beta}} A[j]^{n_j \gamma \beta}  - \delta^{~\gamma}_{\alpha}  \right).
\end{eqnarray}
Varying with respect to $\overline{A[i]_{n_i \alpha \beta}}$ gives the canonical one-site DMRG equations:
\begin{equation}
H_{\text{eff}}[i]^{n_i \alpha \beta}_{\quad \tilde{n}_i \tilde{\alpha} \tilde{\beta}} A[i]^{\tilde{n}_i \tilde{\alpha} \tilde{\beta}} = \lambda A[i]^{n_i \alpha \beta} \label{DMRGeffHamEqPRB88}
\end{equation}
in terms of the effective Hamiltonian \cite{ChanB805292C}. By bringing the MPS into canonical forms of which the left- and right-normalization conditions above are examples, the gauge freedom can be (partially) removed. For the left- and right-normalization conditions, the remaining gauge freedom is a unitary rotation ($G$ unitary). All gauge freedom can be removed by bringing the MPS into Vidal's canonical form \cite{PhysRevLett.91.147902}.

\newpage

The DMRG equations are site-based mean-field equations. There is self-consistency because the effective Hamiltonian in Eq. \eqref{DMRGeffHamEqPRB88}, which determines the site matrices of a particular site, depends on the site matrices of the other sites \cite{ChanB805292C, 2011arXiv1103.2155K, Schollwock201196}. In DMRG, the effective Hamiltonian hence plays the role of Fock operator \cite{ChanB805292C}. Since both of them act locally (respectively, on one site and one orbital), it might be worthwhile to explore Rayleigh-Schr\"odinger perturbation theory analogs for DMRG in the future, such as M\o{}ller-Plesset perturbation theory \cite{helgaker2, PhysRev.46.618, ChanB805292C}.

Note that in practice the two-site DMRG algorithm is used to optimize an MPS. The two-site algorithm is more robust against local minima, and when symmetry is imposed it provides a natural way to distribute the bond dimension $D$ over the symmetry sectors. After the two-site algorithm has converged, a few one-site DMRG sweeps allow to make the MPS fully self-consistent. This can be compared to HF, where the optimal SD is found by gradient methods \cite{RingAndSchuck} or by direct inversion of iterative subspaces \cite{Pulay1980393} for stability reasons. The DMRG and HF solutions satisfy respectively Eqs. \eqref{DMRGeffHamEqPRB88} and \eqref{HFequationsPRB88}, irrespective of the optimization scheme.

\vspace{0.7cm}
\hspace{-\parindent}\textbf{V. THE MPS TANGENT SPACE}
\vspace{0.3cm}

\hspace{-\parindent}\textbf{A. A redundant parameterization}
\vspace{0.3cm}

Flatten the site matrices $A[i]^{n_i}$ to a column $\mathbf{A}$ with entries $\left( A[i]^{n_i} \right)_{\alpha,\beta} = A^{i n_i \alpha \beta} = A^{\mu}$, and consider a small variation $A^{\mu} = A_0^{\mu} + B^{\mu}$. The wave function can then be expanded as
\begin{equation}
\ket{\Phi} = \ket{\Phi^0} + B^{\mu} \ket{\Phi^0_{\mu}} + \frac{1}{2} B^{\mu} B^{\nu} \ket{\Phi^0_{\mu \nu}} + ... \label{wavefunctionexpansionPRB88}
\end{equation}
with first-order tangent space $\ket{\Phi^0_{\mu}} = \partial_{\mu} \ket{\Phi^0} = \frac{\partial \ket{\Phi^0}}{\partial A^{\mu}}$ and second-order tangent space $\ket{\Phi^0_{\mu \nu}} = \partial_{\mu} \partial_{\nu} \ket{\Phi^0}$. Note that the summation convention was used. Each order of MPS tangent space contains all lower orders, e.g., $A_0^{\mu} \ket{\Phi^0_{\mu}} = L \ket{\Phi^0}$ and $A_0^{\mu} B^{\nu} \ket{\Phi^0_{\mu \nu}} = (L-1) B^{\mu} \ket{\Phi^0_{\mu}} $ \cite{2011arXiv1103.2155K, PhysRevLett.107.070601, 2012arXiv1210.7710H}.

The tangent vectors $\ket{\Phi^0_{\mu}}$ are redundant, as the MPS manifold has dimension $\text{dim} \mathbb{T}$, and there are $\text{dim} \mathbb{A}$ such vectors. The metric, or overlap matrix $S_{\mu\nu} = \braket{\overline{\Phi_{\mu}^0} \mid \Phi_{\nu}^0}$, is therefore not invertible. In Sec.~V C, $\text{dim} \mathbb{T}$ explicit linear combinations of the vectors $\ket{\Phi^0_{\mu}}$ are given, so that the overlap in this new basis is the unit matrix, and $\ket{\Phi^0}$ is orthogonal to this new basis. Remember that variations in the direction of $\ket{\Phi^0}$ only cause norm or phase changes of the \textit{Ansatz}, but do not change the physical state. This new basis is then a nonredundant parameterization of the MPS tangent space.

\vspace{0.7cm}
\hspace{-\parindent}\textbf{B. Hamiltonian sparsity}
\vspace{0.3cm}

The Hamiltonian \eqref{HamiltonianSecondQuantizationPRB88} is sparse, as it consists of a sum of one- and two-particle interactions. When it acts on a certain SD, the result lies in the space spanned by the given SD and its single and double OV excitations. This is immediataly clear by changing the single particle basis in Eq. \eqref{HamiltonianSecondQuantizationPRB88} from $\hat{b}_k^{\dagger}$ to the SD orbitals $\hat{a}^{\dagger}_j$.

A typical lattice Hamiltonian can be considered sparse too, as it consists of a sum of one- and two-site operators. It is sparse in site space instead of particle space. Let us focus on the one-dimensional Hubbard model \cite{Hubbard26111963}:
\begin{equation}
\hat{H} = - \sum\limits_{\sigma, i=1}^{L-1} \left( \hat{a}^{\dagger}_{i \sigma} \hat{a}_{i+1 \sigma} + \hat{a}^{\dagger}_{i+1 \sigma} \hat{a}_{i \sigma} \right) + U \sum\limits_{i=1}^{L} \hat{n}_{i \uparrow} \hat{n}_{i \downarrow}. \label{HubbardModelPRB88}
\end{equation}
Consider its action on an MPS. Let $\mu_i$ be a shorthand for $(n_i,\alpha,\beta)$, or $\mu$ restricted to site $i$. The Hamiltonian connects the MPS to a part of its double tangent space:
\begin{equation}
\hat{H} \ket{\Phi^0} \propto C^{\mu_i \nu_{i+1}} \ket{\Phi^0_{\mu_i \nu_{i+1}}}.
\end{equation}
It might hence be worthwhile to construct the site-space analog of the particle Fock space \cite{Dirac01031927, Fock2ndquant}. A new second quantization should be constructed, based on the MPS reference instead of the HF orbitals.

\vspace{0.7cm}
\hspace{-\parindent}\textbf{C. A nonredundant parameterization}
\vspace{0.3cm}

A nonredundant parameterization of the MPS tangent space was first presented by Dorando \textit{et al.} \cite{dorando:184111} in DMRG projector terminology. Haegeman \textit{et al.} \cite{PhysRevLett.107.070601} provided a construction in the language of the MPS wave function and the corresponding manifold. To present the relationship between the two, here we describe the tangent space construction in projector terms, but by using the explicit MPS representation of the projectors.

Consider an MPS where all left-renormalized basis states at boundary $i-1$,
\begin{equation}
\ket{L_{\alpha}^{i-1}} = \sum\limits_{\{n_j:j<i\}} \left[ A^{n_1}[1] ... A^{n_{i-1}}[i-1] \right]_{\alpha} \ket{n_1 ... n_{i-1}} ,
\end{equation}
are orthonormal and all right-renormalized basis states at boundary $i$,
\begin{equation}
\ket{R_{\beta}^{i}} = \sum\limits_{\{n_j:j>i\}} \left[ A^{n_{i+1}}[i+1] ... A^{n_L}[L] \right]_{\beta} \ket{n_{i+1} ... n_L} ,
\end{equation}
are orthonormal. In the DMRG algorithm, a renormalization transformation is constructed to reduce the direct product of $\ket{L_{\alpha}^{i-1}}$ (size $D_{i-1}$) and $\ket{n_i}$ (size $d$) to a new left-renormalized basis at boundary $i$ (size $D_i \leq dD_{i-1}$). This renormalization transformation is a projection, represented by the site matrices of site $i$:
\begin{equation}
\sum\limits_{\alpha n_i} \left( A[i]^{n_i} \right)_{\alpha ,\beta} \ket{L_{\alpha}^{i-1}} \ket{n_i}.
\end{equation}
The projection onto the $dD_{i-1}-D_i$ discarded states from the direct product space, defines the nonredundant tangent space. We now explain the explicit construction of the nonredundant tangent space as provided by Dorando \textit{et al.} \cite{dorando:184111} in MPS terminology. Consider the QR-decomposition of the projector:
\begin{equation}
\left( A[i]^{n_i} \right)_{\alpha, \beta} = A[i]_{(\alpha n_i),\beta} = \sum\limits_{\gamma} Q[i]_{(\alpha n_i),\gamma} R_{\gamma,\beta}. \label{QRdecompeqPRB88}
\end{equation}
$Q[i]$ contains $D_i$ orthonormal columns of size $d D_{i-1}$. Its left null space is spanned by $d D_{i-1} - D_i$ vectors. This allows to construct the $dD_{i-1} \times (d D_{i-1} - D_i)$ matrix $\tilde{Q}[i]$, so that $\left[ Q[i] \tilde{Q}[i] \right]$ is unitary. A part of the nonredundant tangent space can then be parameterized by the matrix $x[i]$ with dimensions $\left( dD_{i-1}-D_i \right) \times D_i$:
\begin{equation}
\sum\limits_{\alpha \beta n_i} \left( \tilde{Q}[i]^{n_i} x[i] \right)_{\alpha,\beta} \ket{L_{\alpha}^{i-1}} \ket{n_i} \ket{R_{\beta}^{i}}.
\end{equation}
If the renormalized basis states $\ket{L_{\alpha}^{i-1}}$, $\ket{R_{\beta}^{i}}$ are not orthonormal, their overlap has to be taken into account. It was Haegeman \textit{et al.} \cite{PhysRevLett.107.070601} who first presented the parameterization in that case:
\begin{equation}
\sum\limits_{\alpha \beta n_i} \left( l[i-1]^{-\frac{1}{2}} \tilde{Q}[i]^{n_i} x[i] r[i]^{-\frac{1}{2}} \right)_{\alpha, \beta} \ket{L_{\alpha}^{i-1}} \ket{n_i} \ket{R_{\beta}^{i}} \label{JuthoTangentSpacePRB88}
\end{equation}
with $l[i-1]$ the density matrix of the left renormalized states $\ket{L_{\alpha}^{i-1}}$ and $r[i]$ the density matrix of the right renormalized states $\ket{R_{\beta}^{i}}$. The QR-decomposition of Eq. \eqref{QRdecompeqPRB88} is now performed on $l[i-1]^{\frac{1}{2}} A[i]^{n_i}$ instead of on $A[i]^{n_i}$. The complete nonredundant tangent space is formed by doing this construction for the projector on each site. Combine all matrices $x[i]$ to a column $\mathbf{x}$ of length $\text{dim} \mathbb{T}$. By writing the construction in Eq. \eqref{JuthoTangentSpacePRB88} as $B^{\mu}(\mathbf{x}) \ket{\Phi^0_{\mu}}$, with
\begin{equation}
B^{n_i}(\mathbf{x})[i] = l[i-1]^{-\frac{1}{2}} \tilde{Q}^{n_i}[i] x[i] r[i]^{-\frac{1}{2}}, \label{BmxExplicitPRB88}
\end{equation}
one possibility for a nonredundant tangent space basis of dimension $\text{dim} \mathbb{T}$ is immediately obtained:
\begin{equation}
\ket{\Phi_k^T} = \frac{\partial}{\partial x^k} B^{\mu}(\mathbf{x}) \ket{\Phi^0_{\mu}}.
\end{equation}
Note that this provides a construction of $\ket{\Phi^T_k}$ as a linear combination of $\ket{\Phi^0_{\mu}}$. Any tangent vector can be constructed by taking a complex linear combination of these $\text{dim} \mathbb{T}$ vectors: $x^k \ket{\Phi^T_k} = B^{\mu}(\mathbf{x}) \ket{\Phi^0_{\mu}}$. Because of the construction of $\tilde{Q}^{n_i}[i]$, these vectors are orthogonal to $\ket{\Phi^0}$: $\braket{\overline{\Phi^0} \mid \Phi^0_{\mu}  } B^{\mu}(\mathbf{x}) = 0$. The metric of the parameterization in Eq. \eqref{BmxExplicitPRB88} is the unit matrix: $\overline{B^{\mu}(\mathbf{x})} S_{\mu \nu} B^{\nu}(\mathbf{y}) = \mathbf{x}^{\dagger}\mathbf{y}$ \cite{PhysRevLett.107.070601, 2012arXiv1210.7710H}. Analogous results have been obtained in a different context \cite{MPSmanifoldSchneider, Uschmajew2013133}.

For an SD written as an MPS ($D=1$ and $d=2$), the nonredundant tangent space vectors correspond to the addition (removal) of an electron to (from) the system.

\vspace{0.7cm}
\hspace{-\parindent}\textbf{VI. THE THOULESS THEOREM FOR MPS}
\vspace{0.3cm}

The operators $\hat{B}^{\dagger}_{vo}$ link an SD $\ket{\Psi^0}$ to its nonredundant tangent space $\hat{B}^{\dagger}_{vo} \ket{\Psi^0}$. Exponentiation of these operators led to the Thouless theorem. Here, we present the MPS counterpart.

\newpage

\hspace{-\parindent}\textbf{A. Proposal}
\vspace{0.3cm}

For the sake of simplicity, we use a part of the gauge freedom to work with a left-canonical MPS. The left-normalization condition in Eq. \eqref{left-can-indexPRB88} then holds for all sites. This implies $\forall i: l[i] = I_{D_i}$. Consider the following matrix notation for site matrices: $C[i]$ has entries $(C[i])_{(\alpha n_i),\beta}$, i.e., the row-index of the matrices $C[i]$ contains the physical index $n_i$. The left-normalization condition then becomes
\begin{equation}
A[i]^{\dagger} A[i] = I_{D_i}. \label{left-can-eqPRB88}
\end{equation}
Because of the construction of $\tilde{Q}[i]$, the site matrices $B(\mathbf{x})[i]$ are left-orthogonal to the site matrices $A[i]$:
\begin{equation}
B(\mathbf{x})[i]^{\dagger} A[i] = 0. \label{B-left-can-eqPRB88}
\end{equation}
This allows to propose the MPS counterpart of the Thouless theorem:
\begin{equation}
A(\mathbf{x},\overline{\mathbf{x}})[i] = \exp{\left( B(\mathbf{x})[i] A_0[i]^{\dagger} - A_0[i] B(\mathbf{x})[i]^{\dagger} \right)} A_0[i] , \label{ThoulessMPSPRB88}
\end{equation}
where now $A_0[i]$ is used for $A[i]$ to clearly mark the difference with $A(\mathbf{x},\overline{\mathbf{x}})[i]$. The matrix in the exponential is anti-Hermitian, and the transformation in Eq. \eqref{ThoulessMPSPRB88} is therefore unitary. As the $A_0[i]$ site matrices were left-normalized, the $A(\mathbf{x},\overline{\mathbf{x}})[i]$ site matrices are also left-normalized. An MPS built with the $A(\mathbf{x},\overline{\mathbf{x}})[i]$ site matrices,
\begin{equation}
\ket{\Phi(\mathbf{x},\overline{\mathbf{x}})} = \sum\limits_{\left\{ n_i \right\}} A(\mathbf{x},\overline{\mathbf{x}})[1]^{n_1} ... A(\mathbf{x},\overline{\mathbf{x}})[L]^{n_L} \ket{n_1 n_2 ... n_L} , \label{Phi_x_thouless_to_mpsPRB88}
\end{equation}
is hence still left-canonical and therefore normalized. For $\mathbf{x}=\mathbf{0}$, $\ket{\Phi(\mathbf{x},\overline{\mathbf{x}})} = \ket{\Phi^0}$. The tangent space of this MPS parameterization is familiar too:
\begin{equation}
\left. \frac{\partial}{\partial x^k} \ket{\Phi(\mathbf{x},\overline{\mathbf{x}})} \right|_{\mathbf{x}=\mathbf{0}} = \ket{\Phi^T_k} ,
\end{equation}
which can be easily checked by using Eqs. \eqref{left-can-eqPRB88} and \eqref{B-left-can-eqPRB88}. $\ket{\Phi(\mathbf{x},\overline{\mathbf{x}})}$ is therefore an explicit nonredundant parameterization of the MPS manifold in the neighbourhood of $\ket{\Phi^0}$.

\vspace{0.7cm}
\hspace{-\parindent}\textbf{B. Global validity}
\vspace{0.3cm}

Here we show that Eq. \eqref{Phi_x_thouless_to_mpsPRB88} is a global parameterization of the MPS manifold, or that any MPS with bond dimensions $D_i$ can be generated from $\ket{\Phi^0}$ (which has the same bond dimensions). This implies that we can optimize over the parameters $\mathbf{x}$ instead of over $\mathbf{A}$ to find an energy minimum.

For a specific site index $i$, the parameterization $A(\mathbf{x},\overline{\mathbf{x}})[i]$ of Eq. \eqref{ThoulessMPSPRB88} is a Grassmann manifold with matrix dimensions $dD_{i-1} \times D_i$. Define $\mathbf{y}$ by $x[i] = y[i]r[i]^{\frac{1}{2}}$ to obtain
\begin{equation}
\tilde{A}(\mathbf{y},\overline{\mathbf{y}}) = A(\mathbf{x},\overline{\mathbf{x}})[i] = \exp{\left( \tilde{Q}[i] y Q[i]^{\dagger} - Q[i] y^{\dagger} \tilde{Q}[i]^{\dagger} \right)} Q[i]. \label{GrassmannFormThoulessMPSPRB88}
\end{equation}
Note the close analogy to Eq. \eqref{counterpartThoulessPRB88}. We show in Appendix that Eq. \eqref{GrassmannFormThoulessMPSPRB88} represents a Grassmann manifold. Note that we assume that the density matrix $r[i]$ is nonsingular (i.e., $r[i]^{-\frac{1}{2}}$ exists) for the construction of the nonredundant tangent space in Eq. \eqref{BmxExplicitPRB88}. For a left-canonical MPS, the Schmidt values are the positive square roots of the eigenvalues of $r[i]$. The condition of nonsingular density matrices $r[i]$ is therefore equal to having all Schmidt values of $\ket{\Phi^0}$ nonzero.  This is a condition for the global validity of Thouless's theorem for MPS.

Now give a normalized MPS $\ket{\Phi^1}$, with the only restriction that it has the same bond dimensions as $\ket{\Phi^0}$. We will prove by construction that
\begin{equation}
\exists \mathbf{x} : \left|\braket{\overline{\Phi^1} \mid \Phi(\mathbf{x},\overline{\mathbf{x}})}\right| = 1.
\end{equation}
(1) Set $i = 1$.
(2) Use a part of the gauge freedom at boundary $i$ to bring the site matrices $A_1[i]$ of $\ket{\Phi^1}$ in left-normalized form: $A^L_1[i]$.
(3) Find $x_1[i]$ so that the columns of $A(\mathbf{x_1},\overline{\mathbf{x_1}})[i]$ and the columns of $A^L_1[i]$ span the same space, which is always possible because $A(\mathbf{x},\overline{\mathbf{x}})[i]$ is a Grassmann manifold.
(4) Use the remaining gauge freedom at boundary $i$, i.e., a unitary transformation $U_{D_i}$, to enforce $A(\mathbf{x_1},\overline{\mathbf{x_1}})[i] = A^{\text{exact}}_1[i] = A^L_1[i] U_{D_i}$.
(5) If $i<L$, set $i = i+1$ and go to 2.

When the construction is finished, all parameters of $\mathbf{x_1}$ are assigned, and the gauge freedom in $\ket{\Phi^1}$ was used to write $\ket{\Phi^1}$ exactly as $\ket{\Phi(\mathbf{x_1},\overline{\mathbf{x_1}})}$, i.e., $\forall i: A^{\text{exact}}_1[i] = A(\mathbf{x_1},\overline{\mathbf{x_1}})[i]$. See Refs. \cite{MPSmanifoldSchneider} and \cite{Uschmajew2013133} on the diffeomorphism between a finite chain MPS manifold and a product manifold of Grassmann manifolds. This concludes the proof that Eq. \eqref{Phi_x_thouless_to_mpsPRB88} can represent any MPS with the same bond dimensions, as long as $\ket{\Phi^0}$ does not have any vanishing Schmidt values. Note that the theorem guarantees a solution $\mathbf{x_1}$, but does not guarantee that this solution is unique, in analogy with the discussion in Sec.~II D.

\vspace{0.7cm}
\hspace{-\parindent}\textbf{C. The double tangent space}
\vspace{0.3cm}

To get a better understanding of the MPS double tangent space, consider the second order term of $A(\mathbf{x},\overline{\mathbf{x}})[i]$:
\begin{eqnarray}
& & A(\mathbf{x},\overline{\mathbf{x}})[i] - A_0[i] - B(\mathbf{x})[i] \nonumber\\
& = & - \frac{1}{2} A_0[i] B(\mathbf{x})[i]^{\dagger}B(\mathbf{x})[i] + \mathcal{O}(x^3) \nonumber\\
& = & - \frac{1}{2} A_0[i] r[i]^{-\frac{1}{2}} x[i]^{\dagger} x[i] r[i]^{-\frac{1}{2}} + \mathcal{O}(x^3).
\end{eqnarray}
The expansion of $\ket{\Phi(\mathbf{x},\overline{\mathbf{x}})}$ up to second order then consists of the following.
(1) The MPS reference $\ket{\Phi(\mathbf{0},\mathbf{0})} = \ket{\Phi^0}$.
(2) The tangent space $\left. \frac{\partial}{\partial x^k} \ket{\Phi(\mathbf{x},\overline{\mathbf{x}})} \right|_{\mathbf{x}=\mathbf{0}} = \frac{\partial}{\partial x^k} \ket{\Phi(\mathbf{0},\mathbf{0})} = \ket{\Phi^T_k}$. Note that $\frac{\partial}{\partial \overline{x_k}} \ket{\Phi(\mathbf{0},\mathbf{0})} = 0$. The tangent space consists of all possible connections between the unused basis states from $\ket{L_{\alpha}^{i-1}} \otimes \ket{n_i}$ and $\ket{R_{\beta}^{i}}$.
(3) The nonlocal part of the double tangent space $\frac{\partial^2}{\partial x^k \partial x^l} \ket{\Phi(\mathbf{0},\mathbf{0})} = \ket{\Phi^{T2}_{kl}}$. Note that this term is only nonzero if $x^k$ and $x^l$ correspond to different sites of the MPS chain. This part corresponds to two single excitations on different sites. Also note that $\frac{\partial^2}{\partial \overline{x_k} \partial \overline{x_l}} \ket{\Phi(\mathbf{0},\mathbf{0})} = 0$.
(4) The local part of the double tangent space $\frac{\partial^2}{\partial \overline{x_k} \partial x^l} \ket{\Phi(\mathbf{0},\mathbf{0})} = \ket{\Phi^{T2}_{\overline{k}l}}$. Note that this term is only nonzero if $\overline{x_k}$ and $x^l$ belong to the same site, and correspond to the same row index in the matrix notation $x[i]$. This part consists of all possible connections between the renormalized basis states $\ket{L_{\alpha}^{i}}$ (from $\ket{L_{\alpha}^{i-1}} \otimes \ket{n_i}$) and $\ket{R_{\beta}^{i}}$.

These states are not all mutually orthogonal. Note that the local part of the double tangent space arises because we have considered a unitary variant of the Thouless theorem for MPS. The original (nonunitary) Thouless parameterization for HF depends only on the complex parameters, and not on their complex conjugates.

If two excitation operators in HF try to annihilate an occupied single particle twice, the state is destroyed. The space of double OV excitations therefore consists of the replacement of two different occupied single particles by two different virtual single particles.

The local part of the double tangent space of an MPS can be written as $B^{\mu} \ket{\Phi^0_{\mu}}$, which lies entirely in the space spanned by the MPS reference $\ket{\Phi^0}$ and the nonredundant tangent space vectors $\ket{\Phi^T_k}$. Together with the other two arguments above, this provides a justification to discard this part of the double tangent space without any loss in variational freedom, and to consider only two single excitations acting on different sites, for the double tangent space.

\vspace{0.7cm}
\hspace{-\parindent}\textbf{D. Excitation operators}
\vspace{0.3cm}

The excitation operators for an MPS can be read from the Thouless theorem:
\begin{equation}
\ket{\Phi^T_k} = \hat{B}^{\dagger}_k \ket{\Phi^0} =\left. \frac{\partial}{\partial x^k} \ket{\Phi(\mathbf{x},\overline{\mathbf{x}})} \right|_{\mathbf{x}=\mathbf{0}}. \label{ExcitationOperatorEqPRB88}
\end{equation}
See, e.g., Sec.~IV in Rowe \textit{et al.} \cite{PhysRevA.22.2362} for a discussion on the relationship between the linearized time-dependent variational principle on a general manifold, and the EOM approach to the RPA equations. The operators $\hat{B}^{\dagger}_k$ are obtained by going from the manifold representation based on the virtual space in Eq. \eqref{ThoulessMPSPRB88}, to a representation based on the physical Hilbert space $\ket{\Phi(\mathbf{x},\overline{\mathbf{x}})} = \exp{\left( x^k \hat{B}_k^{\dagger} - \overline{x_k} \hat{B}^k \right)} \ket{\Phi^0}$. When only the first-order tangent space needs to match, $\hat{B}_k^{\dagger} =  \ket{\Phi^T_k}\bra{\overline{\Phi^0}}$ can be used. It will be a challenge to find the $\hat{B}_k^{\dagger}$'s to match the higher order tangent spaces too. Finding an answer to this problem, is closely related to finding a site-space analog of the particle Fock space, based on the MPS reference. From Eq. \eqref{ThoulessMPSPRB88}, it can be understood that this excitation operator projects out the site matrices $A_0[i(k)]$ and replaces them with the tangent space site matrices $\left. \partial_{x_k} B(\mathbf{x})[i(k)] \right|_{\mathbf{x}=\mathbf{0}}$. It adds a single excitation to the vacuum $\ket{\Phi^0}$. In the chosen gauge, a single MPS excitation is localized to one site, just like a single OV excitation of an SD is localized to one orbital. From Eq. \eqref{ThoulessMPSPRB88}, it can also be understood that a deexcitation operator projects out the tangent space site matrices $\left. \partial_{x_k} B(\mathbf{x})[i(k)] \right|_{\mathbf{x}=\mathbf{0}}$ and replaces them with the site matrices $A_0[i(k)]$. Remember that the tangent space metric is the unit matrix for the chosen parameterization, and that the deexcitation projections are hence not only orthogonal to the MPS reference (they destroy the vacuum $\ket{\Phi^0}$) but also orthogonal to other tangent space site matrices:
\begin{eqnarray}
\hat{B}_l \ket{\Phi^0} & = & 0, \\
\hat{B}_l \ket{\Phi^T_k} & = & \delta_{l,k} \ket{\Phi^0}.
\end{eqnarray}
The deexcitation operators of the ket vectors are the excitation operators of the bra vectors:
\begin{equation}
\bra{\overline{\Phi^0}} \hat{B}_l = \bra{\overline{\Phi^T_l}}.
\end{equation}
Consider the commutators $\left[ \hat{B}^{\dagger}_l, \hat{B}^{\dagger}_k \right]$ and $\left[ \hat{B}_l, \hat{B}^{\dagger}_k \right]$. Their general expressions are far from trivial, only their expectation value with respect to the vacuum $\ket{\Phi^0}$ is clear:
\begin{eqnarray}
\braket{\overline{\Phi^0} \mid \left[ \hat{B}^{\dagger}_l, \hat{B}^{\dagger}_k \right] \mid \Phi^0 } & = & 0 , \label{quasi-boson1PRB88}\\
\braket{\overline{\Phi^0} \mid \left[ \hat{B}_l, \hat{B}^{\dagger}_k \right] \mid \Phi^0 } & = & \delta_{k,l}. \label{quasi-boson2PRB88}
\end{eqnarray}
A bosonic algebra for the excitation operators is hence only retrieved when expectation values with respect to the vacuum are taken. The operators $\hat{B}^{\dagger}_k$ are called quasiboson operators.

\vspace{0.7cm}
\hspace{-\parindent}\textbf{VII. OPTIMAL TIME EVOLUTION FOR MPS}
\vspace{0.3cm}

The optimal time evolution of an MPS, which stays within the MPS \textit{Ansatz} space, was derived by means of the TDVP in Refs. \cite{PhysRevLett.107.070601} and \cite{2011arXiv1103.2155K}. Now that we have established the Thouless theorem for MPS, we can rephrase the result as
\begin{equation}
i \hbar \dot{x}^k (\mathbf{x}=\mathbf{0}) = \braket{\overline{\Phi^0} \mid \hat{B}^k \hat{H} \mid \Phi^0}. \label{TDVPquantumLatticesPRB88}
\end{equation}
Also in this case, Eq. \eqref{TDVPquantumLatticesPRB88} can be obtained by inserting $\ket{\Phi(\mathbf{x},\overline{\mathbf{x}})}$ in the time-dependent Schr\"odinger equation, and by projecting the time-dependent equation onto $\hat{B}^{\dagger}_k \ket{\Phi(\mathbf{x},\overline{\mathbf{x}})}$:
\begin{equation}
i \hbar \bra{\overline{\Phi}} \hat{B}^{k} \left( \dot{x}^{l} \frac{\partial}{\partial x^l} + \dot{\overline{x}}_{l} \frac{\partial}{\partial \overline{x_l}} \right) \ket{\Phi} = \bra{\overline{\Phi}} \hat{B}^{k} \left( \hat{H} - E_{\text{MPS}} \right) \ket{\Phi} . \label{toLinearize2PRB88}
\end{equation}
Evaluation for $\mathbf{x}=\mathbf{0}$ yields Eq. \eqref{TDVPquantumLatticesPRB88}. This form of time propagation automatically stays within the MPS \textit{Ansatz} space. No Hamiltonian decompositions or bond dimension truncations are necessary \cite{2011arXiv1103.2155K, PhysRevLett.107.070601}.

\vspace{0.7cm}
\hspace{-\parindent}\textbf{VIII. RPA FOR MPS}
\vspace{0.3cm}

\hspace{-\parindent}\textbf{A. In a redundant parameterization}
\vspace{0.3cm}

One way to obtain the RPA equations for MPS, is to consider the linearized time-dependent equations in the vicinity of a variational minimum, and to project them onto the tangent space of the manifold \cite{2011arXiv1103.2155K, JuthoThesis}. Consider a small time-dependent step around the minimum $A^{\mu}(t) = A_0^{\mu} + B^{\mu}(t)$. The time-dependent equation, its projection onto the tangent space, and its first-order terms become
\begin{eqnarray}
i \hbar \ket{\Phi_{\nu}(\mathbf{A})} \dot{A}^{\nu} & = & \left( \hat{H} - E_{\text{MPS}} \right) \ket{\Phi(\mathbf{A})} ,\\
i \hbar \braket{\overline{\Phi_{\mu}(\mathbf{A})} \mid \Phi_{\nu}(\mathbf{A})} \dot{A}^{\nu} & = & \braket{\overline{\Phi_{\mu}(\mathbf{A})} \mid \hat{H} - E_{\text{MPS}} \mid \Phi(\mathbf{A})} ,\\
i \hbar \braket{\overline{\Phi_{\mu}(\mathbf{A}_0)} \mid \Phi_{\nu}(\mathbf{A}_0)} \dot{B}^{\nu} & = & \braket{\overline{\Phi_{\mu}(\mathbf{A}_0)} \mid \hat{H} - E_{\text{MPS}} \mid \Phi_{\nu}(\mathbf{A}_0)} B^{\nu} \nonumber \\
& + & \braket{\overline{\Phi_{\mu\nu}(\mathbf{A}_0)} \mid \hat{H} - E_{\text{MPS}} \mid \Phi(\mathbf{A}_0) } \overline{B}^{\nu},
\end{eqnarray}
with $E_{\text{MPS}} = \braket{\overline{\Phi^0} \mid \hat{H} \mid \Phi^0}$. By taking a harmonic \textit{Ansatz} for the perturbation $\mathbf{B} = \mathbf{Y} e^{-i \omega t} + \overline{\mathbf{Z}} e^{i \omega t}$, the RPA equations are found:
\begin{equation}
\hbar \omega \left[ \begin{array}{cc} S & 0 \\ 0 & -\overline{S} \end{array} \right] \left( \begin{array}{c} \mathbf{Y} \\ \mathbf{Z} \end{array} \right) = \left[ \begin{array}{cc} H & W \\ \overline{W} & \overline{H} \end{array} \right] \left( \begin{array}{c} \mathbf{Y} \\ \mathbf{Z} \end{array} \right) \label{RPAeqWithOverlapPRB88}
\end{equation}
with $S_{\mu\nu} = \braket{\overline{\Phi_{\mu}^0} \mid \Phi_{\nu}^0}$, $H_{\mu\nu} = \braket{\overline{\Phi_{\mu}^0} \mid \hat{H} - E_{\text{MPS}} \mid \Phi_{\nu}^0}$, and $W_{\mu\nu} = \braket{\overline{\Phi_{\mu\nu}^0} \mid \hat{H} - E_{\text{MPS}} \mid \Phi^0}$. Note that $\| \left( \hat{H} - E_{\text{MPS}} \right) \ket{\Phi^0} \|_2$ becomes smaller when $\ket{\Phi^0}$ becomes a better approximation for the exact ground state. $\| W \|_2$ is hence a measure for the accuracy of the MPS approximation to the exact ground state \cite{JuthoThesis}. A specific eigenvector of Eq. \eqref{RPAeqWithOverlapPRB88} can be obtained in $\mathcal{O}(L D^3)$ complexity \cite{2011arXiv1103.2155K, dorando:184111}.

\vspace{0.7cm}
\hspace{-\parindent}\textbf{B. In a nonredundant parameterization}
\vspace{0.3cm}

By changing the basis from $\ket{\Phi_{\mu}^0}$ to $\ket{\Phi^T_k} = Z^{\mu}_k \ket{\Phi_{\mu}^0}$, with $\frac{\partial}{\partial x^k} B^{\mu}(\mathbf{x}) = Z^{\mu}_k$, the overlap matrix $S$ becomes the unit matrix $I$: $\overline{Z}^{\mu}_k S_{\mu \nu} Z^{\nu}_l = \delta_{kl}$. Analogously, the Hermitian matrix $A$ and the complex symmetric matrix B, both of dimension $\text{dim} \mathbb{T} \times \text{dim} \mathbb{T}$, can be defined as resp. $A_{kl} = \overline{Z}^{\mu}_k H_{\mu \nu} Z^{\nu}_l$ and $B_{kl} = \overline{Z}^{\mu}_k W_{\mu \nu} Z^{\nu}_l$. The RPA equations become
\begin{equation}
\hbar \omega \left[ \begin{array}{cc} I & 0 \\ 0 & -I \end{array} \right] \left( \begin{array}{c} \mathbf{y} \\ \mathbf{z} \end{array} \right) = \left[ \begin{array}{cc} A & B \\ \overline{B} & \overline{A} \end{array} \right] \left( \begin{array}{c} \mathbf{y} \\ \mathbf{z} \end{array} \right) , \label{RPAeqUNITMPSPRB88}
\end{equation}
where $\mathbf{y}$ and $\mathbf{z}$ are now coefficients with respect to $\ket{\Phi^T_k}$. The same result can be obtained by linearizing Eq. \eqref{toLinearize2PRB88}, just as for HF.

The $A$ and $B$ matrices can be constructed explicitly. If only a few excitation energies are desired, it is better to resort to a sweep algorithm, which can be implemented in an existing DMRG code \cite{dorando:184111}. Implementation details of this sweep algorithm will be presented elsewhere \cite{NaokiLRTpaper}.

\vspace{0.7cm}
\hspace{-\parindent}\textbf{C. EOM derivation}
\vspace{0.3cm}

The excitation operators discussed in Sec.~VI D allow for a rederivation of the RPA equations for MPS by means of the EOM. An exact bosonic algebra can be set up by adding correction terms to operators defined in Sec.~VI D, so that $\left[ \hat{B}^{\dagger}_l, \hat{B}^{\dagger}_k \right] = 0$ and $\left[ \hat{B}^l, \hat{B}^{\dagger}_k \right] = \delta_{l,k}$. A justification is given by Eqs. \eqref{quasi-boson1PRB88} and \eqref{quasi-boson2PRB88}. The Hamiltonian can be expanded in these bosonic operators, and RPA coincides with truncating the expansion after second order. Expressions for the RPA correlation energy and wave function follow, just as for HF:
\begin{eqnarray}
E_{\text{cRPA}} & = & - \frac{\text{tr}A}{2} + \sum\limits_{\omega_n>0} \frac{\hbar \omega_n}{2} = - \sum\limits_{n} \hbar \omega_n \sum\limits_{k} \mid z_{n}^{k} \mid^2 , \label{RPAcorrEnergyEq2PRB88}\quad\\
\ket{\text{RPA}} & \propto & e^{\frac{1}{2} \left( \overline{zy^{-1}} \right)^{k;l} \hat{B}^{\dagger}_{k} \hat{B}^{\dagger}_{l}} \ket{\Phi^0}. \label{RPAwavefunction2PRB88}
\end{eqnarray}

\vspace{0.7cm}
\hspace{-\parindent}\textbf{IX. POST-DMRG METHODS}
\vspace{0.3cm}

\hspace{-\parindent}\textbf{A. TDA and Brillouin's theorem}
\vspace{0.3cm}

A preferred tangent basis can be found by searching the eigenstates of the Hamiltonian in the basis $\ket{\Phi^T_k}$. This corresponds to diagonalizing the $A$ matrix of the RPA Eq. \eqref{RPAeqUNITMPSPRB88}. As $\ket{\Phi^0} = \frac{1}{L} A^{\mu}_0 \ket{\Phi^0_{\mu}}$ gave the lowest energy solution for the \textit{Ansatz} $B^{\mu} \ket{\Phi^0_{\mu}}$ and $\ket{\Phi^T_k} \perp \ket{\Phi^0}$, approximations for excited states are found this way. This is the MPS analog of TDA \cite{helgaker2, Tamm, PhysRev.78.382, RingAndSchuck}. For a variational minimum,
\begin{equation}
0 = \frac{\partial E}{ \partial \overline{A^{\mu}_0}} = \frac{\braket{\overline{\Phi_{\mu}^0} \mid \hat{H} \mid \Phi^0}}{\braket{\overline{\Phi^0} \mid \Phi^0}} - \frac{\braket{\overline{\Phi^0} \mid \hat{H} \mid \Phi^0}}{\braket{\overline{\Phi^0} \mid \Phi^0}^2} \braket{\overline{\Phi_{\mu}^0} \mid \Phi^0}.
\end{equation}
If the wave function $\ket{\Phi^0}$ is normalized and only norm- and phase-conserving changes $\hat{B}_k^{\dagger} \ket{\Phi^0} \perp \ket{\Phi^0}$ are considered \cite{PhysRevLett.107.070601, dorando:184111},
\begin{equation}
\braket{\overline{\Phi^T_k} \mid \hat{H} \mid \Phi^0} = \braket{\overline{\Phi^0} \mid \hat{B}^k \hat{H} \mid \Phi^0} = 0.\label{BrillouinEqPRB88}
\end{equation}
This is the MPS analog of Brillouin's theorem \cite{helgaker2, Brillouin}. For MPS, excited momentum eigenstates of translationally invariant systems have previously been approximated in the nonredundant tangent space basis \cite{PhysRevB.85.035130, PhysRevB.85.100408}.

\vspace{0.7cm}
\hspace{-\parindent}\textbf{B. CC and CI}
\vspace{0.3cm}

The Thouless theorem for MPS and Eq. \eqref{RPAwavefunction2PRB88} suggest CC and CI \textit{Ans\"atze} on top of an MPS reference. Consider, for example, the single and double excitations:
\begin{eqnarray}
\ket{\text{CCSD}} & \propto & e^{y^k \hat{B}^{\dagger}_k + \frac{1}{2} z^{kl} \hat{B}^{\dagger}_k \hat{B}^{\dagger}_l} \ket{\Phi^0} ,\\
\ket{\text{CISD}} & \propto & \left( x + y^k \hat{B}^{\dagger}_k + \frac{1}{2} z^{kl} \hat{B}^{\dagger}_k \hat{B}^{\dagger}_l \right) \ket{\Phi^0}.
\end{eqnarray}
With the exposition in Secs. V, VI C, and VI D, we can also propose the following CCSD and CISD \textit{Ans\"atze}:
\begin{eqnarray}
\ket{\text{CCSD}} & \propto & e^{ C^{\mu\nu} \partial_{\mu} \partial_{\nu} } \ket{\Phi^0} ,\\
\ket{\text{CISD}} & \propto & C^{\mu\nu} \ket{\Phi^0_{\mu\nu}} \label{CISDansatzPRB88}
\end{eqnarray}
with $\mathcal{O}\left[\frac{1}{2} (L d D^2)^2\right]$ parameters in the symmetric $C$ tensor. Note that working in the $L$th order tangent space corresponds to the FCI \textit{Ansatz}. For DMRG (HF), the CCSD and CISD \textit{Ans\"atze} can be considered as a way to improve the correlation between two sites (electrons) embedded in an approximate environment given by the zeroth-order MPS (SD). Since the double tangent space can connect sites that are far apart, this enables the CCSD and CISD expressions to directly build in long-range entanglement. If the zeroth-order description fails (static correlation for HF, critical system for DMRG), these \textit{Ans\"atze} will fail too. Also for MPS, CISD is not size-consistent if there are more than two sites in the compound system, whereas CCSD is always size-consistent because of the exponential \textit{Ansatz}.

\vspace{0.7cm}
\hspace{-\parindent}\textbf{X. SYMMETRY-ADAPTED CALCULATIONS}
\vspace{0.3cm}

For large calculations, symmetry-adapted MPS \textit{Ans\"atze} are often used. They allow to search for eigenstates within a specific symmetry sector of the total Hilbert space, and lead to computational advantages in memory and time. An MPS \textit{Ansatz} without symmetry adaptation can yield an approximate eigenstate that breaks the symmetry. Its tangent space then also contains symmetry-broken vectors. RPA-MPS breaks down if a symmetry multiplet of a non-Abelian group is incomplete at a certain MPS boundary. Therefore we use symmetry-adapted MPS \textit{Ans\"atze} for the applications.

\vspace{0.7cm}
\hspace{-\parindent}\textbf{A. Tangent space without symmetry adaptation}
\vspace{0.3cm}

First consider an MPS \textit{Ansatz} without symmetry adaptation. A basis for its nonredundant tangent space, which is at the same time a basis of symmetry eigenvectors, can only be constructed when the MPS reference is an eigenvector of those symmetries. If the MPS reference is a symmetry eigenvector, its tangent space (in general) also contains symmetry eigenvectors that belong to a different irreducible representation (irrep). We provide a simple counting argument.

Consider an MPS with length $L$ even, then the center virtual dimension has to be $d^{\frac{L}{2}}$ to represent a general FCI state \cite{Schollwock201196}. The number of states in its nonredundant tangent space is then
\begin{equation}
\text{dim} \mathbb{T} = \sum\limits_{k=1}^{\frac{L}{2}} \left( d d^k - d^{k-1} \right) d^{k-1} = d^L - 1,
\end{equation}
i.e., the rest of the Hilbert space. Note that these $d^L - 1$ nonredundant tangent space vectors can only be constructed if all Schmidt values are greater than zero \cite{PhysRevLett.107.070601, 2012arXiv1210.7710H}. Suppose that this is the case. The MPS reference and its nonredundant tangent space then span the entire Hilbert space. If the MPS reference transforms according to a particular irrep of the symmetry group of the Hamiltonian, a basis of symmetry eigenvectors can be constructed for its nonredundant tangent space. If the Hilbert space is spanned by symmetry vectors belonging to at least two different irreps, the nonredundant tangent basis then contains symmetry eigenvectors belonging to a different irrep than the MPS reference.

\vspace{0.7cm}
\hspace{-\parindent}\textbf{B. Implications for RPA}
\vspace{0.3cm}

If an MPS \textit{Ansatz} without symmetry adaptation is variationally optimized, it can occur that due to the choice of virtual dimensions a symmetry multiplet of a non-Abelian group [e.g., $\mathsf{SU(2)}$] is incomplete at a certain boundary. From the projector interpretation of the nonredundant tangent space, it can be understood that this may lead to spurious zero energy RPA excitations: replacing an occuring renormalized basis state of the multiplet by one that is not in the renormalized basis, can lead to a state with the same energy and hence a spurious zero energy RPA excitation. For this reason, we have opted to use symmetry-adapted MPS references in this work.

\vspace{0.7cm}
\hspace{-\parindent}\textbf{C. Tangent space of a symmetry-adapted \textit{Ansatz}}
\vspace{0.3cm}

We now discuss the construction of the tangent space of an $\mathsf{SU(2)} \otimes \mathsf{U(1)}$ adapted MPS \textit{Ansatz}. A spin- and particle number-adapted MPS decomposes into Clebsch-Gordan coefficients of the imposed symmetry groups and reduced tensors, due to the Wigner-Eckart theorem \cite{woutersJCP1, sharma:124121, 0295-5075-57-6-852, 1742-5468-2007-10-P10014, PhysRevA.82.050301, 1367-2630-12-3-033029}:
\begin{equation}
A^{s s^z N}_{j_L j_L^z N_L \alpha_L ; j_R j_R^z N_R \alpha_R } = \braket{j_L j_L^z s s^z \mid j_R j_R^z} \delta_{N_L + N, N_R} T^{s N}_{j_L N_L \alpha_L ; j_R N_R \alpha_R }.
\end{equation}
The derivative operator $\frac{\partial}{\partial A^{\mu}}$ in Eq. \eqref{wavefunctionexpansionPRB88} is then replaced by $\frac{\partial}{\partial T^{\kappa}}$. All symmetry imposing Clebsch-Gordan coefficients are still in place, and the tangent space vectors hence have the same symmetry as the MPS reference. The nonredundant tangent space can be constructed in an analogous way as for the case without symmetry adaptation. The entire symmetry sector of the Hilbert space (minus the MPS reference) is retrieved in the nonredundant tangent space, if the virtual dimensions are taken as large as possible. The difference between the tangent spaces with and without symmetry adaptation can be compared to the restricted and unrestricted HF manifolds \cite{PhysRevA.22.2362, PhysRevA.24.673}. For the former only singlet excitations are possible, while for the latter triplet excitations are allowed too, even if the ground state is a singlet.

Note that if a symmetry-adapted MPS is optimized by the imaginary time evolution of Sec.~VII, the distribution of the bond dimensions over the symmetry sectors is fixed. As such an optimization does not lead to an optimal distribution of the bond dimensions, we have used the two-site DMRG algorithm to optimize all the MPS reference wave functions in this work. Henceforth symmetry-adapted will be used as a shorthand for spin- and particle number-adapted.

\vspace{0.7cm}
\hspace{-\parindent}\textbf{XI. THE 1D HUBBARD CHAIN}
\vspace{0.3cm}

In this section, we approximate low-lying eigenstates of the one-dimensional Hubbard chain with open boundary conditions (OBC) [see Eq. \eqref{HubbardModelPRB88}]. The CISD-MPS \textit{Ansatz} of Eq. \eqref{CISDansatzPRB88}, which contains all excitations to the double tangent space, is used to improve on the ground state and to find low-lying excitations. The results are compared with TDA-MPS, which contains all excitations to the single tangent space. With RPA-MPS, we search for the Goldstone mode of a symmetry-broken ground state. In addition, we discuss RPA-MPS correlation energy calculations.

\vspace{0.7cm}
\hspace{-\parindent}\textbf{A. CISD-MPS}
\vspace{0.3cm}

\begin{table}[t]
\centering
\caption{\label{CISDtablePRB88}CISD-MPS improvement on the ground state and low-lying excitation energies for the one-dimensional Hubbard chain with length $L=8$ and OBC. $D$ is the number of $\mathsf{SU(2)}$ multiplets retained at each boundary in the symmetry-adapted MPS reference calculation. The state for which the absolute energy is shown, was chosen as MPS reference.}

\begin{tabular}{l c c c c c c c}
\hline
\hline
$U$ & Quantity & $S$ & $N$ & Exact & TDA-MPS & TDA-MPS & CISD-MPS \\
  &          &   &   & (FCI) & ($D$=3) & ($D$=9) & ($D$=3) \\
\hline
0.1 & E$_0$       & 0             & 8 & -9.319312 & -9.067465         & -9.301264 & -9.315185\\
    & E$_1$-E$_0$ & $\frac{1}{2}$ & 7 &  0.297631 &  0.222150         &  0.285466 &  0.311181\\
    & E$_2$-E$_0$ & $\frac{1}{2}$ & 9 &  0.397631 &  0.322150         &  0.385466 &  0.411181\\
    & E$_3$-E$_0$ & 0             & 6 &  0.611620 &  0.873285         &  0.619417 &  0.629720\\
\hline
1   & E$_0$-E$_1$ & $\frac{1}{2}$ & 7 & -0.022354 & -0.082237         & -0.029340 &  0.011799\\
    & E$_1$       & 0             & 6 & -7.790647 & -7.532068         & -7.780764 & -7.785715\\
    & E$_2$-E$_1$ & 0             & 8 &  0.095814 &  0.543100$^{a}$ &  0.105942 &  0.135785\\
    & E$_3$-E$_1$ & 1             & 6 &  0.517393 &  0.572255$^{a}$ &  0.530944 &  0.542513\\
\hline
10  & E$_0$       & 0             & 4 & -5.187427 & -5.083270         & -5.186955 & -5.187090\\
    & E$_1$-E$_0$ & $\frac{1}{2}$ & 5 &  0.008950 & -0.010314         &  0.009270 &  0.010988\\
    & E$_2$-E$_0$ & 1             & 4 &  0.113988 &  0.127636         &  0.114721 &  0.114984\\
    & E$_3$-E$_0$ & $\frac{1}{2}$ & 5 &  0.189005 &  0.205577$^{a}$ &  0.196828 &  0.192880\\
\hline
100 & E$_0$       & 0             & 4 & -4.805753 & -4.736845         & -4.805615 & -4.805360\\
    & E$_1$-E$_0$ & 1             & 4 &  0.013020 &  0.013672         &  0.013016 &  0.013783\\
    & E$_2$-E$_0$ & 1             & 4 &  0.027045 &  0.022700         &  0.027013 &  0.028034\\
    & E$_3$-E$_0$ & 0             & 4 &  0.034327 &  0.316707$^{a}$ &  0.034327 &  0.035940\\
\hline
\hline
\end{tabular}

{\small $^{a}$ Excitation with different multiplicity. The required FCI excitation is not in the TDA-MPS spectrum.}
\end{table}

The TDA and CISD calculations were done by optimizing a symmetry-adapted MPS reference, with $D$ retained multiplets at each boundary. This reference was then used in TDA and CISD calculations without symmetry constraints. As the symmetry-adapted reference is not necessarily a variational minimum for an MPS \textit{Ansatz} without symmetry constraints, negative excitation energies can occur.

The CISD \textit{Ansatz} in Eq. \eqref{CISDansatzPRB88} leads to a generalized eigenvalue problem,
\begin{equation}
\braket{\overline{\Phi^0_{\kappa \lambda}} \mid \hat{H} \mid \Phi^0_{\mu \nu} } C^{\mu \nu} = E \braket{\overline{\Phi^0_{\kappa \lambda}} \mid \Phi^0_{\mu \nu} } C^{\mu \nu},
\end{equation}
which was solved by multitargeting the lowest states with the Davidson algorithm \cite{Davidson197587}. By decomposing the $C$ tensor, the CISD \textit{Ansatz} can be written as a sum over MPS wave functions:
\begin{equation}
\ket{\text{CISD}} = \sum\limits_{i < j} C^{\mu_i \nu_j} \ket{\Phi^0_{\mu_i \nu_j}} = \sum\limits_{i < j} \sum\limits_{p(i,j)} C^{\mu_i}_{L,p} C^{\nu_j}_{R,p} \ket{\Phi^0_{\mu_i \nu_j}}. \label{sumoverMPSsPRB88}
\end{equation}
This allows to use standard MPS machinery \cite{Schollwock201196} in the matrix-vector multiplication. Because the sum of several MPS wave functions yields an MPS with a larger bond dimension \cite{Schollwock201196}, this immediately leads to the understanding that the CISD \textit{Ansatz} can introduce extra entanglement.

We chose $L=8$ and four $U$ values: $0.1$, $1$, $10$ and $100$. With increasing $U$, the ground state changes from a collection of quasi-free particles to a highly correlated state. For the latter, HF gives a qualitatively wrong description. For $U=1$, the ground state contains $7$ particles and has spin $\frac{1}{2}$. If a symmetry-broken reference is chosen, the multiplet degeneracy of the excitations is lost. Therefore, we opted for the first singlet state as MPS reference for $U=1$. Although the TDA and CISD calculations were not symmetry-adapted, the multiplet degeneracy was \textit{exactly} retrieved because we started from a symmetry-adapted MPS reference. The first four multiplets for each $U$ value are shown in Table \ref{CISDtablePRB88}.

The TDA-MPS ($D=9$) energies and the CISD-MPS ($D=3$) energies are of roughly the same quality, and improve on the TDA-MPS ($D=3$) energies significantly. The CISD-MPS \textit{Ansatz} contains $\mathcal{O}\left[\frac{1}{2} (L d D^2)^2\right]$ variational parameters to include extra correlation between all pairs of sites on top of the MPS reference, and can be used both to improve on the ground state and to approximate excited states. The TDA-MPS \textit{Ansatz} contains $\mathcal{O}(L d D^2)$ variational parameters, and due to the MPS analog of Brillouin's theorem, it can only be used to approximate excited states. The relative accuracy of the CISD-MPS (D=3) and MPS (D=9) reference state energies changes with $U$. With increasing single-particle nature, the CISD-MPS \textit{Ansatz} performs better for the targeted reference.

For small $D$, not all excited states are retrieved with TDA-MPS. An example is the third excited state for $D=3$ and $U=100$. The targeted state consists of two singlet-triplet excitations, which interact to form a bound singlet state. This is well captured by CISD-MPS ($D=3$) and TDA-MPS ($D=9$), while $E_3$ for TDA-MPS ($D=3$) in Table \ref{CISDtablePRB88} corresponds in fact to a doublet. The MPS ($D=9$) reference has a large enough bond dimension to capture the two excitations in its single tangent space, while the CISD-MPS ($D=3$) \textit{Ansatz} captures the double excitation in the MPS's double tangent space.

\vspace{0.7cm}
\hspace{-\parindent}\textbf{B. RPA-MPS and Goldstone modes}
\vspace{0.3cm}

The $L=8$ and $U=1$ case is an ideal candidate to retrieve a Goldstone mode, because a specific spin-$\frac{1}{2}$ ground-state vector is necessarily a symmetry-broken state. With an MPS reference optimized without any symmetry constraints and $D=16$ (now exceptionally the number of states instead of the number of multiplets), we find one zero-energy solution to the RPA equations, and this solution also satisfies Eq. \eqref{GoldstoneEqPRB88}. This is the Goldstone mode from the symmetry-broken spin-$\frac{1}{2}$ ground state. Zero-energy solutions can also arise for singlet ground states, if the MPS accidently breaks non-Abelian symmetries, as discussed in Sec.~X. This can be avoided by retaining complete multiplets at each boundary, whereas the RPA Goldstone mode for a symmetry-broken ground state will always occur, even for bond dimensions that reproduce the full Hilbert space.

\vspace{0.7cm}
\hspace{-\parindent}\textbf{C. The RPA-MPS correlation energy}
\vspace{0.3cm}

\begin{figure}
\centering
\includegraphics[width=0.70\textwidth]{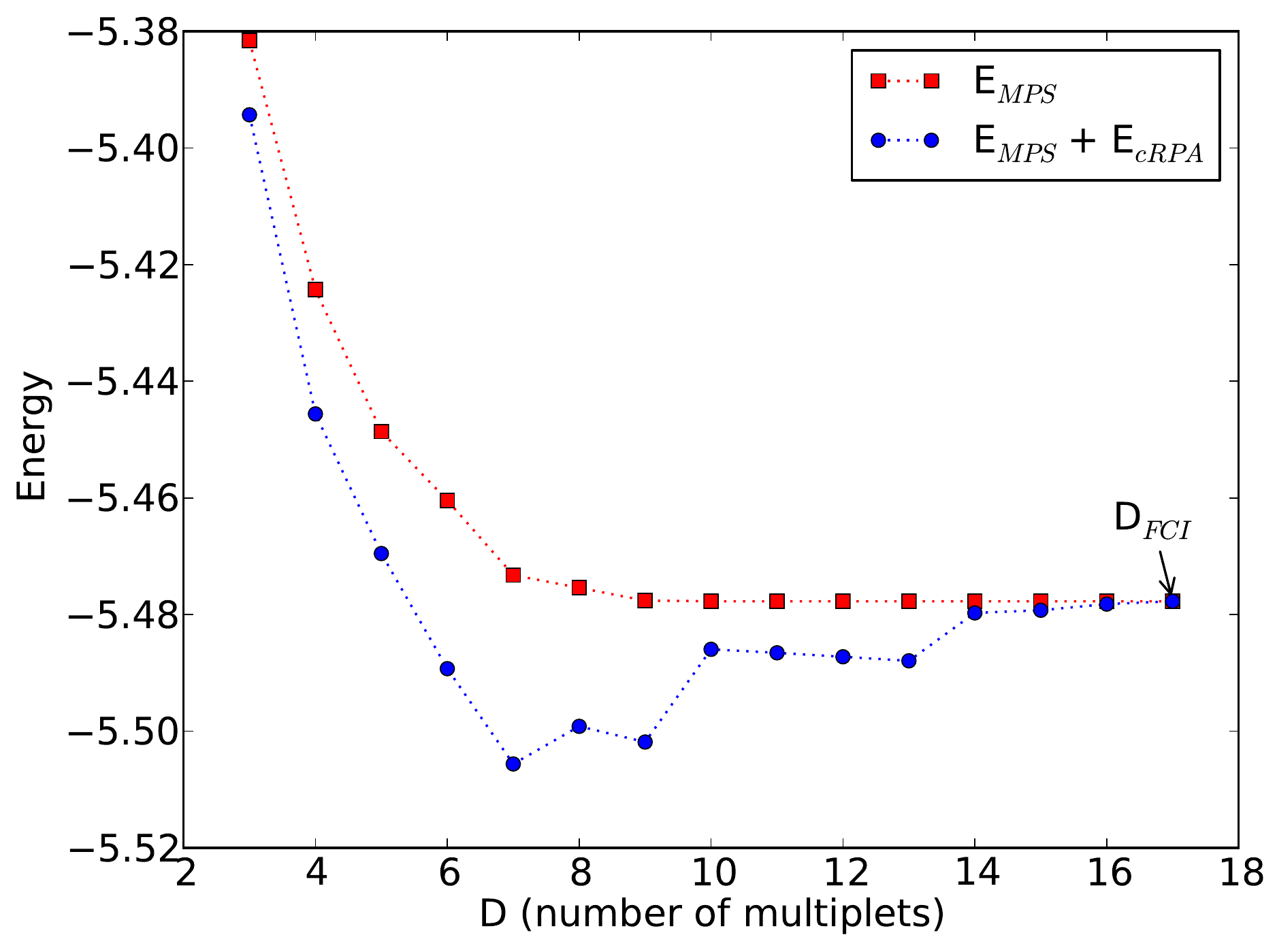}
\caption{\label{RPAcorrFigPRB88} The RPA-MPS correlation energy for the Hubbard chain with OBC, length $L=6$, filled with $N=4$ electrons, in the singlet state, and $U=1$. A symmetry-adapted \textit{Ansatz} was used, with $D$ retained multiplets at each boundary.
}
\end{figure}
We calculated RPA-MPS correlation energies for symmetry-adapted \textit{Ans\"atze}. Remember that only excitations with the same symmetry as the MPS reference are then retrieved. For the Hubbard chain with OBC and length $L=6$, filled with $N=4$ electrons, in the singlet state, and $U=1$, $E_{\text{cRPA}}$ is shown in Fig. \ref{RPAcorrFigPRB88}. From Eq. \eqref{RPAcorrEnergyEq2PRB88}, $2 E_{\text{cRPA}}$ can be interpreted as the mean difference between RPA and TDA excitation energies, multiplied by the number of excitations ($\text{dim} \mathbb{T}$). With increasing $D$, $| E_{\text{cRPA}} |$ first increases because the number of excitations increases. For even larger $D$, the mean difference between the RPA and TDA energies vanishes faster than the number of excitations increases. When the FCI virtual dimensions are reached, the RPA and TDA excitation energies are equal, as the $B$-matrix vanishes, and $E_{\text{cRPA}}$ is exactly zero.

When calculating $E_{\text{cRPA}}$, care has to be taken that the MPS reference is the variational minimum and that no symmetries are broken, such as the multiplet structure at a boundary or, e.g., the $\mathsf{SO(4)}$ symmetry when considering a half-filled Hubbard chain \cite{101142S0217984990000933}. If these conditions are not fulfilled, the RPA-MPS correlation energy breaks down ($| E_{\text{cRPA}} | \gg | E_{\text{MPS}} |$).

\vspace{0.7cm}
\hspace{-\parindent}\textbf{XII. POLYENES}
\vspace{0.3cm}

\begin{figure}
\centering
\includegraphics[width=0.70\textwidth]{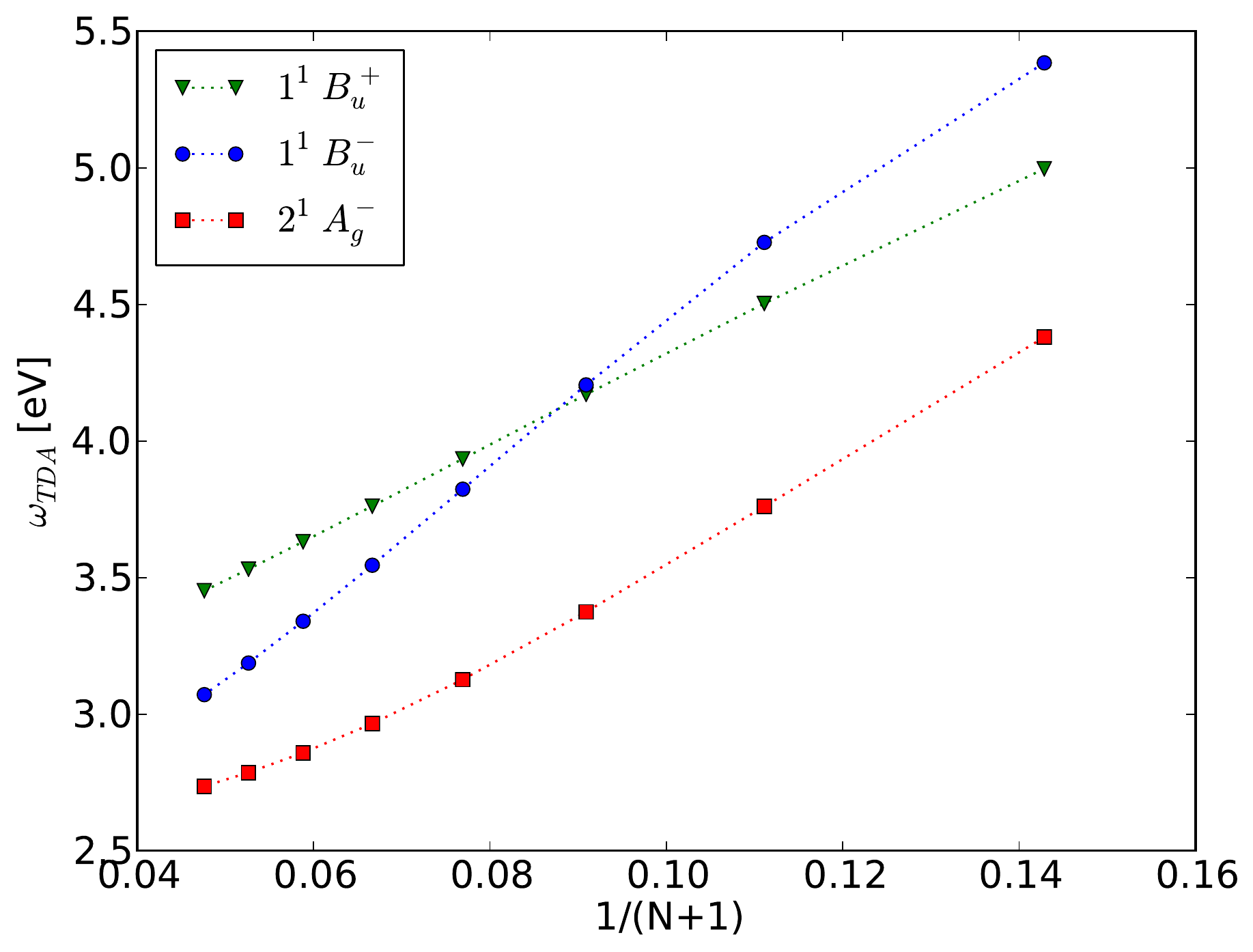}
\caption{\label{PPPTDAPRB88} The first three TDA-MPS excitation energies for a polyene chain with $N$ carbon atoms for which the $\pi$ system was approximated by the long-range PPP Hamiltonian.}
\end{figure}

\begin{figure}
\centering
\includegraphics[width=0.70\textwidth]{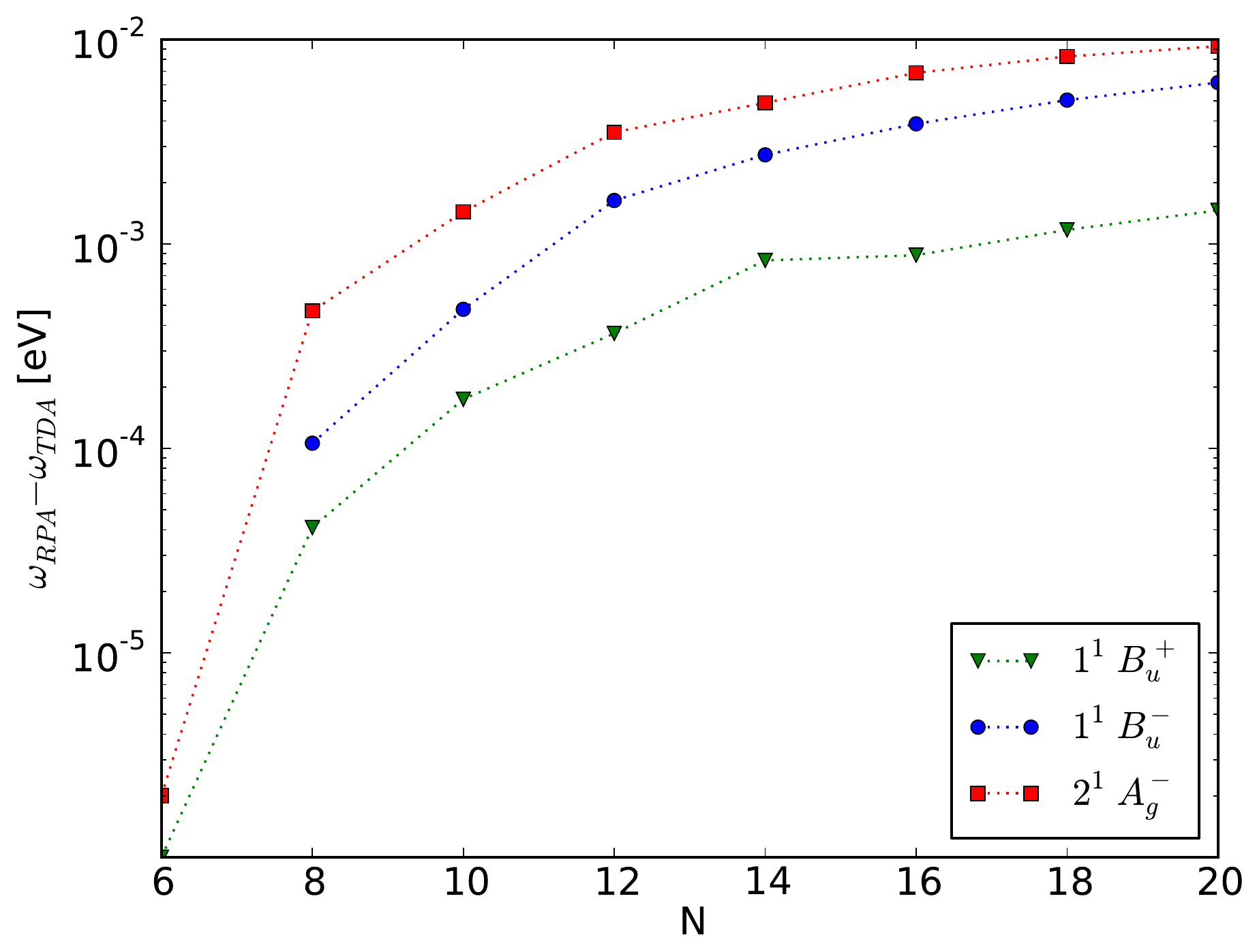}
\caption{\label{PPPTDARPAfigPRB88} The difference between the RPA-MPS and TDA-MPS energies for the first three excitations of a polyene chain with $N$ carbon atoms for which the $\pi$ system was approximated by the long-range PPP Hamiltonian.}
\end{figure}

Polyenes are linear conjugated chains of hydrocarbons:
\begin{equation}
\vspace{-0.15cm}
\text{CH}_2 = \text{CH} - \text{CH} = \text{CH} - \text{CH} = \text{CH}_2.
\end{equation}
Excitations in the $\pi$ system lie in the visible region of the spectrum, and polyenes are therefore important building blocks for light absorption and dyes. They have a long history of use as benchmark systems to test new quantum chemistry excited state methods. The $\pi$ system can be approximated by the long-range Pariser-Parr-Pople (PPP) Hamiltonian, where the two-body terms of the Hamiltonian \eqref{HamiltonianSecondQuantizationPRB88} are approximated by a local Coulomb repulsion:
\begin{equation}
\hat{V} = \frac{1}{2} \sum\limits_{k l \sigma \tau} R^{kl} \hat{n}_{k \sigma} \hat{n}_{l \tau}.
\end{equation}
The Latin letters denote orbitals and the Greek letters spin projections. For our calculations, we used the Ohno parameterization for the electron-electron repulsion $R^{kl}$ \cite{ohnoparameterization}. All Hamiltonian parameters, except $R_{^{\text{single}}_{\text{double}}} = 1.40 \pm 0.05 \AA$, are identical to the ones in Ref. \cite{OhnoChoicePMS}.\\
Many DMRG calculations studying the excited states and response properties of conjugated molecules have been performed, using a parameterized Hamiltonian \cite{Ramasesha1, PhysRevB.54.7598, PhysRevB.55.15368, PhysRevB.56.9298, JCPramaseshajettajetta, PhysRevB.66.035116}. At the \textit{ab initio} level, high-lying excited states have been targeted with the harmonic Davidson adaptation of the DMRG method \cite{dorando:084109}. Frequency-dependent dipole polarizabilities were computed at the \textit{ab initio} level by Dorando \textit{et al.} \cite{dorando:184111} using the TDA-MPS approximation.\\
Using the PPP Hamiltonian, we approximated the first three particle-conserving singlet excitations with the symmetry-adapted RPA-MPS and TDA-MPS methods. We kept $D=20$ retained multiplets at each boundary. The TDA-MPS excitation energies are shown in Fig. \ref{PPPTDAPRB88} as a function of the number of carbon atoms $N$ in the polyene. The symmetry labeling was based on Fig. 10 in Ref. \cite{PhysRevB.36.4337}. The difference between the RPA-MPS and TDA-MPS energies is shown in Fig. \ref{PPPTDARPAfigPRB88}, indicating that the ground state MPS reference is already quite accurate for $D=20$, as the $B$-matrix contributions are small. The RPA-MPS and TDA-MPS excitation energies match better for the higher excitations of Fig. \ref{PPPTDAPRB88}.

\vspace{0.7cm}
\hspace{-\parindent}\textbf{XIII. SUMMARY}
\vspace{0.3cm}

In this work, we attempted to set up a post-DMRG framework by finding the excitation structure of the MPS reference. As a guide, we carefully followed the structure of HF theory and the subsequent post-HF methods, exploiting the fact that both HF and DMRG can be seen as productlike wave functions \cite{ChanB805292C}.

A variational wave-function \textit{Ansatz} can be used in the TIVP to yield self-consistent equations. With the TDVP, optimal time-evolution is found which stays within the \textit{Ansatz} manifold. Linearization of the TDVP around a variational minimum gives the RPA equations. The optimal time-evolution requires a nonredundant parameterization of the \textit{Ansatz}'s tangent space to exclude meaningless variations of the wave function. Occupied-occupied variations in HF theory, as well as variations in the direction of the renormalized DMRG basis states, only lead to norm or phase changes. They do not change the physical state represented by the \textit{Ansatz}.

Exponentiation of the norm- and phase-conserving variations in HF theory, led to the Thouless theorem: a nonredundant parameterization of the entire HF (Grassmann) manifold, generated by the OV excitations of any particular SD. In this work, we have proposed the DMRG counterpart: a nonredundant parameterization of the entire MPS manifold, generated by the norm- and phase-conserving changes of any particular MPS wave function. Just like the norm- and phase-conserving changes of HF theory are generated by replacing an occupied orbital by a virtual orbital, the norm- and phase-conserving changes of an MPS wave function are generated by replacing the occuring renormalized basis states by discarded renormalized basis states. We have proven the MPS counterpart of Thouless's theorem for a general MPS with OBC, for which no Schmidt values vanish.

By identifying the excitation structure of the SD/MPS \textit{Ansatz} by means of the Thouless theorem, the RPA equations can be rederived be means of the EOM. This allows for a bosonic expansion of the Hamiltonian, and the definition of the RPA correlation energy and wave function.

The different orders of tangent space of the Thouless parameterization generate the CI basis. Eigenstates of the Hamiltonian can be approximated in this basis. CIS, or CI with only single excitations, yields again the SD/MPS reference due to Brillouin's theorem, as well as a set of excited states. These excited states are found by diagonalizing the Hamiltonian in the nonredundant tangent basis, or the $A$ matrix of RPA. This method is known as TDA.

When the MPS reference is a good approximation of the true ground state, \newline $\| ( \hat{H} - E_{\text{MPS}} ) \ket{\Phi^0} \|_2$ becomes small, and the B-matrix contributions of RPA vanish. TDA and RPA then lead to the same excitation energies. The RPA wave function suggests a size-consistent CC \textit{Ansatz} on top of the reference wave function.

The ideas presented in this paper are illustrated with proof-of-principle calculations of CISD-MPS improvements on the ground state, TDA-MPS, RPA-MPS and CISD-MPS excitation energies, an RPA-MPS Goldstone mode, and the RPA-MPS correlation energy. In contrast to HF, the MPS reference gives also in the highly correlated regime of the Hubbard model a qualitatively good description, and variational post-DMRG methods such as TDA-MPS and CISD-MPS give numerically relevant results. For an MPS with small bond dimensions, two correlated single excitations are not always retrieved in the tangent space, and the CISD-MPS \textit{Ansatz} is a better choice then.

Recently, we learned about Ref. \cite{PhysRevB.88.075133}, which presents RPA-MPS calculations and new multisite excitation \textit{Ans\"atze} for uniform MPS.

\vspace{0.7cm}
\hspace{-\parindent}\textbf{ACKNOWLEDGEMENTS}
\vspace{0.3cm}

This research was supported by the Research Foundation Flanders (S.W.) and the National Science Foundation Grant No. SI2-SSE:1265277 (G.C.). The authors would like to thank Jutho Haegeman, Frank Verstraete, and Stijn De Baerdemacker for the many stimulating conversations.

\vspace{0.7cm}
\hspace{-\parindent}\textbf{APPENDIX: EXPLICIT GRASSMANN MANIFOLD PARAMETERIZATION}
\vspace{0.3cm}

The proof given here is inspired by the proof for the unitary counterpart of Thouless's theorem for HF, given in Rowe \textit{et al.} \cite{PhysRevA.22.2362}. Give a unitary $m \times n$ matrix $Q$ with $m>n$, i.e., $Q^{\dagger}Q = I_n$, and a second unitary matrix $U$, of the same form as $Q$. Form a unitary $m\times (m-n)$ matrix $\tilde{Q}$ so that $[Q \tilde{Q}] [Q \tilde{Q}]^{\dagger} = I = [Q \tilde{Q}]^{\dagger} [Q \tilde{Q}]$. For the parameterization
\begin{equation}
\tilde{A}(\mathbf{y},\overline{\mathbf{y}}) = \exp{\left( \tilde{Q} y Q^{\dagger} - Q y^{\dagger} \tilde{Q}^{\dagger} \right)} Q \label{appendixEqPRB88}
\end{equation}
with $y$ an $(m-n) \times n$ matrix containing the complex variables and $\mathbf{y}$ the corresponding flattened column, there exists at least one $\mathbf{y_u}$ so that the columns of $\tilde{A}(\mathbf{y_u},\overline{\mathbf{y_u}})$ and the columns of $U$ (denoted by $\mathbf{u_k}$) span the same space. We will provide a proof by construction.

(1) The matrix $M = U^{\dagger} Q Q^{\dagger} U$ is a hermitian positive semidefinite matrix. There exists a unitary transformation to rotate the basis $\mathbf{u_i}$ to $\mathbf{v_i}$, so that $\mathbf{v_k}^{\dagger} Q Q^{\dagger} \mathbf{v_j} = \delta_{kj} n_k^2$.

(2) Write $\mathbf{v_i}$ in terms of $\mathbf{q_k}$ and $\tilde{\mathbf{q}}_\mathbf{k}$, the columns of $Q$ and $\tilde{Q}$: $\mathbf{v_i} = \sum_k \alpha_{ik} \mathbf{q_k} + \sum_l \beta_{il} \tilde{\mathbf{q}}_\mathbf{l}$. From the previous step we know that $\delta_{ij} = \mathbf{v_i}^{\dagger} \mathbf{v_j} = n_i^2 \delta_{ij} + \sum_l \overline{\beta_{il}} \beta_{jl}$.

(3) If $n_i \neq 0$, define $\mathbf{r_i}$ by $n_i \mathbf{r_i} = \sum_k \alpha_{ik} \mathbf{q_k}$. If $n_i \neq 1$, define $\tilde{\mathbf{r}}_{\mathbf{i}}$ by $(1 - n_i^2)^{\frac{1}{2}} \tilde{\mathbf{r}}_{\mathbf{i}} = \sum_l \beta_{il} \tilde{\mathbf{q}}_{\mathbf{l}}$. Note that if, e.g., $2n>m$, a number of $n_i$ will certainly be 1, and the corresponding vectors $\tilde{\mathbf{r}}_{\mathbf{i}}$ cannot be constructed.

(4) From the previous steps, it follows that the vectors $\{ \mathbf{r_i}$, $\tilde{\mathbf{r}}_{\mathbf{i}} \}$ are orthonormal. Complete both sets with additional vectors, so that they span the same space as the columns of, respectively, $Q$ and $\tilde{Q}$. If the matrix $R$ contains $\mathbf{r_i}$ in its columns and $\tilde{R}$ contains $\tilde{\mathbf{r}}_{\mathbf{i}}$ in its columns, a unitary transformation $P$ links $R$ and $Q$ by $R = QP$ and a unitary transformation $\tilde{P}$ links $\tilde{R}$ and $\tilde{Q}$ by $\tilde{R} = \tilde{Q} \tilde{P}$.

(5) If $n_i \neq 1$, consider $\mathbf{w_i} = \exp{\left( \gamma_i ( \tilde{\mathbf{r}}_{\mathbf{i}} \mathbf{r_i}^{\dagger} - \mathbf{r_i} \tilde{\mathbf{r}}_{\mathbf{i}}^{\dagger} ) \right)} \mathbf{r_i} = \cos{(\gamma_i)} \mathbf{r_i} + \sin{(\gamma_i)} \tilde{\mathbf{r}}_{\mathbf{i}}$. Assign $0 \leq \gamma_i \leq \frac{\pi}{2}$ so that $\cos{(\gamma_i)} = n_i$. It then follows that $\mathbf{w_i} = \mathbf{v_i}$. Note that if $n_i=1$, $\gamma_i$ would have been 0, and that the exponential in front of $\mathbf{r_i}$ then becomes the identity. So it poses no problem that the corresponding vectors $\tilde{\mathbf{r}}_{\mathbf{i}}$ cannot be constructed.

(6) If $\gamma$ is regarded as a diagonal matrix containing the $\gamma_i$ values, the singular value decomposition of $y_u$ is given by $y_u = \tilde{P} \gamma P^{\dagger}$. This can be confirmed by writing the exponential expression for $\mathbf{w_i}$ in terms of $Q$ and $\tilde{Q}$.

This concludes the construction of the complex $(m-n) \times n$ matrix $y_u$. Eq. \eqref{appendixEqPRB88} hence represents a Grassmann manifold.

{\color{blue}{
\noindent\makebox[\linewidth]{\rule{\textwidth}{0.4pt}}

\vspace{-0.40cm}

\noindent\makebox[\linewidth]{\rule{\textwidth}{0.4pt}}
}}

\newpage

\section{Remarks}
It is proposed in Ref. \cite{PhysRevB.88.075122} to construct the site-space analog of the particle Fock space. Operators such as the effective Hamiltonian $H_{\text{eff}}[i]$ and the excitation operator $\hat{B}_k^{\dagger} = \frac{\partial}{\partial x^k}B^{\mu}(\mathbf{x}) \frac{\partial}{\partial A^{\mu}}$ (which act on virtual indices) should then be written in terms of Fock space operators (which act solely on physical indices).

Several theorems treat the existence and the support of these Fock space operators for uniform MPSs \cite{PerezGarcia2011832.2011833, PhysRevB.88.075133, PhysRevLett.111.080401}. Uniform MPSs represent translationally invariant states in the TD limit. The support of a Fock space operator is the number of neighbouring physical indices on which it acts. For finite lattices such operators can also be constructed, but for virtual dimension $D$ they act on $\mathcal{O}(D^4)$ physical indices \cite{PerezGarcia2011832.2011833}. For most practical purposes, this is the whole Hilbert space.

The excitation ansatz
\begin{equation}
\sum_k x^k \hat{B}_k^{\dagger} \ket{\Phi^0}
\end{equation}
in terms of (general) Fock space operators $\hat{B}_k^{\dagger}$ is known historically as the Feynman-Bijl ansatz \cite{PhysRev.94.262, Bijl1941655} or the single-mode approximation \cite{PhysRevLett.60.531, PhysRevB.33.2481}.

The single-site excitation ansatz
\begin{equation}
B^{\mu} \ket{\Phi^0_{\mu}}
\end{equation}
of DMRG-TDA is local in the sense that is only able to capture particle excitations
\begin{equation}
\hat{a}^{\dagger}_{j\sigma} \hat{a}_{i\sigma} \ket{\Phi^0}
\end{equation}
for which $i$ and $j$ lie sufficiently close, $|i-j| \leq \mathcal{O}(\ln(D))$ \cite{NaokiLRTpaper}. However, for such $i$ and $j$, it captures particle excitations of higher rank as well:
\begin{equation}
\hat{a}^{\dagger}_{j\uparrow} \hat{a}^{\dagger}_{j\downarrow} \hat{a}_{i\downarrow} \hat{a}_{i\uparrow} \ket{\Phi^0}.
\end{equation}

The DMRG-CISD ansatz has also been used to determine the phase shift of scattering momentum eigenfunctions in one-dimensional spin chains \cite{2013arXiv1312.6793V}.

\chapter{Projector Monte Carlo with matrix product states} \label{DMC-MPS-chapter}
\begin{chapquote}{Eric Hoffer}
Creativity is the ability to introduce order into the randomness of nature.
\end{chapquote}

\section{Introduction}
The two most prevalent types of quantum Monte Carlo (MC) are variational MC \cite{PhysRev.138.A442} and projector or diffusion MC \cite{Grimm1971134, PhysRevD.27.1304}. In this chapter projector MC is introduced for matrix product states, in complete analogy with constrained path quantum MC and its phase-free extension for Slater determinants \cite{PhysRevLett.74.3652, PhysRevB.55.7464, PhysRevLett.90.136401}. This method can hence be seen as a new rung on the post-DMRG ladder.

Projector MC is introduced in section \ref{PMCintroSection}. The sign problem in fermionic systems can be removed with the constrained path method, which is discussed in section \ref{SignProbSec}. In sections \ref{PMCintroSection} and \ref{SignProbSec}, no specification of the wavefunction ansatz is made. Three specific flavours of projector MC for MPS wavefunctions are proposed in section \ref{MPSPMCsec}. For the auxiliary field variant, the projector decomposition is complex-valued and the sign problem becomes a phase problem. A strategy to eliminate the phase problem is given in section \ref{PhaseProbSec}. Some (preliminary) results for the three flavours are discussed in section \ref{resultsDMC}.

\section{Projector Monte Carlo} \label{PMCintroSection}
Consider a hermitian operator $\hat{K}$. Its dominant eigenstate $\ket{\Psi^{*}}$ with largest eigenvalue in magnitude $\lambda$ can be obtained by repeated application of $\hat{K}$ on a state $\ket{\Psi^{(0)}}$, if this state has nonzero overlap with $\ket{\Psi^{*}}$: 
\begin{equation}
\lim\limits_{n \rightarrow \infty}(\hat{K})^n \ket{\Psi^{(0)}} = \lim\limits_{n \rightarrow \infty} \ket{\Psi^{*}} \lambda^n \braket{\Psi^{*} \mid \Psi^{(0)}}.
\end{equation}
To find the ground state of a Hamiltonian $\hat{H}$, possible choices for $\hat{K}$ are $e^{- \delta\tau \hat{H}}$ or $(1 - \delta\tau \hat{H})$. For the latter, the positive time step $\delta \tau$ is bounded by
\begin{equation}
| 1 - \delta\tau E_0 | > | 1 - \delta\tau E_{\text{max}}|, \label{LinearProjectorBoundedness}
\end{equation}
with $E_0$ and $E_{\text{max}}$ respectively the minimum and maximum algebraic eigenvalues of $\hat{H}$.

At each MC time step $n$, the wavefunction is represented by an ensemble of walkers:
\begin{equation}
\ket{\Psi^{(n)}} = (\hat{K})^n \ket{\Psi^{(0)}} \approx \sum_{\phi} \ket{\phi}. \label{Ch7ReferenceForConclusion1}
\end{equation}
These walkers can be, for example, real-space coordinates \cite{Grimm1971134}, SDs \cite{PhysRevLett.74.3652, PhysRevB.55.7464, PhysRevLett.90.136401}, or MPSs.

The operator $\hat{K}$ is decomposed into a probability distribution function (PDF) $P(\mathbf{x})$ and the operators $\hat{B}(\mathbf{x})$ \cite{Grimm1971134, PhysRevLett.74.3652}:
\begin{equation}
\hat{K} = \sum_{\mathbf{x}} P(\mathbf{x}) \hat{B}(\mathbf{x}). \label{origPropagator}
\end{equation}
For the method to be successful, the action of $\hat{B}(\mathbf{x})$ on a walker should not increase its complexity. Real-space coordinates should just change positions. SDs should rotate into SDs. The virtual dimension of MPS walkers should not grow.

At each MC time step $n$, for each walker $\ket{\phi}$, an $\mathbf{x}$ is sampled with probability $P(\mathbf{x})$, and the walker is accordingly updated:
\begin{equation}
\ket{\phi'} = \hat{B}(\mathbf{x}) \ket{\phi}. \label{walkerUpdateEquation}
\end{equation}
After a sufficient amount of MC time steps, the ensemble stochastically represents $\ket{\Psi^*}$.

\section{The sign problem} \label{SignProbSec}
If everything is real-valued, there is sign symmetry in the sense that $\pm \ket{\Psi^{*}}$ are equivalent solutions. The two ensembles of walkers
\begin{equation}
\pm \left( \sum_{\phi} \ket{\phi} \right)
\end{equation}
then represent the targeted state equally well. In projector MC, the walkers are propagated independently, which is the source of the sign problem. Define the nodal plane $\mathcal{N}_{*}$ \cite{PhysRevLett.74.3652, PhysRevB.55.7464}:
\begin{equation}
\ket{\phi} \in \mathcal{N}_{*} \iff \braket{\Psi^{*} \mid \phi} = 0.
\end{equation}
If $\ket{\phi}$ can cross $\mathcal{N}_{*}$ to reach $-\ket{\phi}$ by successive application of some operators $\hat{B}(\mathbf{x})$, then $\pm \ket{\phi}$ will occur with equal probability after infinite MC time. Estimators, such as the projected energy
\begin{equation}
E^{(n)}_P = \frac{\sum_{\phi} \braket{\Psi_P \mid \hat{H} \mid \phi}}{\sum_{\phi} \braket{\Psi_P \mid \phi}},
\end{equation}
then suffer from a decaying signal-to-noise ratio because both the numerator and the denominator vanish. For fermionic systems, this generally cannot be avoided. An exception is the half-filled Hubbard model, for which a special decomposition \eqref{origPropagator} can be constructed which avoids the sign problem \cite{PhysRevLett.74.3652, PhysRevB.55.7464}.

The signal can be recovered by constraining the paths of the walkers with a trial wavefunction $\ket{\Psi_T}$ \cite{PhysRevLett.74.3652, PhysRevB.55.7464}. The projector for walker $\ket{\phi}$ is then changed to
\begin{equation}
\hat{K}_{\phi} = \sum_{\mathbf{x}} P(\mathbf{x}) \frac{\min \left\{0, \braket{\Psi_T \mid \hat{B}(\mathbf{x}) \mid \phi } \right\}}{\braket{\Psi_T \mid \phi }} \hat{B}(\mathbf{x}) = N_{\phi} \sum_{\mathbf{x}} \widetilde{P}_{\phi}(\mathbf{x}) \hat{B}(\mathbf{x}), \label{CPprojector}
\end{equation}
where $N_{\phi}$ is a constant so that $\widetilde{P}_{\phi}(\mathbf{x})$ is a normalized PDF. There is now importance sampling with respect to the overlap $\braket{\Psi_T \mid \phi}$, and the paths of the walkers are constrained to one side of the \textit{trial} nodal plane $\mathcal{N}_{T}$.

With the constrained path projector \eqref{CPprojector}, the walkers become weighted:
\begin{equation}
\sum_{\phi} w_{\phi} \ket{\phi}. \label{WeightedWalkerEquaion}
\end{equation}
The walkers $\ket{\phi}$ are updated according to Eq. \eqref{walkerUpdateEquation}, with $\mathbf{x}$ drawn from $\widetilde{P}_{\phi}(\mathbf{x})$. The nonnegative weigths $w_{\phi}$ absorb the PDF normalization constant $N_{\phi}$:
\begin{equation}
w_{\phi'} = N_{\phi} w_{\phi}.
\end{equation}
Due to the importance sampling, a weighted walker $w_{\phi} \ket{\phi}$ now contains the factor $\braket{\Psi_T \mid \phi}$, which is not present if the original projector \eqref{origPropagator} is used. To relate the ensemble of weighted walkers to the state $\ket{\Psi^{(n)}}$, this factor has to be removed. The state $\ket{\Psi^{(n)}}$ and the projected energy $E^{(n)}_T$ are then:
\begin{eqnarray}
\ket{\Psi^{(n)}} & \propto & \sum_{\phi} \frac{ w_{\phi} \ket{\phi} }{ \braket{\Psi_T \mid \phi} }, \label{newEnsemble}\\
E^{(n)}_T & = & \frac{\sum_{\phi} w_{\phi} \frac{ \braket{\Psi_T \mid \hat{H} \mid \phi} }{ \braket{\Psi_T \mid \phi} } }{\sum_{\phi} w_{\phi}}. \label{projectedEnergy}
\end{eqnarray}
The norm of $\ket{\phi}$ cancels in Eqs. \eqref{newEnsemble} and \eqref{projectedEnergy}. Normalization of $\ket{\phi}$ hence does not change $\ket{\Psi^{(n)}}$ or $E^{(n)}_T$. Care has to be taken that the overlap remains positive: $\braket{\Psi_T \mid \phi} > 0$.

The constrained path method eliminates the sign problem, but it introduces a systematic bias. The magnitude of the systematic bias depends on how good the trial wavefunction $\ket{\Psi_T}$ represents $\ket{\Psi^{*}}$. Once a walker hits the nodal plane $\mathcal{N}_{*}$, it will never contribute to statistics anymore \cite{PhysRevLett.74.3652, PhysRevB.55.7464}:
\begin{equation}
\braket{\Psi^* \mid \phi} = 0 \Rightarrow \forall n: \braket{\Psi^* \mid (\hat{K})^n \mid \phi} = 0.
\end{equation}
For exact $\ket{\Psi_T}$, the systematic bias therefore vanishes. With real-space coordinates as walker wavefunctions, the constrained path method is known as the fixed-node approximation.

Population control is used to duplicate walkers with large weights and to eliminate walkers with small weights. $\lfloor w_{\phi} + u \rfloor$ copies are made of walker $\ket{\phi}$, with $u$ drawn from the uniform PDF on $\left[ 0,1 \right[$. The weights of the copies are set to 1. Stochastically, this population control does not change the wavefunction $\ket{\Psi^{(n)}}$.

\section{Projector MC with matrix product states} \label{MPSPMCsec}
The walkers $\ket{\phi}$ and the trial wavefunction $\ket{\Psi_T}$ are typically of the same ansatz type, because the overlap $\braket{\Psi_T \mid \phi}$ and the expectation value $\braket{\Psi_T \mid \hat{H} \mid \phi}$ can then be evaluated cheaply. This is also the case for MPS wavefunctions.

Whereas an SD trial wavefunction provides a single variational approximation to the ground state, an MPS allows to systematically improve the approximation to the ground state by increasing the virtual dimension. This allows to assess the systematic bias due to the constrained path method.

In this chapter, spin lattice Hamiltonians will be studied:
\begin{equation}
\hat{H} = \frac{1}{2} \sum\limits_{i,j} J_{ij} \hat{\vec{S}}_i \cdot \hat{\vec{S}}_j + h \sum\limits_i \hat{S}_i^z. \label{SpinHamChapDMC}
\end{equation}
From a chemical Hamiltonian, a spin-$\frac{1}{2}$ lattice can be obtained by considering a half-filled system ($N=L$) and by taking the limit of large local repulsion \cite{PhysRev.115.2}. Each orbital is then singly occupied. The only local degrees of freedom $\ket{n_i}$ which remain are $\ket{\uparrow}$ and $\ket{\downarrow}$.
 
\subsection{Sampling the matrix product operator}\label{SamplingMPO}
\begin{figure}
\centering
\includegraphics[width=0.275\textwidth]{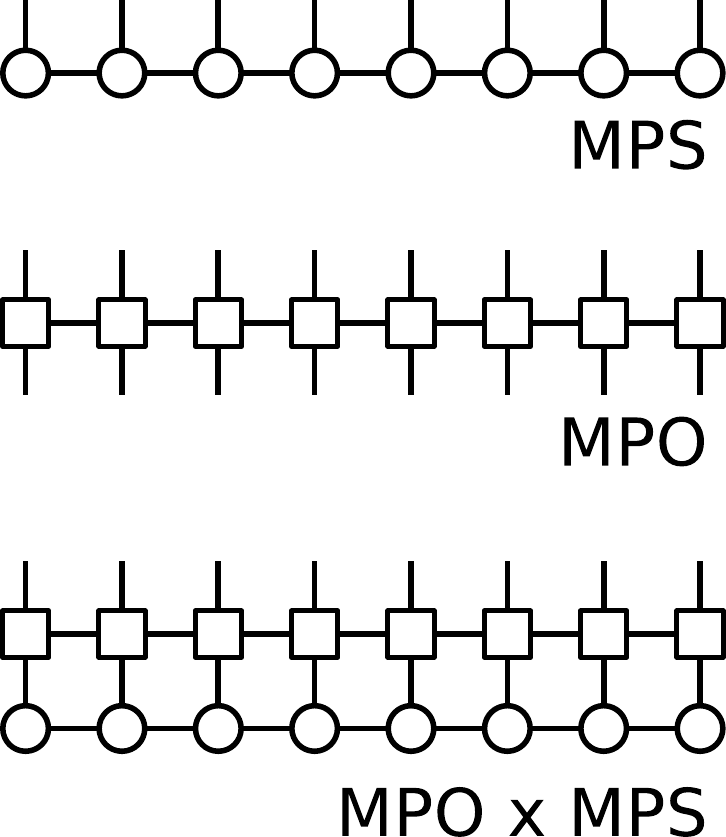}
\caption{\label{MPO-plot} Matrix product operators. Tensors are represented by circles or squares, physical indices by open lines, and virtual indices by connected lines. The graphs hence represent how the FCI tensor is decomposed into an MPS and how the Hamiltonian is decomposed into an MPO. The application of a Hamiltonian on a state yields another state. An MPS decomposition of the latter can be easily obtained from the MPO and MPS decompositions of resp. the Hamiltonian and the initial state.}
\end{figure}

In complete analogy to the MPS construction in section \ref{sec1p4-in-chap1}, the Hamiltonian can be decomposed into a matrix product operator (MPO):
\begin{eqnarray}
& \braket{n_1 n_2 ... n_L \mid \hat{H} \mid \widetilde{n}_1 \widetilde{n}_2 ... \widetilde{n}_L } = H_{\widetilde{n}_1 \widetilde{n}_2 ... \widetilde{n}_L}^{n_1 n_2 ... n_L} \\
& = \sum\limits_{\{ \beta_k \}} \left( M[1]^{n_1}_{\widetilde{n}_1} \right)_{\beta_1} \left( M[2]^{n_2}_{\widetilde{n}_2} \right)_{\beta_1 ; \beta_2} \left( M[3]^{n_3}_{\widetilde{n}_3} \right)_{\beta_2 ; \beta_3} ... \left( M[L]^{n_L}_{\widetilde{n}_L} \right)_{\beta_{L-1}}.
\end{eqnarray}
Except for sites 1 and $L$, the MPO tensors have rank 4. The pictorial representation of an MPS was already introduced in Fig. \ref{TNS-plot}. For an MPO, the pictorial representation is shown in Fig. \ref{MPO-plot}. The MPO tensors are represented by squares, physical indices by open lines, and virtual indices by connected lines. The application of a Hamiltonian on a state yields another state. An MPS decomposition of the latter can be easily obtained from the MPO and MPS decompositions of respectively the Hamiltonian and the initial state, as illustrated in Fig. \ref{MPO-plot}. The virtual dimension of the result is the product of the initial MPS and MPO virtual dimensions. The MPO of the Hamiltonian \eqref{SpinHamChapDMC} can be constructed analytically \cite{Schollwock201196}.

The set of operators $\hat{B}(\mathbf{x})$ should not increase the complexity of the MPS walkers, i.e. their virtual dimension should not increase. One way to obtain such operators $\hat{B}(\mathbf{x})$ is by constructing the MPO of the projector $\hat{K}$ and by sampling the virtual dimension at each virtual bond. Analogously to the gauge invariance of an MPS (see section \ref{TheMPSansatzSectionInChapterTwo}), the MPO has gauge invariance as well. Which gauge should be chosen, or should the gauge itself be sampled?

We will consider the MPO of $\hat{K} = (1-\delta\tau \hat{H} )$ here. For certain gauge choices, there are only a polynomial amount of nonzero terms which can be obtained by sampling its virtual bonds. For example:
\begin{eqnarray}
& \hat{I}_1 \hat{I}_2 \hat{I}_3 \hat{I}_4 \hat{I}_5 \hat{I}_6 \hat{I}_7 \hat{I}_8, \\
-\delta \tau \frac{J_{24}}{2} & \hat{I}_1 \hat{S}^{+}_2 \hat{I}_3 \hat{S}^{-}_4 \hat{I}_5 \hat{I}_6 \hat{I}_7 \hat{I}_8, \\
-\delta \tau \frac{J_{24}}{2} & \hat{I}_1 \hat{S}^{-}_2 \hat{I}_3 \hat{S}^{+}_4 \hat{I}_5 \hat{I}_6 \hat{I}_7 \hat{I}_8, \\
-\delta \tau J_{24} & \hat{I}_1 \hat{S}^z_2 \hat{I}_3 \hat{S}^z_4 \hat{I}_5 \hat{I}_6 \hat{I}_7 \hat{I}_8, ~ ...
\end{eqnarray}
In this example there are at most $1 + \frac{3L(L-1)}{2}$ nonzero operator strings. If certain elements $J_{ij}$ vanish, there will be even less nonzero terms. Instead of working with the local operators $(\hat{S}^{+}, \hat{S}^{-}, \hat{S}^z)$, any unitary rotation among them can be used as well. With $\hat{\vec{S}} = (\hat{S}^{x}, \hat{S}^y, \hat{S}^z)$:
\begin{equation}
\hat{\vec{S}}_i \cdot \hat{\vec{S}}_j = \left( \hat{\vec{S}}_i \mathbf{U} \right) \cdot \left( \mathbf{U}^{\dagger} \hat{\vec{S}}_j \right).
\end{equation}
The choice $(\hat{S}^{+}, \hat{S}^{-}, \hat{S}^z)$ might be unfortunate because the operators $\hat{S}^{+}$ and $\hat{S}^{-}$ are always singular, while the operators $(\hat{S}^{x}, \hat{S}^{y}, \hat{S}^z)$ are nonsingular for half-odd-integer spin. For the spin-$\frac{1}{2}$ lattice studied in section \ref{resultsDMC}, we even have
\begin{equation}
\hat{S}^x \hat{S}^x  = \hat{S}^y \hat{S}^y = \hat{S}^z \hat{S}^z = \frac{1}{4} \hat{I}.
\end{equation}
The individual walker paths are then always reversible. In section \ref{resultsDMC}, \texttt{MPO S$^+$S$^-$} and \texttt{MPO S$^x$S$^x$} will denote the special gauge choices of $\hat{K} = (1-\delta\tau \hat{H} )$ with resp. the local operators $(\hat{S}^{+}, \hat{S}^{-}, \hat{S}^z)$ and $(\hat{S}^{x}, \hat{S}^y, \hat{S}^z)$. One should remember that for both choices, there are at most $1 + \frac{3L(L-1)}{2}$ nonzero operator strings. Also note that while $\hat{S}^y$ is complex-valued, the product $\hat{S}^y_i\hat{S}^y_j$ is always real-valued. The constrained path method can hence be employed.

\subsection{Trotter decomposition}
In this subsection, we study the projector $\hat{K} = e^{-\delta\tau \hat{H}}$. Consider its Trotter-Suzuki decomposition \cite{Trotter, Suzuki}:
\begin{equation}
e^{- \delta\tau \hat{H}} = \left( e^{- \frac{h \delta\tau}{2} \sum_i \hat{S}_i^z} \right) \left( \prod_{i<j} e^{- \delta\tau J_{ij} \hat{\vec{S}}_i \cdot \hat{\vec{S}}_j } \right)\left( e^{- \frac{h \delta\tau}{2} \sum_i \hat{S}_i^z} \right) + \mathcal{O}(\delta\tau^2). \label{TrotterDecompEquation}
\end{equation}
The first and last factors on the right-hand side (RHS) can be written as a single string of local operators with the Baker-Campbell-Hausdorff formula \cite{Campbell01111896, Campbell01111897} because spin operators acting on different sites commute:
\begin{equation}
e^{- \frac{h \delta\tau}{2} \sum_i \hat{S}_i^z} \equiv \prod_i \left( e^{- \frac{h \delta\tau}{2} \hat{S}_i^z} \right). \label{BCHequalityResult}
\end{equation}
The other factors on the RHS act on a local Hilbert space of size $(2S+1)^2$, i.e. two sites in the lattice. They can be considered two-site projectors, and can be decomposed into an MPO as well:
\begin{equation}
\braket{ n_i n_j \mid e^{- \delta\tau J_{ij} \hat{\vec{S}}_i \cdot \hat{\vec{S}}_j } \mid \widetilde{n}_i \widetilde{n}_j } = K_{\widetilde{n}_i \widetilde{n}_j}^{n_i n_j} = \sum\limits_{\beta} \left( M[i]^{n_i}_{\widetilde{n}_i} \right)_{\beta} \left( M[j]^{n_j}_{\widetilde{n}_j} \right)_{\beta}.
\end{equation}
The virtual dimension of such an MPO is $(2S+1)^2$. For each factor in Eq. \eqref{TrotterDecompEquation}, the (single) virtual bond $\beta$ can sampled. We have considered two sampling schemes. In the first, an SVD is performed on $K_{\widetilde{n}_i \widetilde{n}_j}^{n_i n_j}$, which defines the MPO decomposition as:
\begin{eqnarray}
K_{\widetilde{n}_i \widetilde{n}_j}^{n_i n_j} & = & K_{(n_i \widetilde{n}_i) ; (n_j \widetilde{n}_j)} = \sum_{\beta} U_{(n_i \widetilde{n}_i) ; \beta} \lambda_{\beta} V_{\beta ; (n_j \widetilde{n}_j)} \nonumber \\
& = & \sum_{\beta} \left( U_{(n_i \widetilde{n}_i) ; \beta} \sqrt{\lambda_{\beta}} \right) \left( \sqrt{\lambda_{\beta}} V_{\beta ; (n_j \widetilde{n}_j)} \right) = \sum\limits_{\beta} \left( M[i]^{n_i}_{\widetilde{n}_i} \right)_{\beta} \left( M[j]^{n_j}_{\widetilde{n}_j} \right)_{\beta}.
\end{eqnarray}
For the spin-$\frac{1}{2}$ lattice studied in section \ref{resultsDMC}, $(2S+1)^2 = 4$ and an SVD yields the MPO
\begin{equation}
e^{- \delta\tau J \hat{\vec{S}}_i \cdot \hat{\vec{S}}_j } = \left( \frac{ 3e^{-J\delta\tau / 4} +  e^{3J\delta\tau / 4} }{4}  \right) \hat{I}_i \hat{I}_j + \left( e^{-J\delta\tau / 4} - e^{3J\delta\tau / 4} \right) \left( \frac{\hat{S}^{+}_i \hat{S}^{-}_j + \hat{S}^{-}_i \hat{S}^{+}_j}{2} + \hat{S}^{z}_i \hat{S}^{z}_j \right).
\end{equation}
This sampling scheme is called \texttt{Trotter Simple} in section \ref{resultsDMC}. Other gauges for the two-site MPO decomposition can be chosen as well, and the gauge can even be sampled by introducing additional stochastic variables:
\begin{eqnarray}
K_{\widetilde{n}_i \widetilde{n}_j}^{n_i n_j} & = & \sum\limits_{\beta = 1}^d \left( M[i]^{n_i}_{\widetilde{n}_i} \right)_{\beta} \left( M[j]^{n_j}_{\widetilde{n}_j} \right)_{\beta} = \sum\limits_{\beta = 1}^d \hat{M}_i^{\beta} \hat{M}_j^{\beta} = \hat{\vec{M}}_i \cdot \hat{\vec{M}}_j \nonumber \\
& = & d \int d\vec{n} P(\vec{n}) \left( \vec{M}_i \cdot \vec{n} \right) \left( \vec{n} \cdot \vec{M}_j \right)
\end{eqnarray}
where $P(\vec{n})$ is the uniform PDF on the $(d-1)$-sphere with unit radius. A discrete grid of points can also be set up. Suppose that with $\vec{n} = (n_1, n_2, ... ,n_{d})$:
\begin{eqnarray}
\forall \beta, n_i \quad & : & \quad P(n_1, n_2, ... , n_{\beta}, ..., n_{d}) = P(n_1, n_2, ... ,-n_{\beta}, ..., n_{d}), \label{discreteCondition1}\\
\forall i,j,n_k \quad & : & \quad P(n_1, ... , n_{i}, ...,n_{j}, ..., n_{d}) = P(n_1, ... , n_{j}, ...,n_{i}, ..., n_{d}).\label{discreteCondition2}
\end{eqnarray}
Then
\begin{equation}
\int dn_1 ... dn_{i-1} dn_{i+1} ... dn_{d} P(n_1, ..., n_d) = P_i(n_i) = P_1(n_i).
\end{equation}
$P_i(n)$ is hence independent of the index $i$. With $N^{-1} = \int dn P_1(n) n^2$:
\begin{eqnarray}
& \hat{\vec{M}}_i \cdot \hat{\vec{M}}_j = N \int dn P_{1}(n) n^2 \sum_{\beta} \hat{M}_i^{\beta} \hat{M}_j^{\beta} = N \int d\vec{n} P(\vec{n}) \sum_{\beta} n^2_{\beta} \hat{M}_i^{\beta} \hat{M}_j^{\beta} \nonumber \\
& = N \int d\vec{n} P(\vec{n}) \sum_{\beta \gamma} \left( n_{\beta} \hat{M}_i^{\beta} \right) \left( n_{\gamma} \hat{M}_j^{\gamma} \right) = N \int d\vec{n} P(\vec{n}) \left( \hat{\vec{M}}_i \cdot \vec{n} \right) \left( \vec{n} \cdot \hat{\vec{M}}_j \right),
\end{eqnarray}
where Eq. \eqref{discreteCondition1} was used to go from the first to the second line.

Now consider the points
\begin{equation}
x_i = \text{erf}^{-1}\left( \frac{2i+1}{N_{\text{points}}} - 1 \right) \qquad i = 0,1,...,N_{\text{points}}-1.
\end{equation}
from which a discrete set of points on the $(d-1)$-sphere with unit radius can be built:
\begin{equation}
\vec{n}_{ij...l} = \frac{(x_i, x_j, ..., x_l)}{\sqrt{x_i^2 + x_j^2 + ... + x_l^2}}.
\end{equation}
If the PDF is only nonzero at these points, and uniform in these points, Eqs. \eqref{discreteCondition1} and \eqref{discreteCondition2} are valid. This choice of discrete points on the unit sphere was inspired by Marsaglia's sphere picking algorithm \cite{Marsaglia} and will be called \texttt{Trotter Sphere($N_{\text{points}}$)} in section \ref{resultsDMC}.

Without sampling the MPO gauge, it can occur that only discrete points in the walker ansatz space can be reached. For the spin-$\frac{1}{2}$ lattice studied in section \ref{resultsDMC}, this is the case when the local operators $(\hat{I}, \hat{S}^x, \hat{S}^y, \hat{S}^z)$ are used. Action of these operators on an MPS tensor yield the results in Table \ref{ErgodicityTable}. The two MPS site-matrices can be swapped and a relative minus sign can be introduced. For the entire lifetime of this walker, each MPS tensor can only reach four possible states (up to a global factor).

\begin{table}
\centering
\caption{\label{ErgodicityTable} Action of the operators $(\hat{I},\hat{S}^x,\hat{S}^y,\hat{S}^z)$ on an MPS tensor of a spin-$\frac{1}{2}$ lattice. The two MPS site-matrices $\left( \mathbf{M^{\uparrow}}, \mathbf{M^{\downarrow}} \right) = \left( \mathbf{A}, \mathbf{B} \right)$ can be swapped and a relative minus sign can be introduced.}
\begin{tabular}{c c c}
  \hline
  \hline
  $\hat{O}$ & \quad & Action of $\hat{O}$ on $\left( \mathbf{M^{\uparrow}}, \mathbf{M^{\downarrow}} \right) = \left( \mathbf{A}, \mathbf{B} \right)$ \\
  \hline
$\hat{I}$   & \quad & $\left( \mathbf{A}, \mathbf{B} \right)$ \\
$\hat{S^x}$ & \quad & $\frac{1}{2} \left( \mathbf{B}, \mathbf{A} \right)$ \\
$\hat{S^y}$ & \quad & $\frac{i}{2} \left( -\mathbf{B}, \mathbf{A} \right)$ \\
$\hat{S^z}$ & \quad & $\frac{1}{2} \left( \mathbf{A}, -\mathbf{B} \right)$ \\
  \hline
  \hline
\end{tabular}
\end{table}

\subsection{Auxiliary fields}
Auxiliary field quantum MC \cite{PhysRevLett.74.3652, PhysRevB.55.7464, PhysRevLett.90.136401} also provides a way to decompose $\hat{K} = e^{- \delta\tau \hat{H}}$ into operators $\hat{B}(\mathbf{x})$ which do not increase the virtual dimension of the MPS walkers. Consider the spectral decomposition of the symmetric matrix $J_{ij}$ in the Hamiltonian \eqref{SpinHamChapDMC}:
\begin{equation}
J_{ij} = \sum_k V_{ik} \gamma_k (V^T)_{kj}.
\end{equation}
With the operators
\begin{eqnarray}
\hat{v}^x_k & = & \sum_i \hat{S}_i^x V_{ik} \sqrt{-\gamma_k},\\
\hat{v}^y_k & = & \sum_i \hat{S}_i^y V_{ik} \sqrt{-\gamma_k},\\
\hat{v}^z_k & = & \sum_i \hat{S}_i^z V_{ik} \sqrt{-\gamma_k},
\end{eqnarray}
the Hamiltonian \eqref{SpinHamChapDMC} can be rewritten as:
\begin{equation}
\hat{H} = h \sum_i \hat{S}_i^z - \sum\limits_{w,k} \frac{(\hat{v}_k^w)^2}{2} = h \sum_i \hat{S}_i^z - \frac{\hat{\vec{v}}^2}{2},
\end{equation}
with $\hat{\vec{v}} = (v_1^x, v_1^y, v_1^z, v_2^x, ...)$. With this quadratic form for the two-site interaction, a Hubbard-Stratonovich transformation \cite{Stratonovich,PhysRevLett.3.77} yields:
\begin{eqnarray}
P(\vec{x}) & = & \frac{e^{-\vec{x}^2/2}}{(2 \pi)^{3L/2}}, \label{Porb_AFQMC} \\
\hat{B}(\vec{x}) & = & \exp{\left(- \frac{h \delta\tau}{2} \sum_i \hat{S}_i^z \right)}  \exp{\left(\sqrt{\delta\tau} \vec{x} \cdot \hat{\vec{v}} \right) }  \exp{\left(- \frac{h \delta\tau}{2} \sum_i \hat{S}_i^z \right)}, \label{BX_AFQMC} \\
e^{-\delta\tau \hat{H}} & = & \int d\vec{x} P(\vec{x}) \hat{B}(\vec{x}) + \mathcal{O}(\delta\tau^2). \label{Int_QFQMC}
\end{eqnarray}
The first and last factors on the RHS of Eq. \eqref{BX_AFQMC} yield a single string of local operators (see Eq. \eqref{BCHequalityResult}). The middle factor can be decomposed as:
\begin{equation}
\exp{\left(\sqrt{\delta\tau} \vec{x} \cdot \hat{\vec{v}} \right) } \equiv \prod_{i} \exp{\left( \sum_w \hat{S}_{i}^w ~ \sum_k V_{ik} x_k^w \sqrt{-\gamma_k \delta\tau} \right) }. \label{asdgjsdidfncdraggg}
\end{equation}
Note that Eq. \eqref{asdgjsdidfncdraggg} also corresponds to a single string of local operators, for the same reasons as in Eq. \eqref{BCHequalityResult}. For each MPS walker, the auxiliary field $\vec{x}$ is sampled from $P(\vec{x})$ and the walker is propagated with $\hat{B}(\vec{x})$. This does not increase the virtual dimension of the MPS walker. Because the projector decomposition in Eqs. \eqref{Porb_AFQMC}-\eqref{Int_QFQMC} is complex-valued, the fermion sign problem is now a phase problem.

\section{The phase problem} \label{PhaseProbSec}
For complex-valued parameterizations, there is phase symmetry in the sense that $e^{i \theta} \ket{\Psi^*}$ (with $\theta  \in [0,2\pi[$) are equivalent solutions. The ensembles of walkers
\begin{equation}
e^{i \theta} \left( \sum_{\phi} \ket{\phi} \right) \qquad , \qquad \theta \in [0,2\pi[
\end{equation}
then represent the targeted state equally well. In projector MC, the walkers are propagated independently, which is the source of the phase problem. The constrained path method (see section \ref{SignProbSec}) does not resolve the phase problem, and a different strategy is needed. In this section, Zhang's proposal for auxiliary field quantum MC is reviewed \cite{PhysRevLett.90.136401}. The notation is again independent of the specific walker ansatz type, and relies only on the Hubbard-Stratonovich transformation.

The projector in Eq. \eqref{Int_QFQMC} does not change with the following translation:
\begin{equation}
\hat{K} = \int d\vec{x} P(\vec{x} - \vec{y}) \hat{B}(\vec{x} - \vec{y}).
\end{equation}
Importance sampling with respect to the overlap $\braket{\Psi_T \mid \phi}$ with a trial wavefunction $\ket{\Psi_T}$ is again introduced:
\begin{eqnarray}
\hat{K}_{\phi} & = & \int d\vec{x} P(\vec{x} - \vec{y}) \hat{B}(\vec{x} - \vec{y}) \frac{\braket{\Psi_T \mid \hat{B}(\vec{x} - \vec{y}) \mid \phi}}{\braket{\Psi_T \mid \phi}} \nonumber\\
& = & \int d\vec{x} P(\vec{x}) \hat{B}(\vec{x} - \vec{y}) W_{\phi}(\vec{x},\vec{y}), \label{PhaseFreeProp}\\
W_{\phi}(\vec{x},\vec{y}) & = & \frac{\braket{\Psi_T \mid \hat{B}(\vec{x} - \vec{y}) \mid \phi}}{\braket{\Psi_T \mid \phi}} e^{\vec{x}\cdot\vec{y} - \vec{y}\cdot\vec{y}/2}.
\end{eqnarray}
The wavefunction is then again represented by Eq. (\ref{newEnsemble}). The vector $\vec{y}$ is chosen in order to minimize the average fluctuations of $\ln W_{\phi}(\vec{x},\vec{y})$ for variations in $\vec{x}$ \cite{PhysRevLett.90.136401}:
\begin{equation}
\vec{y}_{\phi}^0 = - \sqrt{\delta\tau} \frac{\braket{\Psi_T \mid \hat{\vec{v}} \mid \phi}}{\braket{\Psi_T \mid \phi}} + \mathcal{O}(\delta\tau).
\end{equation}
$W_{\phi}(\vec{x},\vec{y}_{\phi}^0)$ then becomes (approximately) independent of $\vec{x}$:
\begin{equation}
W_{\phi}(\vec{x},\vec{y}_{\phi}^0) \approx W^0_{\phi} = \exp\left[ -\delta\tau \frac{\braket{\Psi_T \mid \hat{H} \mid \phi}}{\braket{\Psi_T \mid \phi}} \right]. \label{weightFuncvtion}
\end{equation}

The walkers are again weighted (see Eq. \eqref{WeightedWalkerEquaion}). The weights absorb the $\vec{x}$-independent prefactor $W^0_{\phi}$ of the projector \eqref{PhaseFreeProp}:
\begin{equation}
w_{\phi'} = W^0_{\phi} w_{\phi}.
\end{equation}
The auxiliary field $\vec{x}$ is sampled from $P(\vec{x})$, and the walker is updated accordingly:
\begin{equation}
\ket{\phi'} = \hat{B}(\vec{x} - \vec{y}_{\phi}^0) \ket{\phi}.
\end{equation}

For an exact $\ket{\Psi_T}$, the walker energy in Eq. (\ref{weightFuncvtion}) is real-valued. To adjust for approximate $\ket{\Psi_T}$, the walker energy is replaced by its real part in Eqs. (\ref{projectedEnergy}) and (\ref{weightFuncvtion}):
\begin{eqnarray}
W^0_{\phi} & = & \exp\left[ -\delta\tau \Re \frac{\braket{\Psi_T \mid \hat{H} \mid \phi}}{\braket{\Psi_T \mid \phi}} \right], \\
E^{(n)}_T & = & \frac{\sum_{\phi} w_{\phi} \Re \frac{\braket{\Psi_T \mid \hat{H} \mid \phi} }{\braket{\Psi_T \mid \phi} } }{\sum_{\phi} w_{\phi}}.
\end{eqnarray}

Thus far only importance sampling was introduced. The phase problem can be eliminated by changing $W^0_{\phi}$ to \cite{PhysRevLett.90.136401, AFQMC_JCP_recent}:
\begin{equation}
\widetilde{W}^0_{\phi} = W^0_{\phi} \max(0, \cos(\Delta\theta)) \qquad \Delta\theta = \Im \ln \frac{\braket{\Psi_T \mid \hat{B}(\vec{x} - \vec{y}_{\phi}^0) \mid \phi}}{\braket{\Psi_T \mid \phi}}.
\end{equation}
This introduces a systematic bias for approximate $\ket{\Psi_T}$. The bias vanishes when $\ket{\Psi_T}$ becomes exact. For real-valued parametrizations, this phase-free approach reduces to the constrained path method of section \ref{SignProbSec}.

\section{Results and discussion} \label{resultsDMC}

In this section, some (preliminary) results are presented for the abovementioned methods: \texttt{MPO S$^x$S$^x$}, \texttt{MPO S$^+$S$^-$}, \texttt{Trotter Simple}, \texttt{Trotter Sphere($N_{\text{points}}$)}, and \texttt{AFQMC}. 

\subsection{The studied system}
The spin-$\frac{1}{2}$ Heisenberg model \cite{PhysRev.115.2} was studied on a $4\times 4$ torus (square lattice with periodic boundary conditions). Only the nearest-neighbour coupling $J=1$ is then nonzero. No magnetic field is present: $h=0$.

The results were obtained with $D_T = D_W$, i.e. equal trial ($D_T$) and walker ($D_W$) virtual dimensions. However, it should be mentioned that for $2 \leq D_W \leq D_T$, the results are not noticably influenced by the specific choice of $D_W$. The number of walkers was taken to be $N_W = 1000$. This number does not influence the outcome, only the statistical fluctations on the outcome. The time step was taken to be $\delta\tau = 0.01$. This time step influences the error of expansions such as Eq. \eqref{TrotterDecompEquation}. It should also be chosen small enough so that Eq. \eqref{LinearProjectorBoundedness} remains valid.

\begin{figure}
\centering
\includegraphics[width=0.70\textwidth]{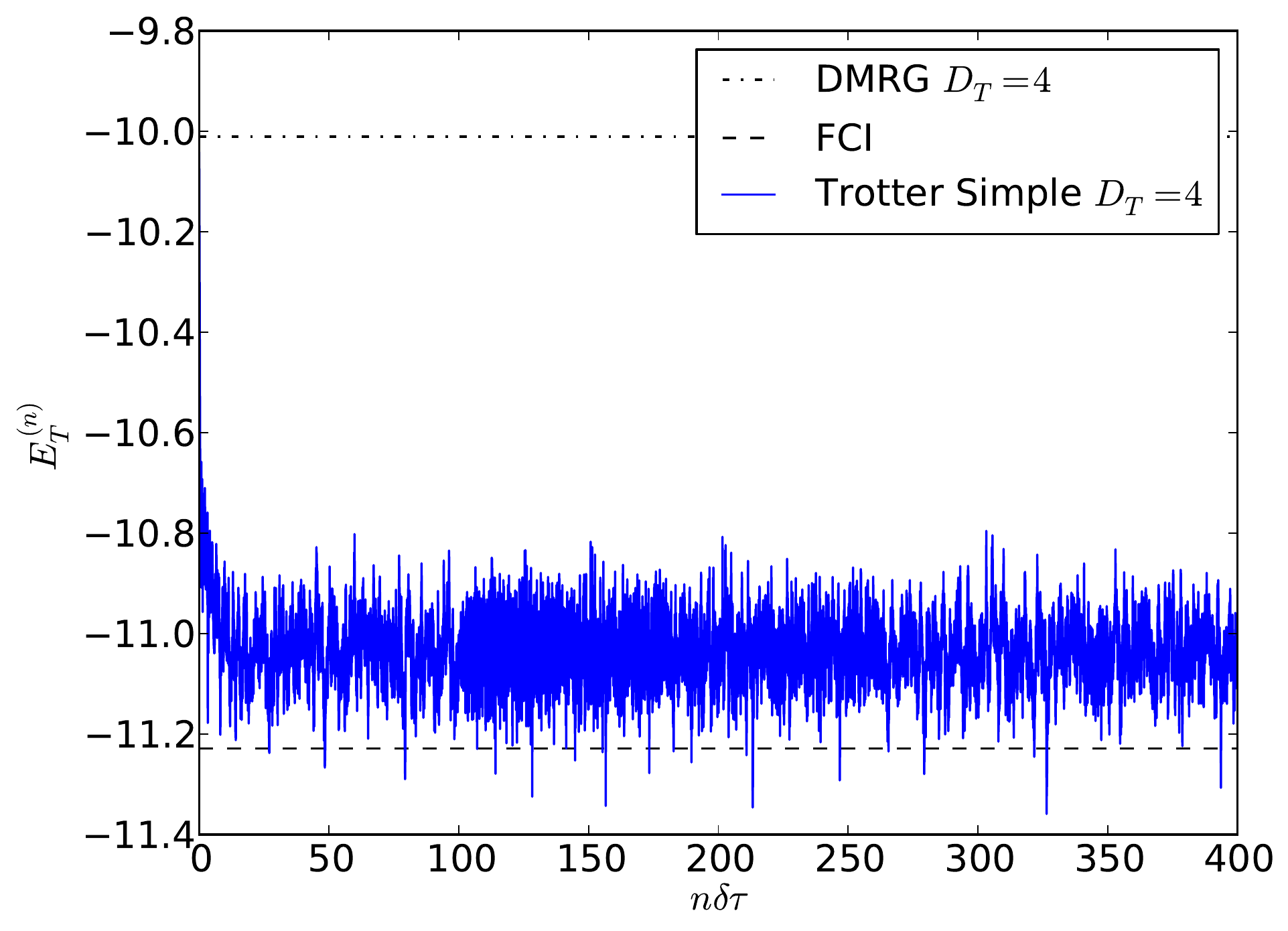}
\caption{\label{MPSQMC-TimeEvo} MC time evolution of the projected energy. An ensemble of $N_W=1000$ walkers was propagated with the \texttt{Trotter Simple} projector with $\delta\tau=0.01$, for the spin-$\frac{1}{2}$ Heisenberg model on the $4 \times 4$ torus. The virtual dimension of the trial and walker wavefunctions was $D_T = D_W = 4$.}
\end{figure}

\subsection{Statistical error}

\begin{figure}
\centering
\includegraphics[width=0.70\textwidth]{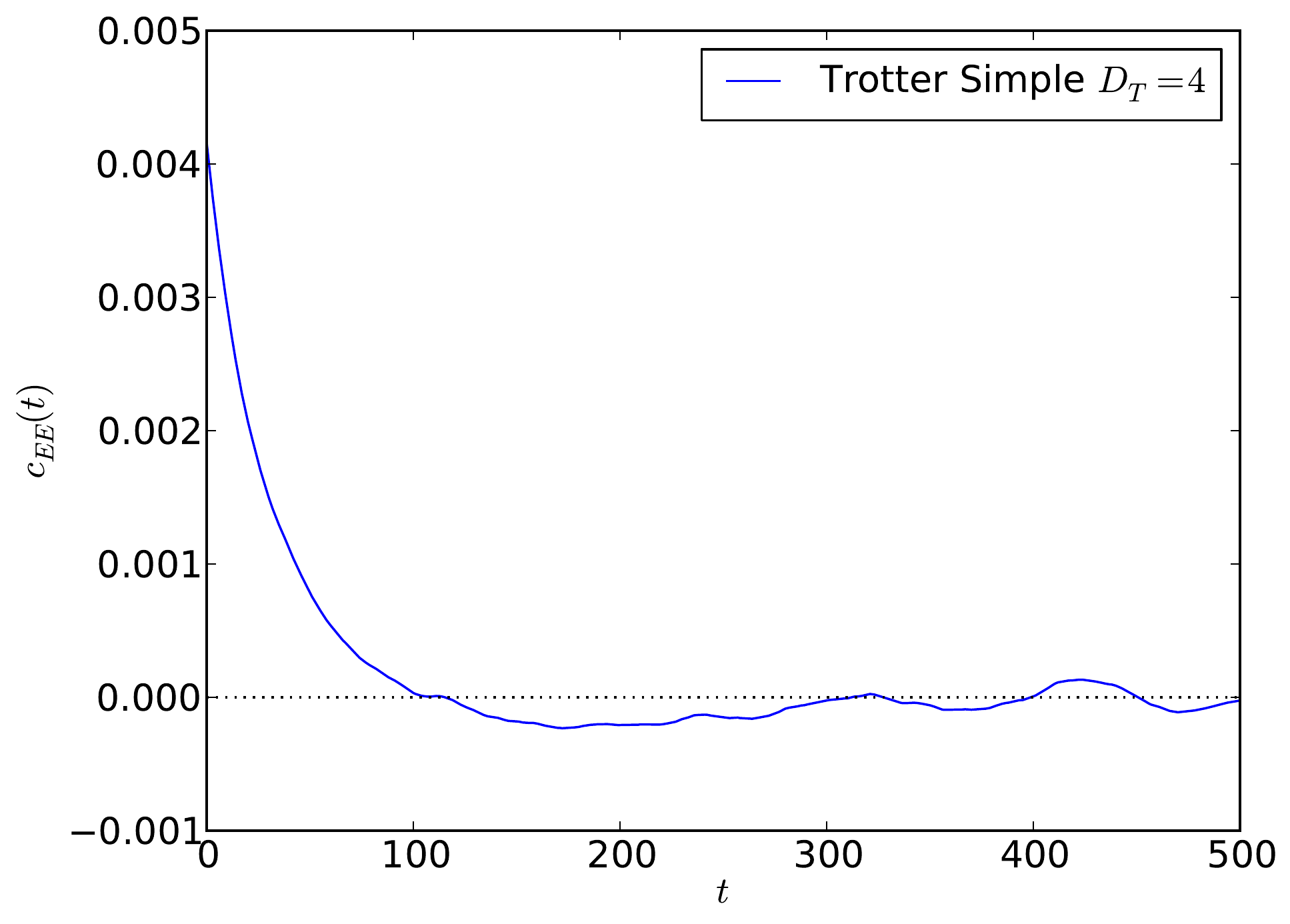}
\caption{\label{MPSQMC-Autocorr} Autocorrelation function of the projected energies in Fig. \ref{MPSQMC-TimeEvo}.}
\end{figure}

\begin{figure}
\centering
\includegraphics[width=0.70\textwidth]{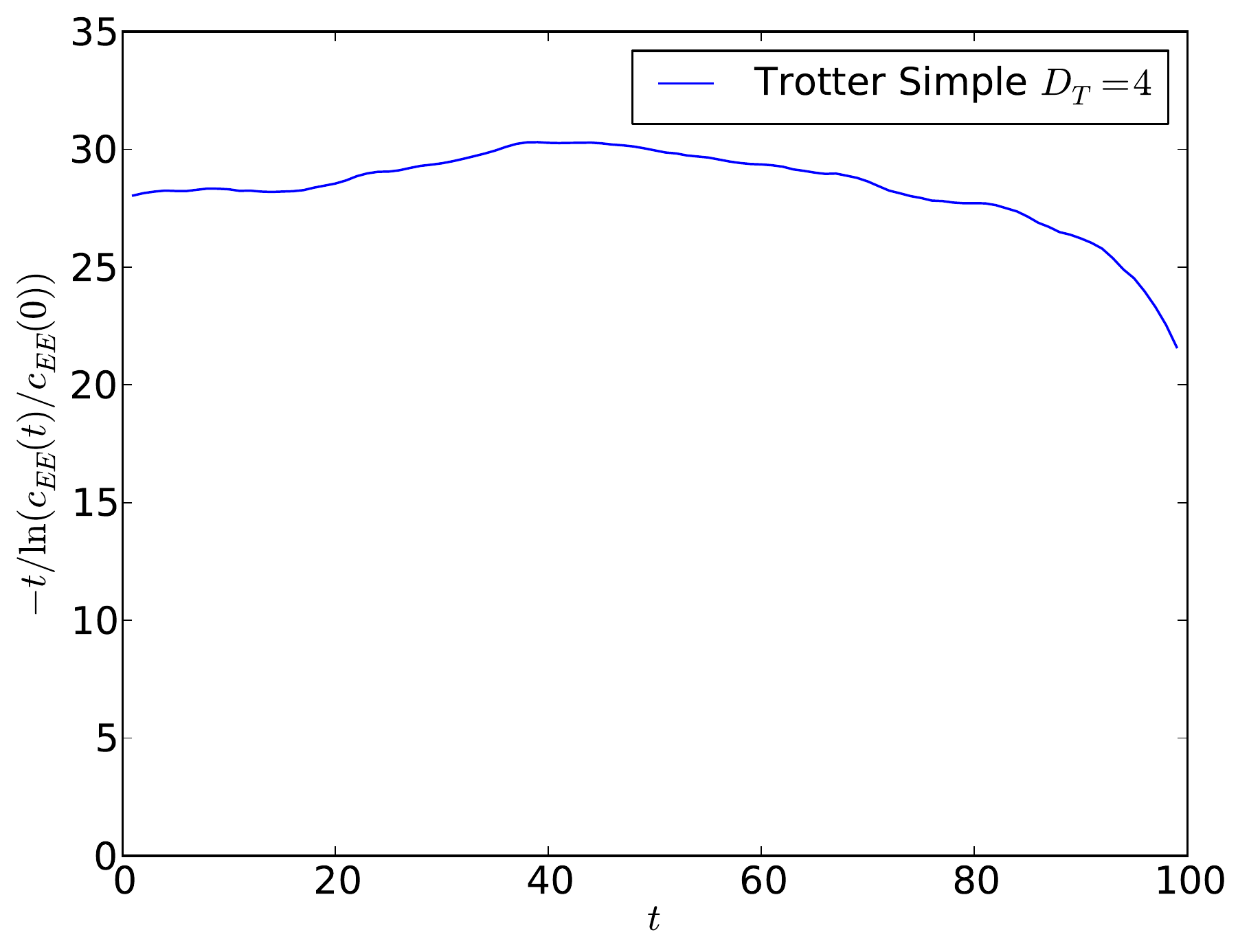}
\caption{\label{MPSQMC-Autocorr2} Estimation of $\tau_{\text{corr}}$ from the autocorrelation function in Fig. \ref{MPSQMC-Autocorr}.}
\end{figure}

\begin{figure}
\centering
\includegraphics[width=0.70\textwidth]{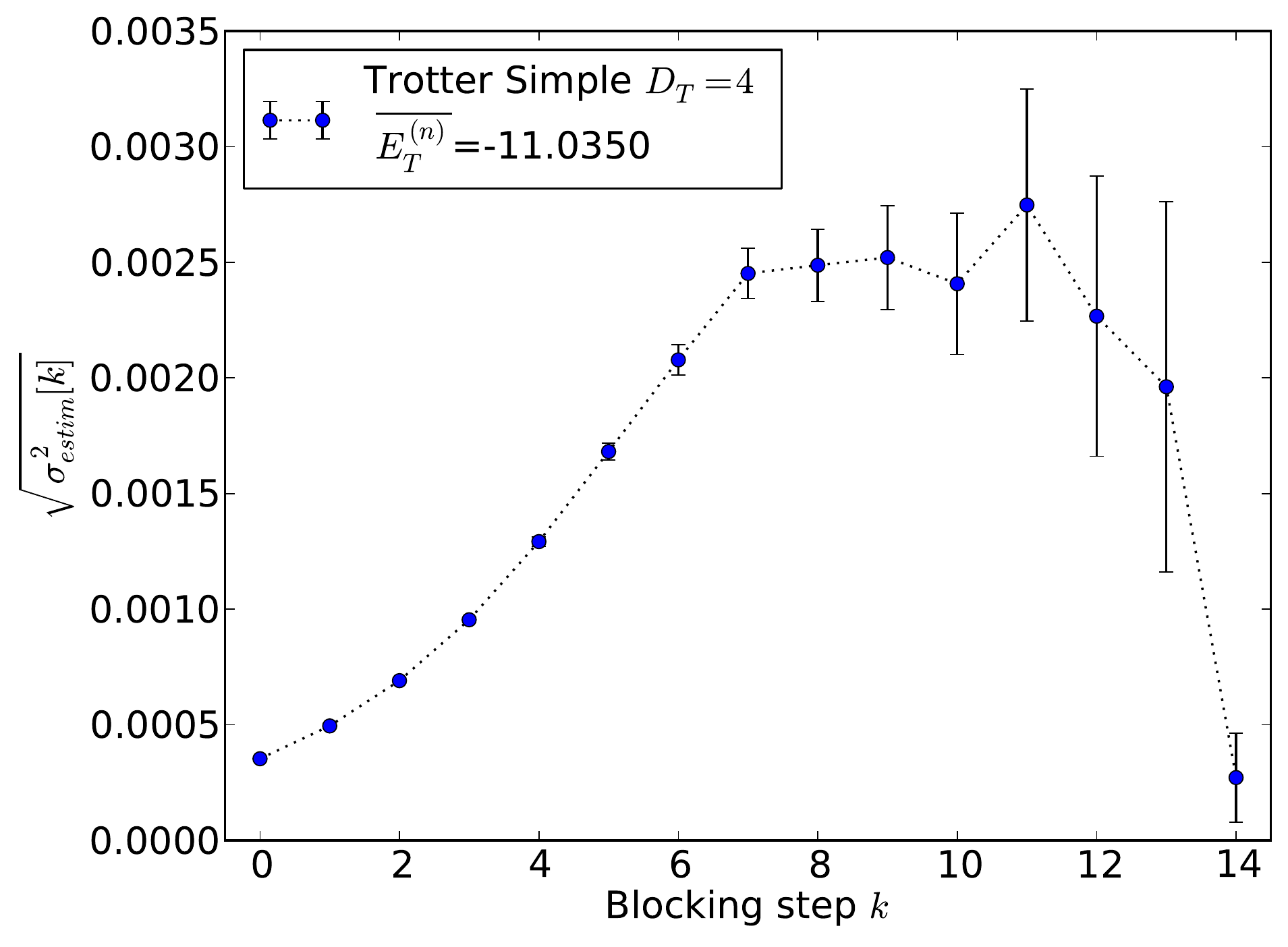}
\caption{\label{MPSQMC-Blocking} Illustration of the blocking method for the projected energies in Fig. \ref{MPSQMC-TimeEvo}.}
\end{figure}

An example of the MC time evolution of the projected energy is shown in Fig. \ref{MPSQMC-TimeEvo}. The projected energy is not variational. After an initial transition period, $E_T^{(n)}$ fluctuates around a mean value. This initial transition period is removed for what follows. The samples in an MC time series $\{ E_1 E_2 ... E_n \}$ can be correlated \cite{Thijssen}. Consider their mean:
\begin{equation}
m = \frac{1}{n} \sum\limits_{k=1}^{n} E_n. \label{MeanValue}
\end{equation}
The autocorrelation function
\begin{equation}
c_{EE}(t) = \frac{1}{n-t} \sum\limits_{k=1}^{n-t} (E_k-m)(E_{k+t}-m) ~ \propto ~ \exp{ \left( - \frac{t}{\tau_{\text{corr}}} \right) }
\end{equation}
allows to estimate the correlation time $\tau_{\text{corr}}$ (measured in number of MC steps), see Figs. \ref{MPSQMC-Autocorr} and \ref{MPSQMC-Autocorr2}.
Two samples can only be considered independent if they are separated by several correlation times $\tau_{\text{corr}}$ in MC time. The statistical error on Eq. \eqref{MeanValue} can be estimated as \cite{Thijssen}
\begin{equation}
\sigma_{\text{stat}}(m) = \sqrt{\frac{2 \tau_{\text{corr}} c_{EE}(0)}{n-1}} \approx \sqrt{ \frac{ 2 \times 28 \times 4.2~10^{-3} }{4.1~10^5} } = 2.4~10^{-3}. \label{FirstNumericalEstimateMCstatError}
\end{equation}
Another way to estimate the statistical error on Eq. \eqref{MeanValue} is the blocking method \cite{BlockingMethod}. Start with blocking step $k=0$. The variance of the (possibly correlated) samples is
\begin{equation}
\sigma^2_{\text{estim}}[k] = \frac{c_0}{n-1}.
\end{equation}
With the blocking transformation:
\begin{eqnarray}
n & \leftarrow & \frac{n}{2}, \\
E_i & \leftarrow & \frac{E_{2i} + E_{2i+1}}{2}, \\
m & \leftarrow & m, \\
k & \leftarrow & k+1,
\end{eqnarray}
the sample size is halved. After sufficient blocking steps the samples become independent, and $\sigma^2_{\text{estim}}[k]$ becomes flat with respect to additional blocking steps $k$. For even more blocking steps, the sample population becomes too small, and $\sigma^2_{\text{estim}}[k]$ becomes noisy. When $\sigma^2_{\text{estim}}[k]$ becomes flat with $k$, the samples $E_i$ are independent gaussian variables (due to the central limit theorem). An error estimate can hence be made of the statistical error \cite{BlockingMethod}:
\begin{equation}
\sigma_{\text{stat}}(m) \approx \sqrt{\sigma^2_{\text{estim}}[k]} \left( 1 \pm \frac{1}{\sqrt{2(n-1)}} \right).
\end{equation}
The blocking method is illustrated in Fig. \ref{MPSQMC-Blocking} for the projected energies in Fig. \ref{MPSQMC-TimeEvo}. It yields $\sigma_{\text{stat}}(m) \approx 2.5~10^{-3}$, which complies well with Eq. \eqref{FirstNumericalEstimateMCstatError}.

\subsection{Systematic error}

\begin{figure}
\centering
\includegraphics[width=0.70\textwidth]{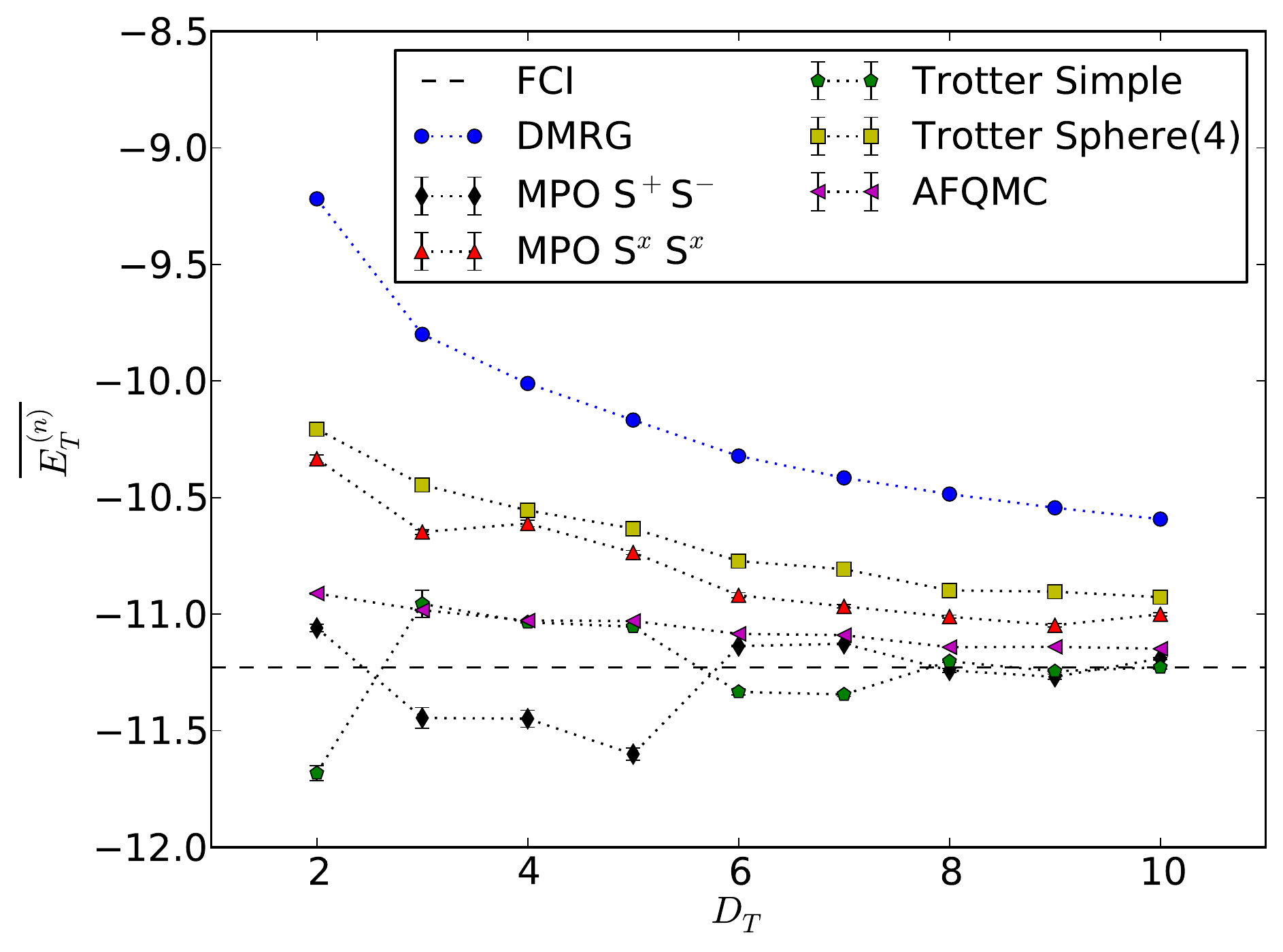}
\caption{\label{MPSQMC-Summary} In the contrained path and phase-free methods, the systematic error due to the trial wavefunction $\ket{\Psi_T}$ can be estimated and/or removed by increasing its virtual dimension $D_T$.}
\end{figure}

The systematic error cannot be estimated by using a single trial wavefunction $\ket{\Psi_T}$. With increasing $D_T$, the optimized trial wavefunction $\ket{\Psi_T}$ becomes a better approximation of the true ground state $\ket{\Psi^*}$, which allows to estimate and/or remove the systematic bias. This is illustrated in Fig. \ref{MPSQMC-Summary}.

For the methods \texttt{MPO S$^x$S$^x$}, \texttt{Trotter Sphere(4)}, and \texttt{AFQMC}, in which the sampled local operators are nonsingular, the systematic bias becomes systematically smaller with increasing $D_T$. Of these three methods, $\texttt{AFQMC}$ performs significantly better. For the methods \texttt{MPO S$^+$S$^-$} and \texttt{Trotter Simple}, in which the sampled local operators $\hat{S}^+$ and $\hat{S}^-$ are singular, the systematic error shows no clear behaviour with increasing $D_T$. The mean projected energies $\overline{E_T^{(n)}}$ are however slightly better than those obtained with \texttt{AFQMC} for larger $D_T$. The systematic error can only be removed if its behaviour as a function of $D_T$ can be predicted. The statistical errors are also the smallest for \texttt{AFQMC}. Based on our exploratory calculations, \texttt{AFQMC} therefore seems the most promising method.

For \texttt{AFQMC} and \texttt{Trotter}, there is also a systematic error due to the finite time step $\delta\tau$. This systematic error can again be estimated and/or removed by considering a few different time steps.

\subsection{Outlook}
We have used \texttt{AFQMC} with MPS walkers to study the $J_1$-$J_2$ model \cite{PhysRevB.86.024424} on larger lattices, in order to investigate the true potential of the method (see Ref. \cite{MPS-AFQMC}). We have observed that the walker virtual dimension can be kept constant at $D_W=2$, without loss of accuracy. Projector MC therefore scales as $\mathcal{O}(D_{T}^2)$ (at least after the $\mathcal{O}(D_{T}^3)$ DMRG ground state calculation has been performed). As can be observed from Fig. \ref{MPSQMC-Summary}, the \texttt{AFQMC} energies are significantly better than the DMRG energies, and the DMRG algorithm would require significantly larger virtual dimensions to yield results of the same accuracy. At the same time, the use of an MPS trial wavefunction allows for a systematic improvement of the nodal plane (fixed-node approximation) by increasing its virtual dimension $D_T$, which allows to estimate and remove the corresponding systematic bias.

DMRG is a very efficient ground state algorithm for the MPS ansatz. For other TNSs, the ground state optimization algorithms are currently less efficient, and it is likely that projector MC can also yield a performance gain for these ansatzes.

\chapter{Summary, conclusions and outlook}
\begin{chapquote}{Arnold H. Glasow}
Happy is the person who knows what to remember of the past, what to enjoy in the present, and what to plan for in the future.
\end{chapquote}

\section{Summary and conclusions} \label{FirstSectionConclusions-sec}
Quantum chemistry tries to predict molecular structure and corresponding energy differences at experimental accuracy. Not all physical effects are relevant for this prediction. The nonrelativistic quantum mechanical description of electrostatically interacting electrons and nuclei is usually sufficient. The electronic motion decouples from the nuclear motion due to the different mass scales of both. Molecular structure prediction therefore boils down to molecular electronic structure calculations. To make calculations feasible, a finite single-particle basis set is introduced, which transforms the Schr\"odinger equation into an algebraic eigenvalue equation.

With $L$ spin-independent spatial orbitals and second quantization, this algebraic equation can be formulated as the diagonalization of the Hamiltonian
\begin{equation*}
\hat{H}_e = E_0 + \sum\limits_{ij} (i | \hat{T} | j ) \sum\limits_{\sigma} \hat{a}_{i\sigma}^{\dagger} \hat{a}_{j\sigma} + \frac{1}{2} \sum\limits_{ijkl} (ij | \hat{V} | kl ) \sum\limits_{\sigma\tau} \hat{a}_{i\sigma}^{\dagger} \hat{a}_{j\tau} ^{\dagger} \hat{a}_{l\tau} \hat{a}_{k\sigma} \tag{\ref{QC-ham}}
\end{equation*}
in the occupation number basis 
\begin{equation*}
\ket{n_{1\uparrow} n_{1\downarrow} n_{2\uparrow} ... n_{L \uparrow} n_{L \downarrow}} = \left( \hat{a}^{\dagger}_{1 \uparrow} \right)^{n_{1\uparrow}} \left( \hat{a}^{\dagger}_{1 \downarrow} \right)^{n_{1\downarrow}} \left( \hat{a}^{\dagger}_{2\uparrow} \right)^{n_{2\uparrow}} ...  \left( \hat{a}^{\dagger}_{L\uparrow} \right)^{n_{L \uparrow}} \left( \hat{a}^{\dagger}_{L\downarrow} \right)^{n_{L \downarrow}}  \ket{-}. \tag{\ref{occupation-number-representation-spin-orbs}}
\end{equation*}
Greek letters denote electron spin projections and Latin letters spin-independent spatial orbitals. With $\ket{n_i} = \ket{n_{i\uparrow}n_{i\downarrow}}$, the exact solution can be written as
\begin{equation*}
\ket{\Psi} = \sum_{\{n_{j} \}} C^{n_{1} n_{2} ... n_{L}} \ket{n_{1} n_{2} ... n_{L}}. \tag{\ref{asdfhasdgjhsajcucahsgasgsjkkskkkk}}
\end{equation*}
For $N$ electrons in $L$ orbitals, the number of variables in this solution grows as $\binom{2L}{N}$, i.e. faster than polynomially in $L$. Approximate solution methods are required. The density matrix renormalization group (DMRG) uses a matrix product state (MPS) ansatz to approximate the $C$-tensor in Eq. \eqref{asdfhasdgjhsajcucahsgasgsjkkskkkk} as a contracted product of matrices
\begin{equation*}
C^{n_{1} n_{2} ... n_{L}} = \sum_{\{ \alpha_k \}} A[1]^{n_{1}}_{\alpha_1} A[2]^{n_{2}}_{\alpha_1 ; \alpha_2} A[3]^{n_{3}}_{\alpha_2 ; \alpha_3} ... A[L-1]^{n_{L-1}}_{\alpha_{L-2} ; \alpha_{L-1}} A[L]^{n_{{L}}}_{\alpha_{L-1}}, \tag{\ref{MPS-A-tensor}}
\end{equation*}
with
\begin{equation*}
 \text{dim}\left( \alpha_j \right) = \min \left( 4^j , 4^{L-j}, D \right). \tag{\ref{MPSapproxSizes}}
\end{equation*}
$D$ is called the virtual dimension of the MPS (with open boundary conditions) and controls the size of the corner of the Hilbert space which can approximated by Eq. \eqref{MPS-A-tensor}.

In chapter \ref{DMRG-QC-chapter}, the DMRG algorithm for quantum chemistry is discussed. DMRG is a renormalization group flow for increasing many-body Hilbert spaces. It can also be formulated as the variational optimization of an MPS. In the thermodynamic limit, the virtual dimension truncation of an MPS results in exponentially decaying correlation functions:
\begin{equation*}
C_{\text{MPS}}(\Delta x) \propto e^{-\alpha \Delta x}, \tag{\ref{MPSexpodecaycorrfuncchap2}}
\end{equation*}
for two sites which are separated by a distance $\Delta x$ on the one-dimensional DMRG lattice. This type of correlation function is typical for ground states of noncritical (gapped) one-dimensional Hamiltonians. For such systems, DMRG works extremely well. The Schmidt decomposition
\begin{equation*}
\ket{\Psi} = \sum_{ij} C_{ij} \ket{A_i} \ket{B_j} = \sum_{ijk} U_{ik} \sigma_k V^{\dagger}_{kj} \ket{A_i} \ket{B_j} = \sum\limits_k \sigma_k \ket{\widetilde{A}_k} \ket{\widetilde{B}_k} \tag{\ref{Schmidt-decomp}}
\end{equation*}
for the bipartition of the one-dimensional lattice in two semi-infinite halves then has a fast-decaying Schmidt spectrum $\sigma_k$, and its truncation is then a good approximation. In quantum chemistry, the active orbital spaces of interest are often far from one-dimensional. DMRG can still be of use, but larger virtual dimensions $D$ are then required.

The gauge freedom of an MPS allows to formulate the simultaneous variational optimization of two neighbouring MPS tensors in Eq. \eqref{MPS-A-tensor}, the so-called micro-iteration, as a numerically stable standard Hermitian eigenvalue problem:
\begin{equation}
\mathbf{H}[i]^{\text{eff}} \mathbf{B}[i] = E_i \mathbf{B}[i]. \tag{\ref{effHameq}}
\end{equation}
Once the lowest energy state of this eigenvalue equation is found, it is decomposed with a singular value decomposition. This decomposition can be related to the Schmidt decomposition of the orbital space. The optimal approximation is obtained by keeping the $D$ largest Schmidt values. This micro-iteration is performed repeatedly at stepwise changing positions in the one-dimensional DMRG lattice, during the so-called sweeps or macro-iterations. DMRG can hence be regarded as a self-consistent mean-field theory in the lattice sites. $\mathbf{H}[i]^{\text{eff}}$ is too large to be fully constructed and only its action on a particular guess $\mathbf{B}[i]$ is available as a function. In order to perform this multiplication efficiently, renormalized operators as well as complementary renormalized operators are constructed. This limits the total cost per macro-iteration to $\mathcal{O}(L^4D^2+L^3D^3)$ in computational time, $\mathcal{O}(L^2D^2)$ in memory, and $\mathcal{O}(L^3D^2)$ in disk. If certain two-body matrix elements $(ij | \hat{V} | kl)$ can be numerically neglected, for example in one-dimensional systems, this cost can be reduced. The use of symmetry reduces this cost as well.

Due to the underlying MPS ansatz, the energies obtained with Eq. \eqref{effHameq} are upper bounds to the exact ground state energy. With increasing virtual dimension $D$, lower energies are obtained, and several successful extrapolation schemes have been proposed. The fundamental difference of the MPS ansatz with a configuration interaction expansion is revealed by taking the Hartree-Fock single-particle states as the orthonormal basis for Eq. \eqref{occupation-number-representation-spin-orbs}. An MPS does not restrict the rank of possible particle excitations relative to the Hartree-Fock reference, but correlates them instead. A configuration interaction expansion restricts the particle-excitation rank, but leaves the allowed excitations uncorrelated. The MPS ansatz is size-consistent for proper orbital orderings.

A renormalization group flow such as DMRG depends on the specific setup. The two-site algorithm is less lickely to get stuck in local minima than its one-site counterpart. Adding noise or perturbative corrections during the initial sweeps helps to reintroduce lost quantum numbers. A good starting guess is also important, as well as the orbital choice and ordering. Thus far, several rules of thumb have been established regarding the latter. For elongated molecules such as polyenes, it is best to use an orthonormal basis of localized orbitals, and to place them according to the molecular topology on the one-dimensional DMRG lattice. The exchange matrix $K_{ij} = (ij | \hat{V} | ji)$ directly reflects the overlap and the distance between localized orbitals, and can be used to order them. Based on the two-orbital mutual information, it was observed that for compact molecules the orbitals are best grouped per irreducible representation (irrep) of the molecular point group, and that bonding and anti-bonding irreps should be placed adjacent. The question regarding the orbital choice and ordering is currently only partially answered, and the author believes that further research in this direction can improve the DMRG algorithm significantly.

DMRG can currently handle active spaces up to 40 electrons in 40 orbitals. It is therefore ideal to replace the full configuration interaction solver in the complete active space self-consistent field method (DMRG-SCF). This allows to capture static correlation in the active space. Dynamic correlation can be added subsequently by perturbation theory (DMRG-CASPT2), a multireference configuration interaction expansion (DMRG-MRCI), or an exponential ansatz (inspired by coupled-cluster theory) such as canonical transformation theory (DMRG-CT). Due to its ability to handle rather large active spaces, DMRG is ideal to tackle large $\pi$-conjugated systems as well as transition metal clusters. Many properties have been studied in a wide variety of systems, and DMRG is a continuously growing field within quantum chemistry.

The symmetry group of the Hamiltonian \eqref{QC-ham} contains $\mathsf{SU(2)}$ spin symmetry, $\mathsf{U(1)}$ particle-number symmetry, and the molecular point-group symmetry $\mathsf{P}$. This symmetry group can be used to reduce the dimensionality of the exact diagonalization problem. The Hamiltonian does not connect states which belong to different irreps, or to different rows of the same irrep. By choosing a basis of symmetry eigenvectors, the Hamiltonian becomes block diagonal, and each block can be diagonalized separately.

In chapter \ref{SYMM-chapter}, we discuss how to construct MPS wavefunctions which are symmetry eigenstates. In our code, we only use the abelian point groups with real-valued character tables:
\begin{equation*}
\mathsf{P} \in \{ C_1, C_i, C_2, C_s, D_2, C_{2v}, C_{2h}, D_{2h} \}. \tag{\ref{PointGroupsInCheMPS2}}
\end{equation*}
The spatial orbitals are then constructed so that they transform according to a particular irrep $I$ of $\mathsf{P}$. To ensure that an MPS is a symmetry eigenstate, the MPS tensors should be irreducible tensor operators of the symmetry group. When the local and virtual basis states are labelled according to the symmetry, it follows from the Wigner-Eckart theorem that each MPS tensor then factorizes into Clebsch-Gordan coefficients and a reduced tensor:
\begin{equation*}
A[i]_{(j_L j_L^z N_L I_L \alpha_L) ; (j_R j_R^z N_R I_R \alpha_R)}^{(s s^z N I)} = \braket{j_L j_L^z s s^z \mid j_R j_R^z} \delta_{N_L + N, N_R} \delta_{I_L \otimes I, I_R} T[i]^{(s N I)}_{(j_L N_L I_L \alpha_L);(j_R N_R I_R \alpha_R)}. \tag{\ref{CheMPS2_WE_MPS}}
\end{equation*}
In this equation, $s$ and $j$ denote the total spin, $s^z$ and $j^z$ the spin projection, $N$ the particle number, and $I$ the point group irrep. Eq. \eqref{CheMPS2_WE_MPS} results in block-sparsity and information compression. If a Clebsch-Gordan coefficient is zero, the corresponding block in the $T[i]$-tensor does not have to be allocated. For the symmetry sector $(j_L N_L I_L)$, there are $D(j_L N_L I_L) = \text{size}(\alpha_L)$ reduced virtual basis states in the $T[i]$-tensor, which effectively represent $(2j_L+1)D(j_L N_L I_L)$ virtual basis states in the $A[i]$-tensor. This block-sparsity and compression lead to reductions in memory and computational time.

The desired global symmetry $(S_G , N_G , I_G)$ can be imposed with the singlet-embedding strategy. The leftmost MPS tensor in the chain then only has the trivial symmetry sector in its left virtual boundary, with reduced virtual dimension 1. The rightmost MPS tensor then only has the symmetry sector $(S_G , N_G , I_G )$ in its right virtual boundary, with reduced virtual dimension 1.

The operators
\begin{equation*}
\hat{b}^{\dagger}_{c \gamma} = \hat{a}^{\dagger}_{c \gamma} \tag{\ref{creaannih1--}}
\end{equation*}
\begin{equation*}
\hat{b}_{c \gamma} = (-1)^{\frac{1}{2}-\gamma}\hat{a}_{c -\gamma} \tag{\ref{creaannih2--}}
\end{equation*}
for orbital $c$ correspond to resp. the $(s=\frac{1}{2}, s^z=\gamma, N=1, I_c)$ row of irrep $(s=\frac{1}{2}, N=1, I_c)$ and the $(s=\frac{1}{2}, s^z=\gamma, N=-1, I_c)$ row of irrep $(s=\frac{1}{2}, N=-1, I_c)$. $\hat{b}^{\dagger}$ and $\hat{b}$ are hence both doublet irreducible tensor operators. This fact permits exploitation of the Wigner-Eckart theorem also for operators and (complementary) renormalized operators. Contracting terms of the type \eqref{CheMPS2_WE_MPS} and \eqref{creaannih1--}-\eqref{creaannih2--} can be done by implicitly summing over the common multiplets and recoupling the local, virtual and operator spins. (Complementary) renormalized operators then formally consist of terms containing Clebsch-Gordan coefficients and reduced tensors. In our code, only the reduced tensors need to be calculated, and Wigner 3-j symbols or Clebsch-Gordan coefficients are never used.

Due to the abelian point group symmetry $\mathsf{P}$, the two-body matrix elements $(ij | \hat{V} | kl)$ of the Hamiltonian \eqref{QC-ham} are only nonzero if $I_i \otimes I_j = I_k \otimes I_l$. If $\mathsf{P}$ is nontrivial, this considerably reduces the number of terms in the construction of complementary renormalized operators, and in the multiplication in Eq. \eqref{effHameq}. The exploitation of non-abelian spatial symmetries is also discussed in chapter \ref{SYMM-chapter}.

In chapter \ref{HCHAIN-chapter}, one-dimensional hydrogen chains were studied. Although the Coulomb interaction is nonlocal, these chains require only a small virtual dimension. Due to the mutual screening of nuclei and electrons, the Coulomb interaction can be considered effectively local in this system, which renders it an ideal test case for DMRG.

Equidistant hydrogen chains,
\begin{equation*}
\vcenter{\hbox{\setlength{\unitlength}{1cm}
\begin{picture}(6.7,1)
\put(0.1,0.375){H}
\put(0.4,0.5){\line(1,0){0.5}}
\put(1.0,0.375){H}
\put(1.3,0.5){\line(1,0){0.5}}
\put(1.9,0.375){H}
\put(2.2,0.5){\line(1,0){0.5}}
\put(2.8,0.375){H}
\put(3.1,0.5){\line(1,0){0.5}}
\put(3.7,0.375){H}
\put(4.0,0.5){\line(1,0){0.5}}
\put(4.6,0.375){H}
\put(4.9,0.5){\line(1,0){0.5}}
\put(5.5,0.375){H}
\put(5.8,0.5){\line(1,0){0.5}}
\put(6.4,0.375){H}
\put(0.5,0.6){$R$}
\put(1.4,0.6){$R$}
\put(2.3,0.6){$R$}
\put(3.2,0.6){$R$}
\put(4.1,0.6){$R$}
\put(5.0,0.6){$R$}
\put(5.9,0.6){$R$}
\end{picture}}} , \tag{\ref{JCP136equallySpacedFormula}}
\end{equation*}
exhibit a large amount of static correlation for large internuclear distances $R$. Band theory predicts that this system is a conductor, while for large internuclear distances the electrons are localized on the nuclei. The system is therefore a Mott insulator. When the internuclear distance decreases, the system goes trough a metal-insulator transition. The electrons become more delocalized, and as the phase transition is approached (from the insulator region), the reduced Schmidt spectrum starts to decay more slowly. This can be explained by the excitation gap, which closes in the metallic regime. In this regime, two other quantities simultaneously diverge: the static dipole polarizability per electron and the fluctuation of the dipole moment per electron. The rate of divergence with increasing system size was obtained for both quantities at several internuclear distances. Both metallic and insulating regions were observed.

The equidistant hydrogen chain cannot exist due to the Peierls instability, and it dimerizes into $H_2$ molecules. Consider the one-dimensional dimerized hydrogen chain,
\begin{equation*}
\vcenter{\hbox{\setlength{\unitlength}{1cm}
\begin{picture}(6.7,1)
\put(0.1,0.375){H}
\put(0.4,0.5){\line(1,0){0.5}}
\put(1.0,0.375){H}
\put(1.3,0.45){......}
\put(1.9,0.375){H}
\put(2.2,0.5){\line(1,0){0.5}}
\put(2.8,0.375){H}
\put(3.1,0.45){......}
\put(3.7,0.375){H}
\put(4.0,0.5){\line(1,0){0.5}}
\put(4.6,0.375){H}
\put(4.9,0.45){......}
\put(5.5,0.375){H}
\put(5.8,0.5){\line(1,0){0.5}}
\put(6.4,0.375){H}
\put(0.5,0.6){$R_f$}
\put(1.4,0.6){$R$}
\put(2.3,0.6){$R_f$}
\put(3.2,0.6){$R$}
\put(4.1,0.6){$R_f$}
\put(5.0,0.6){$R$}
\put(5.9,0.6){$R_f$}
\end{picture}}} , \tag{\ref{JCP136jetajetaDimerizedHchain}}
\end{equation*}
with intramolecular distance $R_f$ and intermolecular distance $R$. For fixed intramolecular distance (with $R_f < R$) the static correlation remains roughly the same, and only the electron delocalization changes with $R$. We have studied the response of this system to an external electric field by means of the longitudinal static dipole polarizability $\alpha_{zz}$ and second hyperpolarizability $\gamma_{zzzz}$, obtained with the finite-field method. In the insulating regime, both quantities eventually grow linearly with chain length, and the values per molecule can be extrapolated to the thermodynamic limit. We have observed that the minimal basis set STO-6G does not suffice. Inclusion of an extra $s$-orbital per atom (6-31G basis) improved the results significantly. The additional inclusion of three extra $p$-orbitals (6-31G(d,p) basis) had only a minor effect. Coupled cluster theory with single and double excitations, and with triple excitations in perturbation (CCSD(T)), is at present the golden standard of quantum chemistry. To study the performance of other approximative methods, CCSD(T) is often used as reference. In our study, the DMRG results are indistinguishable from exact diagonalization, and they were used to assess the performance of CCSD(T) to calculate the static dipole second hyperpolarizability. As the electrons become more delocalized, i.e. with decreasing intermolecular distance, the performance of CCSD(T) decays. With increasing electron delocalization, an increasing amount of electrons are involved in the response to an electric field, and CCSD(T) can only correlate a limited amount of electrons. The power law
\begin{equation*}
 \gamma(M) \propto M^{a(M)} \tag{\ref{JCP136theScalingRelationForTheSecondHyperpol}}
\end{equation*}
is often proposed for the increase of $\gamma_{zzzz}$ with the number of molecules $M$ in the chain. The parameter $a(M)$ varies only slowly with $M$. For the systems and basis sets for which $a(M)$ was nearly equal to one for the largest chains under consideration, we were able to accurately extrapolate the longitudinal static second hyperpolarizability to the thermodynamic limit. The polarizability could be accurately extrapolated to the thermodynamic limit for all the systems and basis sets we have studied.

In chapter \ref{C2-chapter}, we studied the carbon dimer. Its $\pi$-bonds are of the charge-shift type, and it has recently been debated whether the carbon dimer has a quadruple bond. The ground state has significant multireference character, and many crossings and avoided crossings occur between its low-lying states. The carbon dimer is therefore a good system to test the capabilities of our DMRG code. We have obtained the twelve lowest bond dissociation curves to $0.01~mE_h$ accuracy in the cc-pVDZ basis. Due to the $\mathsf{SU(2)}$, $\mathsf{U(1)}$, and abelian $\mathsf{P}$ symmetries in our code, these states were mainly resolved by targeting different symmetry sectors of the many-body Hilbert space. The particle number was always $N=12$, and $^S\mathsf{D_{2 h}}$ symmetry (right) was used to target the original $^S\mathsf{D_{\infty h}}$ symmetry (left):
\begin{equation*}
X^1\Sigma_g^+ ; B^1 \Delta_g ; B'^1\Sigma_g^+ ~ \rightarrow ~ ^1A_g \tag{\ref{CPCmeuh1}}
\end{equation*}
\begin{equation*}
c^3\Sigma_u^+ ; 1^3\Delta_u ; 2^3\Sigma_u^+ ~ \rightarrow ~ ^3B_{1u} \tag{\ref{CPCmeuh2}}
\end{equation*}
\begin{equation*}
C^1\Pi_g ~ \rightarrow ~ ^1B_{2g} \tag{\ref{CPCfirstOfThed2hIrreps}}
\end{equation*}
\begin{equation*}
A^1\Pi_u ~ \rightarrow ~ ^1B_{2u} \tag{\ref{CPCmeuh3}}
\end{equation*}
\begin{equation*}
1^1\Sigma_u^- ~ \rightarrow ~ ^1A_u \tag{\ref{CPCmeuh4}}
\end{equation*}
\begin{equation*}
b^3 \Sigma_g^- ~ \rightarrow ~ ^3B_{1g} \tag{\ref{CPCmeuh5}}
\end{equation*}
\begin{equation*}
d^3\Pi_g ~ \rightarrow ~ ^3B_{2g} \tag{\ref{CPCmeuh6}}
\end{equation*}
\begin{equation*}
a^3\Pi_u ~ \rightarrow ~ ^3B_{2u}. \tag{\ref{CPClastOfThed2hIrreps}}
\end{equation*}
Within the $^1A_g$ and $^3B_{1u}$ sectors, two excited states were calculated by projecting out lower-lying eigenstates. Their $\mathsf{D_{\infty h}}$ symmetries were distinguished by looking at the relative phases of the coefficients of two Slater determinants. We have also estimated the importance of core and core-valence correlation by comparing DMRG(36o,12e) and DMRG-SCF(34o,8e) calculations in the cc-pCVDZ basis. In the DMRG-SCF calculations, the two $1s$ orbitals of the carbon atoms were kept doubly occupied. The non-parallellity error was of the order of $2~mE_h$, while the energy difference between both curves was about $75~mE_h$. To capture the core and core-valence correlation, as well as the correct core dynamics in the united atom limit (small internuclear distance), the basis set should be augmented with extra orbitals for the closed shells.

DMRG can resolve excited states by projecting out lower-lying eigenstates or by targeting a specific energy. These are state-specific DMRG algorithms, because the whole virtual basis is used to represent one single eigenstate. In state-averaged DMRG, the virtual basis is constructed to target several eigenstates at once. Linear response theory for DMRG (DMRG-LRT) can be used as well to find excited states. DMRG-LRT is discussed in chapter \ref{Thouless-chapter}.

In that chapter, we explore the analogy between Hartree-Fock theory and DMRG. Both methods can be formulated as the variational optimization of a wavefunction ansatz, a Slater determinant for Hartree-Fock theory and an MPS for DMRG. The time-independent variational principle yields self-consistent mean-field equations for the particles in Hartree-Fock theory, and for the lattice sites in DMRG. The gauge invariance of the wavefunction can be used to simplify the self-consistent mean-field equations to standard eigenvalue problems. In Hartree-Fock theory, the gauge freedom can be used to construct single-particle states which are eigenvectors of the Fock operator. In DMRG, the gauge freedom can be used to construct MPS site tensors which are eigenvectors of the effective Hamiltonian.

The time-dependent variational principle generates time-evolution equations for the wavefunction, which stay within the ansatz space. Linearization of these equations near a variational minimum leads to the random-phase approximation (RPA). A small time step connects a wavefunction with its first order tangent space. For Hartree-Fock theory, this space is spanned by the single-particle excitations, which correspond to the replacement of occupied orbitals with virtual orbitals. For DMRG, this space is spanned by the single-site excitations, which correspond to the replacement of retained virtual basis states with discarded ones. Exponentiation of the single-particle excitations leads to the Thouless theorem for Hartree-Fock theory, an explicit nonredundant parameterization for the entire manifold of Slater determinants. We have proven the DMRG counterpart in chapter \ref{Thouless-chapter}. Expansion of the Thouless theorem leads to the configuration interaction expansion. For Hartree-Fock theory, the first-order terms yield the configuration interaction with singles (CIS), also called the Tamm-Dancoff approximation (TDA). The same names are used for DMRG: DMRG-CIS or DMRG-TDA. A variational optimization in this tangent space yields approximate excited states. For Hartree-Fock theory, the second-order terms yield the configuration interaction with singles and doubles (CISD). The same name is used for DMRG: DMRG-CISD. A variational optimization in this space yields both an improved description of the ground state, as well as approximate excited states.

We have performed DMRG-CISD calculations for the one-dimensional Hubbard chain. For small virtual dimensions, the variational DMRG-CISD ground state energy is significantly lower than the DMRG result, and the DMRG-CISD excitation energies are also better than the DMRG-TDA energies. Delocalized two-site excitations are not captured by DMRG-TDA, but these can be retrieved by DMRG-CISD. This has to be compared with Hartree-Fock theory, where TDA does not capture two-particle excitations, while CISD does. We have also successfully calculated low-lying singlet excited states of polyenes with both DMRG-TDA and DMRG-RPA. The $\pi$-system of these molecules was parameterized with the Pariser-Parr-Pople Hamiltonian.

DMRG-RPA is able to retrieve the Goldstone boson for a ground state which breaks a continuous symmetry. We have performed a proof-of-principle calculation for a spin doublet ground state of the one-dimensional Hubbard model.

In chapter \ref{DMC-MPS-chapter}, projector Monte Carlo for MPS wavefunctions is discussed. The eigenvector $\ket{\Psi^*}$ with largest eigenvalue in magnitude of a Hermitian operator $\hat{K}$ can be retrieved by repeated application of $\hat{K}$ on an initial wavefunction $\ket{\Psi^{(0)}}$. At each Monte Carlo time step $n$, the wavefunction is approximated by an ensemble of walkers:
\begin{equation*}
\ket{\Psi^{(n)}} = (\hat{K})^n \ket{\Psi^{(0)}} \approx \sum_{\phi} \ket{\phi}. \tag{\ref{Ch7ReferenceForConclusion1}}
\end{equation*} 
The operator $\hat{K}$ is decomposed into a probability distribution function $P(\mathbf{x})$ and a set of operators $\hat{B}(\mathbf{x})$:
\begin{equation*}
\hat{K} = \sum_{\mathbf{x}} P(\mathbf{x}) \hat{B}(\mathbf{x}). \tag{\ref{origPropagator}}
\end{equation*}
For the method to be successful, the operators $\hat{B}(\mathbf{x})$ are chosen so that they do not increase the complexity of the walkers. At each Monte Carlo time step $n$, for each walker $\ket{\phi}$, an $\mathbf{x}$ is drawn from $P(\mathbf{x})$, and the walker is updated with $\hat{B}(\mathbf{x})$. With this stochastic propagation of the ensemble, an approximation for $\ket{\Psi^*}$ is obtained. When the walkers are Slater determinants, a possible (and often used) decomposition for $\hat{K} = e^{-\delta\tau \hat{H}}$ can obtained with the Hubbard-Stratonovich transformation.

For real-valued (complex-valued) parameterizations, the fermion sign (phase) problem can be removed by constraining the walker paths with a trial wavefunction $\ket{\Psi_T}$. This introduces a systematic bias into the Monte Carlo propagation, which has an effect on estimators such as, for example, the projected energy. The magnitude of this systematic bias depends on how good $\ket{\Psi_T}$ represents $\ket{\Psi^*}$. For an exact trial wavefunction, the systematic bias vanishes.

The walkers and the trial wavefunction are typically of the same ansatz type, as this allows to calculate overlaps and expectation values cheaply. The advantage of MPS trial wavefunctions over Slater determinants is that they can be systematically improved by increasing their virtual dimension. When the corresponding systematic bias shows a clear decreasing trend with increasing virtual dimension, it might be possible to estimate and remove this bias.

In chapter \ref{DMC-MPS-chapter}, we have studied the spin-$\frac{1}{2}$ Heisenberg model on a $4 \times 4$ torus. Three decompositions \eqref{origPropagator} were tested: (1) sampling the virtual bonds of the matrix product operator for $\hat{K}$; (2) performing a Trotter decomposition of $\hat{K} = e^{-\delta\tau \hat{H}}$, and sampling the virtual bonds of the corresponding two-site matrix product operators; and (3) performing a Hubbard-Stratonovich transformation (for spin systems), which yields a set of auxiliary fields to sample. We have observed that for fixed gauge choices of the matrix product operator, it is best to avoid singular local operators such as $\hat{S}^+$, as then no clear behaviour of the systematic bias was obtained with increasing virtual dimension. Of the three tested methods, the auxiliary-field quantum Monte Carlo variant seems the most promising.

\section{Outlook}
The DMRG algorithm is well understood by means of the underlying MPS wavefunction. This allows to assess DMRG with concepts from quantum information theory. The large-$D$ regime of DMRG, when the solution becomes quasi-exact, has been explored extensively. Accurate extrapolation schemes are known for the evolution of the variational energy with increasing virtual dimension $D$, or with decreasing discarded weight. The use of symmetry to reduce the computational cost is also well understood. Most progress in the large-$D$ regime can still be made in the orbital choice and ordering.

The question regarding the optimal orbital choice and ordering can be addressed with the two-orbital mutual information. As its gradient and Hessian with respect to orbital rotations can be evaluated efficiently, it might be worthwhile to test whether a Newton-Raphson optimization of this information measure can yield extra rules of thumb.

Planned future improvements of our DMRG code include the optimization of the DMRG-SCF loop around the DMRG algorithm, an MPI implementation of the DMRG algorithm, and the inclusion of dynamic correlation on top of the DMRG-SCF loop.

We are currently performing DMRG-SCF calculations on the oxo-Mn(salen) complex. A large active space ($\geq$ 40 orbitals) was selected around the Fermi level, and a subsequent intermediate-$D$ DMRG calculation was used to find approximate natural orbitals and corresponding occupation numbers. Based on this initial calculation, a smaller active space was selected and optimized with DMRG-SCF. The active space of 17 orbitals from Gagliardi \textit{et al.} \cite{Gagliardi} was thereby augmented with two extra $\pi$-orbitals in the $\pi$-conjugated backbone. A total of ten $\pi$-orbitals is then explicitly correlated for this backbone consisting of ten atoms (six carbon, two nitrogen, and two oxygen atoms). We also found that the $3d_{x^2-y^2}$ orbital of manganese interacts with the in-plane $\pi$-orbitals of the oxygen and nitrogen atoms of this backbone, which required the inclusion of three additional orbitals in the active space. By adding dynamic correlation (and relativistic effects), DMRG should then be able to settle the discussion on the relative energies of the singlet, triplet and quintet states of this complex.

Planned applications for the future include a more thorough study of the metal-insulator transition in equidistant hydrogen chains, an assessment of the correlation between the transition metal atoms and the $\pi$-conjugated backbone in metal-organic frameworks (and its implications for DFT studies), and tackling the active space of the bucky ball (60 electrons in 60 $\pi$-orbitals) with the MPI implementation of our DMRG code.

The exploration of post-DMRG methods has only started recently, and there is much room for improvement. We have performed promising auxiliary-field quantum Monte Carlo calculations with MPS walkers. The extension to other tensor network states still has to be explored. The calculation of low-lying excitations with DMRG-TDA fails if such an excitation involves two orbitals which lie far apart on the one-dimensional DMRG lattice. Low-rank decompositions of the DMRG-CISD ansatz should allow to resolve this problem.

For one-dimensional chemical systems such as all-trans polyenes, a uniform MPS ansatz can be proposed directly in the thermodynamic limit. Because an MPS only captures exponentially decaying correlation functions (see Eq. \eqref{MPSexpodecaycorrfuncchap2}), there is a cutoff distance beyond which the required density correlations $\braket{\hat{n}_i \hat{n}_{i+k}}$ for the Coulomb interaction numerically factorize to $\braket{\hat{n}_i} \braket{\hat{n}_{i+k}}$. In conjunction with the ideas introduced in section \ref{QS-DMRG-section}, this might render the uniform MPS a workable ansatz for ab initio quantum chemistry.

\bibliographystyle{unsrtnat}
\bibliography{PhDbibliography}

\chapter*{Nederlandstalige samenvatting}
\addcontentsline{toc}{chapter}{Nederlandstalige samenvatting}
\markboth{{Nederlandstalige samenvatting}}{{Nederlandstalige samenvatting}}
\begin{chapquote}{Eric van der Steen}
Een boek: het stoffelijk overschot van een idee.
\end{chapquote}

Kwantumchemie probeert moleculaire structuur en de corresponderende energieverschillen te voorspellen tot op experimentele nauwkeurigheid. Niet alle fysische effecten zijn daarvoor van belang. De niet-relativistische kwantummechanische beschrijving van elektrostatisch interagerende elektronen en kernen is doorgaans voldoende. De elektronische beweging ontkoppelt van de kernbeweging door hun sterk verschillende massas. De voorspelling van moleculaire structuur komt daarom neer op de berekening  van elektronische structuur. Om berekeningen mogelijk te maken wordt een eindige \'e\'endeeltjesbasis ge\"introduceerd die de Schr\"odingervergelijking transformeert in een algebra\"ische eigenwaardevergelijking.

Met $L$ spinonafhankelijke ruimtelijke orbitalen en tweede kwantisatie kan deze algebra\"ische vergelijking geformuleerd worden als de diagonalisatie van de Hamiltoniaan
\begin{equation*}
\hat{H}_e = E_0 + \sum\limits_{ij} (i | \hat{T} | j ) \sum\limits_{\sigma} \hat{a}_{i\sigma}^{\dagger} \hat{a}_{j\sigma} + \frac{1}{2} \sum\limits_{ijkl} (ij | \hat{V} | kl ) \sum\limits_{\sigma\tau} \hat{a}_{i\sigma}^{\dagger} \hat{a}_{j\tau} ^{\dagger} \hat{a}_{l\tau} \hat{a}_{k\sigma} \tag{\ref{QC-ham}}
\end{equation*}
in de bezettingsgetalbasis
\begin{equation*}
\ket{n_{1\uparrow} n_{1\downarrow} n_{2\uparrow} ... n_{L \uparrow} n_{L \downarrow}} = \left( \hat{a}^{\dagger}_{1 \uparrow} \right)^{n_{1\uparrow}} \left( \hat{a}^{\dagger}_{1 \downarrow} \right)^{n_{1\downarrow}} \left( \hat{a}^{\dagger}_{2\uparrow} \right)^{n_{2\uparrow}} ...  \left( \hat{a}^{\dagger}_{L\uparrow} \right)^{n_{L \uparrow}} \left( \hat{a}^{\dagger}_{L\downarrow} \right)^{n_{L \downarrow}}  \ket{-}. \tag{\ref{occupation-number-representation-spin-orbs}}
\end{equation*}
Griekse letters duiden de spinprojecties van de elektronen aan en Latijnse letters de spinonafhankelijke ruimtelijke orbitalen. Met $\ket{n_i} = \ket{n_{i\uparrow}n_{i\downarrow}}$ kan de exacte oplossing geschreven worden als
\begin{equation*}
\ket{\Psi} = \sum_{\{n_{j} \}} C^{n_{1} n_{2} ... n_{L}} \ket{n_{1} n_{2} ... n_{L}}. \tag{\ref{asdfhasdgjhsajcucahsgasgsjkkskkkk}}
\end{equation*}
Voor $N$ elektronen in $L$ orbitalen groeit het aantal variabelen in deze oplossing als $\binom{2L}{N}$, dus sneller dan polynomiaal in $L$. Benaderende oplossingsmethoden zijn derhalve nodig. De dichtheidsmatrixrenormalisatiegroep (DMRG) gebruikt als aanzet een matrixproducttoestand (MPS) om de $C$-tensor in vgl. \eqref{asdfhasdgjhsajcucahsgasgsjkkskkkk} te benaderen als een gecontraheerd product van matrices
\begin{equation*}
C^{n_{1} n_{2} ... n_{L}} = \sum_{\{ \alpha_k \}} A[1]^{n_{1}}_{\alpha_1} A[2]^{n_{2}}_{\alpha_1 ; \alpha_2} A[3]^{n_{3}}_{\alpha_2 ; \alpha_3} ... A[L-1]^{n_{L-1}}_{\alpha_{L-2} ; \alpha_{L-1}} A[L]^{n_{{L}}}_{\alpha_{L-1}}, \tag{\ref{MPS-A-tensor}}
\end{equation*}
met
\begin{equation*}
 \text{dim}\left( \alpha_j \right) = \min \left( 4^j , 4^{L-j}, D \right). \tag{\ref{MPSapproxSizes}}
\end{equation*}
De variabele $D$ wordt de virtuele dimensie van de MPS (met open grenscondities) genoemd en controleert de grootte van de ``hoek'' van de Hilbertruimte die kan benaderd worden door vgl. \eqref{MPS-A-tensor}.

In hoofdstuk \ref{DMRG-QC-chapter} wordt het DMRG-algoritme besproken. DMRG is een renormalisatiegroep voor groeiende veeldeeltjes-Hilbertruimten. Het kan ook geformuleerd worden als de variationele optimalisatie van een MPS. In de thermodynamische limiet zorgt de truncatie van de virtuele dimensie van een MPS voor exponentieel afnemende correlatiefuncties:
\begin{equation*}
C_{\text{MPS}}(\Delta x) \propto e^{-\alpha \Delta x}, \tag{\ref{MPSexpodecaycorrfuncchap2}}
\end{equation*}
voor twee roosterplaatsen die zich op een afstand $\Delta x$ op het \'e\'endimensionaal DMRG-rooster van elkaar bevinden. Dit type van correlatiefunctie is typisch voor grondtoestanden van niet-kritische \'e\'endimensionale Hamiltonianen. Voor zulke systemen werkt DMRG zeer goed. De Schmidt-decompositie
\begin{equation*}
\ket{\Psi} = \sum_{ij} C_{ij} \ket{A_i} \ket{B_j} = \sum_{ijk} U_{ik} \sigma_k V^{\dagger}_{kj} \ket{A_i} \ket{B_j} = \sum\limits_k \sigma_k \ket{\widetilde{A}_k} \ket{\widetilde{B}_k} \tag{\ref{Schmidt-decomp}}
\end{equation*}
voor de bipartitie van het \'e\'endimensionaal rooster in twee halfoneindige delen heeft dan een snel afnemend Schmidt-spectrum $\sigma_k$. De truncatie ervan is bijgevolg een goede benadering. In kwantumchemie zijn de actieve orbitaalruimtes waarin men ge\"interesseerd is vaak allesbehalve \'e\'endimensionaal. DMRG kan in dat geval nog steeds nuttig zijn, maar grotere virtuele dimensies $D$ zijn dan nodig.

De golffunctievrijheid van een MPS laat toe om de simultane variationele optimalisatie van twee naburige MPS-tensoren in vgl. \eqref{MPS-A-tensor}, de zogenaamde micro-iteratie, te formuleren als een numeriek stabiel standaard Hermitisch eigenwaardeprobleem:
\begin{equation}
\mathbf{H}[i]^{\text{eff}} \mathbf{B}[i] = E_i \mathbf{B}[i]. \tag{\ref{effHameq}}
\end{equation}
Eens de laagste energietoestand van dit eigenwaardeprobleem gevonden is, wordt het ontleed met een singuliere-waardendecompositie. Deze decompositie kan gerelateerd worden aan de Schmidt-decompositie van de orbitaalruimte. De optimale benadering wordt bekomen door de $D$ grootste Schmidt-waarden te behouden. Deze micro-iteratie wordt herhaaldelijk uitgevoerd op stapsgewijs veranderende lokaties in het \'e\'endimensionaal DMRG-rooster, tijdens de zogenaamde macro-iteraties. DMRG kan dus gezien worden als een zelfconsistente gemiddeld-veldtheorie in de roosterplaatsen. $\mathbf{H}[i]^{\text{eff}}$ is te groot om volledig te construeren, en enkel het matrix-vectorproduct met een gegeven vector $\mathbf{B}[i]$ is beschikbaar als een functie. Om dit product effici\"ent uit te voeren, worden gerenormaliseerde operatoren, alsook hun complement, geconstrueerd. Dit limiteert de totale kost per macro-iteratie tot $\mathcal{O}(L^4D^2+L^3D^3)$ in rekentijd, $\mathcal{O}(L^2D^2)$ in werkgeheugen, en $\mathcal{O}(L^3D^2)$ in harde-schijfgeheugen. Als bepaalde tweedeeltjesmatrixelementen $(ij | \hat{V} | kl)$ numeriek verwaarloosd kunnen worden, bvb. in \'e\'endimensionale systemen, kan deze kost gereduceerd worden. Het gebruik van symmetrie reduceert deze kost ook.

Door de onderliggende MPS-aanzet zijn de energie\"en die bekomen worden met vgl. \eqref{effHameq} bovengrenzen voor de exacte grondtoestandsenergie. Met toenemende virtuele dimensie $D$ worden lagere energie\"en bekomen, en verschillende succesvolle extrapolatieschemas zijn gekend. Het fundamenteel verschil tussen een MPS-aanzet en een configuratie-interactie-expansie kan men begrijpen door Hartree-Fock-\'e\'endeeltjestoestanden als de orthonormale basis voor vgl. \eqref{occupation-number-representation-spin-orbs} te nemen. Een MPS beperkt de deeltjesexcitatiegraad t.o.v. de Hartree-Fock-referentie (HF-referentie) niet, maar in de plaats daarvan worden de excitaties gecorreleerd. Een configuratie-interactie-expansie beperkt de deeltjesexcitatiegraad, maar correleert de toegelaten excitaties niet. De MPS-aanzet is grootte-consistent, bij een goede orbitaalvolgorde.

Het resultaat van een renormalisatiegroep zoals DMRG hangt af van de details. Het algoritme met twee roosterplaatsen raakt minder snel vast in lokale minima dan zijn tegenhanger met \'e\'en roosterplaats. Door ruis of perturbatieve correcties toe te voegen gedurende de initi\"ele macro-iteraties kunnen verdwenen kwantumgetallen opnieuw ge\"introduceerd worden. Een goed startpunt is ook belangrijk, alsook de orbitaalkeuze en -ordening. Er zijn verschillende vuistregels in omloop. Voor uitgerekte molecules zoals polyenen is het best om gelokaliseerde orthonormale orbitalen te gebruiken, en ze te plaatsen op het \'e\'endimensionaal DMRG-rooster volgens de moleculaire topologie. De uitwisselingsmatrix $K_{ij} = (ij | \hat{V} | ji)$ weerspiegelt de overlapping en de afstand tussen gelokaliseerde orbitalen op een directe manier, en kan gebruikt worden om ze te ordenen. Op basis van de wederzijdse informatie tussen twee orbitalen werd er waargenomen dat voor compacte moleculen de orbitalen best gegroepeerd worden per irreducibele representatie (irrep) van de moleculaire puntgroep, en dat bindende en antibindende irreps best naast elkaar geplaatst worden. Het probleem van de orbitaalkeuze en -ordening is op dit moment slechts deels opgelost, en verder onderzoek in deze richting kan het DMRG-algoritme significant verbeteren.

Op dit moment kan DMRG actieve ruimtes van 40 elektronen in 40 orbitalen behandelen. Het is daarom uitermate geschikt om de exacte oplossingsmethode te vervangen in de zelfconsistente veldmethode voor de volledige actieve ruimte (DMRG-SCF). Dit laat toe om statische correlatie in de actieve ruimte te vatten. Dynamische correlatie kan achteraf toegevoegd worden door perturbatietheorie (DMRG-CASPT2), een configuratie-interactie-expansie voor multireferentietoestanden (DMRG-MRCI), of een exponenti\"ele aanzet (ge\"inspireerd door de theorie van gekoppelde clusters) zoals canonische transformatietheorie (DMRG-CT). Door de mogelijkheid om grote actieve ruimtes te behandelen is DMRG ideaal om systemen met een groot geconjugeerd $\pi$-systeem of met transitiemetalen te bestuderen. Een grote vari\"eteit aan eigenschappen en systemen werd reeds bestudeerd met DMRG, en het is een sterk groeiend onderzoeksdomein binnen de kwantumchemie.

De symmetriegroep van de Hamiltoniaan \eqref{QC-ham} bestaat uit $\mathsf{SU(2)}$ spinsymmetrie, $\mathsf{U(1)}$ deeltjesaantalsymmetrie, en de moleculaire puntgroepsymmetrie $\mathsf{P}$. Deze symmetriegroep kan gebruikt worden om het exacte diagonalisatieprobleem te vereenvoudigen. De Hamiltoniaan verbindt geen toestanden die tot verschillende irreps of tot verschillende rijen van \'e\'enzelfde irrep behoren. Door een basis van symmetrievectoren te kiezen wordt de Hamiltoniaan blokdiagonaal. Elk blok kan dan apart gediagonaliseerd worden.

In hoofdstuk \ref{SYMM-chapter} wordt uitgelegd hoe een MPS geconstrueerd kan worden die tevens een eigenvector is van de symmetrie. In onze code worden enkel de abelse puntgroepen met re\"eelwaardige karaktertabellen gebruikt:
\begin{equation*}
\mathsf{P} \in \{ C_1, C_i, C_2, C_s, D_2, C_{2v}, C_{2h}, D_{2h} \}. \tag{\ref{PointGroupsInCheMPS2}}
\end{equation*}
De ruimtelijke orbitalen worden geconstrueerd zodat ze transformeren volgens een bepaalde irrep $I$ van $\mathsf{P}$. Om te verzekeren dat een MPS een eigenvector is van de symmetrie, moeten de MPS-tensoren irreducibele tensoroperatoren zijn. Als de lokale en virtuele basistoestanden eigentoestanden zijn van de symmetrie, volgt uit het Wigner-Eckart theorema dat elke MPS-tensor factoriseert in Clebsch-Gordan-co\"effici\"enten en een gereduceerde tensor:
\begin{equation*}
A[i]_{(j_L j_L^z N_L I_L \alpha_L) ; (j_R j_R^z N_R I_R \alpha_R)}^{(s s^z N I)} = \braket{j_L j_L^z s s^z \mid j_R j_R^z} \delta_{N_L + N, N_R} \delta_{I_L \otimes I, I_R} T[i]^{(s N I)}_{(j_L N_L I_L \alpha_L);(j_R N_R I_R \alpha_R)}. \tag{\ref{CheMPS2_WE_MPS}}
\end{equation*}
In deze vgl. stellen $s$ en $j$ de totale spin voor, $s^z$ en $j^z$ de spinprojectie, $N$ het deeltjesaantal, en $I$ de irrep van de puntgroep. De Clebsch-Gordan-co\"effici\"enten in vgl. \eqref{CheMPS2_WE_MPS} zorgen ervoor dat slechts enkele blokken in de $A[i]$-tensor verschillend van nul zijn. Als een Clebsch-Gordan-co\"effici\"ent nul is, moet de corresponderende blok in de $T[i]$-tensor niet gealloceerd worden. Vgl. \eqref{CheMPS2_WE_MPS} zorgt ook voor informatiecompressie. Voor de symmetriesector $(j_L N_L I_L)$ zijn er $D(j_L N_L I_L) = \text{grootte}(\alpha_L)$ gereduceerde virtuele basistoestanden in de $T[i]$-tensor. Deze stellen $(2j_L+1)D(j_L N_L I_L)$ virtuele basistoestanden voor in de $A[i]$-tensor. De aanwezigheid van nulblokken en de informatiecompressie zorgen voor een reductie in het nodige werkgeheugen en de rekentijd.

De globale symmetrie $(S_G , N_G , I_G)$ kan opgelegd worden met de spin-0 inbeddingstrategie. De meest linkse MPS-tensor heeft dan enkel de triviale symmetriesector in zijn linker virtuele binding, met gereduceerde virtuele dimensie 1. De meest rechtse MPS-tensor heeft dan enkel de symmetriesector $(S_G , N_G , I_G )$ in zijn rechter virtuele binding, met gereduceerde virtuele dimensie 1.

De operatoren
\begin{equation*}
\hat{b}^{\dagger}_{c \gamma} = \hat{a}^{\dagger}_{c \gamma} \tag{\ref{creaannih1--}}
\end{equation*}
\begin{equation*}
\hat{b}_{c \gamma} = (-1)^{\frac{1}{2}-\gamma}\hat{a}_{c -\gamma} \tag{\ref{creaannih2--}}
\end{equation*}
voor orbitaal $c$ corresponderen respectievelijk met de rij $(s=\frac{1}{2}, s^z=\gamma, N=1, I_c)$ van de irrep $(s=\frac{1}{2}, N=1, I_c)$ en de rij $(s=\frac{1}{2}, s^z=\gamma, N=-1, I_c)$ van de irrep $(s=\frac{1}{2}, N=-1, I_c)$. $\hat{b}^{\dagger}$ en $\hat{b}$ zijn daarom beiden irreducibele tensoroperatoren met spin $\frac{1}{2}$. Dit laat toe om het Wigner-Eckart theorema ook voor operatoren en (het complement van) gerenormaliseerde operatoren te gebruiken. Termen van het type \eqref{CheMPS2_WE_MPS} en \eqref{creaannih1--}-\eqref{creaannih2--} kunnen gecontraheerd worden door impliciet over de gemeenschappelijke multipletten te sommeren en door de lokale, virtuele en operatorspins te herkoppelen. (Het complement van) gerenormaliseerde operatoren bestaat dan formeel uit termen die het product zijn van Clebsch-Gordan-co\"effici\"enten en gereduceerde tensoren. In onze code worden enkel de gereduceerde tensoren berekend, en worden Wigner 3-j of Clebsch-Gordan-co\"effici\"enten nooit gebruikt.

Door de abelse puntgroepsymmetrie $\mathsf{P}$ zijn de tweedeeltjesmatrixelementen $(ij | \hat{V} | kl)$ van de Hamiltoniaan \eqref{QC-ham} enkel verschillend van nul als $I_i \otimes I_j = I_k \otimes I_l$. Voor niet-triviale groepen $\mathsf{P}$ reduceert dit het aantal termen in de constructie van het complement van gerenormaliseerde operatoren en in de vermenigvuldiging in vgl. \eqref{effHameq} aanzienlijk. Het gebruik van niet-abelse ruimtelijke symmetrie wordt ook besproken in hoofdstuk \ref{SYMM-chapter}.

In hoofdstuk \ref{HCHAIN-chapter} worden \'e\'endimensionale waterstofketens bestudeerd. Ondanks het feit dat de Coulombinteractie niet lokaal is, hebben deze ketens enkel een kleine virtuele dimensie nodig. Door de wederzijdse elektrostatische afscherming van de elektronen en de kernen krijgt de Coulombinteractie een effectieve korte dracht, wat deze ketens tot een ideaal testgeval voor DMRG maakt.

Equidistante waterstofketens, 
\begin{equation*}
\vcenter{\hbox{\setlength{\unitlength}{1cm}
\begin{picture}(6.7,1)
\put(0.1,0.375){H}
\put(0.4,0.5){\line(1,0){0.5}}
\put(1.0,0.375){H}
\put(1.3,0.5){\line(1,0){0.5}}
\put(1.9,0.375){H}
\put(2.2,0.5){\line(1,0){0.5}}
\put(2.8,0.375){H}
\put(3.1,0.5){\line(1,0){0.5}}
\put(3.7,0.375){H}
\put(4.0,0.5){\line(1,0){0.5}}
\put(4.6,0.375){H}
\put(4.9,0.5){\line(1,0){0.5}}
\put(5.5,0.375){H}
\put(5.8,0.5){\line(1,0){0.5}}
\put(6.4,0.375){H}
\put(0.5,0.6){$R$}
\put(1.4,0.6){$R$}
\put(2.3,0.6){$R$}
\put(3.2,0.6){$R$}
\put(4.1,0.6){$R$}
\put(5.0,0.6){$R$}
\put(5.9,0.6){$R$}
\end{picture}}} , \tag{\ref{JCP136equallySpacedFormula}}
\end{equation*}
vertonen veel statische correlatie voor grote internucleaire afstanden $R$. Bandentheorie voorspelt dat dit systeem een geleider is, terwijl voor grote internucleaire afstanden de elektronen gelokaliseerd zijn op de kernen. Het systeem is daarom een Mott-isolator. Als de internucleaire afstand afneemt, ondergaat het systeem een metaal-isolatortransitie. De elektronen worden geleidelijk aan gedelokaliseerd, en naar mate de fasetransitie benaderd wordt (vanuit het isolatorgebied) begint het Schmidt-spectrum trager te vervallen. Dit kan verklaard worden door de excitatie\"energie die nul wordt in het metaalgebied. In dit gebied divergeren ook twee andere grootheden simultaan: de statische dipoolpolariseerbaarheid per elektron en de fluctuatie van het dipoolmoment per elektron. De machten waarmee deze grootheden divergeren met toenemende ketenlengte werden voor verschillende internucleaire afstanden berekenend. Zowel metaalgebieden als isolatorgebieden werden waargenomen.

Equidistante waterstofketens kunnen niet bestaan door de onstabiliteit van Peierls. De keten dimeriseert in $H_2$-moleculen. Beschouw de \'e\'endimensionale gedimeriseerde waterstofketen,
\begin{equation*}
\vcenter{\hbox{\setlength{\unitlength}{1cm}
\begin{picture}(6.7,1)
\put(0.1,0.375){H}
\put(0.4,0.5){\line(1,0){0.5}}
\put(1.0,0.375){H}
\put(1.3,0.45){......}
\put(1.9,0.375){H}
\put(2.2,0.5){\line(1,0){0.5}}
\put(2.8,0.375){H}
\put(3.1,0.45){......}
\put(3.7,0.375){H}
\put(4.0,0.5){\line(1,0){0.5}}
\put(4.6,0.375){H}
\put(4.9,0.45){......}
\put(5.5,0.375){H}
\put(5.8,0.5){\line(1,0){0.5}}
\put(6.4,0.375){H}
\put(0.5,0.6){$R_f$}
\put(1.4,0.6){$R$}
\put(2.3,0.6){$R_f$}
\put(3.2,0.6){$R$}
\put(4.1,0.6){$R_f$}
\put(5.0,0.6){$R$}
\put(5.9,0.6){$R_f$}
\end{picture}}} , \tag{\ref{JCP136jetajetaDimerizedHchain}}
\end{equation*}
met intramoleculaire afstand $R_f$ en intermoleculaire afstand $R$. Voor vaste intramoleculaire afstand (met $R_f < R$) blijft de statische correlatie ruwweg hetzelfde, en enkel de elektrondelokalisatie verandert met $R$. De respons van dit systeem op een statisch elektrisch veld werd bestudeerd door middel van de longitudinale statische dipoolpolariseerbaarheid $\alpha_{zz}$ en de tweede orde dipoolhyperpolariseerbaarheid $\gamma_{zzzz}$. Beide grootheden werden berekend met de eindige-veldmethode. In het isolatorgebied groeien beide grootheden uiteindelijk lineair met ketenlengte, en de waarden per molecule kunnen ge\"extrapoleerd worden tot de thermodynamische limiet. De minimale basis STO-6G is niet voldoende. Door toevoeging van een extra $s$-orbitaal per atoom (6-31G basis) werden significant betere resultaten bekomen. De bijkomende toevoeging van drie extra $p$-orbitalen (6-31G(d,p) basis) had slechts een kleine invloed. De theorie van gekoppelde clusters met enkele en dubbele excitaties, en met driedubbele excitaties in perturbatie (CCSD(T)), is momenteel de gouden standaard van de kwantumchemie. Om de nauwkeurigheid van benaderende methoden te besturen wordt CCSD(T) vaak als referentie gebruikt. Onze DMRG-resultaten zijn niet te onderscheiden van de exacte, en ze werden gebruikt om de nauwkeurigheid van CCSD(T) af te schatten in de berekening van $\gamma_{zzzz}$. Naar mate de elektronen meer gedelokaliseerd zijn (met dalende $R$) neemt de nauwkeurigheid van CCSD(T) af. Met toenemende elektrondelokalisatie is een steeds groter aantal elektronen gecorreleerd, terwijl CCSD(T) enkel een beperkt aantal elektronen kan correleren. De machtwet
\begin{equation*}
 \gamma(M) \propto M^{a(M)} \tag{\ref{JCP136theScalingRelationForTheSecondHyperpol}}
\end{equation*}
wordt vaak vooropgesteld om de toename van $\gamma_{zzzz}$ met het aantal moleculen $M$ te beschrijven. De parameter $a(M)$ varieert traag met $M$. Voor de systemen en basissen waarvoor $a(M)$ al ongeveer \'e\'en was voor de grootste ketens die werden berekend, kon $\gamma_{zzzz}/M$ accuraat ge\"extrapoleerd worden naar de thermodynamische limiet. De grootheid $\alpha_{zz}/M$ kon altijd accuraat ge\"extrapoleerd worden naar de thermodynamische limiet.

In hoofdstuk \ref{C2-chapter} werd het koolstofdimeer bestudeerd. De $\pi$-bindingen van dit dimeer zijn van het ladingsverschuivingstype, en er werd recent gedebatteerd of dit dimeer een vierdubbele binding heeft. De grondtoestand heeft een uitgesproken multireferentiekarakter, en veel kruisingen en vermeden kruisingen geschieden tussen de laaggelegen toestanden. Het koolstofdimeer is bijgevolg een goed systeem om de mogelijkheden van onze code te testen. We hebben de twaalf laagst gelegen bindingsdissociatiecurven tot op $0.01~mE_h$ nauwkeurigheid berekend in de cc-pVDZ basis. Door de $\mathsf{SU(2)}$, $\mathsf{U(1)}$, en abelse $\mathsf{P}$ symmetrie\"en in onze code konden deze toestanden in grote mate ontward worden door in verschillende symmetriesectoren te zoeken. Het deeltjesaantal was steeds $N=12$, en $^S\mathsf{D_{2 h}}$-symmetrie (rechts) werd gebruikt om de originele $^S\mathsf{D_{\infty h}}$-symmetriesectoren (links) te doorzoeken:
\begin{equation*}
X^1\Sigma_g^+ ; B^1 \Delta_g ; B'^1\Sigma_g^+ ~ \rightarrow ~ ^1A_g \tag{\ref{CPCmeuh1}}
\end{equation*}
\begin{equation*}
c^3\Sigma_u^+ ; 1^3\Delta_u ; 2^3\Sigma_u^+ ~ \rightarrow ~ ^3B_{1u} \tag{\ref{CPCmeuh2}}
\end{equation*}
\begin{equation*}
C^1\Pi_g ~ \rightarrow ~ ^1B_{2g} \tag{\ref{CPCfirstOfThed2hIrreps}}
\end{equation*}
\begin{equation*}
A^1\Pi_u ~ \rightarrow ~ ^1B_{2u} \tag{\ref{CPCmeuh3}}
\end{equation*}
\begin{equation*}
1^1\Sigma_u^- ~ \rightarrow ~ ^1A_u \tag{\ref{CPCmeuh4}}
\end{equation*}
\begin{equation*}
b^3 \Sigma_g^- ~ \rightarrow ~ ^3B_{1g} \tag{\ref{CPCmeuh5}}
\end{equation*}
\begin{equation*}
d^3\Pi_g ~ \rightarrow ~ ^3B_{2g} \tag{\ref{CPCmeuh6}}
\end{equation*}
\begin{equation*}
a^3\Pi_u ~ \rightarrow ~ ^3B_{2u}. \tag{\ref{CPClastOfThed2hIrreps}}
\end{equation*}
Binnen de $^1A_g$- en $^3B_{1u}$-sectoren werden twee ge\"exciteerde toestanden berekend door lager gelegen toestanden uit te projecteren. Hun $\mathsf{D_{\infty h}}$-symmetrie werd bepaald door het relatieve faseverschil van de co\"effici\"enten van twee Slaterdeterminanten. We hebben ook het belang van kern- en kern-valentiecorrelatie afgeschat door DMRG(36o,12e) en DMRG-SCF(34o,8e) berekeningen in de cc-pCVDZ basis met elkaar te vergelijken. In de DMRG-SCF berekeningen werden de $1s$ orbitalen van de koolstofatomen dubbel bezet gehouden. De niet-evenwijdigheidsfout was van de orde $2~mE_h$, terwijl het energieverschil tussen beide curven ongeveer $75~mE_h$ bedraagt. Om de kern- en kern-valentiecorrelatie, alsook de correcte kerndynamica voor kleine internucleaire afstanden correct te beschrijven, moet de basis voldoende bewegingsvrijheid toelaten voor de elektronen in de volledig gevulde schillen.

DMRG kan ge\"exciteerde toestanden vinden door lager gelegen toestanden uit te projecteren of door de dichtstbijgelegen toestand bij een vooropgestelde energie te zoeken. Dit zijn toestandspecifieke DMRG-algoritmes omdat de volledige virtuele basis gebruikt wordt om \'e\'en enkele eigentoestand te benaderen. In het toestandsgemiddelde DMRG-algoritme wordt de virtuele basis gebruikt om verschillende eigentoestanden tegelijk te beschrijven. Lineaire responstheorie voor DMRG (DMRG-LRT) kan ook gebruikt worden om ge\"exciteerde toestanden te vinden. DMRG-LRT wordt beschreven in hoofdstuk \ref{Thouless-chapter}.

In dat hoofdstuk wordt de analogie tussen HF-theorie en DMRG onderzocht. Beide methoden kunnen geformuleerd worden als de variationele optimalisatie van een golffunctie-aanzet, een Slater-determinant voor HF-theorie en een MPS voor DMRG. Het tijdsonafhankelijk variationeel principe levert zelfconsistente gemiddeld-veldvergelijkingen op voor de deeltjes in HF-theorie en voor de roosterplaatsen in DMRG. De golffunctievrijheid kan benut worden om de zelfconsistente gemiddeld-veldvergelijkingen te vereenvoudigen tot standaard eigenwaardevergelijkingen. In HF-theorie kan de golffunctievrijheid gebruikt worden om \'e\'endeeltjestoestanden te construeren die eigenvectoren zijn van de Fock-operator. In DMRG kan de golffunctievrijheid gebruikt worden om MPS-tensoren te construeren die eigenvectoren zijn van de effectieve Hamiltoniaan.

Het tijdsafhankelijk variationeel principe genereert vergelijkingen voor de tijdsevolutie van een golffunctie, die binnen de ruimte van de golffunctie-aanzet blijft. Linearisatie van deze vergelijkingen in de buurt van een variationeel minimum leidt tot de willekeurige-fasebenadering (RPA). Een kleine tijdstap verbindt een golffunctie met zijn eerste-orde raakruimte. Voor HF-theorie wordt deze ruimte opgespannen door de \'e\'endeeltjesexcitaties. Deze corresponderen met het vervangen van bezette \'e\'endeeltjestoestanden met onbezette. Voor DMRG wordt deze ruimte opgespannen door de \'e\'enroosterplaatsexcitaties. Deze corresponderen met het vervangen van behouden virtuele basistoestanden met weggegooide. Exponenti\"ering van de \'e\'endeeltjesexcitaties leidt tot het Thouless-theorema voor HF-theorie, een expliciete minimale parametrisatie van de totale golffunctieruimte van Slater-determinanten. In hoofdstuk \ref{Thouless-chapter} wordt de tegenhanger voor DMRG bewezen. De expansie van het Thouless-theorema leidt tot de configuratie-interactie-expansie. Voor HF-theorie leveren de eerste-ordetermen de configuratie-interactie met enkele excitaties (CIS) op, ook wel de Tamm-Dancoff-benadering (TDA) genoemd. Dezelfde namen worden gebruikt voor DMRG: DMRG-CIS of DMRG-TDA. Een variationele optimalisatie in deze raakruimte levert een benadering voor excitaties op. Voor HF-theorie leiden de tweede-ordetermen tot de configuratie-interactie met enkele en dubbele excitaties (CISD). Dezelfde naam wordt gebruikt voor DMRG: DMRG-CISD. Een variationele optimalisatie in deze ruimte levert zowel een betere beschrijving van de grondtoestand op, alsook een benadering voor excitaties.

Er werden DMRG-CISD-berekeningen uitgevoerd voor het \'e\'endimensionaal Hubbard-model. Voor kleine virtuele dimensies is de variationele DMRG-CISD-grondtoestandsener-gie significant lager dan het DMRG-resultaat, en de DMRG-CISD-excitatie\"energie\"en zijn ook beter dan de DMRG-TDA-energie\"en. Gedelokaliseerde excitaties op twee roosterplaatsen worden niet gevat met DMRG-TDA, maar deze kunnen wel teruggevonden worden met DMRG-CISD. Dit moet vergeleken worden met HF-theorie, waar TDA geen tweedeeltjesexcitaties kan beschrijven, maar CISD wel. De laaggelegen spin-0 excitaties van polyenen werden berekend met DMRG-TDA en DMRG-RPA. Het $\pi$-systeem van deze moleculen werd geparametriseerd met de Pariser-Parr-Pople-Hamiltoniaan.

DMRG-RPA kan ook het Goldstone-boson terugvinden voor een grondtoestand die een continue symmetrie breekt. We hebben dit aangetoond voor een grondtoestand van het \'e\'endimensionaal Hubbard-model met spin $\frac{1}{2}$.

In hoofdstuk \ref{DMC-MPS-chapter} wordt projector-Monte Carlo voor MPS-golffuncties besproken. De eigenvector $\ket{\Psi^*}$ van een Hermitische operator $\hat{K}$ met de grootste absolute eigenwaarde kan teruggevonden worden door herhaaldelijk met $\hat{K}$ in te werken op een initi\"ele golffunctie $\ket{\Psi^{(0)}}$. Op elk Monte Carlo-tijdsstip $n$ wordt de golffunctie benaderd door een groep ``wandelaars'':
\begin{equation*}
\ket{\Psi^{(n)}} = (\hat{K})^n \ket{\Psi^{(0)}} \approx \sum_{\phi} \ket{\phi}. \tag{\ref{Ch7ReferenceForConclusion1}}
\end{equation*}
De operator $\hat{K}$ wordt ontbonden in een waarschijnlijkheidsdichtheid $P(\mathbf{x})$ en een verzameling operatoren $\hat{B}(\mathbf{x})$:
\begin{equation*}
\hat{K} = \sum_{\mathbf{x}} P(\mathbf{x}) \hat{B}(\mathbf{x}). \tag{\ref{origPropagator}}
\end{equation*}
Opdat de methode succesvol zou zijn, moeten de operatoren $\hat{B}(\mathbf{x})$ zo gekozen worden dat de complexiteit van de wandelaars niet vergroot. Op elk Monte Carlo-tijdsstip $n$ wordt er voor elke wandelaar $\ket{\phi}$ een $\mathbf{x}$ getrokken uit $P(\mathbf{x})$ en wordt die wandelaar gepropageerd met $\hat{B}(\mathbf{x})$. Met deze stochastische propagatie van de groep wandelaars wordt een benadering voor $\ket{\Psi^*}$ bekomen. Als de wandelaars Slater-determinanten zijn, kan een mogelijke (en vaak gebruikte) ontleding van $\hat{K} = e^{-\delta\tau \hat{H}}$ bekomen worden met de Hubbard-Stratonovich-transformatie. 

Voor re\"eelwaardige (complexwaardige) parametrisaties kan het fermiontekenprobleem (fermionfaseprobleem) opgelost worden door de paden van de wandelaars te beperken met een padbeperkingsgolffunctie $\ket{\Psi_T}$. Dit introduceert een systematische afwijking in de Monte Carlo-propagatie, hetgeen een invloed heeft op geschatte grootheden zoals de geprojecteerde energie. De grootte van deze systematische afwijking hangt af van hoe goed $\ket{\Psi^*}$ benaderd wordt door $\ket{\Psi_T}$. Als deze laatste exact is, verdwijnt de systematische afwijking.

De wandelaars en de padbeperkingsgolffunctie zijn over het algemeen van hetzelfde golffunctie-aanzettype omdat dit toelaat om overlapping en verwachtingswaarden goedkoop te berekenen. Het voordeel van MPS-padbeperkingsgolffuncties over Slater-determinanten is dat ze systematisch verbeterd kunnen worden door hun virtuele dimensie te vergroten. Als de corresponderende systematische afwijking een duidelijke afnemende trend vertoont met groeiende virtuele dimensie, is het mogelijk om de systematische afwijking af te schatten en te verwijderen.

In hoofstuk \ref{DMC-MPS-chapter} wordt het Heisenberg-model met spin $\frac{1}{2}$ bestudeerd op een $4 \times 4$ torus. Drie decomposities \eqref{origPropagator} werden getest: (1) monsters trekken uit de virtuele bindingen van een matrixproductoperator voor $\hat{K}$; (2) een Trotter-decompositie voor $\hat{K} = e^{-\delta\tau \hat{H}}$ opstellen, en monsters trekken uit de virtuele bindingen van de corresponderende tweeroosterplaatsmatrixproductoperatoren; en (3) een Hubbard-Stratonovich-transformatie voor spinsystemen uitvoeren, hetgeen een verzameling hulpvelden oplevert om te bemonsteren. We hebben waargenomen dat voor een vaste keuze van de matrixproductoperatorvrijheid het beter is om singuliere lokale operatoren zoals $\hat{S}^+$ te vermijden, omdat de systematische afwijking dan geen duidelijke trend vertoond met toenemende virtuele dimensie. Van de drie geteste methoden lijkt de variant met hulpvelden (3) het best te werken.

\newpage
\thispagestyle{empty}
\hbox{}
\newpage
\thispagestyle{empty}
\hbox{}

\end{document}